\documentclass[12pt,journal,compsoc]{IEEEtran}
\ifCLASSOPTIONcompsoc
\else
\fi
\ifCLASSINFOpdf
\else
\fi
\begin{document}
%
\title{Covariant local field theory equations following from the relativistic canonical quantum mechanics of arbitrary spin}
%
%
%
%

\author{Volodimir Simulik~\IEEEmembership{}
        ~\IEEEmembership{}
        ~\IEEEmembership{}
\IEEEcompsocitemizethanks{\IEEEcompsocthanksitem V. Simulik is the pripcipal research associate of the Institute of Electron Physics, National Academy of Sciences of Ukraine, Uzhgorod, 21 Universitetska Str. Ukraine, 88000.\protect\\
E-mail: vsimulik@gmail.com
\IEEEcompsocthanksitem }
\thanks{Manuscript received June 28, 2014}}

%
%

\markboth{}%
{Shell \MakeLowercase{\textit{et al.}}: Bare Demo of IEEEtran.cls for Computer Society Journals}
%



\IEEEcompsoctitleabstractindextext{%
\begin{abstract}
The new relativistic equations of motion for the particles with spin s=1, s=3/2, s=2 and nonzero mass have been introduced. The description of the relativistic canonical quantum mechanics of the arbitrary mass and spin has been given. The link between the relativistic canonical quantum mechanics of the arbitrary spin and the covariant local field theory has been found. The manifestly covariant field equations that follow from the quantum mechanical equations, have been considered. The covariant local field theory equations for spin s=(1,1) particle-antiparticle doublet, spin s=(1,0,1,0) particle-antiparticle multiplet, spin s=(3/2,3/2) particle-antiparticle doublet, spin s=(2,2) particle-antiparticle doublet, spin s=(2,0,2,0) particle-antiparticle multiplet  and spin s=(2,1,2,1) particle-antiparticle multiplet have been introduced. The Maxwell-like equations for the boson with spin s=1 and $m>0$ have been introduced as well.
\end{abstract}

}

\maketitle

\IEEEdisplaynotcompsoctitleabstractindextext

%
\IEEEpeerreviewmaketitle

\tableofcontents

\section{Introduction}
%
%

%
%
%
%
\IEEEPARstart{T}{he} start of the relativistic quantum mechanics was given by Paul Dirac
with his well known equation for electron [1]. More precisely, in this 4-component model, the spin s=(1/2,1/2) particle-antiparticle doublet of two fermions was considered (in particular, the electron-positron doublet). Nevertheless, the quantum-mechanical interpretation of the Dirac equation, which should be similar to the physical interpretation of the nonrelativistic Schr$\mathrm{\ddot{o}}$dinger equation, is not evident and is hidden deeply in the Dirac model. In order to visualize the quantum mechanical interpretation of the Dirac equation, transformation to the canonical (quantum-mechanical) representation was suggested [2]. In this Foldy--Wouthuysen (FW) representation of the Dirac equation, the quantum-mechanical interpretation is much more clear. Nevertheless, the direct and evident quantum-mechanical interpretation of the spin s=(1/2,1/2) particle-antiparticle doublet can be fulfilled only within the framework of the relativistic canonical quantum mechanics (RCQM), see, e. g., consideration in [3].

Note that here only the first-order particle and the field equations (together with their canonical nonlocal pseudo-differential representations) are considered. The second order equations (like the Klein--Gordon--Fock equation) are not the subject of this investigation.

Different approaches to the description of the particles of an arbitrary spin can be found in [4--12]. The allusion on the RCQM and the first steps are given in [13]. In the papers [3, 14], this relativistic model for the test case of the spin s=(1/2,1/2) particle-antiparticle doublet is formulated. In [14], this relativistic model is considered as the system of the axioms on the level of the von Neumann monograph [15], where the mathematically well-defined consideration of the nonrelativistic quantum mechanics was given. Furthermore, in  [3] the link between the spin s=(1/2,1/2) particle-antiparticle doublet RCQM and the Dirac theory is given.

Below the same procedure is fulfilled for the spin s=(1,1), s=(1,0,1,0), s=(3/2,3/2), s=(2,2), s=(2,0,2,0) and spin s=(2,1,2,1) RCQM. The corresponding equations, which follow from the RCQM for the covariant local field theory, are introduced.

In other words, the so-called Foldy synthesis [13] of the covariant particle equations is extended here by starting from RCQM, and equations of the covariant local field theory are the final step of such program. The canonical representation of the field equations (the analog of the FW representation) is the intermediate step in this method.

The goals of this paper are as follows: to formulate the foundations of the RCQM of an arbitrary spin and mass, to demonstrate and explain this model on the examples of the spin s=1/2, s=1, s=3/2, s=2 singlets, spin s=(1/2,1/2), s=(1,1), s=(3/2,3/2), s=(2,2) particle-antiparticle doublets, spin s=(1,0) multiplet and spin s=(1,0,1,0), s=(2,0,2,0), s=(2,1,2,1) particle-antiparticle multiplets, to give the link of the spin s=(1,1), s=(1,0,1,0), s=(3/2,3/2), s=(2,2), s=(2,0,2,0), s=(2,1,2,1) RCQM with the Dirac-like equations and the covariant local field theory, to find the corresponding transition operators, to derive new manifestly covariant equations for higher spins, and to derive new equations for electromagnetic theory.

The results are presented on the three linking levels.

$$\mathrm{COVARIANT} \quad \mathrm{LOCAL}$$
$$\mathrm{FIELD} \quad \mathrm{THEORY}\quad \mathrm{OF}$$
$$\mathrm{ARBITRARY} \quad \mathrm{SPIN}$$
$$\uparrow$$
$$\mathrm{CANONICAL} \quad \mathrm{FIELD} \quad \mathrm{THEORY} \quad \mathrm{OF}$$
$$\mathrm{ARBITRARY} \quad \mathrm{SPIN}$$
$$\uparrow$$
$$\mathrm{RELATIVISTIC} \quad \mathrm{CANONICAL}$$
$$\mathrm{QUANTUM} \quad \mathrm{MECHANICS} \quad \mathrm{OF}$$
$$\mathrm{ARBITRARY} \quad \mathrm{SPIN}$$

Here the standard relativistic concepts, definitions and notations
in the form convenient for our consideration are chosen. For example, in the
Minkowski space-time

\begin{equation}
\label{eq1}
\mathrm{M}(1,3)=\{x\equiv(x^{\mu})=(x^{0}=t, \,
\overrightarrow{x}\equiv(x^{j}))\};
\end{equation}
$$\mu=\overline{0,3}, \, j=1,2,3,$$

\noindent $x^{\mu}$ denotes the Cartesian (covariant)
coordinates of the points of the physical space-time in the
arbitrary-fixed inertial reference frame (IRF). We use the system of units with $\hbar=c=1$. The metric tensor is given by

\begin{equation}
\label{eq2}
g^{\mu\nu}=g_{\mu\nu}=g^{\mu}_{\nu}, \, \left(g^{\mu}_{\nu}\right)=\mathrm{diag}\left(1,-1,-1,-1\right);
\end{equation}
$$x_{\mu}=g_{\mu\nu}x^{\mu},$$

\noindent and summation over the twice repeated indices is implied.

\section{Foundations of the relativistic canonical quantum mechanics of
the particle with nonzero mass and arbitrary spin s}

Equation of motion is the Schr$\mathrm{\ddot{o}}$dinger--Foldy
equation

\begin{equation}
\label{eq3}
i\partial_{t}f(x)=\sqrt{m^{2}-\Delta}f(x)
\end{equation}

\noindent for the N-component wave function

\begin{equation}
\label{eq4}
f\equiv \mathrm{column}(f^{1},f^{2},...,f^{\mathrm {N}}), \quad \mathrm {N}=2\mathrm {s}+1.
\end{equation}

Suggestion to call the main equation of the RCQM as the Schr$\mathrm{\ddot{o}}$dinger--Foldy
equation was given in [14] and [3]. Our motivation was as follows. In the papers [13], [16] the two component version of equation (3) is called the Schr$\mathrm{\ddot{o}}$dinger equation. Moreover, the one component version of equation (3) was suggested in [17] and is called in the literature as the spinless Salpeter equation, see, e. g. [18, 19]. Nevertheless, taking into account the L. Foldy's contribution in the construction of RCQM and his proof of the principle of correspondence between RCQM and non-relativistic quantum mechanics, we propose \textit{to call the $N$-component equations of this type as the Schr$\mathrm{\ddot{o}}$dinger-Foldy equations}.

The space of the states is taken as the rigged Hilbert space

\begin{equation}
\label{eq5}
\mathrm{S}^{3,\mathrm {N}}\equiv \mathrm{S}(\mathrm{R}^{3})\times\mathrm{C}^{\mathrm {N}}\subset\mathrm{H}^{3,\mathrm {N}}\subset\mathrm{S}^{3,\mathrm {N}*}.
\end{equation}

\noindent Here $\mathrm{S}^{3,\mathrm {N}}$ is the N-component
Schwartz test function space over the space
$\mathrm{R}^{3}\subset \mathrm{M}(1,3)$ and
$\mathrm{H}^{3,\mathrm {N}}$ is the Hilbert space of the
N-component square-integrable functions over the
$x\in\mathrm{R}^{3}\subset \mathrm{M}(1,3)$

\begin{equation}
\label{eq6}
\mathrm{H}^{3,\mathrm {N}}=\mathrm{L}_{2}(\mathrm{R}^3)\otimes\mathrm{C}^{\otimes \mathrm{N}}=\{f=(f^{\mathrm {N}}):\mathrm{R}^{3}\rightarrow\mathrm{C}^{\otimes \mathrm {N}};
 \end{equation}
$$\int d^{3}x|f(t,\overrightarrow{x})|^{2} <\infty\},$$

\noindent  where $d^{3}x$ is the Lebesgue measure in the space $\mathrm{R}^{3}\subset \mathrm{M}(1,3)$ of the eigenvalues of the position operator $\overrightarrow{x}$ of the Cartesian coordinate of the particle in an arbitrary-fixed IFR. Further, $\mathrm{S}^{3,\mathrm {N}*}$ is the space of the
N-component Schwartz generalized functions. The space
$\mathrm{S}^{3,\mathrm {N}*}$ is conjugated to that of the
Schwartz test functions $\mathrm{S}^{3,\mathrm {N}}$ by the
corresponding topology (see, e. g. [20]).

In general, the mathematical correctness of consideration demands to make the calculations in the space $\mathrm{S}^{3,\mathrm {N}*}$ of generalized functions, i. e. with the application of cumbersome functional analysis. Nevertheless, one can take into account that the Schwartz test function space $\mathrm{S}^{3,\mathrm {N}}$ in the
triple (5) is \textit{kernel}. It means that $\mathrm{S}^{3,\mathrm {N}}$ is dense both in quantum-mechanical space $\mathrm{H}^{3,\mathrm {N}}$ and in the space of generalized functions $\mathrm{S}^{3,\mathrm {N}*}$. Therefore, any physical state $f\in\mathrm{H}^{3,\mathrm {N}}$ can be approximated with an arbitrary precision by the corresponding elements of the Cauchy sequence in $\mathrm{S}^{3,\mathrm {N}}$, which converges to the given $f\in\mathrm{H}^{3,\mathrm {N}}$. Further, taking into account the requirement to measure the arbitrary value of the quantum-mechanical model with non-absolute precision, it means that all concrete calculations can be fulfilled within the Schwartz test function space $\mathrm{S}^{3,\mathrm {N}}$.

Furthermore, the mathematical correctness of the consideration demands to determine the domain of definitions and the range of values for any used operator and for the functions of operators. Note that if the kernel space  $\mathrm{S}^{3,\mathrm {N}}\subset\mathrm{H}^{3,\mathrm {N}}$ is taken as the common domain of definitions of the generating operators $\overrightarrow{x}=(x^{j}), \, \widehat{\overrightarrow{p}}=(\widehat{p}^{j}), \, \overrightarrow{s} \equiv\left(s^{j}\right)=\left(s_{23},s_{31},s_{12}\right)$ of coordinate, momentum and spin, respectively, then this space appears to be also the range of their values. Moreover, the space $\mathrm{S}^{3,\mathrm {N}}$ appears to be the common domain of definitions and values for the set of all below mentioned functions from the 9 operators $\overrightarrow{x}=(x^{j}), \, \widehat{\overrightarrow{p}}=(\widehat{p}^{j}), \, \overrightarrow{s} \equiv\left(s^{j}\right)$ (for example, for the generators $(\widehat{p}_{\mu}, \widehat{j}_{\mu\nu})$ of the irreducible unitary representations of the Poincar$\mathrm{\acute{e}}$ group $\mathcal{P}$ and for different sets of commutation relations). Therefore, in order to guarantee the realization of the principle of correspondence between the results of cognition and the instruments of cognition in the given model, it is sufficient to take the algebra $\mathrm{A}_{\mathrm{S}}$ of the all sets of observables of the given model in the form of converged in $\mathrm{S}^{3,\mathrm {N}}$ Hermitian power series of the 9 generating operators $\overrightarrow{x}=(x^{j}), \, \widehat{\overrightarrow{p}}=(\widehat{p}^{j}), \, \overrightarrow{s} \equiv\left(s^{j}\right)$.

Note that the Schr$\mathrm{\ddot{o}}$dinger--Foldy equation (3) has generalized solutions, which do not belong to the space $\mathrm{H}^{3,4}$ (6). Therefore, the application of the rigged Hilbert space $\mathrm{S}^{3,\mathrm {N}}\subset\mathrm{H}^{3,\mathrm {N}}\subset\mathrm{S}^{3,\mathrm {N}*}$ (5) is necessary.

Some other details of motivations of the choice of the spaces (5), (6) (and all
necessary notations) are given in [3], where  the corresponding
4-component spaces are considered.

The operator of the particle spin is chosen in the complete matrix
form and is associated with the SU(2) group. The orthonormalized
diagonal Cartesian basis, in which the third component of the
spin has the diagonal form, is necessary. The corresponding
generators of the SU(2) group irreducible representations are
chosen to be the spin operators of the corresponding particle
states.

Hence, the spin operator is given as

\begin{equation}
\label{eq7}
\overrightarrow{s} \equiv\left(s^{j}\right)=\left(s_{23},s_{31},s_{12}\right): \, \left[s^{j},s^{l}\right]=i\varepsilon^{jln}s^{n},
\end{equation}

\noindent where $\varepsilon^{jln}$ is the Levi-Civita tensor and
$s^{j}=\varepsilon^{j\ell n}s_{\ell n}$ are the Hermitian
$\mathrm{M}\times \mathrm{M}$ matrices -- the generators of the
M-dimensional representation of the spin group SU(2) (universal
covering of the SO(3)$\subset$SO(1,3) group). Below, in sections
3--6 the fixed concrete representations of the SU(2) group are
associated with the fixed particular spin values.

General solution of the equation of motion (3) is given by

\begin{equation}
\label{eq8}
f(x)= \frac{1}{\left(2\pi\right)^{\frac{3}{2}}}\int d^{3}k e^{-ikx} a^{\mathrm {N}}\left(\overrightarrow{k}\right)\mathrm{d}_{\mathrm {N}},
\end{equation}

\noindent where the following notations

\begin{equation}
\label{eq9}
kx\equiv \omega t -\overrightarrow{k}\overrightarrow{x}, \quad \omega \equiv \sqrt{\overrightarrow{k}^{2}+m^{2}},
\end{equation}

\noindent are used. The orts of the N-dimensional Cartesian basis
have the form

\begin{equation}
\label{eq10}
\mathrm{d}_{1} = \left|
\begin{array}{cccc}
 1 \\
 0 \\
 0 \\
 . \\
 . \\
 . \\
 0 \\
\end{array} \right|, \,
\mathrm{d}_{2} = \left|
\begin{array}{cccc}
 0 \\
 1 \\
 0 \\
 . \\
 . \\
 . \\
 0 \\
\end{array} \right|, \, ...., \,
\mathrm{d}_{\mathrm {N}} = \left|
\begin{array}{cccc}
 0 \\
 0 \\
 0 \\
 . \\
 . \\
 . \\
 1 \\
\end{array} \right|.
\end{equation}

Solution (8) is associated with the stationary complete set of
operators $(\overrightarrow{p}, \, s^{3}=s_{z}, \, g)$ of the
momentum, spin projection and sign of the charge (in the case of
the charge particles). It is easy to see from the sections 3--6
below that for the different N the spin projection operators are
different. The stationary complete set of operators is the set of all functionally independent mutually commuting operators, each of which commutes with the operator of energy (in this model with the operator $\sqrt{m^{2}-\Delta}$).

Interpretation of the amplitudes $a^{\mathrm
{N}}\left(\overrightarrow{k}\right)$ follows from the equations
on the eigenvalues of the operators $(\overrightarrow{p}, \,
s^{3}=s_{z}, \, g)$. The functions $a^{\mathrm
{N}}\left(\overrightarrow{k}\right)$ are the quantum-mechanical
momentum-spin amplitudes of the single particle with
corresponding momentum, spin and charge values (in the case of the
charge particle), respectively.

The relativistic invariance of the model under consideration
requires, as a first step, consideration of its invariance with
respect to the proper ortochronous Lorentz $\mbox{L}_ + ^
\uparrow $ =
SO(1,3)=$\left\{\Lambda=\left(\Lambda^{\mu}_{\nu}\right)\right\}$
and  Poincar$\mathrm{\acute{e}}$ $\mbox{P}_ + ^ \uparrow =
\mbox{T(4)}\times )\mbox{L}_ + ^ \uparrow  \supset \mbox{L}_ + ^
\uparrow$ groups. This invariance in an arbitrary relativistic
model is the implementation of the Einstein's relativity
principle in the special relativity form. Note that the
mathematical correctness requires the invariance mentioned above
to be considered as the invariance with respect to the universal
coverings $\mathcal{L}$ = SL(2,C) and
$\mathcal{P}\supset\mathcal{L}$ of the groups $\mbox{L}_ + ^
\uparrow $ and $\mbox{P}_ + ^ \uparrow $, respectively.

For the group $\mathcal{P}$ we choose real parameters
$a=\left(a^{\mu}\right)\in$M(1,3) and
$\varpi\equiv\left(\varpi^{\mu\nu}=-\varpi^{\nu\mu}\right)$ with
well-known physical meaning. For the standard $\mathcal{P}$
generators $\left(p_{\mu},j_{\mu\nu}\right)$ we use commutation
relations in the manifestly covariant form

$$\left[p_{\mu},p_{\nu}\right]=0, \, \left[p_{\mu},j_{\rho\sigma}\right]=ig_{\mu\rho}p_{\sigma}-ig_{\mu\sigma}p_{\rho},$$
\begin{equation}
\label{eq11}
\left[j_{\mu\nu},j_{\rho\sigma}\right]=-i\left(g_{\mu\rho}j_{\nu\sigma}+g_{\rho\nu}j_{\sigma\mu}+g_{\nu\sigma}j_{\mu\rho}+g_{\sigma\mu}j_{\rho\nu}\right).
\end{equation}

The following assertion should be noted. Not a matter of fact that
non-covariant objects such as the Lebesgue measure $d^{3}x$ are
explored, the model of RCQM of arbitrary spin is a relativistic
invariant in the following sense. The
Schr$\mathrm{\ddot{o}}$dinger--Foldy equation (3) and the set of
its solution $\{f\}$ (8) are invariant with respect to the
irreducible unitary representation of the group $\mathcal{P}$,
the $\mathrm {N} \times \mathrm {N}$ matrix-differential
generators of which are given by the following nonlocal operators

$$\widehat{p}_{0}=\widehat{\omega}\equiv \sqrt{-\Delta+m^{2}}, \quad \widehat{p}_{\ell}=i\partial_{\ell},$$
\begin{equation}
\label{eq12}
\widehat{j}_{\ell n}=x_{\ell}\widehat{p}_{n}-x_{n}\widehat{p}_{\ell}+s_{ln}\equiv \widehat{m}_{\ell n}+s_{\ell n},
\end{equation}

\begin{equation}
\label{eq13}
\widehat{j}_{0 \ell}=-\widehat{j}_{\ell 0}=t\widehat{p}_{\ell}-\frac{1}{2}\left\{x_{\ell},\widehat{\omega}\right\}-\left(\frac{s_{\ell n}\widehat{p}_{n}}{\widehat{\omega}+m} \equiv \breve{s}_{\ell}\right),
\end{equation}

\noindent where the orbital parts of the generators are not
changed under the transition from one spin to another. Under such
transitions only the spin parts (7) of the expressions (12), (13)
are changed. In formulae (12), (13), the SU(2)-spin generators $s
^{\ell n}$ have particular specific forms for each representation
of the SU(2) group (see the examples in sections 3--6 below).

Thus, the irreducible unitary representation of the
Poincar$\mathrm{\acute{e}}$ group $\mathcal{P}$ in the space (5),
with respect to which the Schr$\mathrm{\ddot{o}}$dinger-Foldy
equation (3) and the set of its solution $\{f\}$ (8) are
invariant, is given by a series converges in this space

\begin{equation}
\label{eq14} (a,\varpi)\rightarrow U(a,\varpi)=\exp
(-ia^{0}\widehat{p}_{0}-i\overrightarrow{a}\widehat{\overrightarrow{p}}-\frac{i}{2}\varpi^{\mu\nu}\widehat{j}_{\mu\nu})
\end{equation}

\noindent where the generators $(\widehat{p}^{\mu}, \,
\widehat{j}^{\mu\nu})$ are given in (12), (13) with the arbitrary
values of the SU(2) spins $\overrightarrow{s}=(s^{\ell n})$ (7).

The validity of this assertion is verified by the following three
steps. (i) The calculation that the $\mathcal{P}$-generators (12),
(13) commute with the operator $i\partial _{0}-\widehat{\omega}$
of the Schr$\mathrm{\ddot{o}}$dinger--Foldy equation (3). (ii) The
verification that the $\mathcal{P}$-generators (12), (13) satisfy
the commutation relations (11) of the Lie algebra of the
Poincar$\mathrm{\acute{e}}$ group $\mathcal{P}$. (iii) The proof
that generators (12), (13) realize the spin s(s+1) representation
of this group. Therefore, the Bargman--Wigner classification on
the basis of the corresponding Casimir operators calculation
should be given. These three steps can be made by direct and
non-cumbersome calculations.

The corresponding Casimir operators have the form

\begin{equation}
\label{eq15}
p^{2}=\widehat{p}^{\mu}\widehat{p}_{\mu}=m^{2}\mathrm {I}_{\mathrm {N}},
\end{equation}
\begin{equation}
\label{eq16}
W=w^{\mu}w_{\mu}=m^{2}\overrightarrow{s}^{2}=\mathrm {s}(\mathrm {s}+1)m^{2}\mathrm {I}_{\mathrm {N}},
\end{equation}

\noindent where $\mathrm {I}_{\mathrm {N}}$ is the $\mathrm {N}
\times \mathrm {N}$ unit matrix and s =1/2, 1, 3/2, 2, ...

Below in the next sections the particular examples of spins s
=1/2, 1, 3/2, 2 singlets and spins s=(1,1), (1,0), (1,0,1,0)
multiplets are considered briefly.

The partial cases s =1/2, 1, 3/2, 2 and s=(1,1), (1,0), (1,0,1,0)
can be presented on the level of axiomatic approach [21]. The way
of such consideration is demonstrated in [3] on the test example
of the spin s=(1/2,1/2) particle-antiparticle doublet of fermions.

\section{A brief scheme of the relativistic canonical quantum mechanics of the single spin s=1/2 fermion}

The Schr$\mathrm{\ddot{o}}$dinger--Foldy equation is given by

\begin{equation}
\label{eq17}
i\partial_{t}f(x)=\sqrt{m^{2}-\Delta}f(x), \quad f=\left|
{{\begin{array}{*{20}c}
 f^{1} \hfill  \\
 f^{2} \hfill  \\
\end{array} }} \right|.
\end{equation}

The space of the states is as follows

\begin{equation}
\label{eq18}
\mathrm{S}^{3,2}\subset\mathrm{H}^{3,2}\subset\mathrm{S}^{3,2*}.
\end{equation}

The generators of the SU(2)-spin have an explicit form

\begin{equation}
\label{eq19}
\overrightarrow{s}=\frac{1}{2}\overrightarrow{\sigma}, \quad \left[s^{j},s^{\ell}\right]=i\varepsilon^{j \ell n}s^{n},
\end{equation}

\noindent where $\overrightarrow{\sigma}$ are the standard Pauli matrices

\begin{equation}
\label{eq20}
\sigma^{1}= \left| {{\begin{array}{*{20}c}
 0 \hfill & 1 \hfill \\
 1 \hfill & 0 \hfill \\
\end{array} }} \right|, \quad
\sigma^{2}= \left| {{\begin{array}{*{20}c}
 0 \hfill & -i \hfill \\
 i \hfill & 0 \hfill \\
\end{array} }} \right|, \quad
\sigma^{3}=\left|
{{\begin{array}{*{20}c}
 1 \hfill & 0 \hfill \\
 0 \hfill & { -1} \hfill \\
\end{array} }} \right|.
\end{equation}

The Casimir operator is given by

\begin{equation}
\label{eq21}
\overrightarrow{s}^{2}=\frac{3}{4}\mathrm{I}_{2}=\frac{1}{2}\left(\frac{1}{2}+1\right)\mathrm{I}_{2}, \quad \mathrm{I}_{2}= \left|
{{\begin{array}{*{20}c}
 1 \hfill & 0 \hfill \\
 0 \hfill &  1 \hfill \\
\end{array} }} \right|.
\end{equation}

The general solution of the Schr$\mathrm{\ddot{o}}$dinger--Foldy equation (17) is given by

\begin{equation}
\label{eq22}
f(x)= \frac{1}{\left(2\pi\right)^{\frac{3}{2}}}\int d^{3}k e^{-ikx}\left[a^{-}_{+}(\overrightarrow{k})\mathrm{d}_{1}+a^{-}_{-}(\overrightarrow{k})\mathrm{d}_{2}\right],
\end{equation}

\noindent where notations (9) are used. The orts of the
2-dimensional Cartesian basis have the form

\begin{equation}
\label{eq23}
\mathrm{d}_{1} = \left|
\begin{array}{cccc}
 1 \\
 0 \\
\end{array} \right|, \, \mathrm{d}_{2} = \left|
\begin{array}{cccc}
 0 \\
 1 \\
\end{array} \right|.
\end{equation}

The solution (22) is associated with the stationary complete set of operators $\overrightarrow{p}, \, s^{3}=s_{z}, \, g=-e$ of the momentum, spin projection and sign of the charge of the spin 1/2 fermion, respectively.

The equations on the eigenvalues of the spin projection operator
$s^{3}=\frac{1}{2}\left| {{\begin{array}{*{20}c}
 1 \hfill & 0 \hfill \\
 0 \hfill &  -1 \hfill \\
\end{array} }} \right|$ and momentum $\overrightarrow{p}$ are given by

\begin{equation}
\label{eq24}
s^{3}\mathrm{d}_{1} = \frac{1}{2}\mathrm{d}_{1}, \quad s^{3}\mathrm{d}_{2} = -\frac{1}{2}\mathrm{d}_{2},
\end{equation}
$$\overrightarrow{p} e^{-ikx}\mathrm{d}_{\mathrm{r}}=\overrightarrow{k}\mathrm{d}_{\mathrm{r}}, \quad \mathrm{r} = 1,2.$$

The interpretation of the amplitudes
$a^{-}_{+}(\overrightarrow{k}),\,a^{-}_{-}(\overrightarrow{k})$
follows from equations (24) and similar equations on the eigenvalues of the operator $g=-e$ from
stationary complete set of operators. The functions
$a^{-}_{+}(\overrightarrow{k}),\,a^{-}_{-}(\overrightarrow{k})$
are the quantum-mechanical momentum-spin amplitudes of the
fermion with charge -e and the spin projection eigenvalues +1/2
and -1/2, respectively.

The Schr$\mathrm{\ddot{o}}$dinger--Foldy equation (17) and the
set $\{\mathrm{f}\}$ of its solutions (22) are invariant with
respect to the unitary spin $s=1/2$ representation (14) of the
Poincar$\mathrm{\acute{e}}$ group $\mathcal{P}$. The
corresponding $2 \times 2$ matrix-differential generators are
given by (12), (13), where the spin 1/2 SU(2) generators
$\overrightarrow{s}=(s^{\ell n})$ are given in (19).

The validity of this assertion is verified by the three steps
considered in the previous section after formula (14). The
corresponding Casimir operators have the form

\begin{equation}
\label{eq25}
p^{2}=\widehat{p}^{\mu}\widehat{p}_{\mu}=m^{2}\mathrm{I}_{2},
\end{equation}
\begin{equation}
\label{eq26}
W=w^{\mu}w_{\mu}=m^{2}\overrightarrow{s}^{2}=\frac{1}{2}\left(\frac{1}{2}+1\right)m^{2}\mathrm{I}_{2},
\end{equation}

\noindent where $\mathrm{I}_{2}$ is given in (21).

Hence, above a brief consideration of the RCQM foundations of the
particle with the mass $m>0$ and the spin $s=1/2$ has been given.

\section{A brief scheme of the relativistic canonical quantum mechanics of the single spin s=1 boson}

The Schr$\mathrm{\ddot{o}}$dinger--Foldy equation is given by

\begin{equation}
\label{eq27}
i\partial_{t}f(x)=\sqrt{m^{2}-\Delta}f(x), \quad f=\left|
{{\begin{array}{*{20}c}
 f^{1} \hfill  \\
 f^{2} \hfill  \\
 f^{3} \hfill  \\
\end{array} }} \right|.
\end{equation}

The space of the states is as follows

\begin{equation}
\label{eq28}
\mathrm{S}^{3,3}\subset\mathrm{H}^{3,3}\subset\mathrm{S}^{3,3*}.
\end{equation}

The generators of the SU(2)-spin in the most spread explicit form
are given by

\begin{equation}
\label{eq29}
s^{1}= \frac{1}{\sqrt{2}}\left| {{\begin{array}{*{20}c}
 0 \hfill & 1 \hfill & 0 \hfill\\
 1 \hfill & 0 \hfill & 1 \hfill\\
 0 \hfill & 1 \hfill & 0 \hfill\\
\end{array} }} \right|, \quad
s^{2}= \frac{i}{\sqrt{2}}\left| {{\begin{array}{*{20}c}
 0 \hfill & -1 \hfill & 0 \hfill\\
 1 \hfill & 0 \hfill & -1 \hfill\\
 0 \hfill & 1 \hfill & 0 \hfill\\
\end{array} }} \right|,
\end{equation}

$$s^{3}= \left| {{\begin{array}{*{20}c}
 1 \hfill & 0 \hfill & 0 \hfill\\
 0 \hfill & 0 \hfill & 0 \hfill\\
 0 \hfill & 0 \hfill & -1 \hfill\\
\end{array} }} \right|.$$

It is easy to verify that the commutation relations

\begin{equation}
\label{eq30}
\left[s^{j},s^{\ell}\right]=i\varepsilon^{j \ell n}s^{n}
\end{equation}

\noindent of the SU(2)-algebra are valid.

The Casimir operator for this representation of the SU(2)-algebra
is given by

\begin{equation}
\label{eq31}
\overrightarrow{s}^{2}=2\mathrm{I}_{3}=1\left(1+1\right)\mathrm{I}_{3}, \quad \mathrm{I}_{3}= \left|
{{\begin{array}{*{20}c}
 1 \hfill & 0 \hfill & 0 \hfill \\
 0 \hfill &  1 \hfill & 0 \hfill \\
 0 \hfill &  0 \hfill & 1 \hfill \\
\end{array} }} \right|.
\end{equation}

The general solution of the Schr$\mathrm{\ddot{o}}$dinger--Foldy equation (27) is given by

$$f(x)= \frac{1}{\left(2\pi\right)^{\frac{3}{2}}}\int d^{3}k e^{-ikx} $$
\begin{equation}
\label{eq32}
\left[c^{1}(\overrightarrow{k})\mathrm{d}_{1}+c^{2}(\overrightarrow{k})\mathrm{d}_{2}+c^{3}(\overrightarrow{k})\mathrm{d}_{3}\right],
\end{equation}

\noindent where notations (9) are used. The orts of the
3-dimensional Cartesian basis have the form

\begin{equation}
\label{eq33}
\mathrm{d}_{1} = \left|
\begin{array}{cccc}
 1 \\
 0 \\
 0 \\
\end{array} \right|, \, \mathrm{d}_{2} = \left|
\begin{array}{cccc}
 0 \\
 1 \\
 0 \\
\end{array} \right|, \,
\mathrm{d}_{3} = \left|
\begin{array}{cccc}
 0 \\
 0 \\
 1 \\
\end{array} \right|.
\end{equation}

The solution (32) is associated with the stationary complete set
$\overrightarrow{p}, \, s^{3}=s_{z}$ of the momentum and spin
projection operators of the spin s=1 boson, respectively.

The equations on the eigenvalues of the spin projection operator
$s^{3}= \left| {{\begin{array}{*{20}c}
 1 \hfill & 0 \hfill & 0 \hfill\\
 0 \hfill & 0 \hfill & 0 \hfill\\
 0 \hfill & 0 \hfill & -1 \hfill\\
\end{array} }} \right|$ are given by

\begin{equation}
\label{eq34}
s^{3}\mathrm{d}_{1} = 1\mathrm{d}_{1}, \quad s^{3}\mathrm{d}_{2} = 0,\quad s^{3}\mathrm{d}_{3} = -1 \mathrm{d}_{3}.
\end{equation}

The interpretation of the amplitudes $c^{j}(\overrightarrow{k})$
in (32) follows from equations (34) and similar equations on the
operator $\overrightarrow{p}$ eigenvalues. The functions
$c^{1}(\overrightarrow{k}), \, c^{2}(\overrightarrow{k}), \,
c^{3}(\overrightarrow{k})$ are the quantum-mechanical
momentum-spin amplitudes of the boson with the spin projection
eigenvalues +1, 0 and -1, respectively.

The Schr$\mathrm{\ddot{o}}$dinger--Foldy equation (27) and the
set $\{\mathrm{f}\}$ of its solutions (32) are invariant with
respect to the irreducible unitary spin $s=1$ representation (14)
of the Poincar$\mathrm{\acute{e}}$ group $\mathcal{P}$. The
corresponding $3 \times 3$ matrix-differential generators are
given by (12), (13), whereas the spin 1 SU(2) generators
$\overrightarrow{s}=(s^{\ell n})$ are given in (29).

The validity of this assertion is verified by the three steps
already given in section 2 after formula (14). The corresponding
Casimir operators have the form

\begin{equation}
\label{eq35}
p^{2}=\widehat{p}^{\mu}\widehat{p}_{\mu}=m^{2}\mathrm{I}_{3},
\end{equation}
\begin{equation}
\label{eq36}
W=w^{\mu}w_{\mu}=m^{2}\overrightarrow{s}^{2}=1(1+1)m^{2}\mathrm{I}_{3},
\end{equation}

\noindent where $\mathrm{I}_{3}$ is given in (31).

Hence, above a brief consideration of the RCQM foundations of the
particle with the mass $m>0$ and the spin $s=1$ has been given.

\section{A brief scheme of the relativistic canonical quantum mechanics of the single spin s=3/2 fermion}

The Schr$\mathrm{\ddot{o}}$dinger--Foldy equation is given by

\begin{equation}
\label{eq37}
i\partial_{t}f(x)=\sqrt{m^{2}-\Delta}f(x), \quad f=\left|
{{\begin{array}{*{20}c}
 f^{1} \hfill  \\
 f^{2} \hfill  \\
 f^{3} \hfill  \\
 f^{4} \hfill  \\
\end{array} }} \right|.
\end{equation}

The space of the states is as follows

\begin{equation}
\label{eq38}
\mathrm{S}^{3,4}\subset\mathrm{H}^{3,4}\subset\mathrm{S}^{3,4*}.
\end{equation}

The generators of the SU(2)-spin in the most spread explicit form
are given by

$$s^{1}= \frac{1}{2}\left| {{\begin{array}{*{20}c}
 0 \hfill & \sqrt{3} \hfill & 0 \hfill & 0 \hfill\\
 \sqrt{3} \hfill & 0 \hfill & 2 \hfill & 0 \hfill\\
 0 \hfill & 2 \hfill & 0 \hfill & \sqrt{3} \hfill\\
 0 \hfill & 0 \hfill & \sqrt{3} \hfill & 0 \hfill\\
\end{array} }} \right|,$$

\begin{equation}
\label{eq39}
s^{2}= \frac{i}{2}\left| {{\begin{array}{*{20}c}
 0 \hfill & -\sqrt{3} \hfill & 0 \hfill & 0 \hfill\\
 \sqrt{3} \hfill & 0 \hfill & -2 \hfill & 0 \hfill\\
 0 \hfill & 2 \hfill & 0 \hfill & -\sqrt{3} \hfill\\
 0 \hfill & 0 \hfill & \sqrt{3} \hfill & 0 \hfill\\
\end{array} }} \right|, \quad
\end{equation}

$$s^{3}= \frac{1}{2}\left| {{\begin{array}{*{20}c}
 3 \hfill & 0 \hfill & 0 \hfill & 0 \hfill\\
 0 \hfill & 1 \hfill & 0 \hfill & 0 \hfill\\
 0 \hfill & 0 \hfill & -1 \hfill & 0 \hfill\\
 0 \hfill & 0 \hfill & 0 \hfill & -3 \hfill\\
\end{array} }} \right|.$$

\noindent It is easy to verify that the commutation relations $
\left[s^{j},s^{\ell}\right]=i\varepsilon^{j \ell n}s^{n}$
\noindent of the SU(2)-algebra are valid.

The Casimir operator for this representation of the SU(2)-algebra
is given by

\begin{equation}
\label{eq40}
\overrightarrow{s}^{2}=\frac{15}{4}\mathrm{I}_{4}=\frac{3}{2}\left(\frac{3}{2}+1\right)\mathrm{I}_{4}, \quad \mathrm{I}_{4}= \left|
{{\begin{array}{*{20}c}
 1 \hfill & 0 \hfill & 0 \hfill & 0 \hfill \\
 0 \hfill &  1 \hfill & 0 \hfill & 0 \hfill \\
 0 \hfill &  0 \hfill & 1 \hfill & 0 \hfill \\
 0 \hfill &  0 \hfill & 0 \hfill & 1 \hfill \\
\end{array} }} \right|.
\end{equation}

The general solution of the Schr$\mathrm{\ddot{o}}$dinger--Foldy equation (37) is given by

$$f(x)= \frac{1}{\left(2\pi\right)^{\frac{3}{2}}}\int d^{3}k e^{-ikx}$$
\begin{equation}
\label{eq41}
\left[b^{1}(\overrightarrow{k})\mathrm{d}_{1}+b^{2}(\overrightarrow{k})\mathrm{d}_{2}+b^{3}(\overrightarrow{k})\mathrm{d}_{3}+b^{4}(\overrightarrow{k})\mathrm{d}_{4}\right],
\end{equation}

\noindent where notations (9) are used. The orts of 4-dimensional
Cartesian basis have the form

\begin{equation}
\label{eq42}
\mathrm{d}_{1} = \left|
\begin{array}{cccc}
 1 \\
 0 \\
 0 \\
 0 \\
\end{array} \right|, \, \mathrm{d}_{2} = \left|
\begin{array}{cccc}
 0 \\
 1 \\
 0 \\
 0 \\
\end{array} \right|, \,
\mathrm{d}_{3} = \left|
\begin{array}{cccc}
 0 \\
 0 \\
 1 \\
 0 \\
\end{array} \right|, \,
\mathrm{d}_{4} = \left|
\begin{array}{cccc}
 0 \\
 0 \\
 0 \\
 1 \\
\end{array} \right|.
\end{equation}

The solution (41) is associated with the stationary complete set
$\overrightarrow{p}, \, s^{3}=s_{z}$ of the momentum and spin
projection operators of spin s=3/2 fermion, respectively.

The equations on the spin projection operator $s^{3}=
\frac{1}{2}\left| {{\begin{array}{*{20}c}
 3 \hfill & 0 \hfill & 0 \hfill & 0 \hfill\\
 0 \hfill & 1 \hfill & 0 \hfill & 0 \hfill\\
 0 \hfill & 0 \hfill & -1 \hfill & 0 \hfill\\
 0 \hfill & 0 \hfill & 0 \hfill & -3 \hfill\\
\end{array} }} \right|$ eigenvalues are given by

$$s^{3}\mathrm{d}_{1} = \frac{3}{2}\mathrm{d}_{1}, \, s^{3}\mathrm{d}_{2} = \frac{1}{2}\mathrm{d}_{2},$$
\begin{equation}
\label{eq43}
s^{3}\mathrm{d}_{3} = -\frac{1}{2} \mathrm{d}_{3}, \, s^{3}\mathrm{d}_{4} = -\frac{3}{2}\mathrm{d}_{4}.
\end{equation}

The interpretation of the amplitudes
$b^{\alpha}(\overrightarrow{k})$ in (41) follows from equations
(43) and similar equations on the operator $\overrightarrow{p}$
eigenvalues. The functions $b^{1}(\overrightarrow{k}), \,
b^{2}(\overrightarrow{k}), \, b^{3}(\overrightarrow{k}, \,
b^{4}(\overrightarrow{k})$ are the quantum-mechanical
momentum-spin amplitudes of the fermion with the spin projection
eigenvalues $\frac{3}{2}, \, \frac{1}{2}, \,-\frac{1}{2}, \,
-\frac{3}{2}$, respectively.

The Schr$\mathrm{\ddot{o}}$dinger--Foldy equation (37) and the
set $\{\mathrm{f}\}$ of its solutions (41) are invariant with
respect to the irreducible unitary spin $s=3/2$ representation
(14) of the Poincar$\mathrm{\acute{e}}$ group $\mathcal{P}$. The
corresponding $4 \times 4$ matrix-differential generators are
given by (12). (13), where the spin 3/2 SU(2) generators
$\overrightarrow{s}=(s^{\ell n})$ are given in (39).

The validity of this assertion is verified by the three steps
already explained in section 2 after formula (14). The
corresponding Casimir operators have the form

\begin{equation}
\label{eq44}
p^{2}=\widehat{p}^{\mu}\widehat{p}_{\mu}=m^{2}\mathrm{I}_{4},
\end{equation}
\begin{equation}
\label{eq45}
W=w^{\mu}w_{\mu}=m^{2}\overrightarrow{s}^{2}= \frac{3}{2}\left(\frac{3}{2}+1\right)m^{2}\mathrm{I}_{4},
\end{equation}

\noindent where $\mathrm{I}_{4}$ is given in (40).

Hence, above a brief consideration of the RCQM foundations of the
particle with the mass $m>0$ and the spin $s=3/2$ has been given.

\section{A brief scheme of the relativistic canonical quantum mechanics of the single spin s=2 boson}

The Schr$\mathrm{\ddot{o}}$dinger--Foldy equation is given by

\begin{equation}
\label{eq46}
i\partial_{t}f(x)=\sqrt{m^{2}-\Delta}f(x), \quad f=\left|
{{\begin{array}{*{20}c}
 f^{1} \hfill  \\
 f^{2} \hfill  \\
 f^{3} \hfill  \\
 f^{4} \hfill  \\
 f^{5} \hfill  \\
\end{array} }} \right|.
\end{equation}

The space of the states is as follows

\begin{equation}
\label{eq47}
\mathrm{S}^{3,5}\subset\mathrm{H}^{3,5}\subset\mathrm{S}^{3,5*}.
\end{equation}

The generators of the SU(2)-spin in the most spread explicit form
are given by

$$s^{1}= \frac{1}{2}\left| {{\begin{array}{*{20}c}
 0 \hfill & 2 \hfill & 0 \hfill & 0 \hfill & 0 \hfill \\
 2 \hfill & 0 \hfill & \sqrt{6} \hfill & 0 \hfill & 0 \hfill \\
 0 \hfill & \sqrt{6} & 0 \hfill & \sqrt{6} & 0 \hfill \\
 0 \hfill & 0 \hfill & \sqrt{6} \hfill & 0 \hfill & 2 \hfill \\
 0 \hfill & 0 \hfill & 0 \hfill & 2 \hfill & 0 \hfill \\
\end{array} }} \right|,$$

\begin{equation}
\label{eq48}
s^{2}= \frac{i}{2}\left| {{\begin{array}{*{20}c}
 0 \hfill & -2 \hfill & 0 \hfill & 0 \hfill & 0 \hfill \\
 2 \hfill & 0 \hfill & -\sqrt{6} \hfill & 0 \hfill & 0 \hfill \\
 0 \hfill & \sqrt{6} & 0 \hfill & -\sqrt{6} & 0 \hfill \\
 0 \hfill & 0 \hfill & \sqrt{6} \hfill & 0 \hfill & -2 \hfill \\
 0 \hfill & 0 \hfill & 0 \hfill & 2 \hfill & 0 \hfill \end{array} }} \right|,
\end{equation}

$$s^{3}= \left| {{\begin{array}{*{20}c}
 2 \hfill & 0 \hfill & 0 \hfill & 0 \hfill & 0 \hfill \\
 0 \hfill & 1 \hfill & 0 \hfill & 0 \hfill & 0 \hfill \\
 0 \hfill & 0 \hfill & 0 \hfill & 0 \hfill & 0 \hfill \\
 0 \hfill & 0 \hfill & 0 \hfill & -1 \hfill & 0 \hfill \\
 0 \hfill & 0 \hfill & 0 \hfill & 0 \hfill & -2 \hfill \\
\end{array} }} \right|.$$

\noindent It is easy to verify that the commutation relations $
\left[s^{j},s^{\ell}\right]=i\varepsilon^{j \ell n}s^{n}$
\noindent of the SU(2)-algebra are valid.

The Casimir operator for this representation of the SU(2)-algebra
is given by

\begin{equation}
\label{eq49}
\overrightarrow{s}^{2}=6\mathrm{I}_{5}=2\left(2+1\right)\mathrm{I}_{5},
\end{equation}

\noindent where $\mathrm{I}_{5}$ is the $5 \times 5$ unit matrix.

The general solution of the Schr$\mathrm{\ddot{o}}$dinger--Foldy equation (46) is given by

$$f(x)= \frac{1}{\left(2\pi\right)^{\frac{3}{2}}}\int d^{3}k e^{-ikx}$$
\begin{equation}
\label{eq50}
\left[g^{1}(\overrightarrow{k})\mathrm{d}_{1}+g^{2}(\overrightarrow{k})\mathrm{d}_{2}+...+g^{5}(\overrightarrow{k})\mathrm{d}_{5}\right],
\end{equation}

\noindent where notations (9) are used. The orts of the
5-dimensional Cartesian basis have the form

$$\mathrm{d}_{1} = \left|
\begin{array}{cccc}
 1 \\
 0 \\
 0 \\
 0 \\
 0 \\
\end{array} \right|, \, \mathrm{d}_{2} = \left|
\begin{array}{cccc}
 0 \\
 1 \\
 0 \\
 0 \\
 0 \\
\end{array} \right|,$$

\begin{equation}
\label{eq51}
\mathrm{d}_{3} = \left|
\begin{array}{cccc}
 0 \\
 0 \\
 1 \\
 0 \\
 0 \\
\end{array} \right|, \,
\mathrm{d}_{4} = \left|
\begin{array}{cccc}
 0 \\
 0 \\
 0 \\
 1 \\
 0 \\
\end{array} \right| \,
\mathrm{d}_{5} = \left|
\begin{array}{cccc}
 0 \\
 0 \\
 0 \\
 0 \\
 1 \\
\end{array} \right|.
\end{equation}

The solution (50) is associated with the stationary complete set
$\overrightarrow{p}, \, s^{3}=s_{z}$ of the momentum and spin
projection operators of the spin 2 boson, respectively.

The equations on the spin projection operator $s^{3}= \left|
{{\begin{array}{*{20}c}
 2 \hfill & 0 \hfill & 0 \hfill & 0 \hfill & 0 \hfill \\
 0 \hfill & 1 \hfill & 0 \hfill & 0 \hfill & 0 \hfill \\
 0 \hfill & 0 \hfill & 0 \hfill & 0 \hfill & 0 \hfill \\
 0 \hfill & 0 \hfill & 0 \hfill & -1 \hfill & 0 \hfill \\
 0 \hfill & 0 \hfill & 0 \hfill & 0 \hfill & -2 \hfill \\
\end{array} }} \right|$ eigenvalues are given by

$$s^{3}\mathrm{d}_{1} = 2\mathrm{d}_{1}, \, s^{3}\mathrm{d}_{2} = 1\mathrm{d}_{2},$$
\begin{equation}
\label{eq52}
s^{3}\mathrm{d}_{3} = 0, \, s^{3}\mathrm{d}_{4} = -1\mathrm{d}_{4}, \, s^{3}\mathrm{d}_{5} = -2\mathrm{d}_{5}.
\end{equation}

The interpretation of the amplitudes
$g^\mathrm{A}(\overrightarrow{k}), \, \mathrm{A}=\overline{1,5}$
in (50) follows from the equations (52) and similar equations on
the operator $\overrightarrow{p}$ eigenvalues. The functions
$g^{1}(\overrightarrow{k}), \, g^{2}(\overrightarrow{k}), \,
g^{3}(\overrightarrow{k}, \, g^{4}(\overrightarrow{k}), \, \,
g^{5}(\overrightarrow{k})$ are the quantum-mechanical
momentum-spin amplitudes of the boson with the spin $s=2$, mass
$m>0$ and with the spin projection eigenvalues $2, \, 1, \, 0, \,
-1, \, -2$, respectively.

The Schr$\mathrm{\ddot{o}}$dinger--Foldy equation (46) and the
set $\{\mathrm{f}\}$ of its solutions (50) are invariant with
respect to the irreducible unitary spin $s=2$ representation (14)
of the Poincar$\mathrm{\acute{e}}$ group $\mathcal{P}$. The
corresponding $5 \times 5$ matrix-differential generators are
given by (12), (13), whereas the spin 2 SU(2) generators
$\overrightarrow{s}=(s^{\ell n})$ are given in (48).

The validity of this assertion is verified by the three steps,
which already are given in section 2 after the formula (14). The
corresponding Casimir operators have the form

\begin{equation}
\label{eq53}
p^{2}=\widehat{p}^{\mu}\widehat{p}_{\mu}=m^{2}\mathrm{I}_{5},
\end{equation}
\begin{equation}
\label{eq54}
W=w^{\mu}w_{\mu}=m^{2}\overrightarrow{s}^{2}= 2\left(2+1\right)m^{2}\mathrm{I}_{5},
\end{equation}

\noindent where $\mathrm{I}_{5}$ is the $5 \times 5$ unit matrix.

Hence, above a brief consideration of the RCQM foundations of the
particle with the mass $m>0$ and the spin $s=2$ has been given.

\section{The relativistic canonical quantum mechanics of the spin s=(1/2,1/2) particle-antiparticle doublet}

The RCQM of arbitrary spin can be formulated on the level of modern axiomatic approaches to the quantum field theory. Below in this section the demonstration on the principal example of the spin s=(1/2,1/2) particle-antiparticle doublet ($e^{-}e^{+}$-doublet in partial case) is given.

The \textit{axioms of the model} are formulated on the level of correctness of von Neuman's monograph [15]. Requirements of such physically verified principles as \textit{the principle of relativity with respect to the tools of cognition (PRTC)}, \textit{principle of heredity (PH)} with both classical mechanics of single mass point and nonrelativistic quantum mechanics (and \textit{the principle of correspondence (PC)} with these theories), and also \textit{the Einstein principle of relativity (EPR)}, are taken into consideration. The last principle requires first of all \textit{the special relativity (SR)} to be taken into account.

The \textit{ basic axioms} of the model (in the brief consideration) as the mathematical assertions have the form of the following statements.

\textbf{On the space of states}. The space of states of isolated  spin s=(1/2,1/2) particle-antiparticle doublet in an arbitrarily-fixed inertial frame of reference (IFR) in its  $\overrightarrow{x}$-realization is the Hilbert space

\begin{equation}
\label{eq55}
\mathrm{H}^{3,4}=\mathrm{L}_{2}(\mathrm{R}^3)\otimes\mathrm{C}^{\otimes4}=\{f=(f^{\alpha}):\mathrm{R}^{3}\rightarrow\mathrm{C}^{\otimes4},
 \end{equation}
$$\int d^{3}x|f(t,\overrightarrow{x})|^{2} <\infty\},$$

\noindent of complex-valued 4-component square-integrable functions of $x\in\mathrm{R}^{3}\subset \mathrm{M}(1,3)$ (similarly, in momentum,  $\overrightarrow{p}$-realization). Here  $\overrightarrow{x}$ and $\overrightarrow{p}$ are the operators of canonically conjugated dynamical variables of the spin s=(1/2,1/2) particle-antiparticle doublet, and the vectors $f$, $\tilde{f}$  in $\overrightarrow{x}$- and $\overrightarrow{p}$-realizations are linked by the 3-dimensional Fourier transformation (the variable $t$ is the parameter of time-evolution).

\textbf{The mathematical correctness of the consideration} demands the application of the rigged Hilbert space $\mathrm{S}^{3,4}\subset\mathrm{H}^{3,4}\subset\mathrm{S}^{3,4*}$, where the Schwartz test function space $\mathrm{S}^{3,4}$, which is the verified tool of the PRTC implementation, is the kernel (i. e., it is dense both in $\mathrm{H}^{3,4}$ and in the space $\mathrm{S}^{3,4*}$ of the generalized Schwartz functions). Such application allows us to perform, without any loss of generality, all necessary calculations in the space $\mathrm{S}^{3,4}$ on the level of correct differential and integral calculus. The more detailed consideration is given in section 2.

\textbf{On the time evolution of the state vectors}. The time dependence of the state vectors $f\in \mathrm{H}^{3,4}$  (time t is the parameter of evolution) is given either in the integral form by the unitary operator

\begin{equation}
\label{eq56}
u\left(t_{0},t\right)=\exp \left[-i \widehat{\omega}(t-t_{0})\right]; \quad \widehat{\omega}\equiv\sqrt{-\Delta +m^{2}},
\end{equation}

\noindent (below $t_{0}=t$ is put), or  in the differential form by the Schr$\mathrm{\ddot{o}}$dinger--Foldy equation of motion

\begin{equation}
\label{eq57}
(i\partial_{0}- \widehat{\omega})f(x)=0, \quad f=\left|
{{\begin{array}{*{20}c}
 f^{1} \hfill  \\
 f^{2} \hfill  \\
 f^{3} \hfill  \\
 f^{4} \hfill  \\
 \end{array} }} \right|.
\end{equation}

\noindent Here the operator $\widehat{\omega}\equiv\sqrt{-\Delta +m^{2}}$  is the relativistic analog of the energy operator (Hamiltonian) of nonrelativistic quantum mechanics. The Minkowski space-time M(1,3) is pseudo Euclidean with metric $g =\mathrm{diag}(+1,-1,-1,-1)$.

Thus, for the fermionic spin s=(1/2,1 /2) particle-antiparticle doublet the system of two 2-component equations $(i\partial_{0}- \widehat{\omega})f(x)=0$ and $(i\partial_{0}- \widehat{\omega})f(x)=0$ is used. Therefore, the corresponding Schr$\mathrm{\ddot{o}}$dinger--Foldy equation is given by (57), where the 4-component wave function is the direct sum of the particle and antiparticle wave functions, respectively. Due to the historical tradition of the physicists the antiparticle wave function is put in the down part of the 4-column.

The general solution of the Schr$\mathrm{\ddot{o}}$dinger--Foldy equation of motion (57) has the form

$$f(x)= \left|
{{\begin{array}{*{20}c}
 f_{\mathrm{part}} \hfill  \\
 f_{\mathrm{antipart}} \hfill  \\
\end{array} }} \right| =\frac{1}{\left(2\pi\right)^{\frac{3}{2}}}\int d^{3}k e^{-ikx}$$
\begin{equation}
\label{eq58}
\left[a^{-}_{+}(\overrightarrow{k})\mathrm{d}_{1}+a^{-}_{-}(\overrightarrow{k})\mathrm{d}_{2}+a^{+}_{-}(\overrightarrow{k})\mathrm{d}_{3}+a^{+}_{+}(\overrightarrow{k})\mathrm{d}_{4}\right],
\end{equation}

\noindent where the orts $\left\{\mathrm{d}_{\alpha}\right\}$ of the Cartesian basis are given in (42) and the notations (9) are used.

Moreover, the form (57) of the equation of motion means that the information about equal and positive masses of the particle and antiparticle is inserted into the model.

The pseudo-differential (non-local) operator

\begin{equation}
\label{eq59}
\widehat{\omega} \equiv \sqrt{\widehat{\overrightarrow{p}}^{2} + m^2} =\sqrt { - \Delta + m^2}\geq m>0;
\end{equation}
$$\widehat{\overrightarrow{p}}\equiv(\widehat{p}^{j})=-i\nabla, \quad \nabla\equiv(\partial_{\ell}),$$

\noindent is determined alternatively either in the form of the power series

\begin{equation}
\label{eq60}
\widehat{\omega}=m\sqrt{1-\widehat{B}}\equiv 1-\frac{1}{2}\widehat{B}+\frac{1\cdot 2}{2\cdot 3}\widehat{B}^{2}-..., \quad \widehat{B}=\frac{\Delta}{m^{2}},
\end{equation}

\noindent or in the integral form

\begin{equation}
\label{eq61}
(\widehat{\omega}f)(t,\overrightarrow{x})=\frac{1}{(2\pi)^{\frac{3}{2}}}\int d^{3}k e^{i\overrightarrow{k}\overrightarrow{x}} \widetilde{\omega}\widetilde{f}(t,\overrightarrow{k});
\end{equation}
$$\widetilde{\omega}\equiv \sqrt{\overrightarrow{k}^{2}+m^{2}}, \quad \widetilde{f}\in \widetilde{\mathrm{H}}^{3,4},$$

\noindent where $f$ and $\widetilde{f}$ are linked by the 3-dimensional Fourier transformations

\begin{equation}
\label{eq62}
f(t,\overrightarrow{x})=\frac{1}{(2\pi)^{\frac{3}{2}}}\int d^{3}k e^{i\overrightarrow{k}\overrightarrow{x}}\widetilde{f}(t,\overrightarrow{k})\Leftrightarrow \widetilde{f}(t,\overrightarrow{k})=
\end{equation}
$$\frac{1}{(2\pi)^{\frac{3}{2}}}\int d^{3}k e^{-i\overrightarrow{k}\overrightarrow{x}}\widetilde{f}(t,\overrightarrow{x}),$$

\noindent (in (62) $\overrightarrow{k}$ belongs to the spectrum $\mathrm{R}^{3}_{\vec{k}}$ of the operator $\widehat{\overrightarrow{p}}$, and the parameter $t\in (-\infty,\infty)\subset\mathrm{M}(1,3)$).

Note that the space of states (55) is invariant with respect to the Fourier transformation (62). Therefore, both $\overrightarrow{x}$-realization (55) and $\overrightarrow{k}$-realization $\widetilde{\mathrm{H}}^{3,4}$ for the doublet states space are suitable for the purposes of our consideration. In the $\overrightarrow{k}$-realization the Schr$\mathrm{\ddot{o}}$dinger--Foldy equation has the algebraic-differential form

\begin{equation}
\label{eq63}
i\partial_{t}\widetilde{f}(t,\overrightarrow{k})=\sqrt { \overrightarrow{k}^{2} + m^2}\widetilde{f}(t,\overrightarrow{k});
\end{equation}
$$\overrightarrow{k}\in\mathrm{R}^{3}_{\vec{k}}, \quad \widetilde{f}\in \widetilde{\mathrm{H}}^{3,4}.$$

\noindent Below in the places, where misunderstanding is impossible, the symbol "tilde" is omitted.

\textbf{On the fundamental dynamical variables}. The dynamical variable $\overrightarrow{x}\in \mathrm{R}^{3}\subset$M(1,3) (as well as the variable $\overrightarrow{k}\in \mathrm{R}_{\vec{k}}^{3}$) represents the external degrees of freedom of the spin s=(1/2,1/2) particle-antiparticle doublet. The spin $\overrightarrow{s}$ of the spin s=(1/2,1/2) particle-antiparticle doublet is the first in the list of the carriers of the internal degrees of freedom. Taking into account the Pauli principle and the fact that experimentally positron is observed as the mirror reflection of an electron, the operators of  the charge sign  and the spin of the s=(1/2,1/2) particle-antiparticle doublet are taken in the form

\begin{equation}
\label{eq64} g\equiv-\gamma^0 = \left| {{\begin{array}{*{20}c}
 -\mathrm{I}_{2} \hfill & 0 \hfill \\
 0 \hfill & \mathrm{I}_{2} \hfill \\
\end{array} }} \right|, \quad \overrightarrow{s} =
\frac{1}{2}\left| {{\begin{array}{*{20}c}
 \overrightarrow{\sigma} \hfill  0 \hfill \\
 0 -C\hfill\overrightarrow{\sigma}\hfill C \\
\end{array} }} \right|,
\end{equation}

\noindent where $\overrightarrow{\sigma}$ are the standard Pauli matrices (20), $C$ is the operator of complex conjugation, the operator of involution in $\mathrm{H}^{3,4}$, $\mathrm{I}_{2}$ is explained in (21).

In the choice of the spin (64) the principle of correspondence and heredity with the FW representation is used, where the particle-antiparticle doublet spin operator is given by

\begin{equation}
\label{eq65} \overrightarrow{s}_{\mathrm{FW}} =
\frac{1}{2}\left| {{\begin{array}{*{20}c}
 \overrightarrow{\sigma}  \hfill   0  \hfill \\
 0  \hfill \overrightarrow{\sigma} \hfill  \\
\end{array} }} \right|.
\end{equation}

\noindent The link between the spins (64) and (65) is given by the transformation operator $v$ from [3] (note that transformation $v$ is valid for the case of antiHermitian form of spins (64), (65)).

The spin matrices (64), (65) satisfy the commutation relations $\left[s^{j},s^{l}\right]=i\varepsilon^{jln}s^{n}, \, \varepsilon^{123}=+1,$ (7) of the algebra of SU(2) group, where $\varepsilon^{jln}$ is the Levi-Civita tensor and $s^{j}=\varepsilon^{j\ell n}s_{\ell n}$ are the Hermitian $\mathrm{4}\times \mathrm{4}$ matrices (64), (65) -- the generators of a 4-dimensional reducible representation of the spin group SU(2) (universal covering of the SO(3)$\subset$SO(1,3) group).

The Casimir operator for the RCQM representation of SU(2) spin given in (64) has the form

\begin{equation}
\label{eq66} \overrightarrow{s}^{2}=\frac{3}{4}\mathrm{I}_{4}=\frac{1}{2}(\frac{1}{2}+1)\mathrm{I}_{4},
\end{equation}

\noindent where $\mathrm{I}_{4}$ is $4 \times 4$ unit matrix.

\textbf{On the external and internal degrees of freedom}. The coordinate $\overrightarrow{x}$ (as an operator in $\mathrm{H}^{3,4})$ is an analog of the discrete index of generalized coordinates $q\equiv(q_{1}, q_{2}, ...)$ in non-relativistic quantum mechanics of the finite number degrees of freedom. In other words the coordinate $\overrightarrow{x}\in \mathrm{R}^{3}\subset$M(1,3) is the continuous carrier of the external degrees of freedom of a multiplet (the similar consideration was given in [23]). The coordinate operator together with the operator $\widehat{\overrightarrow{p}}$ determines the operator $m_{ln}=x_{l}\widehat{p}_{n}-x_{n}\widehat{p}_{l}$ of an orbital angular momentum, which also is connected with the external degrees of freedom.

However, the RCQM doublet has the additional characteristics such as the spin operator $\overrightarrow{s}$ (64), which is the carrier of the internal degrees of freedom of this multiplet. The set of generators $(\widehat{p}_{\mu}, \widehat{j}_{\mu\nu})$ (12), (13) of the main dynamical variables (formulae (73) below) of the doublet are the functions of the following basic set of 9 functionally independent operators

\begin{equation}
\label{eq67}
\overrightarrow{x}=(x^{j}), \, \widehat{\overrightarrow{p}}=(\widehat{p}^{j}), \, \overrightarrow{s} \equiv\left(s^{j}\right)=\left(s_{23},s_{31},s_{12}\right).
\end{equation}

Note that $\overrightarrow{s}$ commutes both with $(\overrightarrow{x},\widehat{\overrightarrow{p}})$ and with the operator $i\partial_{t}-\sqrt { - \Delta + m^2}$ of the Schr$\mathrm{\ddot{o}}$dinger--Foldy equation (57). Thus, for the free doublet the external and internal degrees of freedom are independent. Therefore, 9 operators (67) in $\mathrm{H}^{3,4}$, which have the univocal physical sense, are the \textit{generating} operators not only for the 10 $\mathcal{P}$ generators $(\widehat{p}_{\mu}, \widehat{j}_{\mu\nu})$ (12), (13) but also for other operators of any experimentally observable quantities of the doublet.

\textbf{On the algebra of observables}. Using the operators of canonically conjugated coordinate $\overrightarrow{x}$ and momentum $\overrightarrow{p}$ (where $\left[x^{j},\widehat{p}^{\ell}\right]=i\delta^{j \ell}, \quad \left[x^{j},x^{\ell}\right]=\left[\widehat{p}^{j},\widehat{p}^{\ell}\right]=0,$ in $\mathrm{H}^{3,4}$), being completed by the operators $\overrightarrow{s}$ (64) and $g$, we construct the algebra of observables (according to the PH) as the Hermitian functions of 10 ($\overrightarrow{x}, \, \overrightarrow{p} \, \overrightarrow{s}, \, -\gamma^{0}$) generating elements of the algebra.

\textbf{On the relativistic invariance of the theory}. This invariance (implementation of the SR) is ensured by the proof of the invariance of the Schr$\mathrm{\ddot{o}}$dinger--Foldy equation (57) with respect to the unitary representation (14) of the universal covering $\mathcal{P}\supset\mathcal{L}$=SL(2,C) of the proper ortochronous Poincar$\mathrm{\acute{e}}$ group $\mbox{P}_ + ^
\uparrow = \mbox{T(4)}\times )\mbox{L}_ + ^ \uparrow  \supset \mbox{L}_ + ^ \uparrow$. Here $\mathcal{L}$ = SL(2,C) is the universal covering of proper ortochronous Lorentz group $\mbox{L}_ + ^ \uparrow $.

The generators of the fermionic ($\mathcal{P}^{\mathrm{f}}$) representation of the group $\mathcal{P}$, with respect to which the Schr$\mathrm{\ddot{o}}$dinger--Foldy equation (57) is invariant, are given by (12), (13) in the $\overrightarrow{x}$-realization of the space $\mathrm{H}^{3,4}$ (55) and by

\begin{equation}
\label{eq68}
p_{0}=\omega, \, p_{\ell}=k_{\ell}, \, \widetilde{j}_{\ell n}=\widetilde{x}_{\ell}k_{n}-\widetilde{x}_{n}k_{\ell}+s_{\ell n};
\end{equation}
$$(\omega\equiv\sqrt{\overrightarrow{k}^{2}+m^{2}}, \, \widetilde{x}_{\ell}=-i\widetilde{\partial}_{\ell}, \, \widetilde{\partial}_{\ell}\equiv \frac{\partial}{\partial k^{\ell}}),$$
\begin{equation}
\label{eq69}
\widetilde{j}_{0 \ell}=-\widetilde{j}_{\ell 0}=t k_{l}-\frac{1}{2}\left\{\widetilde{x}_{\ell},\omega \right\}-\left(\frac{s_{\ell n}k_{n}}{\omega+m} \equiv \breve{\widetilde{s}}_{\ell}\right),
\end{equation}

\noindent in the momentum $\overrightarrow{k}$-realization $\widetilde{\mathrm{H}}^{3,4}$ of the doublet states space, respectively. The explicit form of the spin terms $s_{\ell n}$ in the expressions (12), (13), (68), (69), which is used for the spin s=(1/2,1/2) particle-antiparticle doublet, is given in the definition (64).

Despite the manifestly non-covariant forms (12), (13), (68), (69) of the $\mathcal{P}^{\mathrm{f}}$-generators, they satisfy the commutation relations of the $\mathcal{P}$ algebra in the manifestly covariant form (11).

The $\mathcal{P}^{\mathrm{f}}$-representation of the group $\mathcal{P}$ in the space $\mathrm{H}^{3,4}$ (55) is given by a converged in this space exponential series (14) or, in the momentum space $\widetilde{\mathrm{H}}^{3,4}$, by a corresponding exponential series given in terms of the generators (68), (69).

The corresponding Casimir operators have the form

\begin{equation}
\label{eq70}
p^{2}=\widehat{p}^{\mu}\widehat{p}_{\mu}=m^{2}\mathrm{I}_{4},
\end{equation}
\begin{equation}
\label{eq71}
W=w^{\mu}w_{\mu}=m^{2}\overrightarrow{s}^{2}=\frac{1}{2}\left(\frac{1}{2}+1\right)m^{2}\mathrm{I}_{4},
\end{equation}

\noindent where $\overrightarrow{s}$ is given in (64) and $\mathrm{I}_{4}$ is $4\times 4$ unit matrix.

Note that the modern definition of $\mathcal{P}$ invariance (or $\mathcal{P}$ symmetry) of the equation of motion (57) in $\mathrm{H}^{3,4}$ is given by the following assertion, see, e. g. [22]. \textit{The set} $\mathrm{F}\equiv\left\{f\right\}$ \textit{of all possible solutions of the equation (57) is invariant with respect to the} $\mathcal{P}^{\mathrm{f}}$-\textit{representation of the group} $\mathcal{P}$, \textit{if for arbitrary solution} $f$ \textit{and arbitrarily-fixed parameters} $(a,\varpi)$ \textit{the assertion}

\begin{equation}
\label{eq72}
(a,\varpi)\rightarrow U(a,\varpi)\left\{f\right\}=\left\{f\right\}\equiv\mathrm{F}
\end{equation}

\noindent \textit{is valid}. Furthermore, the assertion (72) is ensured by the fact that (as it is easy to verify) all the $\mathcal{P}$-generators (12), (13) commute with the operator $i\partial_{t}-\sqrt { - \Delta + m^2}$ of the equation (57).

In spite of the fact that in RCQM many manifestly noncovariant objects are used, \textit{the model under consideration is relativistic invariant in the sense of the definition given above}.

\textbf{On the main and additional conservation laws}.

Similarly to the nonrelativistic quantum mechanics \textit{the conservation laws are found in the form of quantum-mechanical mean values of the operators, which commute with the operator of the equation of motion}.

The important physical consequence of the assertion about the relativistic invariance is the fact that 10 integral dynamical variables of the doublet

\begin{equation}
\label{eq73}
(P_{\mu}, \, J_{\mu\nu}) \equiv \int d^{3}x f^{\dag}(t,\overrightarrow{x})(\widehat{p}_{\mu}, \, \widehat{j}_{\mu\nu})f(t,\overrightarrow{x})=
\end{equation}
$$\mathrm{Const}$$

\noindent do not depend on time, i. e. they are the constants of motion for this doublet.

Note that the external and internal degrees of freedom  for the free spin s=(1/2,1/2) partical-antipartical doublet are independent. Therefore, the operator $\overrightarrow{s}$ (64) commutes not only with the operators $\widehat{\overrightarrow{p}}, \overrightarrow{x}$, but also with the orbital part $\widehat{m}_{\mu\nu}$ of the total angular momentum operator. And both operators $\overrightarrow{s}$ and $\widehat{m}_{\mu\nu}$ commute with the operator $i\partial_{t}-\sqrt { - \Delta + m^2}$ of the equation (57). Therefore, besides the 10 main (consequences of the 10 Poincar$\mathrm{\acute{e}}$ generators) conservation laws (73), 12 additional constants of motion exist for the free spin s=(1/2,1/2) partical-antipartical doublet. These additional conservation laws are the consequences of the operators of the following observables:

$$s_{j}, \, \breve{s}_{\ell}=\frac{s_{\ell n}\widehat{p}_{n}}{\widehat{\omega}+m}, \, \widehat{m}_{\ell n}=x_{l}\widehat{p}_{n}-x_{n}\widehat{p}_{\ell},$$
\begin{equation}
\label{eq74}
\widehat{m}_{0 \ell}=-\widehat{m}_{l0}=t\widehat{p}_{\ell}-\frac{1}{2}\left\{x_{\ell},\widehat{\omega}\right\},
\end{equation}

\noindent where $s_{j}=s_{\ell n}$ are given in (64).

Thus, the following assertions can be proved. In the space $\mathrm{H}^{\mathrm{A}}=\left\{A\right\}$ of the quantum-mechanical amplitudes the 10 main conservation laws (73) have the form

\begin{equation}
\label{eq75}
(P_{\mu},J_{\mu\nu})=\int d^{3}k A^{\dag}(\overrightarrow{k})(\widetilde{p}_{\mu},\widetilde{j}_{\mu\nu})A(\overrightarrow{k}),
\end{equation}
$$A(\overrightarrow{k})\equiv \left|
{{\begin{array}{*{20}c}
 a^{-}_{\mathrm{r}}\\
 a^{+}_{\acute{\mathrm{r}}}\\
\end{array} }} \right|,$$

\noindent where the density generators of $\mathcal{P}^{\mathrm{A}}$, $(\widetilde{p}_{\mu},\widetilde{j}_{\mu\nu})$ of (75) are given by

\begin{equation}
\label{eq76}
\widetilde{p}_{0}=\omega, \, \widetilde{p}_{l}=k_{l}, \, \widetilde{j}_{ln}=\widetilde{x}_{l}k_{n}-\widetilde{x}_{n}k_{l}+s_{ln};
\end{equation}
$$(\widetilde{x}_{l}=-i\frac{\partial}{\partial k^{l}}),$$
\begin{equation}
\label{eq77}
\widetilde{j}_{0l}=-\widetilde{j}_{l0}=-\frac{1}{2}\left\{\widetilde{x}_{l},\omega \right\}-(\breve{\widetilde{s}}_{l}\equiv\frac{s_{ln}k_{n}}{\omega+m}).
\end{equation}

\noindent In the formula (75) $A(\overrightarrow{k})\equiv \left|
{{\begin{array}{*{20}c}
 a^{-}_{\mathrm{r}}\\
 a^{+}_{\acute{\mathrm{r}}}\\
\end{array} }} \right|$ is a 4-column of amplitudes $a^{-}_{+}(\overrightarrow{k}),\,a^{-}_{-}(\overrightarrow{k}), \, a^{+}_{-}(\overrightarrow{k}), \, a^{+}_{+}(\overrightarrow{k})$, where $\mathrm{r}=(+.-). \, \acute{\mathrm{r}}=(-,+)$.

Note that the operators (75)--(77) satisfy the Poincar$\mathrm{\acute{e}}$ commutation relations in the manifestly covariant form (11).

It is evident that the 12 additional conservation laws

$$(M_{\mu \nu}, \, S_{\ell n}, \, \breve{S}_{\ell}) \equiv$$
\begin{equation}
\label{eq78}
\int d^{3}x f^{\dag}(t,\overrightarrow{x})(\widehat{m}_{\mu \nu}, \, s_{\ell n}, \, \breve{s}_{\ell})f(t,\overrightarrow{x})
\end{equation}

\noindent generated by the operators (74), are the separate terms in the expressions (75)--(77) of principal (main) conservation laws.

\textbf{On the stationary complete sets of operators}. Let us consider now the outstanding role of the different \textit{complete sets} of operators from the algebra of observables $\mathrm{A}_{\mathrm{S}}$. If one does not appeal to the complete sets of operators, then the solutions of the the Schr$\mathrm{\ddot{o}}$dinger-Foldy equation (57) are linked directly only with the Sturm-Liouville problem for the energy operator (59). In this case one comes to so-called "degeneration" of solutions. Recall that for an arbitrary complete sets of operators the notion of degeneration is absent in the Sturm-Liouville problem (see, e.g., [21]): only one state vector corresponds to any one point of the common spectrum of a complete set of operators. To wit,  for a comp;ete set of operators there is a one to one correspondence between any point of the common spectrum and an eigenvector.

The \textit{stationary complete sets} play the special role among the complete sets of operators. Recall that the stationary complete set is the set of all functionally independent mutually commuting operators, each of which commutes with the operator of energy (in our case with the operator (59)). The examples of the stationary complete sets in $\mathrm{H}^{3,4}$ are given by $(\widehat{\overrightarrow{p}}, \, s_{z}\equiv s^{3}, \, g)$, $(\overrightarrow{p}, \, \overrightarrow{s}\cdot\overrightarrow{p}, \, g)$, ets. The set $(\overrightarrow{x}, \, s_{z}, \, g)$ is an example of non-stationary complete set. The $\overrightarrow{x}$-realization (55) of the space $\mathrm{H}^{3,4}$ and of quantum-mechanical Schr$\mathrm{\ddot{o}}$dinger-Foldy equation (57) are related just to this complete set.

For the goals of this paper the stationary complete set $(\widehat{\overrightarrow{p}}, \, s_{z}\equiv s^{3}, \, g)$ is chosen. The equations on eigenvectors and eigenvalues of the operators of this stationary complete set have the form

\begin{equation}
\label{eq79} \widehat{\overrightarrow{p}}e^{-ikx}\mathrm{d}_{\alpha} =
\overrightarrow{k}e^{-ikx}\mathrm{d}_{\alpha}, \quad \alpha =1,2,3,4,
\end{equation}
$$s^{3}\mathrm{d}_{1} = \frac{1}{2}\mathrm{d}_{1}, \, s^{3}\mathrm{d}_{2} = -\frac{1}{2}\mathrm{d}_{2},$$
\begin{equation}
\label{eq80}
s^{3}\mathrm{d}_{3} = -\frac{1}{2} \mathrm{d}_{3}, \, s^{3}\mathrm{d}_{4} = \frac{1}{2}\mathrm{d}_{4},
\end{equation}
\begin{equation}
\label{eq81} g\mathrm{d}_{1} = -\mathrm{d}_{1}, \,
g\mathrm{d}_{2} = -\mathrm{d}_{2}, \,  g\mathrm{d}_{3} =
\mathrm{d}_{3}, \, g\mathrm{d}_{4} = \mathrm{d}_{4},
\end{equation}

\noindent where the Cartesian orts $\{\mathrm{d}_{\alpha}\}$ are
given in (42).

The interpretation of the amplitudes in the general solution (58) follows from equations (79)--(81). Thus, the functions $a^{-}_{+}(\overrightarrow{k}), \, a^{-}_{-}(\overrightarrow{k})$ are the momentum-spin amplitudes of the particle (e. g., electron) with the momentum $\widehat{\overrightarrow{p}}$, sign of the charge ($-e$) and spin projections ($\frac{1}{2}, \, -\frac{1}{2}$), respectively. Further, the functions $a^{+}_{-}(\overrightarrow{k}), \, a^{+}_{+}(\overrightarrow{k})$ in (58) are the momentum-spin amplitudes of the antiparticle (e. g., positron) with the momentum $\widehat{\overrightarrow{p}}$, sign of the charge ($+e$) and spin projections ($-\frac{1}{2}, \, \frac{1}{2}$), respectively.

Thus, the conclusion about the fermionic spin s=(1/2,1/2) features of solution (58) (i. e. the interpretation of the solution (58)) follows from the equations (79)--(81) and the above given interpretation of the amplitudes.

\textbf{On the solutions of the Schr$\mathrm{\ddot{o}}$dinger-Foldy equation}. Let us consider the Schr$\mathrm{\ddot{o}}$dinger-Foldy equation (57) general solution  related to the stationary complete sets $(\widehat{\overrightarrow{p}}, \, s_{z}\equiv s^{3}, \, g)$, where $s^{3}$ is given in (64). The fundamental solutions of the equation (57), which are the eigen solutions of this stationary complete sets, are given by the relativistic de Broglie waves:

\begin{equation}
\label{eq82}
\varphi_{\vec{k}\alpha}(t,\overrightarrow{x})=\frac{1}{(2\pi)^{\frac{3}{2}}}e^{-i\omega t + i\vec{k}\vec{x} }\mathrm{D}_{\alpha}, \quad \mathrm{D}_{\alpha}=(\delta^{\beta}_{\alpha}),
\end{equation}
$$\alpha = \mathrm{r},\acute{\mathrm{r}}, \quad \mathrm{r}=1,2, \, \acute{\mathrm{r}}=3,4,$$

\begin{equation}
\label{eq83}
\mathrm{D}_{\mathrm{r}}\equiv \left| {{\begin{array}{*{20}c}
 \mathrm{d}_{\mathrm{r}}\\
 0\\
\end{array} }} \right|, \, \mathrm{D}_{\acute{\mathrm{r}}}\equiv \left|
{{\begin{array}{*{20}c}
 0\\
 \mathrm{d}_{\acute{\mathrm{r}}}\\
\end{array} }} \right|,
\end{equation}
$$\mathrm{d}_{1}=\mathrm{d}_{3}=\left| {{\begin{array}{*{20}c}
 1\\
 0\\
\end{array} }} \right|,\, \mathrm{d}_{2}=\mathrm{d}_{4}=\left|
{{\begin{array}{*{20}c}
 0\\
 1\\
\end{array} }} \right|,$$

\noindent where the Cartesian orts $\mathrm{D}_{\alpha}$ are the common eigenvectors for the operators $(s_{z}, \, g)$.

Vectors (82) are the generalized solutions of the equation (57). These solutions do not belong to the quantum-mechanical space $\mathrm{H}^{3,4}$, i. e. they are not realized in the nature. Nevertheless, the solutions (82) are the complete orthonormalized orts in the rigged Hilbert space $\mathrm{S}^{3,4}\subset\mathrm{H}^{3,4}\subset\mathrm{S}^{3,4*}$. In symbolic form the conditions of orthonormalisation and completeness are given by

\begin{equation}
\label{eq84}
\int d^{3}x\varphi^{\dag}_{\vec{k}\alpha}(t,\overrightarrow{x})\varphi_{\vec{k}^{\prime}\alpha^{\prime}}(t,\overrightarrow{x})=\delta(\overrightarrow{k}-\overrightarrow{k}^{\prime})\delta_{\alpha\alpha^{\prime}},
\end{equation}

\begin{equation}
\label{eq85}
\int d^{3}k\sum_{\alpha=1}^{4}\varphi^{\beta}_{\vec{k}\alpha}(t,\overrightarrow{x})\varphi^{*\beta^{\prime}}_{\vec{k}\alpha}(t,\overrightarrow{x}^{\prime})=\delta(\overrightarrow{x}-\overrightarrow{x}^{\prime})\delta_{\beta\beta^{\prime}}.
\end{equation}

\noindent The functional forms of these conditions are omitted because of their bulkiness.

In the rigged Hilbert space $\mathrm{S}^{3,4}\subset\mathrm{H}^{3,4}\subset\mathrm{S}^{3,4*}$ an arbitrary solution of the equation (57) can be decomposed in terms of fundamental solutions (82). Furthermore, for the solutions $f\in\mathrm{S}^{3,4}\subset\mathrm{H}^{3,4}$ the expansion

\begin{equation}
\label{eq86}
f(t,\overrightarrow{x})=\frac{1}{(2\pi)^{\frac{3}{2}}}\int d^{3}xe^{-ikx}[a^{-}_{\mathrm{r}}(\overrightarrow{k})\mathrm{D}_{\mathrm{r}}+a^{+}_{\acute{\mathrm{r}}}(\overrightarrow{k})\mathrm{D}_{\acute{\mathrm{r}}}^{+}],
\end{equation}
$$kx \equiv \omega t- \overrightarrow{k}\overrightarrow{x}, \quad \omega \equiv \sqrt{\overrightarrow{k}^{2}+m^{2}},$$

\noindent is, (i) mathematically well-defined in the framework of the standard differential and integral calculus, (ii) if in the expansion (86) a state $f\in\mathrm{S}^{3,4}\subset\mathrm{H}^{3,4}$, then the amplitudes $(a_{\alpha})=(a^{-}_{\mathrm{r}}, \, a^{+}_{\acute{\mathrm{r}}})$ in (86) belong to the set of the Schwartz test functions over $\mathrm{R}^{3}_{\vec{k}}$. Therefore, they have the unambiguous physical sense of the amplitudes of probability distributions over the eigen values of the stationary complete sets $(\widehat{\overrightarrow{p}}, \, s_{z}, \, g)$. Moreover, the complete set of quantum-mechanical amplitudes unambiguously determine the corresponding representation of the space $\mathrm{H}^{3,4}$ (in this case it is the $(\overrightarrow{k}, \, s_{z}, \, g)$-representation), which vectors have the harmonic time dependence

\begin{equation}
\label{eq87}
\widetilde{f}(t,\overrightarrow{k})=e^{-i\omega t}A(\overrightarrow{k}),
\end{equation}
$$A(\overrightarrow{k})\equiv \mathrm{column}(a^{-}_{+},\,a^{-}_{-},\,a^{+}_{-},\,a^{+}_{+}),$$

\noindent i. e. are the states with the positive sign of the energy $\widetilde{\omega}$.

The similar assertion is valid for the expansions of the states $f\in\mathrm{H}^{3,4}$ over the basis states, which are the eigen vectors of an arbitrary stationary complete sets. Therefore, the corresponding representation of the space $\mathrm{H}^{3,4}$, which is related to such expansions, is often called as the generalized Fourier transformation.

By the way, the $\overrightarrow{x}$-realization (55) of the states space is associated with the non-stationary complete set of operators $(\overrightarrow{x}, \, s_{z}, \, g)$. Therefore, the amplitudes $f^{\alpha}(t,\overrightarrow{x})=\mathrm{D}^{\dag}_{\alpha}f(t,\overrightarrow{x})=U(t)f(0,\overrightarrow{x})$ of the probability distribution over the eigen values of this complete set depend on time $t$ non-harmonically.

\textbf{On the Clifford--Dirac algebra}.

The Clifford--Dirac algebra of the $\gamma$-matrices must be introduced into the FW representations. The reasons are as follows.

Part of the Clifford--Dirac algebra operators are directly related to the spin 1/2 doublet operators $(\frac{1}{2}\gamma^{2}\gamma^{3}, \,\frac{1}{2}\gamma^{3}\gamma^{1}, \,\frac{1}{2}\gamma^{1}\gamma^{2})$ (in the anti-Hermitian form). In the FW representation for the spinor field [2] these spin operators commute with the Hamiltonian and with the operator of the equation of motion $i\partial _{0}-\gamma^{0}\widehat{\omega}$. In the Pauli--Dirac representation these operators do not commute with the Dirac equation operator. Only the sums of the orbital operators and such spin operators commute with the Diracian. So \textit{if we want to relate the orts $\gamma^{\mu}$ of the Clifford--Dirac algebra with the actual spin we must introduce this algebra into the FW representation}.

In the quantum-mechanical representation (i. e. in the space $\left\{\mathrm{f}\right\}$ of the solutions (58) of the Schr$\mathrm{\ddot{o}}$dinger--Foldy equation(57)) the $\gamma$-matrices are obtained by the transformation $v$ given in the formula (89) below.

Moreover, we use the generalized Clifford--Dirac algebra over the field of real numbers. This algebra was introduced in Refs. [24--28]. The use of 29 orts of this \textit{proper extended real Clifford--Dirac algebra} gives the additional possibilities in comparison with only 16 elements of the standard Clifford--Dirac algebra, see, e. g., [24--28].

The definitions of spin matrices (64) de facto determine a so-called "quantum-mechanical" representation of the Dirac matrices

\begin{equation}
\label{eq88}
\bar{\gamma}^{\hat{\mu}}: \, \bar{\gamma}^{\hat{\mu}}\bar{\gamma}^{\hat{\nu}}+\bar{\gamma}^{\hat{\nu}}\bar{\gamma}^{\hat{\mu}}=2g^{\hat{\mu}\hat{\nu}}; \, \bar{\gamma}^{-1}_{0}=\bar{\gamma}_{0}, \, \bar{\gamma}^{-1}_{l}=-\bar{\gamma}_{l},
\end{equation}
$$g^{\hat{\mu}\hat{\nu}} \equiv (+----), \, \hat{\mu}=0,1,2,3,4.$$

\noindent The matrices $\bar{\gamma}^{\mu}$ (88) of this representation are linked with the Dirac matrices $\gamma^{\hat{\mu}}$ in the standard Pauli-Dirac (PD) representation:

$$\bar{\gamma}^{0}=\gamma^{0}, \, \bar{\gamma}^{1}=\gamma^{1}C, \, \bar{\gamma}^{2}=\gamma^{0}\gamma^{2}C, \, \bar{\gamma}^{3}=\gamma^{3}C,$$
$$\bar{\gamma}^{4}=\gamma^{0}\gamma^{4}C;$$
\begin{equation}
\label{eq89}
\bar{\gamma}^{\hat{\mu}}=v\bar{\gamma}^{\hat{\mu}}v, \quad v \equiv\left|
{{\begin{array}{*{20}c}
 \mathrm{I}_{2} \hfill & 0 \hfill \\
 0 \hfill &  C\mathrm{I}_{2} \hfill \\
\end{array} }} \right|=v^{-1},
\end{equation}

\noindent where the standard Dirac matrices $\gamma^{\hat{\mu}}$ are given by

\begin{equation}
\label{eq90}
\gamma^0 = \left| {{\begin{array}{*{20}c}
 \mathrm{I}_{2} \hfill & 0 \hfill \\
 0 \hfill & -\mathrm{I}_{2} \hfill \\
\end{array} }} \right|, \quad \gamma ^k = \left| {{\begin{array}{*{20}c}
 0 \hfill & {\sigma ^k} \hfill \\
 { - \sigma ^k} \hfill & 0 \hfill \\
\end{array} }} \right|,
\end{equation}
$$\gamma^{4} \equiv \gamma^{0}\gamma^{1}\gamma^{2}\gamma^{3}, \quad \gamma^{0}\gamma^{1}\gamma^{2}\gamma^{3}\gamma^{4}=-\mathrm{I}_{4},$$

\noindent $C$ is the operator of complex conjugation.

Note that in the terms of $\bar{\gamma}^{\mu}$ matrices (89) the RCQM spin operator (64) has the form

\begin{equation}
\label{eq91}
\overrightarrow{s}=\frac{i}{2}(\bar{\gamma}^{2}\bar{\gamma}^{3}, \, \bar{\gamma}^{3}\bar{\gamma}^{1}, \, \bar{\gamma}^{1}\bar{\gamma}^{2}).
\end{equation}

\noindent Therefore, the complete analogy with the particle-antiparticle doublet spin in the FW representation exists

\begin{equation}
\label{eq92} \overrightarrow{s}_{\mathrm{FW}} =
\frac{i}{2}(\gamma^{2}\gamma^{3}, \, \gamma^{3}\gamma^{1}, \, \gamma^{1}\gamma^{2}).
\end{equation}

The $\bar{\gamma}^{\mu}$ matrices (89) together with the matrix $\bar{\gamma}^{4}\equiv \bar{\gamma}^{0}\bar{\gamma}^{1}\bar{\gamma}^{2}\bar{\gamma}^{3}$, imaginary unit $i\equiv\sqrt{-1}$ and operator $C$ of complex conjugation in $\mathrm{H}^{3,4}$ generate the quantum-mechanical representations of the extended real Clifford-Dirac algebra and proper extended real Clifford-Dirac algebra, which were put into consideration in [24] (see also [25--28]).

Remind that in [24--28] for the purposes of finding links between the fermionic and bosonic states not 5 (as in (88)--(90)) but 7 generating $\gamma$ matrices were used. In addition to $\gamma^{1}, \, \gamma^{2}, \, \gamma^{3}, \, \gamma^{4}$ matrices from (90) 3 new $\gamma$ matrices were introduced. Therefore, the set of 7 generating $\gamma$ matrices is given by

\begin{equation}
\label{eq93} \gamma^{1}, \, \gamma^{2}, \, \gamma^{3}, \, \gamma^{4}, \, \gamma^{5} \equiv \gamma^{1}\gamma^{3}C, \, \gamma^{6} \equiv i\gamma^{1}\gamma^{3}C,
\end{equation}
$$\gamma^{7} \equiv i\gamma^{0}, \, \gamma^{1}\gamma^{2}\gamma^{3}\gamma^{4}\gamma^{5}\gamma^{6}\gamma^{7}=\mathrm{I}_{4}.$$

Here, in the quantum-mechanical representation these $\gamma$ matrices (in the terms of standard $\gamma$ matrices) have the form

\begin{equation}
\label{eq94} \bar{\gamma}^{1}, \, \bar{\gamma}^{2}, \, \bar{\gamma}^{3}, \, \bar{\gamma}^{4} \equiv \bar{\gamma}^{0}\bar{\gamma}^{1}\bar{\gamma}^{2}\bar{\gamma}^{3}, \, \bar{\gamma}^{5} \equiv \gamma^{1}\gamma^{3}C,
\end{equation}
$$\bar{\gamma}^{6} \equiv -i\gamma^{2}\gamma^{4}C, \, \bar{\gamma}^{7} \equiv i, \,\bar{\gamma}^{1}\bar{\gamma}^{2}\bar{\gamma}^{3}\bar{\gamma}^{6}\bar{\gamma}^{5}\bar{\gamma}^{6}\bar{\gamma}^{7}=\mathrm{I}_{4},$$

\noindent and satisfy the anticommutation relations in the following form

\begin{equation}
\label{eq95}
\bar{\gamma}^{\mathrm{A}}\bar{\gamma}^{\mathrm{B}}+\bar{\gamma}^{\mathrm{B}}\bar{\gamma}^{\mathrm{A}}=2\delta^{\mathrm{A}\mathrm{B}}, \quad \mathrm{A}=\overline{1,7}.
\end{equation}

\noindent The Clifford--Dirac anticommutation relations for the matrices (93) are similar.

The $\gamma$ matrices (93), (94) generate the 29 dimensional representation  of the proper extended real Clifford--Dirac algebra SO(8), which was introduced in [24--28]. Both the fundamental representation $S^{\mathrm{A}\mathrm{B}}=\frac{1}{4}[\gamma^{\mathrm{A}}\gamma^{\mathrm{B}}]$ and the RCQM representation $\bar{S}^{\mathrm{A}\mathrm{B}}=\frac{1}{4}[\bar{\gamma}^{\mathrm{A}}\bar{\gamma}^{\mathrm{B}}]$ are generated by the matrices (93) and (94), respectively. On the structure, subalgebras and different representations of the extended real Clifford--Dirac algebra see, e. g., in [28].

The additional possibilities, which are open by the additional 29 orts of algebra SO(8) in comparison with 16 orts of the standard Clifford--Dirac algebra, are principal in description of the Bose states in the framework of the Dirac theory [24--28]. The algebra SO(8) includes two independent SU(2) subalgebras ($\gamma^{2}\gamma^{3}, \, \gamma^{3}\gamma^{1}, \, \gamma^{1}\gamma^{2}$) and ($\gamma^{5}\gamma^{6}, \, \gamma^{6}\gamma^{4}, \, \gamma^{4}\gamma^{5}$), when the standard Clifford--Dirac algebra includes only one given by the elements ($\gamma^{2}\gamma^{3}, \, \gamma^{3}\gamma^{1}, \, \gamma^{1}\gamma^{2}$).

Therefore, the algebra SO(8) should be taken as the Clifford--Dirac algebra of RCQM.

\textbf{Dynamic and kinematic aspects of the relativistic invariance}. Consider briefly some detalizations of the relativistic invariance of the Schr$\mathrm{\ddot{o}}$dinger--Foldy equation (57). Note that for the free spin s=(1/2,1/2) particle-antiparticle doublet the equation (57) has one and the same explicit form in arbitrary-fixed IFR (its set of solutions is one and the same in every IFR). Therefore, the algebra of observables and the conservation laws (as the functionals of the free spin s=(1/2,1/2) particle-antiparticle doublet states) have one and the same form too. This assertion explains the dynamical sense of the $\mathcal{P}$ invariance (the invariance with respect to the dynamical symmetry group $\mathcal{P}$).

Another, kinematic, aspect of the $\mathcal{P}$ invariance of the RQCM model has the following physical sense. Note at first that any solution of the Schr$\mathrm{\ddot{o}}$dinger--Foldy equation (57) is determined by the concrete given set of the amplitudes $\left\{A\right\}$. It means that if $f$ with the fixed set of amplitudes $\left\{A\right\}$ is the state of the doublet in some arbitrary IFR, then for the observer in the $(a, \, \varpi)$-transformed $\mathrm{IFR}^{\prime}$ this state $f^{\prime}$ is determined by the amplitudes $\left\{A^{\prime}\right\}$. The last ones are received from the given $\left\{A\right\}$ by the unitary  $\mathcal{P}^{\mathrm{A}}$ -transformation

\begin{equation}
\label{eq96}
\mathcal{P}^{\mathrm{A}}: \, (a,\varpi)\rightarrow \widetilde{U}(a,\varpi)=\exp (-ia^{\mu}\widetilde{p}_{\mu}-\frac{i}{2}\varpi^{\mu\nu}\widetilde{j}_{\mu\nu}),
\end{equation}

\noindent where $(\widetilde{p}_{\mu}, \, \widetilde{j}_{\mu\nu})$ are given in (68), (69).

\textbf{The axiom on the mean value of the operators of observables}. Note that any apparatus can not fulfill the absolutely precise measurement of a value of the physical quantity having continuous spectrum. Therefore, the customary quantum-mechanical axiom about the possibility of "precise" measurement, for example, of the coordinate (or another quantity with the continuous spectrum), which is usually associated with the corresponding "reduction" of the wave-packet, can be revisited. This assertion for the values with the continuous spectrum can be replaced by the axiom that only the mean value of the operator of observable (or the corresponding complete set of observables) is the experimentally observed for $\forall f\in\mathrm{H}^{3,4}$. Such axiom, without any loss of generality of consideration, unambiguously justifies the using of the subspace $\mathrm{S}^{3,4}\subset\mathrm{H}^{3,4}$ as an approximative space of the physically realizable states of the considered object. This axiom as well does not enforce the application of the conception of the ray in $\mathrm{H}^{3,4}$ (the set of the vectors $e^{i\alpha}f$ with an arbitrary-fixed real number $\alpha$) as the state of the object. Therefore, the mapping $(a, \, \varpi)\rightarrow U(a, \, \varpi)$ in the formula (96) and in the formula (14) for the $\mathcal{P}$-representations in $\mathrm{S}^{3,4}\subset\mathrm{H}^{3,4}$ is an unambiguous. Such axiom actually removes the problem of the wave packet "reduction", which discussion started from the well-known von Neumann monograph [15]. Therefore, the subjects of the discussions of all "paradoxes" of quantum mechanics, a lot of attention to which was paid in the past century, are removed also.

The important conclusion about the RCQM is as follows. The consideration of all aspects of this model is given on the basis of using only such conceptions and quantities, which have the direct relation to the experimentally observable physical quantities of this "elementary" physical system.

\textbf{On the principles of heredity and the correspondence}. The explicit forms (73)--(78) of the main and additional conservation laws demonstrate evidently that the model of RCQM satisfies the principles of the heredity and the correspondence with the non-relativistic classical and quantum theories. The deep analogy between RCQM and these theories for the physical system with the finite number degrees of freedom (where the values of the free dynamical conserved quantities are additive) is also evident.

Our new way of the Dirac equation derivation (section 9 below) is started from the RCQM of the spin s=(1/2,1/2) doublet, which is formulated in this section above.

\textbf{On the physical interpretation}. The physical interpretation always is the final step in the arbitrary model of the physical reality formulation.

The above considered model of physical reality is called the canonical quantum mechanics due to the principles of the heredity and correspondence with nonrelativistic Schr$\mathrm{\ddot{o}}$dinger quantum mechanics [15].

The above considered canonical quantum mechanics is called the relativistic canonical quantum mechanics due to its invariance with respect to the corresponding representations of the Poincar$\mathrm{\acute{e}}$ group $\mathcal{P}$.

The above considered RCQM describes the spin s = (1/2,1/2) particle-antiparticle doublet due to the corresponding eigenvalues in the equations (79)--(81) for the stationary complete set of operators and the explicit forms of the Casimir operators (70), (71) of the corresponding Poincar$\mathrm{\acute{e}}$ group $\mathcal{P}$ representation, with respect to which the dynamical equation of motion is invariant.

\section{Briefly on the second quantization}

Finally, consider briefly the program of the canonical quantization of the RCQM model. Note that the expression for the total energy $P_{0}$ plays a special role in the procedure of a so called "second quantization". In the RCQM doublet model, as it is evident from the expression for the $P_{0}$ in (75) in terms of the charge sign-momentum-spin amplitudes

\begin{equation}
\label{eq97}
P_{0}=\int d^{3}k\omega\left(\left|a^{-}_{\mathrm{r}}(\overrightarrow{k})\right|^{2}+\left|a^{+}_{\acute{\mathrm{r}}}(\overrightarrow{k})\right|^{2}\right)\geq m>0,
\end{equation}

\noindent the energy is positive. The same assertion is valid for the amplitudes related to the arbitrary-fixed stationary complete set of operators. Furthermore, the operator $\widehat{P}_{0}$ of the quantized energy corresponding to expression (97) is a positive-valued operator. The explicit form of the operator $\widehat{P}_{0}$ follows from the expression (97) after the anticommutation quantization of the amplitudes

\begin{equation}
\label{eq98}
 \left\{\widehat{a}_{\alpha}(\overrightarrow{k}),\widehat{a}^{\dag}_{\beta}(\overrightarrow{k})\right\}=\delta_{\alpha\beta}\delta\left(\overrightarrow{k}-\overrightarrow{k}^{\prime}\right)
\end{equation}

\noindent (other operators anticommute) and their substitution $a^{\mp}\rightarrow \widehat{a}^{\mp}$ into the formula (97). Note that the quantized amplitudes determine the Fock space $\mathcal{H}^{\mathrm{F}}$ (over the quantum-mechanical space $\mathrm{H}^{3,4}$). Moreover, the operators of dynamical variables $\widehat{P}_{\mu}, \, \widehat{J}_{\mu\nu}$ in $\mathcal{H}^{\mathrm{F}}$, which are expressed according to formulae (75) in terms of  the operator amplitudes $\widehat{a}_{\alpha}(\overrightarrow{k}), \, \widehat{a}^{\dag}_{\beta}(\overrightarrow{k})$, automatically have the form of "normal products" and satisfy the commutation relations (11) of the $\mathcal{P}$ group in the Fock space $\mathcal{H}^{\mathrm{F}}$. Operators $\widehat{P}_{\mu}, \, \widehat{J}_{\mu\nu}$ determine the corresponding unitary representation in $\mathcal{H}^{\mathrm{F}}$. Other details are not the subject of this paper.

\section{Derivation of the Foldy--Wouthuysen and the standard Dirac equations from the relativistic canonical quantum mechanics}

At first, the FW and the standard Dirac equations are proved to be the direct and unambiguous consequences of the Schr$\mathrm{\ddot{o}}$dinger--Foldy equation (57) and of the RCQM model. Further, consider briefly the physical and mathematical sense of three different models of the fermionic doublet (RCQM, the FW model, the Dirac model) at the example of comparison of the group-theoretical approaches to these three models.

Thus, below we search for the infallible links between the RCQM, the FW model and the Dirac model of the fermionic doublet. The links between the objects of equations (and between the associated operators of the algebra of observables) of these three models are under consideration. The following mathematical and physical assertions should be taken into account in order to perform such a program.

The Poincar$\mathrm{\acute{e}}$ group $\mathcal{P}$ is the group of real parameters $(a, \, \varpi)$, i. e. it is a real Lie group. Therefore, as a matter of fact the prime (anti-Hermitian) generators

\begin{equation}
\label{eq99}
 (p_{\mu}, \, j_{\mu\nu})^{\mathrm{prime}} \equiv (p_{\mu}^{\mathrm{pr}}, \, j_{\mu\nu}^{\mathrm{pr}}) \equiv (-ip_{\mu}, \, -ij_{\mu\nu})
\end{equation}

\noindent are the generators of the group $\mathcal{P}$, and not the operators $(p_{\mu}, \, j_{\mu\nu})$, which were used in all the above formulae. In terms of prime generators (99) the $\mathcal{P}$-representation in the space $\mathrm{H}^{3,4}$ is given by the formula

\begin{equation}
\label{eq100}
 (a,\varpi)\rightarrow U(a,\varpi)=\exp(a^{\mu}p_{\mu}^{\mathrm{pr}}+\frac{1}{2}\varpi^{\mu\nu}j_{\mu\nu}^{\mathrm{pr}}),
\end{equation}

\noindent and the commutation relations of the Lie algebra of the group $\mathcal{P}$ in manifestly covariant form are given by

$$\left[p_{\mu}^{\mathrm{pr}},p_{\nu}^{\mathrm{pr}}\right]=0, \quad \left[p_{\mu}^{\mathrm{pr}},j_{\rho\sigma}^{\mathrm{pr}}\right]=g_{\mu\rho}p_{\sigma}-g_{\mu\sigma}p_{\rho},$$
\begin{equation}
\label{eq101}
\left[j_{\mu\nu}^{\mathrm{pr}},j_{\rho\sigma}^{\mathrm{pr}}\right]=-g_{\mu\rho}j_{\nu\sigma}^{\mathrm{pr}}-g_{\rho\nu}j_{\sigma\mu}^{\mathrm{pr}}-g_{\nu\sigma}j_{\mu\rho}^{\mathrm{pr}}-g_{\sigma\mu}j_{\rho\nu}^{\mathrm{pr}}.
\end{equation}

It was demonstrated [24--28] that the prime (anti-Hermitian) generators play a special role in the group-theoretical approach to the quantum theory and symmetry analysis of the corresponding equations. It is due to the anti-Hermitian generators of the groups under consideration that in [24--28] the additional bosonic properties of the FW and Dirac equations have been found. The mathematical correctness of appealing to the anti-Hermitian generators is considered in details in Refs. [29, 30].

Below in this section only the prime (anti-Hermitian) $\mathcal{P}$-generators and the prime operators of all energy-momentum an angular momentum quantities are used. Therefore, after this \textit{warning} in the consideration below \textbf{the notation "prime" or "pr" of the corresponding operators is omitted}.

Moreover, we rewrite the explicit forms of the Schr$\mathrm{\ddot{o}}$dinger--Foldy, the FW and the Dirac equations in the terms of anti-Hermitian operators.

The link between the Schr$\mathrm{\ddot{o}}$dinger--Foldy equation (57) and the FW equation [2] is given by the operator $v$

\begin{equation}
\label{eq102}
 v=\left|
{{\begin{array}{*{20}c}
 \mathrm{I}_{2} \hfill & 0 \hfill \\
 0 \hfill &  C\mathrm{I}_{2} \hfill \\
\end{array} }} \right|; \quad v^{2}=\mathrm{I}_{4}, \quad \mathrm{I}_{2}=\left|
{{\begin{array}{*{20}c}
 1 \hfill & 0 \hfill \\
 0 \hfill & { 1} \hfill \\
\end{array} }} \right|,
\end{equation}

\noindent (on the existence of such an operator see in the formulae (89)), $C$ is the operator of complex conjugation, the operator of involution in the space $\mathrm{H}^{3,4}$. For the non-singular operator $v$ from (102) the equality

\begin{equation}
\label{eq103}
v\left(\partial_{0}+i\widehat{\omega}\right)v=\partial_{0}+i\gamma^{0}\widehat{\omega}
\end{equation}

\noindent is valid. It means that after the transformation $f\rightarrow \phi =vf$ the Schr$\mathrm{\ddot{o}}$dinger--Foldy equation (57) becomes the FW equation

\begin{equation}
\label{eq104}
\left(\partial_{0}+i\gamma^{0}\widehat{\omega}\right)\phi(x)=0; \quad \phi = vf \equiv \left|
{{\begin{array}{*{20}c}
 \varphi_{-} \hfill  \\
 \varphi_{+}^{*} \hfill  \\
\end{array} }} \right|\in \mathrm{H}^{3,4}.
\end{equation}

\noindent The inverse transformation is also valid

\begin{equation}
\label{eq105}
 f=v\phi.
\end{equation}

The operator $v$ (102) transforms arbitrary operator $q$ of the RCQM into the operator $Q$ in the FW representation for the spinor field \textit{and vice versa}:

\begin{equation}
\label{eq106}
 Q=vqv\leftrightarrow q=vQv.
\end{equation}

\noindent The only warning is that formula (106) is valid only for the anti-Hermitian operators! It means that in order to avoid mistakes one should apply this formula only for the prime (anti-Hermitian) energy-momentum, angular momentum and spin quantities. Especially if one uses the operator (102) for the transformation of the commutation relations.

\textbf{The Foldy-Wouthuysen model of the fermionic doublet}. Comparison of the RCQM and the FW model of the fermionic doublet is the content of the paper [16]. Here some additional important features of the FW model are considered.

The quantum-mechanical sense of the object $\phi$ is the following. The equation (104) is a system of two 2-component equations

\begin{equation}
\label{eq107}
(\partial_{0}+i\widehat{\omega})\varphi_{-}(x)=0, \quad (\partial_{0}-i\widehat{\omega})\varphi_{+}^{*}(x)=0.
\end{equation}

\noindent The first equation is the equation for the wave function $\varphi_{-}$ of the particle (electron) and the second is the one for the function $\varphi_{+}^{*}$ being the complex conjugate to the wave function $\varphi_{+}$ of the antiparticle (positron).

Under the transformations (106) the $\mathcal{P}^{\mathrm{f}}$-generators (12), (13) \textit{(taken with the spin term (64) and in the prime anti-Hermitian form)} become the prime $\mathcal{P}^{\mathrm{\phi}}$-generators ($\mathcal{P}$-symmetries of the FW equation (104))

\begin{equation}
\label{eq108}
\widehat{p}_{0}=-i\gamma_{0}\widehat{\omega}, \, \widehat{p}_{\ell}=\partial_{\ell}, \, \widehat{j}_{\ell n}=x_{\ell}\partial_{n}-x_{n}\partial_{\ell}+\widehat{s}_{\ell n}\equiv
\end{equation}
$$\widehat{m}_{\ell n}+\widehat{s}_{\ell n},$$
\begin{equation}
\label{eq109}
\widehat{j}_{0\ell}=t\partial_{\ell}+\frac{i}{2}\gamma_{0}\left\{x_{l},\widehat{\omega}\right\}+\gamma_{0}\frac{\widehat{s}_{\ell n}\widehat{p}_{n}}{\widehat{\omega}+m},
\end{equation}

\noindent where

\begin{equation}
\label{eq110}
\widehat{s}_{\ell n}=\widehat{s}^{\ell n}\equiv\frac{1}{4}\left[\gamma^{\ell},\gamma^{n}\right],
\end{equation}

\noindent and $\gamma^{\mu}$ are the standard Dirac matrices (90) (in the Pauli--Dirac representation).

Therefore, the $\mathcal{P}^{\mathrm{\phi}}$-generators (108), (109) of the $\mathcal{P}^{\mathrm{\phi}}$-representation in $\mathrm{H}^{3,4}$ (as well as the operators $q^{\mathrm{\phi}}=vq^{\mathrm{f}}v$) of the algebra of all observable physical quantities in the FW model of the fermionic doublet) are the functions generated by the 10 operators

\begin{equation}
\label{eq111}
\overrightarrow{x}=(x^{j})\in \mathrm{R}^{3}, \, \widehat{\overrightarrow{p}}=-\nabla, \, \widehat{\overrightarrow{s}}\equiv (\widehat{s}_{23},\widehat{s}_{31},\widehat{s}_{12}), \, \gamma^{0},
\end{equation}

\noindent where $\widehat{s}_{\ell n}$ are given in (110). The physical sense of these operators (as well as of the functions $q^{\mathrm{\phi}}$ from them) follows from the physical sense of the corresponding quantum-mechanical operators ($\overrightarrow{x}, \, \overrightarrow{p} \, \overrightarrow{s}, \, \gamma^{0}$) (and $q^{\mathrm{f}}$), which are verified by the principles of the heredity and correspondence with non-relativistic quantum and classical theories.

The solution (58) of the Schr$\mathrm{\ddot{o}}$dinger--Foldy equation (57), which is associated with the stationary complete set of operators (momentum -- sign of the charge -- spin projection $\bar{s}_{\mathrm{z}}$) is transformed into the solution

$$\phi(x)=\frac{1}{(2\pi)^{\frac{3}{2}}}\int
d^{3}k\{e^{-ikx}[a^{-}_{+}(\overrightarrow{k})\mathrm{d}_{1}+a^{-}_{-}(\overrightarrow{k})\mathrm{d}_{2}]+$$
\begin{equation}
\label{eq112}
e^{ikx}[a^{*+}_{-}(\overrightarrow{k})\mathrm{d}_{3}+a^{*+}_{+}(\overrightarrow{k})\mathrm{d}_{4}]\}.
\end{equation}

\noindent of the FW equation (104), where
$a^{-}_{+}(\overrightarrow{k}),\,a^{-}_{-}(\overrightarrow{k})$
are the same as in (58) and $a^{*+}_{-}(\overrightarrow{k}),\,
a^{*+}_{+}(\overrightarrow{k})$ are the complex conjugated
functions to the quantum-mechanical amplitudes
$a^{+}_{-}(\overrightarrow{k}),\,a^{+}_{+}(\overrightarrow{k})$
of the standard quantum-mechanical sense, which was considered in
section 7.

Further, similarly to the RCQM in the FW model the additional
conservation laws also exist together with the main 10
Poincar$\mathrm{\acute{e}}$ conservation quantities

$$(\widehat{p}_{\mu},\widehat{j}_{\mu\nu})^{\mathrm{\phi}} \rightarrow (P_{\mu},J_{\mu\nu})^{\mathrm{\phi}}=$$
\begin{equation}
\label{eq113} \int d^{3}x
\phi^{\dag}(x)i(\widehat{p}_{\mu},\widehat{j}_{\mu\nu})^{\mathrm{\phi}}\phi(x).
\end{equation}

\noindent The 12 additional conservation laws, which were
considered in section 7, also exist and can be very easy
calculated here. Naturally, due to non-unitarity of the operator
$v$ from (102) the explicit form of the conservation laws (113) does
not coincide with the quantum-mechanical quantities (73). It is
evident from the expression (113) in the terms of
quantum-mechanical amplitudes

\begin{equation}
\label{eq114} (P_{\mu},J_{\mu\nu})^{\mathrm{\phi}}=\int d^{3}k
A^{\mathrm{\phi}\dag}(\overrightarrow{k})(\widetilde{p}_{\mu},\widetilde{j}_{\mu\nu})^{\mathrm{\phi}}A^{\mathrm{\phi}}(\overrightarrow{k}),
\end{equation}

\noindent where $A^{\mathrm{\phi}}(\overrightarrow{k})$ has the form $A^{\mathrm{\phi}}(\overrightarrow{k})\equiv \left|
\begin{array}{cccc}
 a^{-}_{+} \\
 a^{-}_{-} \\
a^{*+}_{-} \\
a^{*+}_{+} \\
\end{array} \right|,$

\noindent $(\widetilde{p}_{\mu},\widetilde{j}_{\mu\nu})^{\mathrm{\phi}}$ are given by

\begin{equation}
\label{eq115} \widetilde{p}_{0}=\gamma^{0}\omega, \,
\widetilde{p}_{\ell}=\gamma^{0}k_{\ell}, \, \widetilde{j}_{\ell
n}=\widetilde{x}_{\ell}k_{n}-\widetilde{x}_{n}k_{\ell}+
\widehat{s}_{\ell n},
\end{equation}

\begin{equation}
\label{eq116} \widetilde{j}_{0\ell}=-\widetilde{j}_{\ell
0}=-\frac{1}{2}\left\{\widetilde{x}_{\ell},\omega
\right\}+\gamma^{0}(\breve{\widetilde{s}}_{\ell}\equiv\frac{\widehat{s}_{\ell
n}k_{n}}{\omega+m}),
\end{equation}

\noindent and the definitions $(\widetilde{x}_{\ell}=-i\widetilde{\partial}_{\ell}, \, \widetilde{\partial}_{\ell}\equiv \frac{\partial}{\partial k^{\ell}}), \, \omega \equiv \sqrt{\overrightarrow{k}^{2}+m^{2}}$ are used.

For example, the total energy of the field $\phi$, instead of the
expression (97) in RCQM, has the form

\begin{equation}
\label{eq117} P_{0}=\int d^{3}k \omega
\left(a^{*-}_{\mathrm{r}}(\overrightarrow{k})a^{-}_{\mathrm{r}}(\overrightarrow{k})-a^{+}_{\acute{\mathrm{r}}}(\overrightarrow{k})a^{*+}_{\acute{\mathrm{r}}}(\overrightarrow{k})\right),
\end{equation}

\noindent which is not positively defined. In this sense, the FW
model (whose quantum-mechanical content is unambiguous) in fact is
not the quantum-mechanical model for the spin s=(1/2,1/2)
particle-antiparticle doublet. Therefore, in the procedure of
"canonical quantization" of the field $\phi$ on the basis of
anticommutation relations (98) an \textit{additional axiom} is
applied in this model for the quantized field $\widehat{\phi}$.
According to this axiom the definition of the operators of the
physical quantities in the Fock space $\mathcal{H}^{\mathrm{F}}$
is extended by taking them only in the form of "normal products"
with respect to the operator amplitudes
$\widehat{a}^{-}_{\mathrm{r}}(\overrightarrow{k}), \,
\widehat{a}^{*-}_{\mathrm{r}}(\overrightarrow{k}), \,
\widehat{a}^{+}_{\acute{\mathrm{r}}}(\overrightarrow{k}), \,
\widehat{a}^{*+}_{\acute{\mathrm{r}}}(\overrightarrow{k})$. It is
easy to verify that the operators of 10 main conserved quantities
of the "quantized field" $\widehat{\phi}$ in the form of normal
products coincide with the corresponding expressions in the
"second quantized" RCQM model of the Fermionic doublet

\begin{equation}
\label{eq118}
:(\widehat{P}_{\mu},\widehat{J}_{\mu\nu})^{\mathrm{\phi}} :=\int
d^{3}k
\widehat{A}^{\dag}(\overrightarrow{k})(\widetilde{p}_{\mu},\widetilde{j}_{\mu\nu})\widehat{A}(\overrightarrow{k}),
\end{equation}

\noindent where $\widehat{A}(\overrightarrow{k})\equiv \left|
\begin{array}{cccc}
 \widehat{a}^{-}_{+} \\
 \widehat{a}^{-}_{-} \\
\widehat{a}^{+}_{-} \\
\widehat{a}^{+}_{+} \\
\end{array} \right|$ and $(\widetilde{p}_{\mu},\widetilde{j}_{\mu\nu})$ are given in (76), (77).

\textbf{The Dirac model of the fermionic doublet}. Taking into account the above consideration and using the well-known from [2] transition operators $V^{\pm}$

$$\phi\rightarrow\psi=V^{+}\phi, \, \psi\rightarrow\phi=V^{-}\psi,$$
\begin{equation}
\label{eq119} V^{\pm}\equiv\frac{\pm
i\gamma^{l}\partial_{l}+\widehat{\omega}+m}{\sqrt{2\widehat{\omega}(\widehat{\omega}+m)}},
\end{equation}

\noindent we find the resulting operator

$$W=V^{+}v, \, W^{-1}=vV^{-}:$$
\begin{equation}
\label{eq120}
 f\rightarrow\psi=Wf, \, \psi\rightarrow f= W^{-1}\psi,
\end{equation}

\noindent $WW^{-1}=W^{-1}W=1,$ which transforms (one-to-one) all quantities of RCQM model into the corresponding quantities of the Dirac model and vice versa. For example,

\begin{equation}
\label{eq121}
 W\left(\partial_{0}+i\widehat{\omega}\right)W^{-1}=\partial_{0}+iH_{\mathrm{D}}; \quad H_{\mathrm{D}}\equiv \overrightarrow{\alpha}\cdot \overrightarrow{p}+\beta m.
\end{equation}

\noindent It means that quantum-mechanical
Schr$\mathrm{\ddot{o}}$dinger--Foldy equation (57) (in the
anti-Hermitian form
$\left(\partial_{0}+i\widehat{\omega}\right)f(t,\overrightarrow{x})=0$)
is transformed into the Dirac equation in the
Schr$\mathrm{\ddot{o}}$dinger form

$$\left(\partial_{0}+i\widehat{\omega}\right)f(t,\overrightarrow{x})=0 \rightarrow$$
\begin{equation}
\label{eq122} \left[\partial_{0}+i(\overrightarrow{\alpha}\cdot
\overrightarrow{p}+\beta m)\right]\psi(t,\overrightarrow{x})=0.
\end{equation}

Furthermore,

\begin{equation}
\label{eq123}
 W(\widehat{p}_{\mu},\widehat{j}_{\mu\nu})W^{-1}=(\breve{p}_{\mu},\breve{j}_{\mu\nu})^{\mathrm{Dirac} } ,
\end{equation}

\noindent where

\begin{equation}
\label{eq124} \breve{p}_{0}=-iH_{\mathrm{D}}, \,
\breve{p}_{\ell}=\partial_{\ell}, \, \breve{j}_{\ell
n}=x_{\ell}\partial_{n}-x_{n}\partial_{\ell}+\widehat{s}_{\ell
n}\equiv
\end{equation}
$$\widehat{m}_{\ell n}+\widehat{s}_{\ell n},$$
\begin{equation}
\label{eq125}
\breve{j}_{0\ell}=t\partial_{\ell}+\frac{i}{2}\left\{x_{l},H_{\mathrm{D}}\right\}=t\partial_{\ell}-x_{\ell}\breve{p}_{0}+\widehat{s}_{0
\ell};
\end{equation}
$$\widehat{s}_{\mu\nu}\equiv\frac{1}{4}\left[\gamma_{\mu},\gamma_{\nu}\right].$$

Note that the generators of the \textit{local}
$\mathcal{P}^{\mathrm{\psi}}$-representation (the standard
$\mathcal{P}$-algebra of the Dirac equation (122) invariance)
have the form
\begin{equation}
\label{eq126} p_{\mu}= \partial _{\mu}, \quad j_{\mu\nu}=
x_{\mu}\partial _{\nu}-x_{\nu}\partial_{\mu}+\widehat{s}_{\mu\nu}.
\end{equation}

\noindent Moreover, in the manifold
${\psi}\subset\mathrm{H}^{3,4}$ of the solutions of the Dirac
equation (122), the operators (126) coincide with the
$\mathcal{P}^{\mathrm{I}}$-generators (124), (125). Therefore, the
$\mathcal{P}^{\mathrm{\psi}}$-representation determined by the
generators (124), (125) is called induced. As a consequence of
this fact, for example, the following equalities are valid for
the 10 main dynamical variables of the field $\psi$

$$(p_{\mu}, j_{\mu\nu})\rightarrow(P_{\mu},J_{\mu\nu})^{\mathrm{\psi}}\equiv$$
\begin{equation}
\label{eq127} \int d^{3}x \overline{\psi}i(p_{\mu},
j_{\mu\nu})\psi = \int d^{3}x \overline{\psi}i(\breve{p}_{\mu},
\breve{j}_{\mu\nu})\psi =
\end{equation}
$$(P_{\mu},J_{\mu\nu})^{\mathrm{\phi}},$$

\noindent where the conservation laws
$(P_{\mu},J_{\mu\nu})^{\mathrm{\phi}}$ are given by formula
(113) and in terms of the amplitudes
$a^{-}_{\mathrm{r}}(\overrightarrow{k}), \,
a^{*-}_{\mathrm{r}}(\overrightarrow{k}), \,
a^{+}_{\acute{\mathrm{r}}}(\overrightarrow{k}), \,
a^{*+}_{\acute{\mathrm{r}}}(\overrightarrow{k})$ are given by formula (114). This means that the procedure of canonical
quantization of the field $\psi$ is reduced to the corresponding
procedure of the field $\phi$ quantization.

Note that the $\mathcal{P}$-operators (126) are the functions
of 14 independent "generated" operators $x_{\mu}, \
\partial_{\mu}, \, \widehat{s}_{\mu\nu}$. Further, the
$\mathcal{P}$-generators (124), (125) are the functions of 12
independent operators $x_{\ell}, \ \partial_{\ell}, \,
\widehat{s}_{\mu\nu}$. Nevertheless, only the operator
$\overrightarrow{p}=-\nabla$ has the physical sense of the
quantum-mechanical Fermi doublet momentum operator among the
above mentioned 14 independent operators. As proven in [2],
the operators $\overrightarrow{x}=(x^{\ell})$ and
$\widehat{s}_{\mu\nu}$ essentially used in the
constructions (124)--(127), have no physical sense of the
quantum-mechanical operators of the Fermionic doublet coordinate
and the SU(2)-spin. This fact evidently demonstrates the validity of
the assertion that the standard Dirac local model (in the ordinary version) is not the
\textit{quantum-mechanical model} of the fermionic doublet at all.

The axioms of section 7 eventually need to be reconciled with three levels of description used in this paper: RCQM, FW and Dirac equations. Nevertheless, this interesting problem cannot be considered in few pages. The simple level comparison of RCQM, FW and Dirac models given above is very brief and not complete. Only the main features are considered briefly. The extended version will be given in next paper. The readers of this paper can compare the axioms of RCQM (section 7) with the main principles of the Dirac model given in B. Thaller's monograph [31] on the high mathematical level.

\section{On the quantum-mechanical interpretation of the Dirac equation and the Foldy synthesis}

The incomplete but maximum possible quantum-mechanical interpretation of the Dirac equation is
based on the two main assertions. These assertions are similar to
those considered in the RCQM in sections 2--7.

The first one is about the explicit form of the stationary complete
set of operators and their equations on eigenvectors and
eigenvalues. The second one is the Bargman-Wigner analysis of the
representations of the Poincar$\mathrm{\acute{e}}$ group
$\mathcal{P}$, with respect to which the Dirac equation is
invariant. Based on this two assertions incomplete but maximum possible quantum-mechanical interpretation of the Dirac equation is given below in this section.

It is well known that ordinary group-theoretical and
quantum-mechanical interpretation of the Dirac equation is not
completed. For example, the spin (110) (both in the Hermitian (65)=(92) and
anti-Hermitian (110) forms) is not the spin operator in the Dirac
model. The operator (92) does not commute with the Dirac
Hamiltonian (121). This, and few other reasons given in [2], lead
to the transition in [2] to another representation of the Dirac
equation. In this canonical FW representation the spin operator
(65)=(92) commutes with the FW Hamiltonian
$\gamma^{0}\widehat{\omega}$ (104).

It follows from the above facts that the equations on
eigenvectors and eigenvalues of the spin operator
$s^{\ell
n}\equiv\frac{i}{4}\left[\gamma^{\ell},\gamma^{n}\right]$ (65)=(92) are not
correct in the Dirac model as well. The well-defined spin operator
for the Dirac model is known from [2]. This nonlocal operator is
given by
$\overrightarrow{s}_{\mathrm{D}}=\overrightarrow{s}-\frac{\overrightarrow{\gamma}\times
\nabla}{2\widehat{\omega}}+\frac{\nabla \times
(\overrightarrow{s}\times
\nabla)}{\widehat{\omega}(\widehat{\omega}+m)}$. In [32], we
demonstrated that for the Dirac model the equations on eigenvectors
and eigenvalues of this nonlocal spin are valid.

Thus, the possible quantum-mechanical interpretation of the Dirac
equation (as the equation for the spin s=(1/2,1/2)
particle-antiparticle doublet) will be given in this section
below. After that in next sections the interpretation procedure
developed here will be used to interprete the new higher
spin multiplet equations of covariant local field theory.

Consider first the Dirac equation on \textbf{the level of the
FW representation}. The operator $v$ (102) according to transition (104) transforms solution (58) of the
Schr$\mathrm{\ddot{o}}$dinger--Foldy equation (57) for the spin
s=(1/2,1/2) particle-antiparticle doublet into solution (112), or into that

$$\phi(x)=\frac{1}{(2\pi)^{\frac{3}{2}}}\int
d^{3}k\{e^{-ikx}[a^{1}(\overrightarrow{k})\mathrm{d}_{1}+a^{2}(\overrightarrow{k})\mathrm{d}_{2}]+$$
\begin{equation}
\label{eq128}
e^{ikx}[a^{*3}(\overrightarrow{k})\mathrm{d}_{3}+a^{*4}(\overrightarrow{k})\mathrm{d}_{4}]\},
\end{equation}

\noindent of the Dirac equation in the FW representation (104). Note that the amplitudes $a^{\alpha}(\overrightarrow{k})$ in (128) are the same as in solution $f(x)= \frac{1}{\left(2\pi\right)^{\frac{3}{2}}}\int d^{3}k e^{-ikx}[a^{1}(\overrightarrow{k})\mathrm{d}_{1}+a^{2}(\overrightarrow{k})\mathrm{d}_{2}+a^{3}(\overrightarrow{k})\mathrm{d}_{3}+a^{4}(\overrightarrow{k})\mathrm{d}_{4}]$ of the Schr$\mathrm{\ddot{o}}$dinger--Foldy equation (57) and the orts $\left\{\mathrm{d}_{\alpha}\right\}$ of the Cartesian basis are given by (42). Note that transformation $v$ (102), (104) does not change the orts $\left\{\mathrm{d}_{\alpha}\right\}$.

Further, the operator $V^{\pm}$ from (119) according to the
transformation $\phi\rightarrow\psi=V^{+}\phi$ transforms solution (128) of the FW equation (104) into that

$$\psi(x)=V^{+}\phi(x)= \frac{1}{\left(2\pi\right)^{\frac{3}{2}}}\int d^{3}k$$
\begin{equation}
\label{eq129}
\left[e^{-ikx}a^{\mathrm{r}}(\overrightarrow{k})\mathrm{v}^{-}_{\mathrm{r}}(\overrightarrow{k})+e^{ikx}a^{*\mathrm{\check{r}}}(\overrightarrow{k})\mathrm{v}^{+}_{\mathrm{\check{r}}}(\overrightarrow{k})\right],
\end{equation}
$$\mathrm{r}=(1,2), \, \mathrm{\check{r}}=(3,4),$$

\noindent of the Dirac equation (122). The amplitudes
$a^{\alpha}(\overrightarrow{k})$ in (129) are the same as in (128)
and in the corresponding RCQM. The basis vectors are changed and
now have the form

\begin{equation}
\label{eq130} \mathrm{v}^{-}_{1}(\overrightarrow{k}) = N\left|
\begin{array}{cccc}
 \tilde{\omega}+m \\
 0 \\
 k^{3} \\
 k^{1}+ik^{2} \\
\end{array} \right|, \,
\mathrm{v}^{-}_{2}(\overrightarrow{k}) = N\left|
\begin{array}{cccc}
 0 \\
 \tilde{\omega}+m \\
 k^{1}-ik^{2} \\
 -k^{3} \\
\end{array} \right|,
\end{equation}
$$\mathrm{v}^{+}_{3}(\overrightarrow{k}) = N\left|
\begin{array}{cccc}
 k^{3} \\
 k^{1}+ik^{2} \\
 \tilde{\omega}+m \\
 0 \\
\end{array} \right|, \,
\mathrm{v}^{+}_{4}(\overrightarrow{k}) = N\left|
\begin{array}{cccc}
  k^{1}-ik^{2} \\
 -k^{3} \\
  0 \\
 \tilde{\omega}+m \\
\end{array} \right|,$$

\noindent where

\begin{equation}
\label{eq131} N\equiv
\frac{1}{\sqrt{2\tilde{\omega}(\tilde{\omega}+m)}}, \quad
\tilde{\omega}\equiv \sqrt{\overrightarrow{k}^{2}+m^{2}}.
\end{equation}

\noindent Thus, the orts
$\mathrm{v}^{\pm}_{\alpha}(\overrightarrow{k})$ are the standard
4-component Dirac spinors. Their conditions of orthonormalization
and completeness are well known, see, e. g. [33].

Note that on this step of transition from the FW equation to the Dirac equation the appealing to the anti-Hermitian operators (99), (101) (and corresponding equations of motion (104), (122)) is not necessary. Therefore, contrary to the transition from the Schr$\mathrm{\ddot{o}}$dinger--Foldy to the FW equation, here the ordinary Hermitian operators and corresponding equations can be used

\begin{equation}
\label{eq132} i\partial_{0}-\gamma^{0}\widehat{\omega} \stackrel{V^{\pm}}{\leftrightarrow} i\partial_{0}-(\overrightarrow{\alpha}\cdot
\overrightarrow{p}+\beta m),
\end{equation}

\noindent where the transition operator $V^{\pm}$ is given in (119). Nevertheless, in the direct transition $W$ (120)--(123) from the RCQM to the Dirac model the anti-Hermitian operators are necessary (because they are necessary on the first step in transition from the Schr$\mathrm{\ddot{o}}$dinger--Foldy to the FW equation).

Thus, the transition from the RCQM to the covariant local Dirac theory is better to fulfill in two steps. (i) The transition $v$ (102)--(104) from the Schr$\mathrm{\ddot{o}}$dinger--Foldy to the FW equation, (ii) the transition from the FW to the Dirac equation. In this step by step procedure, on the second step the ordinary Hermitian operators are updated.

Further, in the FW representation the spin (65)=(92) is found from the
spin (64)=(91) in RCQM on the basis of the transformation $v$ (102), (103) (for the anti-Hermitian spin operators). The equations on
the eigenvectors and eigenvalues of the FW's spin projection
operator are given by

$$s^{3}\mathrm{d}_{1}=\frac{1}{2}\mathrm{d}_{1}, \, s^{3}\mathrm{d}_{2}=-\frac{1}{2}\mathrm{d}_{2},$$
\begin{equation}
\label{eq133} s^{3}\mathrm{d}_{3}=\frac{1}{2}\mathrm{d}_{3}, \,
s^{3}\mathrm{d}_{4}=-\frac{1}{2}\mathrm{d}_{4},
\end{equation}

\noindent  of course for the Hermitian spin $s^{\ell
n}\equiv\frac{i}{4}\left[\gamma^{\ell},\gamma^{n}\right]$ (65)=(92). These equations are also the consequences of the equations $s^{3}\mathrm{d}_{1}=\frac{1}{2}\mathrm{d}_{1}, \, s^{3}\mathrm{d}_{2}=-\frac{1}{2}\mathrm{d}_{2}, \, s^{3}\mathrm{d}_{3}=-\frac{1}{2}\mathrm{d}_{3}, \, s^{3}\mathrm{d}_{4}=\frac{1}{2}\mathrm{d}_{4}$ (80) in RCQM and the transformation $v$ (102)--(104). Note that the direct quantum-mechanical interpretation of the amplitudes $a^{\alpha}(\overrightarrow{k})$ in solution (128) should be taken not from equations (133) but from the corresponding above mentioned quantum-mechanical equations (80).

The relativistic invariance of the FW equation (104) follows
from the relativistic invariance of the
Schr$\mathrm{\ddot{o}}$dinger--Foldy equation (57) and the
transformation $v$ (102), (103) (for the anti-Hermitian operators).
The FW equation (104) and the set $\left\{\phi\right\}$ of its
solutions (128) are invariant with respect to the reducible
unitary fermionic representation (14) of the
Poincar$\mathrm{\acute{e}}$ group $\mathcal{P}$. The
corresponding $4 \times 4$ matrix-differential fermionic
generators are given by

$$p^{0}=\gamma^{0}\widehat{\omega}\equiv \gamma^{0}\sqrt{-\Delta+m^{2}}, \quad p^{\ell}=-i\partial_{\ell},$$
\begin{equation}
\label{eq134} j^{\ell n}=x^{\ell}p^{n}-x^{n}p^{\ell}+s^{\ell
n}\equiv m^{\ell n}+s^{\ell n},
\end{equation}
$$j^{0 \ell}=-j^{\ell 0}=x^{0}p^{\ell}-\frac{1}{2}\gamma^{0}\left\{x^{\ell},\widehat{\omega}\right\} +\gamma^{0}\frac{(\overrightarrow{s}\times \overrightarrow{p})^{\ell}}{\widehat{\omega}+m},$$

\noindent where the SU(2) spin matrices have the form $\overrightarrow{s}=(s^{23},s^{31},s^{12})=s_{\ell n}=s^{\ell
n}\equiv\frac{i}{4}\left[\gamma^{\ell},\gamma^{n}\right]$ (in the anti-Hermitian form the generators (134) are given by (108)--(110)).

The explicit form of the generators (134) is the consequence of the transformation $v$ (102), (103) and generators (12), (13)
 with the SU(2) spin matrices (64)=(91). Recall that the transformation $v$ (102), (103) is fulfilled only for the anti-Hermitian operators like (108)--(110).

Not a matter of fact that generators (134) are well-known (see, e. g., the formulae (D-64)--(D-67) in [13]), this consideration is important for the construction of the method of transition from the nonlocal RCQM to the covariant local field theory. The construction of such method is one of the goals of this paper.

The Casimir operator for the SU(2) spin $s^{\ell
n}\equiv\frac{i}{4}\left[\gamma^{\ell},\gamma^{n}\right]$ (65)=(92) has the form

\begin{equation}
\label{eq135} \overrightarrow{s}^{2}=
\frac{3}{4}\mathrm{I}_{4}=\frac{1}{2}\left(\frac{1}{2}+1\right)\mathrm{I}_{4},
\end{equation}

\noindent where $\mathrm{I}_{4}$ is the $4 \times 4$ unit matrix. This operator also can be found by the transition $v$ (102), (103) from the corresponding Casimir operator in the RCQM.

It is easy to prove by the direct verification that generators (134) commute with the operator
$i\partial_{0}-\gamma^{0}\widehat{\omega}$ of the FW equation and
satisfy the commutation relations (11) of the Lie algebra of the
Poincar$\mathrm{\acute{e}}$ group $\mathcal{P}$. The
corresponding Casimir operators are given by

\begin{equation}
\label{eq136} p^{2}=p^{\mu}p_{\mu}=m^{2}\mathrm{I}_{4},
\end{equation}
\begin{equation}
\label{eq137} W=w^{\mu}w_{\mu}=m^{2}\overrightarrow{s}^{2}=
\frac{1}{2}\left(\frac{1}{2}+1\right)m^{2}\mathrm{I}_{4},
\end{equation}

Recall that the group theoretical interpretation of the FW
equation is based on equations (133) and the explicit form of
the Casimir operators (136), (137). The last one is given by the
Bargman--Wigner analysis. Due to these two facts one can come to a
conclusion that the Dirac equation in the FW representation (104)
describes the field (the fermionic spin s=(1/2,1/2)
particle-antiparticle doublet) with the spin s=1/2 and $m>0$.

Consider now the Dirac equation on \textbf{the level of standard Dirac representation}. Thus, on the basis of the FW transformation $V^{\pm}$ (119) from the FW equation (104) the Dirac equation

\begin{equation}
\label{eq138} [i\partial_{0}-(\overrightarrow{\alpha}\cdot
\overrightarrow{p}+\beta m)]\psi(x)=0
\end{equation}

\noindent is obtained

\begin{equation}
\label{eq139} V^{+}(i\partial_{0}-\gamma^{0}\widehat{\omega})V^{-}
= i\partial_{0}-(\overrightarrow{\alpha}\cdot
\overrightarrow{p}+\beta m),
\end{equation}
$$\overrightarrow{\alpha}\equiv \gamma^{0}\overrightarrow{\gamma}, \, \beta \equiv \gamma^{0}.$$

The well-defined spin operator for the Dirac equation is found
from the FW spin
$s^{\ell
n}\equiv\frac{i}{4}\left[\gamma^{\ell},\gamma^{n}\right]$ (65)=(92) on the
basis of the formula

\begin{equation}
\label{eq140}
V^{+}\widehat{q}_{\mathrm{FW}}V^{-}=\widehat{q}_{\mathrm{D}},
\end{equation}

\noindent which is valid for the arbitrary operators
$\widehat{q}_{\mathrm{FW}}$ and $\widehat{q}_{\mathrm{D}}$ in the
FW and standard Dirac representations, respectively. Therefore,
the spin
$\widehat{s}_{\mathrm{D}}=V^{+}\widehat{s}_{\mathrm{FW}}V^{-}$
has the form

\begin{equation}
\label{eq141}
\overrightarrow{s}_{\mathrm{D}}=\overrightarrow{s}-\frac{\overrightarrow{\gamma}\times
\nabla}{2\widehat{\omega}}+\frac{\nabla \times
(\overrightarrow{s}\times
\nabla)}{\widehat{\omega}(\widehat{\omega}+m)}
\end{equation}

\noindent and satisfies the SU(2) commutation relations (30).

It is easy to verify that spin (141) commutes with the operator
$i\partial_{0}-(\overrightarrow{\alpha}\cdot
\overrightarrow{p}+\beta m)$ of the equation (138). Moreover, it
is easy to verify that the Dirac spinors (130) satisfy the equations
on eigenvectors and eigenvalues of the spin operator (141) in the
similar to equations (133) form

$$s^{3}_{\mathrm{D}}\mathrm{v}^{-}_{1}(\overrightarrow{k})=\frac{1}{2}\mathrm{v}^{-}_{1}(\overrightarrow{k}), \, s^{3}_{\mathrm{D}}\mathrm{v}^{-}_{2}(\overrightarrow{k})= -\frac{1}{2}\mathrm{v}^{-}_{2}(\overrightarrow{k}),$$
\begin{equation}
\label{eq142}
s^{3}_{\mathrm{D}}\mathrm{v}^{+}_{1}(\overrightarrow{k})=\frac{1}{2}\mathrm{v}^{+}_{1}(\overrightarrow{k}),
\, s^{3}_{\mathrm{D}}\mathrm{v}^{+}_{2}(\overrightarrow{k})=
-\frac{1}{2}\mathrm{v}^{+}_{2}(\overrightarrow{k}).
\end{equation}

\noindent Furthermore, equations (142) follow from equations (133) and the transformation $V^{\pm}$ (119).

The Dirac equation (138) and the set $\{\psi\}$ of its solutions
(129) are invariant with respect to the induced fermionic
representation (14) of the Poincar$\mathrm{\acute{e}}$ group
$\mathcal{P}$. The corresponding $4 \times 4$ matrix-differential
generators are given by

$$p^{0}=H_{\mathrm{D}}=\overrightarrow{\alpha}\cdot
\overrightarrow{p}+\beta m, \quad p^{\ell}=-i\partial_{\ell},$$
\begin{equation}
\label{eq143} j^{\ell n}=x^{\ell}p^{n}-x^{n}p^{\ell}+s^{\ell
n}\equiv m^{\ell n}+s^{\ell n},
\end{equation}
$$j^{0 \ell}=x^{0}p^{\ell}-\frac{1}{2}\left\{x^{\ell},H_{\mathrm{D}}\right\} =x^{0}p^{\ell}-x^{\ell}p^{0}+s^{0\ell
},$$

\noindent whereas the SU(2) spin matrices are given by

\begin{equation}
\label{eq144} \overrightarrow{s}=(s^{23},s^{31},s^{12})=s_{\ell
n}=s^{\ell
n}\equiv\frac{i}{4}\left[\gamma^{\ell},\gamma^{n}\right]
\end{equation}

\noindent and the matrices of the spinor
$\mathrm{L}^{\uparrow}_{+}=$ SO(1,3) representation of the Lorentz
group have the form

\begin{equation}
\label{eq145} s^{\mu
\nu}\equiv\frac{i}{4}\left[\gamma^{\mu},\gamma^{\nu}\right].
\end{equation}

In the set of solutions $\{\psi\}$ (129) of the Dirac equation
(138), realization (143) coincides with the covariant
one

\begin{equation}
\label{eq146} p^{\mu}= i\partial ^{\mu}, \quad j^{\mu
\nu}=x^{\mu}p^{\nu}-x^{\nu}p^{\mu}+s^{\mu \nu},
\end{equation}

\noindent in which the generators have the form of the local Lie
operators.

For the goals of this manuscript the important conclusion is as follows. The realization (143) is the direct consequence of generators (134). Generators (143) are found from those (134) by the FW transformation $V^{\pm}$ (119). Moreover, generators (143) are the consequence of the RCQM generators (12), (13) with the spin members (64)=(91). The operator of such resulting transformation is given by $W$ (120)--(123) and is valid only for the anti-Hermitian operators.

It is easy to verify by the direct calculations that both generators (143) and generators (146) commute with the operator $i\partial_{0}-(\overrightarrow{\alpha}\cdot
\overrightarrow{p}+\beta m)$ of the Dirac equation (138). Furthermore, both sets of operators (143) and (146) satisfy the commutation relations (11) of the Lie algebra of the
Poincar$\mathrm{\acute{e}}$ group $\mathcal{P}$.

The results of the Casimir operators calculation for realization (143) coincide with the calculations of the invariant operators (136), (137). It is the consequence of the link $V^{\pm}$ (119) between the sets of generators (134) and (143). For realization (146) given by the local Lie generators, the Casimir operators have the form

\begin{equation}
\label{eq147} p^{2}=p^{\mu}p_{\mu}=\partial ^{\mu}\partial_{\mu}\mathrm{I}_{4},
\end{equation}
\begin{equation}
\label{eq148} W=w^{\mu}w_{\mu}=m^{2}\overrightarrow{s}^{2}=
\frac{1}{2}\left(\frac{1}{2}+1\right)\partial ^{\mu}\partial_{\mu}\mathrm{I}_{4}.
\end{equation}

In the set of solutions $\{\psi\}$ (129) of the Dirac equation (138) (more exactly, in the set of solutions of the Klein--Gordon--Fock equation) the Casimir operators (147), (148) coincide with the invariant operators (136), (137) for the sets of generators (143).

The important conclusion from this section is the group-theoretical and quantum-mechanical interpretation of the Dirac equation. It is shown that its relation to the spinor field (spin s = (1/2,1/2) particle-antiparticle doublet) with the spin s=(1/2,1/2) and $m>0$ follows from equations (142) and the Bargman--Wigner analysis of the fermionic representations (143), (146) on the basis of the Casimir operators (136), (137) and (147), (148), respectively. Only such incomplete (in comparison with the interpretation given in section 7) fermionic quantum-mechanical interpretation of the Dirac equation is possible. The important external step in this interpretation is given in the section 9 and here. It is the direct link between the Dirac model and the spin s=(1/2,1/2) particle-antiparticle doublet RCQM, where the quantum-mechanical interpretation is complete and evident.

In our papers [24--28] after the construction of the bosonic spin s=(1,0) representations of the Poincar$\mathrm{\acute{e}}$ group
$\mathcal{P}$, with respect to which the Dirac equation (138) is invariant, the similar procedure of interpretation was used. On this basis in [24--28], the relation of the Dirac equation to the description of bosonic field (spin s=(1,0) particle multiplet) of spins s=1, s=0 and $m>0$ was proved. Similarly to the above given fermionic interpretation such bosonic interpretation of the Dirac model is incomplete as well. And again the important external step in such interpretation is the direct link between the spin s=(1,0), or spin s=(1,1), RCQM and the corresponding Dirac, or Dirac-like equations. This link is given in this paper below. Recal that in the framework of corresponding RCQM the quantum-mechanical interpretation is complete and evident. Below in the next sections the methods of interpretation developed here are applied to other covariant models of higher-spin particle-antiparticle doublets.

Second important conclusion from  sections 7, 9 and this one is the evident demonstration of the link between the nonlocal RCQM of the spin s =(1/2,1/2) particle-antiparticle doublet and the covariant local spinor Dirac model. In other words, in the section 9 and here the derivation of the Dirac equation from the RCQM Schr$\mathrm{\ddot{o}}$dinger--Foldy
equation for the spin s=(1/2,1/2) particle-antiparticle doublet is given (a brief review of other methods of the Dirac equation derivation see in [3]).

Below in similar way the Dirac-like equations for other particle-antiparticle doublets of higher spins are derived. The start from the corresponding models of RCQM is given as well. Thus, here the foundations of the synthesis of covariant particle equations on the basis of the start from the RCQM are formulated. Hence, the author's small addition to the Foldy's synthesis of covariant particle equations is suggested.

\section{A brief scheme of the relativistic canonical quantum mechanics of the 6-component spin s=(1,1) particle-antiparticle bosonic doublet}

For the bosonic spin s=(1,1) particle-antiparticle doublet the
system of two 3-component equations (27) should be used.
Therefore, the corresponding Schr$\mathrm{\ddot{o}}$dinger--Foldy
equation is given by

\begin{equation}
\label{eq149} (i\partial_{0}- \widehat{\omega})f(x)=0, \quad
f=\left| {{\begin{array}{*{20}c}
 f^{1} \hfill  \\
 f^{2} \hfill  \\
 f^{3} \hfill  \\
 f^{4} \hfill  \\
 f^{5} \hfill  \\
 f^{6} \hfill  \\
\end{array} }} \right|,
\end{equation}

\noindent where the pseudo-differential operator
$\widehat{\omega}$ is given in (59)--(61). In (149), the
6-component wave function is the direct sum of the particle and
antiparticle wave functions, whereas the particle wave function
has the form (32). According to the quantum-mechanical tradition
the antiparticle wave function is put into the bottom part of the
6-column.

Therefore, the general solution of the
Schr$\mathrm{\ddot{o}}$dinger--Foldy equation (149) has the form

$$f(x)= \left|
{{\begin{array}{*{20}c}
 f_{\mathrm{part}} \hfill  \\
 f_{\mathrm{antipart}} \hfill  \\
\end{array} }} \right| =\frac{1}{\left(2\pi\right)^{\frac{3}{2}}}\int d^{3}k e^{-ikx}$$
\begin{equation}
\label{eq150}
\left[c^{1}(\overrightarrow{k})\mathrm{d}_{1}+c^{2}(\overrightarrow{k})\mathrm{d}_{2}+...+c^{5}(\overrightarrow{k})\mathrm{d}_{5}+c^{6}(\overrightarrow{k})\mathrm{d}_{6}\right],
\end{equation}

\noindent where the orts of the 6-component Cartesian basis are
given by

$$\mathrm{d}_{1} = \left|
\begin{array}{cccc}
 1 \\
 0 \\
 0 \\
 0 \\
 0 \\
 0 \\
\end{array} \right|, \, \mathrm{d}_{2} = \left|
\begin{array}{cccc}
 0 \\
 1 \\
 0 \\
 0 \\
 0 \\
 0 \\
\end{array} \right|,
\, \mathrm{d}_{3} = \left|
\begin{array}{cccc}
 0 \\
 0 \\
 1 \\
 0 \\
 0 \\
 0 \\
\end{array} \right|,$$
\begin{equation}
\label{eq151} \mathrm{d}_{4} = \left|
\begin{array}{cccc}
 0 \\
 0 \\
 0 \\
 1 \\
 0 \\
 0 \\
\end{array} \right|, \,
\mathrm{d}_{5} = \left|
\begin{array}{cccc}
 0 \\
 0 \\
 0 \\
 0 \\
 1 \\
 0 \\
\end{array} \right| \,
\mathrm{d}_{6} = \left|
\begin{array}{cccc}
 0 \\
 0 \\
 0 \\
 0 \\
 0 \\
 1 \\
\end{array} \right|.
\end{equation}

Hence, the space of the states in the RCQM of the spin s=(1,1)
particle-antiparticle doublet is the rigged Hilbert space

\begin{equation}
\label{eq152}
\mathrm{S}^{3,6}\subset\mathrm{H}^{3,6}\subset\mathrm{S}^{3,6*},
\end{equation}

\noindent i. e. is the direct sum of two spaces (28).

Thus, in the model under consideration information about the
positive and equal masses of the particle and antiparticle is
inserted. Further, information that the observer sees the
antiparticle as the mirror reflection of the particle is also
inserted. Therefore, the charge of the antiparticle should be
opposite in sign to that of the particle, and the spin projection
of the antiparticle should be opposite in sign to the spin
projection of the particle.

Therefore, according to these principles, the corresponding
SU(2)-spin generators are taken in the form

$$s^{1}= \frac{1}{\sqrt{2}}\left| {{\begin{array}{*{20}c}
 0 \hfill & 1 \hfill & 0 \hfill & 0 \hfill & 0 \hfill  & 0 \hfill \\
 1 \hfill & 0 \hfill & 1 \hfill & 0 \hfill & 0 \hfill & 0 \hfill \\
 0 \hfill & 1 \hfill & 0 \hfill & 0 \hfill & 0 \hfill & 0 \hfill \\
 0 \hfill & 0 \hfill & 0 \hfill & 0 \hfill & -1 \hfill & 0 \hfill\\
 0 \hfill & 0 \hfill & 0 \hfill & -1 \hfill & 0 \hfill & -1 \hfill\\
 0 \hfill & 0 \hfill & 0 \hfill & 0 \hfill & -1 \hfill & 0 \hfill\\
\end{array} }} \right|,$$

\begin{equation}
\label{eq153} s^{2}= \frac{1}{\sqrt{2}}\left|
{{\begin{array}{*{20}c}
 0 \hfill & -i \hfill & 0 \hfill & 0 \hfill & 0 \hfill  & 0 \hfill \\
 i \hfill & 0 \hfill & -i \hfill & 0 \hfill & 0 \hfill & 0 \hfill \\
 0 \hfill & i \hfill & 0 \hfill & 0 \hfill & 0 \hfill & 0 \hfill \\
 0 \hfill & 0 \hfill & 0 \hfill & 0 \hfill & -i \hfill & 0 \hfill\\
 0 \hfill & 0 \hfill & 0 \hfill & i \hfill & 0 \hfill & -i \hfill\\
 0 \hfill & 0 \hfill & 0 \hfill & 0 \hfill & i \hfill & 0 \hfill\\
\end{array} }} \right|,
\end{equation}

$$s^{3}= \left| {{\begin{array}{*{20}c}
 1 \hfill & 0 \hfill & 0 \hfill & 0 \hfill & 0 \hfill  & 0 \hfill \\
 0 \hfill & 0 \hfill & 0 \hfill & 0 \hfill & 0 \hfill & 0 \hfill \\
 0 \hfill & 0 \hfill & -1 \hfill & 0 \hfill & 0 \hfill & 0 \hfill \\
 0 \hfill & 0 \hfill & 0 \hfill & -1 \hfill & 0 \hfill & 0 \hfill\\
 0 \hfill & 0 \hfill & 0 \hfill & 0 \hfill & 0 \hfill & 0 \hfill\\
 0 \hfill & 0 \hfill & 0 \hfill & 0 \hfill & 0 \hfill & 1 \hfill\\
\end{array} }} \right|.$$

\noindent The matrices (153) can also be presented in the
following useful form

\begin{equation}
\label{eq154} \overrightarrow{s}= \left| {{\begin{array}{*{20}c}
 \overrightarrow{s}_{3} \hfill  & 0 \hfill \\
 0 \hfill & -C\overrightarrow{s}_{3}C \hfill \\
 \end{array} }} \right|,
\end{equation}

\noindent where the $3 \times 3$-matrices $\overrightarrow{s}_{3}$
are given in (29) and $C$ is the operator of complex conjugation.

It is easy to verify that for the operators (153), (154) the
commutation relations $
\left[s^{j},s^{\ell}\right]=i\varepsilon^{j \ell n}s^{n}$
\noindent of the SU(2)-algebra are valid.

The Casimir operator for this reducible representation of the
SU(2)-algebra is given by

\begin{equation}
\label{eq155} \overrightarrow{s}^{2}=
2\mathrm{I}_{6}=1\left(1+1\right)\mathrm{I}_{6},
\end{equation}

\noindent where $\mathrm{I}_{6}$ is the $6 \times 6$- unit matrix.

The solution (150) is associated with the stationary complete set
$\overrightarrow{p}, \, s^{3}=s_{z}$ of the momentum and spin
projection operators of the spin s=(1,1) bosonic
particle-antiparticle doublet.

The equations on the momentum operator eigenvalues have the form

\begin{equation}
\label{eq156}  \overrightarrow{p}e^{-ikx}\mathrm{d}_{\mathrm{A}} =
\overrightarrow{k}e^{-ikx}\mathrm{d}_{\mathrm{A}}, \quad \mathrm{A}=\overline{1,6},
\end{equation}

\noindent and the equations on the spin projection operator
$s^{3}=\left| {{\begin{array}{*{20}c}
 1 \hfill & 0 \hfill & 0 \hfill & 0 \hfill & 0 \hfill  & 0 \hfill \\
 0 \hfill & 0 \hfill & 0 \hfill & 0 \hfill & 0 \hfill & 0 \hfill \\
 0 \hfill & 0 \hfill & -1 \hfill & 0 \hfill & 0 \hfill & 0 \hfill \\
 0 \hfill & 0 \hfill & 0 \hfill & -1 \hfill & 0 \hfill & 0 \hfill\\
 0 \hfill & 0 \hfill & 0 \hfill & 0 \hfill & 0 \hfill & 0 \hfill\\
 0 \hfill & 0 \hfill & 0 \hfill & 0 \hfill & 0 \hfill & 1 \hfill\\
\end{array} }} \right|$ eigenvalues are given by

$$s^{3}\mathrm{d}_{1} = \mathrm{d}_{1}, \, s^{3}\mathrm{d}_{2} = 0, \, s^{3}\mathrm{d}_{3} = - \mathrm{d}_{3},$$
\begin{equation}
\label{eq157} s^{3}\mathrm{d}_{4} = - \mathrm{d}_{4}, \,
s^{3}\mathrm{d}_{5} = 0, \, s^{3}\mathrm{d}_{6} = \mathrm{d}_{6}.
\end{equation}

Therefore, the functions
$c^{1}(\overrightarrow{k}),c^{2}(\overrightarrow{k}),c^{3}(\overrightarrow{k})$
in the solution (150) are the momentum-spin amplitudes of the
particle (boson) with the momentum $\overrightarrow{p}$ and spin
projection eigenvalues (+1, 0, -1), respectively, and the
functions
$c^{4}(\overrightarrow{k}),c^{5}(\overrightarrow{k}),c^{6}(\overrightarrow{k})$
are the momentum-spin amplitudes of the antiparticle with the
momentum $\overrightarrow{p}$ and spin projection eigenvalues
(-1, 0, +1), respectively.

Note that for the bosons having charge ($W^{\mp}$ bosons) the stationary complete set includes the sign charge operator and the additional equation on eigenvectors and eigenvalues of this operator is valid. As soon as the $W^{+}$ boson is considered as the antiparticle for the $W^{-}$ boson, the sign charge operator is determined in the form $g=-\Gamma_{6}^{0} = \left|
{{\begin{array}{*{20}c}
 -\mathrm{I}_{3} \hfill &  0 \hfill\\
 0 \hfill & \mathrm{I}_{3}  \hfill\\
 \end{array} }} \right|.$ The corresponding equations on eigenvalues are given by $g\mathrm{d}_{1} = -\mathrm{d}_{1}, \, g\mathrm{d}_{2} = -\mathrm{d}_{2}, \, g\mathrm{d}_{3} = - \mathrm{d}_{3},\, g\mathrm{d}_{4} = \mathrm{d}_{4}, \, g\mathrm{d}_{5} = \mathrm{d}_{5}, \, g\mathrm{d}_{6} =  \mathrm{d}_{6}.$  In this case the functions
$c^{1}(\overrightarrow{k}), \, c^{2}(\overrightarrow{k}), \, c^{3}(\overrightarrow{k})$
in the solution (150) are the charge-momentum-spin amplitudes of the
particle (boson) with the charge $-e$, momentum $\overrightarrow{p}$ and spin
projection eigenvalues (+1, 0, -1), respectively, and the
functions
$c^{4}(\overrightarrow{k}), \, c^{5}(\overrightarrow{k}), \, c^{6}(\overrightarrow{k})$
are the charge-momentum-spin amplitudes of the antiparticle with the
charge $+e$, momentum $\overrightarrow{p}$ and spin projection eigenvalues
(-1, 0, +1), respectively.

The Schr$\mathrm{\ddot{o}}$dinger--Foldy equation (149) and the
set $\{\mathrm{f}\}$ of its solutions (150) are invariant with
respect to the reducible unitary bosonic representation (14) of
the Poincar$\mathrm{\acute{e}}$ group $\mathcal{P}$. The
corresponding $6 \times 6$ matrix-differential generators are
given by (12), (13), whereas the spin s=(1,1) SU(2) generators
$\overrightarrow{s}=(s^{\ell n})$ are given in (153), (154).

The proof of this assertion is fulfilled by the three steps
already given in section 2 after formula (14). The corresponding
Casimir operators have the form

\begin{equation}
\label{eq158}
p^{2}=\widehat{p}^{\mu}\widehat{p}_{\mu}=m^{2}\mathrm{I}_{6},
\end{equation}
\begin{equation}
\label{eq159} W=w^{\mu}w_{\mu}=m^{2}\overrightarrow{s}^{2}=
1\left(1+1\right)m^{2}\mathrm{I}_{6},
\end{equation}

\noindent where $\mathrm{I}_{6}$ is the $6 \times 6$ unit matrix.

Hence, above a brief consideration of the RCQM foundations of the
particle-antiparticle doublet with the mass $m>0$ and the spin
s=(1,1) has been given. In the limit m=0 this model describes the
particular case of the photon-antiphoton doublet.

\section{A brief scheme of the relativistic canonical quantum mechanics of the spin s=(1,0) bosonic multiplet}

In this case both partners of the multiplet are the ordinary
particles. Therefore, specification of the antiparticle, which is
the content of the previous section, is absent. Hence, the model
of the spin s=(1,0) multiplet is constructed as the ordinary
direct sum of the spin s=1 and spin s=0 singlets. The last one is
described by the one component
Schr$\mathrm{\ddot{o}}$dinger--Foldy equation also called the
spinless Salpeter equation [17--19]. As shown in [24--28], this s=(1,0)
multiplet is directly linked with the spin s=(1/2,1/2)
particle-antiparticle doublet (in particular, with the $e^{-}
e^{+}$ doublet).

Thus, the corresponding Schr$\mathrm{\ddot{o}}$dinger--Foldy equation has the form

\begin{equation}
\label{eq160} (i\partial_{0}- \widehat{\omega})f(x)=0, \quad
f=\left| {{\begin{array}{*{20}c}
 f^{1} \hfill  \\
 f^{2} \hfill  \\
 f^{3} \hfill  \\
 f^{4} \hfill  \\
 \end{array} }} \right|,
\end{equation}

\noindent where the s=0 contribution is taken as the $f^{4}$
component of the column and the pseudo-differential operator
$\widehat{\omega}$ is given in (59)--(61). This equation is
similar to that for the spin s=(1/2,1/2) particle-antiparticle
doublet, see its consideration in sections 7, 9, 10. Therefore,
equation (160) should be considered in the same space of the
states

\begin{equation}
\label{eq161}
\mathrm{S}^{3,4}\subset\mathrm{H}^{3,4}\subset\mathrm{S}^{3,4*},
\end{equation}

\noindent where the Schr$\mathrm{\ddot{o}}$dinger--Foldy equation
(57) for the spin s=(1/2,1/2) particle-antiparticle doublet is
determined. Moreover, equation (160) and the rigged Hilbert space
(161) for this consideration are similar to equation (37) and the
space (38) for the spin s=3/2 particle singlet.

The general solution of the Schr$\mathrm{\ddot{o}}$dinger--Foldy
equation (160) is given by

$$f(x)= \frac{1}{\left(2\pi\right)^{\frac{3}{2}}}\int d^{3}k e^{-ikx}$$
\begin{equation}
\label{eq162}
\left[c^{1}(\overrightarrow{k})\mathrm{d}_{1}+c^{2}(\overrightarrow{k})\mathrm{d}_{2}+c^{3}(\overrightarrow{k})\mathrm{d}_{3}+c^{4}(\overrightarrow{k})\mathrm{d}_{4}\right],
\end{equation}

\noindent where notations (9) are used. The orts of
the 4-dimensional Cartesian basis have the form (42).

The generators of the corresponding SU(2)-spin that satisfy the
commutation relations (30) of the SU(2) algebra are given by

$$s^{1}= \frac{1}{\sqrt{2}}\left| {{\begin{array}{*{20}c}
 0 \hfill & 1 \hfill & 0 \hfill & 0 \hfill\\
 1 \hfill & 0 \hfill & 1 \hfill & 0 \hfill\\
 0 \hfill & 1 \hfill & 0 \hfill & 0 \hfill\\
 0 \hfill & 0 \hfill & 0 \hfill & 0 \hfill\\
\end{array} }} \right|,$$

\begin{equation}
\label{eq163} s^{2}= \frac{1}{\sqrt{2}}\left|
{{\begin{array}{*{20}c}
 0 \hfill & -i \hfill & 0 \hfill & 0 \hfill\\
 i \hfill & 0 \hfill & -i \hfill & 0 \hfill\\
 0 \hfill & i \hfill & 0 \hfill & 0 \hfill\\
 0 \hfill & 0 \hfill & 0 \hfill & 0 \hfill\\
\end{array} }} \right|, \quad
\end{equation}

$$s^{3}= \left| {{\begin{array}{*{20}c}
 1 \hfill & 0 \hfill & 0 \hfill & 0 \hfill\\
 0 \hfill & 0 \hfill & 0 \hfill & 0 \hfill\\
 0 \hfill & 0 \hfill & -1 \hfill & 0 \hfill\\
 0 \hfill & 0 \hfill & 0 \hfill & 0 \hfill\\
\end{array} }} \right|.$$

\noindent It is easy to verify that the commutation relations $
\left[s^{j},s^{\ell}\right]=i\varepsilon^{j \ell n}s^{n}$
\noindent are valid.

The Casimir operator for this reducible representation of the
SU(2)-algebra is given by

\begin{equation}
\label{eq164} \overrightarrow{s}^{2}= 2 \left|
{{\begin{array}{*{20}c}
 \mathrm{I}_{3} \hfill & 0 \\
 0 \hfill &  0 \\
\end{array} }} \right| = \left|
{{\begin{array}{*{20}c}
 1(1+1)\mathrm{I}_{3} \hfill & 0 \\
 0 \hfill &  0 \\
\end{array} }} \right|,
\end{equation}

\noindent where $\mathrm{I}_{3}$ is $3\times 3$ unit matrix.

The stationary complete set of operators is given by
$\overrightarrow{p}, \, s^{3}=s_{z}$ The equations on the eigenvalues of the spin projection operator $s^{3}= \left|
{{\begin{array}{*{20}c}
 1 \hfill & 0 \hfill & 0 \hfill & 0 \hfill\\
 0 \hfill & 0 \hfill & 0 \hfill & 0 \hfill\\
 0 \hfill & 0 \hfill & -1 \hfill & 0 \hfill\\
 0 \hfill & 0 \hfill & 0 \hfill & 0 \hfill\\
\end{array} }} \right|$ have the form

$$s^{3}\mathrm{d}_{1} = \mathrm{d}_{1}, \, s^{3}\mathrm{d}_{2} = 0,$$
\begin{equation}
\label{eq165} s^{3}\mathrm{d}_{3} = -\mathrm{d}_{3}, \,
s^{3}\mathrm{d}_{4} = 0.
\end{equation}

The equation on the eigenvalues of the momentum operator
$\overrightarrow{p}$ is the same as in (79).

Interpretation of the amplitudes $c^{\alpha}(\overrightarrow{k})$
in the solution (162) follows from equations (79) and (165). The
functions $c^{1}(\overrightarrow{k}), \,
c^{2}(\overrightarrow{k}), \, c^{3}(\overrightarrow{k}),$ are the
quantum-mechanical momentum-spin amplitudes of the boson with the
spin s=1 and the eigenvalues of the spin projection $1, \, 0,
\,-1$, respectively. The function $c^{4}(\overrightarrow{k})$ is
the amplitude of the spinless boson.

The Schr$\mathrm{\ddot{o}}$dinger--Foldy equation (160) and the
set $\{\mathrm{f}\}$ of its solutions (162) are invariant with
respect to the reducible unitary bosonic representation (14) of
the Poincar$\mathrm{\acute{e}}$ group $\mathcal{P}$. The
corresponding $4 \times 4$ matrix-differential generators are
given by (12), (13), where the spin s=(1,0) SU(2) generators
$\overrightarrow{s}=(s^{\ell n})$ are given in (163).

The proof is fulfilled similarly to that given in section 2
after formula (14). The Casimir operators of this reducible
bosonic representation of the group $\mathcal{P}$ have the form

\begin{equation}
\label{eq166} p^{2}=\widehat{p}^{\mu}\widehat{p}_{\mu}=m^{2}\mathrm{I}_{4},
\end{equation}
\begin{equation}
\label{eq167} W=w^{\mu}w_{\mu}=m^{2}\overrightarrow{s}^{2}=
m^{2}\left| {{\begin{array}{*{20}c}
 1\left(1+1\right)\mathrm{I}_{3} \hfill & 0 \\
 0 \hfill &  0 \\
\end{array} }} \right|,
\end{equation}

\noindent where $\mathrm{I}_{3}$ and $\mathrm{I}_{4}$ are the $3 \times 3$ and $4 \times 4$ unit matrices, respectively.

Hence, above a brief consideration of the RCQM foundations of the
particle multiplet with mass $m>0$ and the spin s=(1,0) is given.

\section{A brief scheme of the relativistic canonical quantum mechanics of the 8-component bosonic spin s=(1,0,1,0) particle-antiparticle multiplet}

The 8-component bosonic spin s=(1,0,1,0) particle-antiparticle
multiplet is constructed as the direct sum of the two spin s=(1,0)
multiplets. The spin s=(1,0) multiplet was considered in the
previous section. The principles of constructing and describing such particle-antiparticle multiplet within the
framework of the RCQM are in a complete analogy with the principles
of the describing and the constructing of the spin s=(1,1)
particle-antiparticle doublet considered in
section 11. Therefore, the details can be omitted.

The most important fact is that here the link with the Dirac-like
equation is similar to that between the spin s=(1/2,1/2)
particle-antiparticle doublet and the standard 4-component Dirac
equation demonstrated in sections 9, 10. Therefore, the
spin s=(1,0,1,0) particle-antiparticle multiplet is of special
interest. It is much more useful than the spin s=(1,0) particle
multiplet.

Thus, the Schr$\mathrm{\ddot{o}}$dinger--Foldy equation has the
form

\begin{equation}
\label{eq168} (i\partial_{0}- \widehat{\omega})f(x)=0, \quad
f=\left| {{\begin{array}{*{20}c}
 f^{1} \hfill  \\
 f^{2} \hfill  \\
 f^{3} \hfill  \\
 f^{4} \hfill  \\
 f^{5} \hfill  \\
 f^{6} \hfill  \\
 f^{7} \hfill  \\
 f^{8} \hfill  \\
\end{array} }} \right|,
\end{equation}

\noindent where the operator $\widehat{\omega}$ is given in
(59)--(61).

The space of the states is as follows

\begin{equation}
\label{eq169}
\mathrm{S}^{3,8}\subset\mathrm{H}^{3,8}\subset\mathrm{S}^{3,8*}.
\end{equation}

The general solution of the Schr$\mathrm{\ddot{o}}$dinger--Foldy
equation (168) is given by

$$f(x)= \left|
{{\begin{array}{*{20}c}
 f_{\mathrm{part}} \hfill  \\
 f_{\mathrm{antipart}} \hfill  \\
\end{array} }} \right| =\frac{1}{\left(2\pi\right)^{\frac{3}{2}}}\int d^{3}k e^{-ikx}$$
\begin{equation}
\label{eq170}
\left[c^{1}(\overrightarrow{k})\mathrm{d}_{1}+c^{2}(\overrightarrow{k})\mathrm{d}_{2}+...+c^{7}(\overrightarrow{k})\mathrm{d}_{5}+c^{8}(\overrightarrow{k})\mathrm{d}_{6}\right],
\end{equation}

\noindent where the orts of the 8-component Cartesian basis have the
form

$$\mathrm{d}_{1} = \left|
\begin{array}{cccc}
 1 \\
 0 \\
 0 \\
 0 \\
 0 \\
 0 \\
 0 \\
 0 \\
\end{array} \right|, \, \mathrm{d}_{2} = \left|
\begin{array}{cccc}
 0 \\
 1 \\
 0 \\
 0 \\
 0 \\
 0 \\
 0 \\
 0 \\
\end{array} \right|,
\, \mathrm{d}_{3} = \left|
\begin{array}{cccc}
 0 \\
 0 \\
 1 \\
 0 \\
 0 \\
 0 \\
 0 \\
 0 \\
\end{array} \right|,\mathrm{d}_{4} = \left|
\begin{array}{cccc}
 0 \\
 0 \\
 0 \\
 1 \\
 0 \\
 0 \\
 0 \\
 0 \\
\end{array} \right|,$$
\begin{equation}
\label{eq171} \mathrm{d}_{5} = \left|
\begin{array}{cccc}
 0 \\
 0 \\
 0 \\
 0 \\
 1 \\
 0 \\
 0 \\
 0 \\
\end{array} \right|, \,
\mathrm{d}_{6} = \left|
\begin{array}{cccc}
 0 \\
 0 \\
 0 \\
 0 \\
 0 \\
 1 \\
 0 \\
 0 \\
\end{array} \right|, \,
\mathrm{d}_{7} = \left|
\begin{array}{cccc}
 0 \\
 0 \\
 0 \\
 0 \\
 0 \\
 0 \\
 1 \\
 0 \\
\end{array} \right|,
\mathrm{d}_{8} = \left|
\begin{array}{cccc}
 0 \\
 0 \\
 0 \\
 0 \\
 0 \\
 0 \\
 0 \\
 1 \\
\end{array} \right|.
\end{equation}

The explicit form of the generators of the corresponding SU(2)-spin that satisfy the commutation relations (30) of the SU(2)
algebra is as follows

\begin{equation}
\label{eq172} \overrightarrow{s}_{8}=\left|
{{\begin{array}{*{20}c}
 \overrightarrow{s} \hfill & 0 \\
 0 \hfill &  -C\overrightarrow{s}C \\
\end{array} }} \right|,
\end{equation}

\noindent where $C\mathrm{I}_{4}$ is the diagonal $4 \times 4$
operator of the complex conjugation and components of $\overrightarrow{s}$ are
given in (163).

The Casimir operator is given by the following $8 \times 8$ diagonal matrix

$$\overrightarrow{s}^{2}= 2\left|
{{\begin{array}{*{20}c}
 \mathrm{I}_{3} \hfill & 0 \hfill & 0 \hfill & 0 \hfill\\
 0 \hfill & 0 \hfill & 0 \hfill & 0 \hfill\\
 0 \hfill & 0 \hfill & \mathrm{I}_{3} \hfill & 0 \hfill\\
 0 \hfill & 0 \hfill & 0 \hfill & 0 \hfill\\
\end{array} }} \right|=$$
\begin{equation}
\label{eq173} \left|
{{\begin{array}{*{20}c}
 1(1+1)\mathrm{I}_{3} \hfill & 0 \hfill & 0 \hfill & 0 \hfill\\
 0 \hfill & 0 \hfill & 0 \hfill & 0 \hfill\\
 0 \hfill & 0 \hfill & 1(1+1)\mathrm{I}_{3} \hfill & 0 \hfill\\
 0 \hfill & 0 \hfill & 0 \hfill & 0 \hfill\\
\end{array} }} \right|,
\end{equation}

\noindent where $\mathrm{I}_{3}$ is the $3\times 3$ unit matrix.

The stationary complete set of operators is given by $\overrightarrow{p}, \, s_{8}^{3}=s_{z}$ and the equations on the eigenvalues of the operator $s_{8}^{3}=s_{z}$ have the form

$$s_{8}^{3}\mathrm{d}_{1} = \mathrm{d}_{1}, \, s_{8}^{3}\mathrm{d}_{2} = 0, \, s_{8}^{3}\mathrm{d}_{3} = -\mathrm{d}_{3}, \,
s_{8}^{3}\mathrm{d}_{4} = 0,$$
\begin{equation}
\label{eq174} s_{8}^{3}\mathrm{d}_{5} = -\mathrm{d}_{5}, \,
s_{8}^{3}\mathrm{d}_{6} = 0, \, s_{8}^{3}\mathrm{d}_{7} =
\mathrm{d}_{7}, \, s_{8}^{3}\mathrm{d}_{8} = 0.
\end{equation}

Therefore, the functions $c^{1}(\overrightarrow{k}), \,
c^{2}(\overrightarrow{k}), \, c^{3}(\overrightarrow{k})$ in
solution (170) are the momentum-spin amplitudes of the massive
boson with the spin s=1 and the spin projection $(1,0,-1)$,
respectively, $c^{4}(\overrightarrow{k})$ is the amplitude of the
spinless boson; $c^{5}(\overrightarrow{k}), \,
c^{6}(\overrightarrow{k}), \, c^{7}(\overrightarrow{k})$ are the
momentum-spin amplitudes of the antiparticle (antiboson) with the
spin s=1 and the spin projection $(-1,0,1)$, respectively,
$c^{8}(\overrightarrow{k})$ is the amplitude of the spinless
antiboson.

The Schr$\mathrm{\ddot{o}}$dinger--Foldy equation (168) (and the
set $\{\mathrm{f}\}$ of its solutions (170)) is invariant with
respect to the reducible unitary bosonic representation (14) of
the Poincar$\mathrm{\acute{e}}$ group $\mathcal{P}$, whose
Hermitian $8 \times 8$ matrix-differential generators are given
by (12), (13), where the spin s=(1,0,1,0) SU(2) generators
$\overrightarrow{s}=(s^{\ell n})$ are given in (172).

The proof is similar to that given in section 2 after formula (14). The Casimir operators of this reducible bosonic spin s=(1,0,1,0) representation of the group $\mathcal{P}$ have the form

\begin{equation}
\label{eq175} p^{2}=\widehat{p}^{\mu}\widehat{p}_{\mu}=m^{2}\mathrm{I}_{8},
\end{equation}
$$W=w^{\mu}w_{\mu}=m^{2}\overrightarrow{s}_{8}^{2}=$$
\begin{equation}
\label{eq176} m^{2}\left| {{\begin{array}{*{20}c}
 1\left(1+1\right)\mathrm{I}_{3} \hfill & 0 \hfill & 0 \hfill & 0 \hfill\\
 0 \hfill & 0 \hfill & 0 \hfill & 0 \hfill\\
 0 \hfill & 0 \hfill & 1\left(1+1\right)\mathrm{I}_{3} \hfill & 0 \hfill\\
 0 \hfill & 0 \hfill & 0 \hfill & 0 \hfill\\
\end{array} }} \right|,
\end{equation}

\noindent where $\mathrm{I}_{8}$ and $\mathrm{I}_{3}$ are the $8\times 8$ and $3\times 3$ unit matrices, respectively.

Thus, above the foundations of the RCQM of the 8-component multiplet of two bosons with the spins s=(1,0) and their antiparticle doublet are considered briefly. It is the basis for the transition to the covariant local field theory of the spin s=(1,0,1,0) particle-antiparticle multiplet given below.

\section{A brief scheme of the relativistic canonical quantum mechanics of the 8-component fermionic spin s=(3/2,3/2) particle-antiparticle doublet}

This model is constructed in complete analogy with the RCQM of the 4-component spin s=(1/2,1/2) particle-antiparticle doublet, which is given in the section 7 in details. Moreover, the principles of constructing and describing such particle-antiparticle multiplet within the
framework of the RCQM are in a complete analogy with the principles
of the describing and the constructing of the spin s=(1,1)
particle-antiparticle doublet considered in
section 11. The difference is only in the dimensions of the corresponding spaces and matrices. Therefore, the details can be omitted. The model below is useful for the $\Sigma$-hyperon description.

The 8-component fermionic spin s=(3/2,3/2) particle-antiparticle
doublet is constructed as the direct sum of the two spin s=3/2
singlets. The spin s=3/2 singlet was considered in the
section 5.

The most important fact is that here the link with the Dirac-like
equation is similar to that between the spin s=(1/2,1/2)
particle-antiparticle doublet and the standard 4-component Dirac
equation demonstrated in sections 9, 10. Therefore, the
spin s=(3/2,3/2) particle-antiparticle doublet is of special
interest.

The Schr$\mathrm{\ddot{o}}$dinger--Foldy equation and the space of states are the same as in the previous section and are already given in (168), (169). The general solution of the equation (168) for the spin s=(3/2,3/2) particle-antiparticle doublet is given by

$$f(x)= \left|
{{\begin{array}{*{20}c}
 f_{\mathrm{part}} \hfill  \\
 f_{\mathrm{antipart}} \hfill  \\
\end{array} }} \right| =$$
\begin{equation}
\label{eq177}
\frac{1}{\left(2\pi\right)^{\frac{3}{2}}}\int d^{3}k e^{-ikx}b^{\mathrm{A}}(\overrightarrow{k})\mathrm{d}_{\mathrm{A}}, \quad \mathrm{A}=\overline{1,8},
\end{equation}

\noindent where the orts of the 8-component Cartesian basis have the
form (171), but the amplitudes $b^{\mathrm{A}}(\overrightarrow{k})$ correspond to the spin s=(3/2,3/2) particle-antiparticle doublet  and differ from the amplitudes $c^{\mathrm{A}}(\overrightarrow{k})$ in solution (170).

The generators of the corresponding SU(2)-spin that satisfy the commutation relations (30) of the SU(2)
algebra are as follows

\begin{equation}
\label{eq178} \overrightarrow{s}_{8}=\left|
{{\begin{array}{*{20}c}
 \overrightarrow{s} \hfill & 0 \\
 0 \hfill &  -C\overrightarrow{s}C \\
\end{array} }} \right|,
\end{equation}

\noindent where $C\mathrm{I}_{4}$ is the diagonal $4 \times 4$
operator of the complex conjugation and the matrices $\overrightarrow{s}$ for the single spin s=3/2 particle are
given in (39). In the explicit form the SU(2) spin operators (178) are given by

$${s}^{1}_{8}= \frac{1}{2}\cdot$$
$$\left| {{\begin{array}{*{20}c}
 0 \hfill & \sqrt{3} \hfill & 0 \hfill & 0 \hfill & 0 \hfill  & 0 \hfill & 0 \hfill  & 0 \hfill \\
 \sqrt{3} \hfill & 0 \hfill & 2 \hfill & 0 \hfill & 0 \hfill & 0 \hfill & 0 \hfill  & 0 \hfill \\
 0 \hfill & 2 \hfill & 0 \hfill & \sqrt{3} \hfill & 0 \hfill & 0 \hfill & 0 \hfill  & 0 \hfill \\
 0 \hfill & 0 \hfill & \sqrt{3} \hfill & 0 \hfill & 0 \hfill & 0 \hfill & 0 \hfill  & 0 \hfill \\
 0 \hfill & 0 \hfill & 0 \hfill & 0 \hfill & 0 \hfill & -\sqrt{3} \hfill & 0 \hfill  & 0 \hfill \\
 0 \hfill & 0 \hfill & 0 \hfill & 0 \hfill & -\sqrt{3} \hfill & 0 \hfill& -2 \hfill  & 0 \hfill \\
 0 \hfill & 0 \hfill & 0 \hfill & 0 \hfill & 0 \hfill & -2 \hfill& 0 \hfill  & -\sqrt{3} \hfill \\
 0 \hfill & 0 \hfill & 0 \hfill & 0 \hfill & 0 \hfill & 0 \hfill& -\sqrt{3} \hfill  & 0 \hfill \\
\end{array} }} \right|,$$

$${s}^{2}_{8}= \frac{i}{2}\cdot$$
\begin{equation}
\label{eq179}\left| {{\begin{array}{*{20}c}
 0 \hfill & -\sqrt{3} \hfill & 0 \hfill & 0 \hfill & 0 \hfill  & 0 \hfill & 0 \hfill  & 0 \hfill \\
 \sqrt{3} \hfill & 0 \hfill & -2 \hfill & 0 \hfill & 0 \hfill & 0 \hfill & 0 \hfill  & 0 \hfill \\
 0 \hfill & 2 \hfill & 0 \hfill & -\sqrt{3} \hfill & 0 \hfill & 0 \hfill & 0 \hfill  & 0 \hfill \\
 0 \hfill & 0 \hfill & \sqrt{3} \hfill & 0 \hfill & 0 \hfill & 0 \hfill & 0 \hfill  & 0 \hfill \\
 0 \hfill & 0 \hfill & 0 \hfill & 0 \hfill & 0 \hfill & -\sqrt{3} \hfill & 0 \hfill  & 0 \hfill \\
 0 \hfill & 0 \hfill & 0 \hfill & 0 \hfill & \sqrt{3} \hfill & 0 \hfill& -2 \hfill  & 0 \hfill \\
 0 \hfill & 0 \hfill & 0 \hfill & 0 \hfill & 0 \hfill & 2 \hfill& 0 \hfill  & -\sqrt{3} \hfill \\
 0 \hfill & 0 \hfill & 0 \hfill & 0 \hfill & 0 \hfill & 0 \hfill& \sqrt{3} \hfill  & 0 \hfill \\
\end{array} }} \right|,
\end{equation}
$${s}^{3}_{8}= \frac{1}{2}\left| {{\begin{array}{*{20}c}
 3 \hfill & 0 \hfill & 0 \hfill & 0 \hfill & 0 \hfill  & 0 \hfill & 0 \hfill  & 0 \hfill \\
 0 \hfill & 1 \hfill & 0 \hfill & 0 \hfill & 0 \hfill & 0 \hfill & 0 \hfill  & 0 \hfill \\
 0 \hfill & 0 \hfill & -1 \hfill & 0 \hfill & 0 \hfill & 0 \hfill & 0 \hfill  & 0 \hfill \\
 0 \hfill & 0 \hfill & 0 \hfill & -3 \hfill & 0 \hfill & 0 \hfill & 0 \hfill  & 0 \hfill \\
 0 \hfill & 0 \hfill & 0 \hfill & 0 \hfill & -3 \hfill & 0 \hfill & 0 \hfill  & 0 \hfill \\
 0 \hfill & 0 \hfill & 0 \hfill & 0 \hfill & 0 \hfill & -1 \hfill& 0 \hfill  & 0 \hfill \\
 0 \hfill & 0 \hfill & 0 \hfill & 0 \hfill & 0 \hfill & 0 \hfill& 1 \hfill  & 0 \hfill \\
 0 \hfill & 0 \hfill & 0 \hfill & 0 \hfill & 0 \hfill & 0 \hfill& 0 \hfill  & 3 \hfill \\
\end{array} }} \right|.$$

The Casimir operator has the form of the following $8 \times 8$ diagonal matrix

\begin{equation}
\label{eq180} \overrightarrow{s}^{2}= \frac{15}{4}
\mathrm{I}_{8}=\frac{3}{2}\left(\frac{3}{2}+1\right)\mathrm{I}_{8},
\end{equation}

\noindent where $\mathrm{I}_{8}$ is the $8\times 8$ unit matrix.

The stationary complete set of operators is given by

$$g=\left|
{{\begin{array}{*{20}c}
 -\mathrm{I}_{4} \hfill & 0 \\
 0 \hfill &  \mathrm{I}_{4} \\
\end{array} }} \right|, \quad p^{j}=-i\partial _{j},$$
\begin{equation}
\label{eq181} {s}^{3}_{8}= \frac{1}{2}\left| {{\begin{array}{*{20}c}
 3 \hfill & 0 \hfill & 0 \hfill & 0 \hfill & 0 \hfill  & 0 \hfill & 0 \hfill  & 0 \hfill \\
 0 \hfill & 1 \hfill & 0 \hfill & 0 \hfill & 0 \hfill & 0 \hfill & 0 \hfill  & 0 \hfill \\
 0 \hfill & 0 \hfill & -1 \hfill & 0 \hfill & 0 \hfill & 0 \hfill & 0 \hfill  & 0 \hfill \\
 0 \hfill & 0 \hfill & 0 \hfill & -3 \hfill & 0 \hfill & 0 \hfill & 0 \hfill  & 0 \hfill \\
 0 \hfill & 0 \hfill & 0 \hfill & 0 \hfill & -3 \hfill & 0 \hfill & 0 \hfill  & 0 \hfill \\
 0 \hfill & 0 \hfill & 0 \hfill & 0 \hfill & 0 \hfill & -1 \hfill& 0 \hfill  & 0 \hfill \\
 0 \hfill & 0 \hfill & 0 \hfill & 0 \hfill & 0 \hfill & 0 \hfill& 1 \hfill  & 0 \hfill \\
 0 \hfill & 0 \hfill & 0 \hfill & 0 \hfill & 0 \hfill & 0 \hfill& 0 \hfill  & 3 \hfill \\
\end{array} }} \right|
,
\end{equation}

\noindent where $g$ is the charge sign operator, $\overrightarrow{p}=(p^{j})$ is the momentum operator and $s_{8}^{3}=s_{z}$ is the operator of the spin (178) projection on the axe $z$. 

The equations on the eigenvalues of the operators $g, \, s_{8}^{3}=s_{z}$ have the form

$$g\mathrm{d}_{1} = -\mathrm{d}_{1}, \,
g\mathrm{d}_{2} = -\mathrm{d}_{2}, \, g\mathrm{d}_{3} =
-\mathrm{d}_{3}, \, g\mathrm{d}_{4} = -\mathrm{d}_{4},$$
\begin{equation}
\label{eq182} g\mathrm{d}_{5} = +\mathrm{d}_{5}, \,
g\mathrm{d}_{6} = +\mathrm{d}_{6}, \, g\mathrm{d}_{7} =
+\mathrm{d}_{7}, \, g\mathrm{d}_{8} = +\mathrm{d}_{8},
\end{equation}

$$s_{8}^{3}\mathrm{d}_{1} = \frac{3}{2}\mathrm{d}_{1}, \, s_{8}^{3}\mathrm{d}_{2} = \frac{1}{2}\mathrm{d}_{2},$$
$$s_{8}^{3}\mathrm{d}_{3} = -\frac{1}{2}\mathrm{d}_{3}, \,
s_{8}^{3}\mathrm{d}_{4} =  -\frac{3}{2}\mathrm{d}_{4},$$
\begin{equation}
\label{eq183} s_{8}^{3}\mathrm{d}_{5} = -\frac{3}{2}\mathrm{d}_{5}, \,
s_{8}^{3}\mathrm{d}_{6} =  -\frac{1}{2}\mathrm{d}_{6},
\end{equation}
$$s_{8}^{3}\mathrm{d}_{7} =
\frac{1}{2}\mathrm{d}_{7}, \, s_{8}^{3}\mathrm{d}_{8} = \frac{3}{2}\mathrm{d}_{8}.$$

\noindent The equations on the eigenvalues of the momentum operator $\overrightarrow{p}$ are similar to (156) (here the Cartesian basis has 8 dimensions).           

Therefore, the functions $b^{1}(\overrightarrow{k}), \,
b^{2}(\overrightarrow{k}), \, b^{3}(\overrightarrow{k}), \, b^{4}(\overrightarrow{k})$ in
solution (177) are the momentum-spin amplitudes of the massive
fermion with the spin s=3/2 and the spin projection $(3/2,1/2,-1/2,-3/2)$,
respectively; $b^{5}(\overrightarrow{k}), \,
b^{6}(\overrightarrow{k}), \, b^{7}(\overrightarrow{k}), \, b^{8}(\overrightarrow{k})$ are the
momentum-spin amplitudes of the antiparticle (antifermion) with the
spin s=3/2 and the spin projection $(-3/2,-1/2,1/2,3/2)$, respectively.

In addition to the bosonic $\mathcal{P}$ invariance, which was considered in the previous section, the Schr$\mathrm{\ddot{o}}$dinger--Foldy equation (168) (and the set $\{\mathrm{f}\}$ of its solutions (177)) is invariant with
respect to the reducible unitary fermionic representation (14) of
the Poincar$\mathrm{\acute{e}}$ group $\mathcal{P}$, whose
Hermitian $8 \times 8$ matrix-differential generators are given
by (12), (13), where the spin s=(3/2,3/2) SU(2) generators
$\overrightarrow{s}=(s^{\ell n})$ are given in (178), (179).

The proof is similar to that given in section 2 after formula (14). The Casimir operators of this reducible fermionic spin s=(3/2,3/2) representation of the group $\mathcal{P}$ have the form

\begin{equation}
\label{eq184} p^{2}=\widehat{p}^{\mu}\widehat{p}_{\mu}=m^{2}\mathrm{I}_{8},
\end{equation}
\begin{equation}
\label{eq185} W=w^{\mu}w_{\mu}=m^{2}\overrightarrow{s}_{8}^{2}=\frac{3}{2}\left(\frac{3}{2}+1\right)m^{2}\mathrm{I}_{8},
\end{equation}

\noindent where $\mathrm{I}_{8}$ is the $8\times 8$ unit matrix.

Thus, above the foundations of the RCQM of the 8-component spin s=(3/2,3/2) particle-antiparticle doublet are considered briefly. It is the basis for the transition to the covariant local field theory of the spin s=(3/2,3/2) particle-antiparticle doublet given below.

\section{A brief scheme of the relativistic canonical quantum mechanics of the 10-component spin s=(2,2) particle-antiparticle bosonic doublet}

This model is constructed in complete analogy with the RCQM of the 4-component spin s=(1/2,1/2) particle-antiparticle doublet (section 7), spin s=(1,1) particle-antiparticle doublet (section 11) and spin s=(3/2,3/2) particle-antiparticle doublet (section 14). The most close analogy is with the model of bosonic spin s=(1,1) particle-antiparticle doublet (section 11).

For the bosonic spin s=(2,2) particle-antiparticle doublet the
system of two 5-component equations (46) should be used.
Therefore, the corresponding Schr$\mathrm{\ddot{o}}$dinger--Foldy
equation is given by

\begin{equation}
\label{eq186} (i\partial_{0}- \widehat{\omega})f(x)=0, \quad
f=\left| {{\begin{array}{*{20}c}
 f^{1} \hfill  \\
 f^{2} \hfill  \\
 f^{3} \hfill  \\
 f^{4} \hfill  \\
 f^{5} \hfill  \\
 f^{6} \hfill  \\
 f^{7} \hfill  \\
 f^{8} \hfill  \\
 f^{9} \hfill  \\
 f^{10} \hfill  \\

\end{array} }} \right|,
\end{equation}

\noindent where the pseudo-differential operator
$\widehat{\omega}$ is given in (59)--(61). In (186), the
10-component wave function is the direct sum of the particle and
antiparticle wave functions, whereas the particle wave function
has the form (50). According to the quantum-mechanical tradition
the antiparticle wave function is put into the bottom part of the
10-column.

Therefore, the general solution of the
Schr$\mathrm{\ddot{o}}$dinger--Foldy equation (186) has the form

$$f(x)= \left|
{{\begin{array}{*{20}c}
 f_{\mathrm{part}} \hfill  \\
 f_{\mathrm{antipart}} \hfill  \\
\end{array} }} \right|=$$
\begin{equation}
\label{eq187}
\frac{1}{\left(2\pi\right)^{\frac{3}{2}}}\int d^{3}k e^{-ikx}g^{\mathrm{A}}(\overrightarrow{k})\mathrm{d}_{\mathrm{A}}, \quad \mathrm{A}=\overline{1,10},
\end{equation}

\noindent where the orts of the 10-component Cartesian basis are
given by

$$\mathrm{d}_{1} = \left|
\begin{array}{cccc}
 1 \\
 0 \\
 0 \\
 0 \\
 0 \\
 0 \\
 0 \\
 0 \\
 0 \\
 0 \\
\end{array} \right|, \, \mathrm{d}_{2} = \left|
\begin{array}{cccc}
 0 \\
 1 \\
 0 \\
 0 \\
 0 \\
 0 \\
 0 \\
 0 \\
 0 \\
 0 \\
\end{array} \right|,
\, \mathrm{d}_{3} = \left|
\begin{array}{cccc}
 0 \\
 0 \\
 1 \\
 0 \\
 0 \\
 0 \\
 0 \\
 0 \\
 0 \\
 0 \\
\end{array} \right|,$$
\begin{equation}
\label{eq188} \mathrm{d}_{4} = \left|
\begin{array}{cccc}
 0 \\
 0 \\
 0 \\
 1 \\
 0 \\
 0 \\
 0 \\
 0 \\
 0 \\
 0 \\
\end{array} \right|, \,
\mathrm{d}_{5} = \left|
\begin{array}{cccc}
 0 \\
 0 \\
 0 \\
 0 \\
 1 \\
 0 \\
 0 \\
 0 \\
 0 \\
 0 \\
\end{array} \right|, \,
\mathrm{d}_{6} = \left|
\begin{array}{cccc}
 0 \\
 0 \\
 0 \\
 0 \\
 0 \\
 1 \\
 0 \\
 0 \\
 0 \\
 0 \\
\end{array} \right|, \,
\mathrm{d}_{7} = \left|
\begin{array}{cccc}
 0 \\
 0 \\
 0 \\
 0 \\
 0 \\
 0 \\
 1 \\
 0 \\
 0 \\
 0 \\
\end{array} \right|,
\end{equation}
$$\mathrm{d}_{8} = \left|
\begin{array}{cccc}
 0 \\
 0 \\
 0 \\
 0 \\
 0 \\
 0 \\
 0 \\
 1 \\
 0 \\
 0 \\
\end{array} \right|, \, 
\mathrm{d}_{9} = \left|
\begin{array}{cccc}
 0 \\
 0 \\
 0 \\
 0 \\
 0 \\
 0 \\
 0 \\
 0 \\
 1 \\
 0 \\
\end{array} \right|, \, 
\mathrm{d}_{10} = \left|
\begin{array}{cccc}
 0 \\
 0 \\
 0 \\
 0 \\
 0 \\
 0 \\
 0 \\
 0 \\
 0 \\
 1 \\
\end{array} \right|.$$

Hence, the space of the states in the RCQM of the spin s=(2,2)
particle-antiparticle doublet is the rigged Hilbert space

\begin{equation}
\label{eq189}
\mathrm{S}^{3,10}\subset\mathrm{H}^{3,10}\subset\mathrm{S}^{3,10*},
\end{equation}

\noindent i. e. is the direct sum of two spaces (47).

Thus, in the model under consideration information about the
positive and equal masses of the particle and antiparticle is
inserted. Further, information that the observer sees the
antiparticle as the mirror reflection of the particle is also
inserted. Therefore, the charge of the antiparticle should be
opposite in sign to that of the particle (in the case of the existence of the charge), and the spin projection
of the antiparticle should be opposite in sign to the spin
projection of the particle.

Therefore, according to these principles (see the sections 7, 11, 14 as well), the corresponding
SU(2)-spin generators are taken in the form

\begin{equation}
\label{eq190} \overrightarrow{s}_{10}=\left|
{{\begin{array}{*{20}c}
 \overrightarrow{s}_{5} \hfill & 0 \\
 0 \hfill &  -C\overrightarrow{s}_{5}C \\
\end{array} }} \right|,
\end{equation}

\noindent where the $5 \times 5$-matrices $\overrightarrow{s}_{5}$
are given in (48) and $C\mathrm{I}_{5}$ is $5 \times 5$ diagonal matrix operator of complex conjugation.

It is easy to verify that for the operators (190) the
commutation relations $
\left[s^{j},s^{\ell}\right]=i\varepsilon^{j \ell n}s^{n}$
\noindent of the SU(2)-algebra are valid.

In the explicit form the SU(2) spin operators (190) are given by \small

$$s^{1}_{10}= \frac{1}{2}\cdot$$
$$\left| {{\begin{array}{*{20}c}
 0 \hfill & 2 \hfill & 0 \hfill & 0 \hfill & 0 \hfill  & 0 \hfill & 0 \hfill  & 0 \hfill & 0 \hfill  & 0 \hfill \\
 2 \hfill & 0 \hfill & \sqrt{6} \hfill & 0 \hfill & 0 \hfill & 0 \hfill & 0 \hfill  & 0 \hfill & 0 \hfill  & 0 \hfill \\
 0 \hfill & \sqrt{6} \hfill & 0 \hfill & \sqrt{6} \hfill & 0 \hfill & 0 \hfill & 0 \hfill  & 0 \hfill & 0 \hfill  & 0 \hfill \\
 0 \hfill & 0 \hfill & \sqrt{6} \hfill & 0 \hfill & 2 \hfill & 0 \hfill & 0 \hfill  & 0 \hfill & 0 \hfill  & 0 \hfill \\
 0 \hfill & 0 \hfill & 0 \hfill & 2 \hfill & 0 \hfill & 0 \hfill & 0 \hfill  & 0 \hfill & 0 \hfill  & 0 \hfill \\
 0 \hfill & 0 \hfill & 0 \hfill & 0 \hfill & 0 \hfill & 0 \hfill& -2 \hfill  & 0 \hfill & 0 \hfill  & 0 \hfill \\
 0 \hfill & 0 \hfill & 0 \hfill & 0 \hfill & 0 \hfill & -2 \hfill& 0 \hfill  & -\sqrt{6} \hfill & 0 \hfill  & 0 \hfill \\
 0 \hfill & 0 \hfill & 0 \hfill & 0 \hfill & 0 \hfill & 0 \hfill& -\sqrt{6} \hfill  & 0 \hfill & -\sqrt{6} \hfill  & 0 \hfill \\
 0 \hfill & 0 \hfill & 0 \hfill & 0 \hfill & 0 \hfill & 0 \hfill& 0 \hfill  & -\sqrt{6} \hfill & 0 \hfill  & -2 \hfill \\
 0 \hfill & 0 \hfill & 0 \hfill & 0 \hfill & 0 \hfill & 0 \hfill& 0 \hfill  & 0 \hfill & -2 \hfill  & 0 \hfill \\
\end{array} }} \right|,$$

$$s^{2}_{10}= \frac{i}{2}\cdot$$
\begin{equation}
\label{eq191} \left| {{\begin{array}{*{20}c}
 0 \hfill & -2 \hfill & 0 \hfill & 0 \hfill & 0 \hfill  & 0 \hfill & 0 \hfill  & 0 \hfill & 0 \hfill  & 0 \hfill \\
 2 \hfill & 0 \hfill & -\sqrt{6} \hfill & 0 \hfill & 0 \hfill & 0 \hfill & 0 \hfill  & 0 \hfill & 0 \hfill  & 0 \hfill \\
 0 \hfill & \sqrt{6} \hfill & 0 \hfill & -\sqrt{6} \hfill & 0 \hfill & 0 \hfill & 0 \hfill  & 0 \hfill & 0 \hfill  & 0 \hfill \\
 0 \hfill & 0 \hfill & \sqrt{6} \hfill & 0 \hfill & -2 \hfill & 0 \hfill & 0 \hfill  & 0 \hfill & 0 \hfill  & 0 \hfill \\
 0 \hfill & 0 \hfill & 0 \hfill & 2 \hfill & 0 \hfill & 0 \hfill & 0 \hfill  & 0 \hfill & 0 \hfill  & 0 \hfill \\
 0 \hfill & 0 \hfill & 0 \hfill & 0 \hfill & 0 \hfill & 0 \hfill& -2 \hfill  & 0 \hfill & 0 \hfill  & 0 \hfill \\
 0 \hfill & 0 \hfill & 0 \hfill & 0 \hfill & 0 \hfill & 2 \hfill& 0 \hfill  & -\sqrt{6} \hfill & 0 \hfill  & 0 \hfill \\
 0 \hfill & 0 \hfill & 0 \hfill & 0 \hfill & 0 \hfill & 0 \hfill& \sqrt{6} \hfill  & 0 \hfill & -\sqrt{6} \hfill  & 0 \hfill \\
 0 \hfill & 0 \hfill & 0 \hfill & 0 \hfill & 0 \hfill & 0 \hfill& 0 \hfill  & \sqrt{6} \hfill & 0 \hfill  & -2 \hfill \\
 0 \hfill & 0 \hfill & 0 \hfill & 0 \hfill & 0 \hfill & 0 \hfill& 0 \hfill  & 0 \hfill & 2 \hfill  & 0 \hfill \\
\end{array} }} \right|,
\end{equation}

\normalsize

$$s^{3}_{10}= \left| {{\begin{array}{*{20}c}
 2 \hfill & 0 \hfill & 0 \hfill & 0 \hfill & 0 \hfill  & 0 \hfill & 0 \hfill  & 0 \hfill & 0 \hfill  & 0 \hfill \\
 0 \hfill & 1 \hfill & 0 \hfill & 0 \hfill & 0 \hfill & 0 \hfill & 0 \hfill  & 0 \hfill & 0 \hfill  & 0 \hfill \\
 0 \hfill & 0 \hfill & 0 \hfill & 0 \hfill & 0 \hfill & 0 \hfill & 0 \hfill  & 0 \hfill & 0 \hfill  & 0 \hfill \\
 0 \hfill & 0 \hfill & 0 \hfill & -1 \hfill & 0 \hfill & 0 \hfill & 0 \hfill  & 0 \hfill & 0 \hfill  & 0 \hfill \\
 0 \hfill & 0 \hfill & 0 \hfill & 0 \hfill & -2 \hfill & 0 \hfill & 0 \hfill  & 0 \hfill & 0 \hfill  & 0 \hfill \\
 0 \hfill & 0 \hfill & 0 \hfill & 0 \hfill & 0 \hfill & -2 \hfill& 0 \hfill  & 0 \hfill & 0 \hfill  & 0 \hfill \\
 0 \hfill & 0 \hfill & 0 \hfill & 0 \hfill & 0 \hfill & 0 \hfill& -1 \hfill  & 0 \hfill & 0 \hfill  & 0 \hfill \\
 0 \hfill & 0 \hfill & 0 \hfill & 0 \hfill & 0 \hfill & 0 \hfill& 0 \hfill  & 0 \hfill & 0 \hfill  & 0 \hfill \\
 0 \hfill & 0 \hfill & 0 \hfill & 0 \hfill & 0 \hfill & 0 \hfill& 0 \hfill  & 0 \hfill & 1 \hfill  & 0 \hfill \\
 0 \hfill & 0 \hfill & 0 \hfill & 0 \hfill & 0 \hfill & 0 \hfill& 0 \hfill  & 0 \hfill & 0 \hfill  & 2 \hfill \\
\end{array} }} \right|.$$

The Casimir operator for this reducible representation of the
SU(2)-algebra is given by 

\begin{equation}
\label{eq192} \overrightarrow{s}^{2}=
6\mathrm{I}_{10}=2\left(2+1\right)\mathrm{I}_{10},
\end{equation}

\noindent where $\mathrm{I}_{10}$ is the $10 \times 10$ unit matrix.

The solution (187) is associated with the stationary complete set
$\overrightarrow{p}, \, s^{3}=s_{z}$ of the momentum and spin
projection operators of the spin s=(2,2) bosonic
particle-antiparticle doublet.

The equations on the momentum operator eigenvalues have the form

\begin{equation}
\label{eq193} \overrightarrow{p}e^{-ikx}\mathrm{d}_{\mathrm{A}} =
\overrightarrow{k}e^{-ikx}\mathrm{d}_{\mathrm{A}}, \quad \mathrm{A}=\overline{1,10},
\end{equation}

\noindent and the equations on the spin projection operator
$s^{3}_{10}$ (191) eigenvalues are given by

$$s^{3}_{10}\mathrm{d}_{1} = 2\mathrm{d}_{1}, \, s^{3}_{10}\mathrm{d}_{2} = \mathrm{d}_{2}, \, s^{3}_{10}\mathrm{d}_{3} = 0, \, s^{3}_{10}\mathrm{d}_{4} = -\mathrm{d}_{4},$$
\begin{equation}
\label{eq194} s^{3}_{10}\mathrm{d}_{5} = - 2\mathrm{d}_{5}, \,
s^{3}_{10}\mathrm{d}_{6} = - 2\mathrm{d}_{6}, \, s^{3}_{10}\mathrm{d}_{7} = -\mathrm{d}_{7},
\end{equation}
$$s^{3}_{10}\mathrm{d}_{8}=0, \, s^{3}_{10}\mathrm{d}_{9} = \mathrm{d}_{9}, \, s^{3}_{10}\mathrm{d}_{10} = 2\mathrm{d}_{10}.$$

Therefore, the functions
$g^{1}(\overrightarrow{k}), \, g^{2}(\overrightarrow{k}), \, g^{3}(\overrightarrow{k}), \, g^{4}(\overrightarrow{k}), \, g^{5}(\overrightarrow{k})$
in the solution (187) are the momentum-spin amplitudes of the
particle (boson) with the momentum $\overrightarrow{p}$ and spin
projection eigenvalues (+2, +1, 0, -1, -2), respectively, and the
functions
$g^{6}(\overrightarrow{k}), \, g^{7}(\overrightarrow{k}), \, g^{8}(\overrightarrow{k}), \, g^{9}(\overrightarrow{k}), \, g^{10}(\overrightarrow{k})$
are the momentum-spin amplitudes of the antiparticle with the
momentum $\overrightarrow{p}$ and spin projection eigenvalues
(-2, -1, 0, +1, +2), respectively.

The Schr$\mathrm{\ddot{o}}$dinger--Foldy equation (186) and the
set $\{\mathrm{f}\}$ of its solutions (187) are invariant with
respect to the reducible unitary bosonic representation (14) of
the Poincar$\mathrm{\acute{e}}$ group $\mathcal{P}$. The
corresponding $10 \times 10$ matrix-differential generators are
given by (12), (13), whereas the spin s=(1,1) SU(2) generators
$\overrightarrow{s}=(s^{\ell n})$ are given in (190), (191).

The proof of this assertion is fulfilled by the three steps
already given in section 2 after formula (14). The corresponding
Casimir operators have the form

\begin{equation}
\label{eq195}
p^{2}=\widehat{p}^{\mu}\widehat{p}_{\mu}=m^{2}\mathrm{I}_{10},
\end{equation}
\begin{equation}
\label{eq196} W=w^{\mu}w_{\mu}=m^{2}\overrightarrow{s}^{2}_{10}=
2\left(2+1\right)m^{2}\mathrm{I}_{10},
\end{equation}

\noindent where $\mathrm{I}_{10}$ is the $10 \times 10$ unit matrix.

Hence, above a brief consideration of the RCQM foundations of the
particle-antiparticle doublet with the mass $m>0$ and the spin
s=(2,2) has been given. In the limit m=0 this model describes the
partial case of the graviton-antigraviton doublet. The hypothesis about the massive graviton and the problems of the gravity in general are out of this consideration.

\section{A brief scheme of the relativistic canonical quantum mechanics of the 12-component spin s=(2,0,2,0) particle-antiparticle bosonic multiplet}

This model is constructed in complete analogy with the RCQM of spin s=(1,0,1,0) particle-antiparticle multiplet (section 13). The 12-component bosonic spin s=(2,0,2,0) particle-antiparticle
multiplet is constructed as the direct sum of the two spin s=(2,0)
multiplets. The principles of constructing and describing such particle-antiparticle multiplet within the
framework of the RCQM are in a complete analogy with the principles
of principles considered in the previous sections. Therefore, the details can be omitted.

The most important fact is that here the link with the Dirac-like
equation is similar to that between the spin s=(1/2,1/2)
particle-antiparticle doublet and the standard 4-component Dirac
equation demonstrated in sections 9, 10. Therefore, the
spin s=(2,0,2,0) particle-antiparticle multiplet is of special
interest. It is much more useful than the spin s=(1,0) or (2,0) particle
multiplets.

Therefore, the corresponding Schr$\mathrm{\ddot{o}}$dinger--Foldy
equation is given by

\begin{equation}
\label{eq197} (i\partial_{0}- \widehat{\omega})f(x)=0, \quad
f=\left| {{\begin{array}{*{20}c}
 f^{1} \hfill  \\
 f^{2} \hfill  \\
 .  \\
 .  \\
 .  \\
 f^{12} \hfill  \\

\end{array} }} \right|,
\end{equation}

\noindent where the pseudo-differential operator
$\widehat{\omega}$ is given in (59)--(61). In (197), the
12-component wave function is the direct sum of the particle and
antiparticle wave functions. According to the quantum-mechanical tradition
the wave function of two antiparticles is put into the bottom part of the
12-column.

Therefore, the general solution of the
Schr$\mathrm{\ddot{o}}$dinger--Foldy equation (197) has the form

$$f(x)= \left|
{{\begin{array}{*{20}c}
 f_{\mathrm{part}} \hfill  \\
 f_{\mathrm{antipart}} \hfill  \\
\end{array} }} \right|=$$
\begin{equation}
\label{eq198}
\frac{1}{\left(2\pi\right)^{\frac{3}{2}}}\int d^{3}k e^{-ikx}g^{\mathrm{B}}(\overrightarrow{k})\mathrm{d}_{\mathrm{B}}, \quad \mathrm{B}=\overline{1,12},
\end{equation}

\noindent where the orts of the 12-component Cartesian basis are
given by

\begin{equation}
\label{eq199} \mathrm{d}_{1} = \left|
\begin{array}{cccc}
 1 \\
 0 \\
 0 \\
 0 \\
 0 \\
 0 \\
 0 \\
 0 \\
 0 \\
 0 \\
 0 \\
 0 \\
\end{array} \right|, \,
\mathrm{d}_{2} = \left|
\begin{array}{cccc}
 0 \\
 1 \\
 0 \\
 0 \\
 0 \\
 0 \\
 0 \\
 0 \\
 0 \\
 0 \\
 0 \\
 0 \\
\end{array} \right|, \quad  ... \, , \quad   
\mathrm{d}_{12} = \left|
\begin{array}{cccc}
 0 \\
 0 \\
 0 \\
 0 \\
 0 \\
 0 \\
 0 \\
 0 \\
 0 \\
 0 \\
 0 \\
 1 \\
\end{array} \right|.
\end{equation}

Hence, the space of the states in the RCQM of the spin s=(2,0,2,0)
particle-antiparticle doublet is the rigged Hilbert space

\begin{equation}
\label{eq200}
\mathrm{S}^{3,12}\subset\mathrm{H}^{3,12}\subset\mathrm{S}^{3,12*},
\end{equation}

\noindent i. e. is the direct sum of 4 corresponding spaces.

The corresponding SU(2)-spin generators are taken in the form

\begin{equation}
\label{eq201} \overrightarrow{s}_{12}=\left|
{{\begin{array}{*{20}c}
 \overrightarrow{s}_{6} \hfill & 0 \\
 0 \hfill &  -C\overrightarrow{s}_{6}C \\
\end{array} }} \right|,
\end{equation}

\noindent where the $6 \times 6$-matrices $\overrightarrow{s}_{6}$
are given by 

$$s^{1}_{6}= \frac{1}{2}\left| {{\begin{array}{*{20}c}
 0 \hfill & 2 \hfill & 0 \hfill & 0 \hfill & 0 \hfill  & 0 \hfill  \\
 2 \hfill & 0 \hfill & \sqrt{6} \hfill & 0 \hfill & 0 \hfill & 0 \hfill  \\
 0 \hfill & \sqrt{6} \hfill & 0 \hfill & \sqrt{6} \hfill & 0 \hfill & 0 \hfill  \\
 0 \hfill & 0 \hfill & \sqrt{6} \hfill & 0 \hfill & 2 \hfill & 0 \hfill  \\
 0 \hfill & 0 \hfill & 0 \hfill & 2 \hfill & 0 \hfill & 0 \hfill  \\
 0 \hfill & 0 \hfill & 0 \hfill & 0 \hfill & 0 \hfill & 0 \hfill  \\
 \end{array} }} \right|,$$
\begin{equation}
\label{eq202} s^{2}_{6}= \frac{i}{2}\left| {{\begin{array}{*{20}c}
 0 \hfill & -2 \hfill & 0 \hfill & 0 \hfill & 0 \hfill  & 0 \hfill  \\
 2 \hfill & 0 \hfill & -\sqrt{6} \hfill & 0 \hfill & 0 \hfill & 0 \hfill  \\
 0 \hfill & \sqrt{6} \hfill & 0 \hfill & -\sqrt{6} \hfill & 0 \hfill & 0 \hfill  \\
 0 \hfill & 0 \hfill & \sqrt{6} \hfill & 0 \hfill & -2 \hfill & 0 \hfill  \\
 0 \hfill & 0 \hfill & 0 \hfill & 2 \hfill & 0 \hfill & 0 \hfill  \\
 0 \hfill & 0 \hfill & 0 \hfill & 0 \hfill & 0 \hfill & 0 \hfill  \\
 \end{array} }} \right|,
\end{equation}
$$s^{3}_{6}= \left| {{\begin{array}{*{20}c}
 2 \hfill & 0 \hfill & 0 \hfill & 0 \hfill & 0 \hfill  & 0 \hfill  \\
 0 \hfill & 1 \hfill & 0 \hfill & 0 \hfill & 0 \hfill & 0 \hfill  \\
 0 \hfill & 0 \hfill & 0 \hfill & 0 \hfill & 0 \hfill & 0 \hfill  \\
 0 \hfill & 0 \hfill & 0 \hfill & -1 \hfill & 0 \hfill & 0 \hfill \\
 0 \hfill & 0 \hfill & 0 \hfill & 0 \hfill & -2 \hfill & 0 \hfill  \\
 0 \hfill & 0 \hfill & 0 \hfill & 0 \hfill & 0 \hfill & 0 \hfill  \\
 \end{array} }} \right|,$$

\noindent and $C\mathrm{I}_{6}$ is $6 \times 6$ diagonal matrix operator of complex conjugation.

It is easy to verify that for the operators (201) the
commutation relations $
\left[s^{j},s^{\ell}\right]=i\varepsilon^{j \ell n}s^{n}$
\noindent of the SU(2)-algebra are valid.

The Casimir operator for this reducible representation of the
SU(2)-algebra is given by 

$$\overrightarrow{s}^{2}_{12}= 6\left|
{{\begin{array}{*{20}c}
 \mathrm{I}_{5} \hfill & 0 \hfill & 0 \hfill & 0 \hfill\\
 0 \hfill & 0 \hfill & 0 \hfill & 0 \hfill\\
 0 \hfill & 0 \hfill & \mathrm{I}_{5} \hfill & 0 \hfill\\
 0 \hfill & 0 \hfill & 0 \hfill & 0 \hfill\\
\end{array} }} \right|=$$
\begin{equation}
\label{eq203} \left|
{{\begin{array}{*{20}c}
 2(2+1)\mathrm{I}_{5} \hfill & 0 \hfill & 0 \hfill & 0 \hfill\\
 0 \hfill & 0 \hfill & 0 \hfill & 0 \hfill\\
 0 \hfill & 0 \hfill & 2(2+1)\mathrm{I}_{5} \hfill & 0 \hfill\\
 0 \hfill & 0 \hfill & 0 \hfill & 0 \hfill\\
\end{array} }} \right|,
\end{equation}

\noindent where $\mathrm{I}_{5}$ is the $5 \times 5$ unit matrix.

The solution (198) is associated with the stationary complete set
$\overrightarrow{p}, \, s^{3}_{12}=s_{z}$ of the momentum and spin
projection operators of the spin s=(2,0,2,0) bosonic
particle-antiparticle multiplet.

The equations on the momentum operator eigenvalues have the form

\begin{equation}
\label{eq204} \overrightarrow{p}e^{-ikx}\mathrm{d}_{\mathrm{B}} =
\overrightarrow{k}e^{-ikx}\mathrm{d}_{\mathrm{B}}, \quad \mathrm{B}=\overline{1,12},
\end{equation}

\noindent and the equations on the spin projection operator
$s^{3}_{12}$ (201) eigenvalues are given by

$$s^{3}_{12}\mathrm{d}_{1} = 2\mathrm{d}_{1}, \, s^{3}_{12}\mathrm{d}_{2} = \mathrm{d}_{2}, \, s^{3}_{12}\mathrm{d}_{3} = 0, \, s^{3}_{12}\mathrm{d}_{4} = -\mathrm{d}_{4},$$
\small
\begin{equation}
\label{eq205} s^{3}_{12}\mathrm{d}_{5} = - 2\mathrm{d}_{5}, \, s^{3}_{12}\mathrm{d}_{6} =0, \, s^{3}_{12}\mathrm{d}_{7} = -2\mathrm{d}_{7},
s^{3}_{12}\mathrm{d}_{8} = -\mathrm{d}_{8},
\end{equation}
\normalsize
$$s^{3}_{12}\mathrm{d}_{9} = 0, \, s^{3}_{12}\mathrm{d}_{10}=\mathrm{d}_{10}, \, s^{3}_{12}\mathrm{d}_{11} = 2\mathrm{d}_{11}, \, s^{3}_{12}\mathrm{d}_{12} =0.$$

Therefore, the functions
$g^{1}(\overrightarrow{k}), \, g^{2}(\overrightarrow{k}), \, g^{3}(\overrightarrow{k}), \, g^{4}(\overrightarrow{k}), \, g^{5}(\overrightarrow{k})$
in the solution (198) are the momentum-spin amplitudes of the
particle (boson) with the momentum $\overrightarrow{p}$, spin s=2 and spin
projection eigenvalues (+2, +1, 0, -1, -2), respectively, $g^{6}(\overrightarrow{k})$ is the amplitude of the spinless boson; the
functions
$g^{7}(\overrightarrow{k}), \, g^{8}(\overrightarrow{k}), \, g^{9}(\overrightarrow{k}), \, g^{10}(\overrightarrow{k}), \, g^{11}(\overrightarrow{k})$
are the momentum-spin amplitudes of the antiparticle (antiboson) with the
momentum $\overrightarrow{p}$, spin s=2 and spin projection eigenvalues
(-2, -1, 0, +1, +2), respectively, $g^{12}(\overrightarrow{k})$ is the amplitude of the spinless antiboson.

The Schr$\mathrm{\ddot{o}}$dinger--Foldy equation (197) and the
set $\{\mathrm{f}\}$ of its solutions (198) are invariant with
respect to the reducible unitary bosonic representation (14) of
the Poincar$\mathrm{\acute{e}}$ group $\mathcal{P}$. The
corresponding $12 \times 12$ matrix-differential generators are
given by (12), (13), whereas the spin s=(2,0,2,0) SU(2) generators
$\overrightarrow{s}=(s^{\ell n})$ are given in (201).

The proof of this assertion is fulfilled by the three steps
already given in section 2 after formula (14). The corresponding
Casimir operators have the form

\begin{equation}
\label{eq206}
p^{2}=\widehat{p}^{\mu}\widehat{p}_{\mu}=m^{2}\mathrm{I}_{12},
\end{equation}
$$W=w^{\mu}w_{\mu}=m^{2}\overrightarrow{s}^{2}_{12}=$$
\begin{equation}
\label{eq207} m^{2}\left|
{{\begin{array}{*{20}c}
 2\left(2+1\right)\mathrm{I}_{5} \hfill & 0 \hfill & 0 \hfill & 0 \hfill\\
 0 \hfill & 0 \hfill & 0 \hfill & 0 \hfill\\
 0 \hfill & 0 \hfill & 2\left(2+1\right)\mathrm{I}_{5} \hfill & 0 \hfill\\
 0 \hfill & 0 \hfill & 0 \hfill & 0 \hfill\\
\end{array} }} \right|,
\end{equation}

\noindent where $\mathrm{I}_{12}$ and $\mathrm{I}_{5}$ are the $12 \times 12$ and $5 \times 5$ unit matrices, respectively.

Hence, above a brief consideration of the RCQM foundations of the
particle-antiparticle multiplet with the mass $m>0$ and the spin
s=(2,0,2,0) has been given. In the limit m=0 this model describes the
partial case of corresponding massless particle-antiparticle multiplet.

\section{A brief scheme of the relativistic canonical quantum mechanics of the 16-component bosonic spin s=(2,1,2,1) particle-antiparticle multiplet}

The 16-component bosonic spin s=(2,1,2,1) particle-antiparticle
multiplet is constructed as the direct sum of the two spin s=(2,1)
multiplets. The principles of constructing and describing such particle-antiparticle multiplet within the
framework of the RCQM are in a complete analogy with the principles
of principles considered in the previous sections. Therefore, the details can be omitted.

The most important fact is that here the link with the Dirac-like
equation is similar to that between the spin s=(1/2,1/2)
particle-antiparticle doublet and the standard 4-component Dirac
equation demonstrated in sections 9, 10. Therefore, the
spin s=(2,1,2,1) particle-antiparticle multiplet is of special
interest. It is much more useful than the spin s=(1,0), (2,0), (2,1) particle
multiplets.

Thus, the Schr$\mathrm{\ddot{o}}$dinger--Foldy equation has the
form

\begin{equation}
\label{eq208} (i\partial_{0}- \widehat{\omega})f(x)=0, \quad
f=\left| {{\begin{array}{*{20}c}
 f^{1} \hfill  \\
 f^{2} \hfill  \\
 .  \\
 .  \\
 .  \\
 f^{16} \hfill  \\
\end{array} }} \right|,
\end{equation}

\noindent where the operator $\widehat{\omega}$ is given in
(59)--(61).

The space of the states is given by

\begin{equation}
\label{eq209}
\mathrm{S}^{3,16}\subset\mathrm{H}^{3,16}\subset\mathrm{S}^{3,16*}.
\end{equation}

The general solution of the Schr$\mathrm{\ddot{o}}$dinger--Foldy
equation (208) is given by

$$f(x)= \left|
{{\begin{array}{*{20}c}
 f_{\mathrm{part}} \hfill  \\
 f_{\mathrm{antipart}} \hfill  \\
\end{array} }} \right|=$$
\begin{equation}
\label{eq210}
\frac{1}{\left(2\pi\right)^{\frac{3}{2}}}\int d^{3}k e^{-ikx}b^{\mathrm{\widetilde{A}}}(\overrightarrow{k})\mathrm{d}_{\mathrm{\widetilde{A}}}, \quad \mathrm{\widetilde{A}}=\overline{1,16},
\end{equation}

\noindent where the orts of the 16-component Cartesian basis have the
form

\begin{equation}
\label{eq211} \mathrm{d}_{1} = \left|
\begin{array}{cccc}
 1 \\
 0 \\
 0 \\
 0 \\
 0 \\
 0 \\
 0 \\
 0 \\
 0 \\
 0 \\
 0 \\
 0 \\
 0 \\
 0 \\
 0 \\
 0 \\
\end{array} \right|, \,
\mathrm{d}_{2} = \left|
\begin{array}{cccc}
 0 \\
 1 \\
 0 \\
 0 \\
 0 \\
 0 \\
 0 \\
 0 \\
 0 \\
 0 \\
 0 \\
 0 \\
 0 \\
 0 \\
 0 \\
 0 \\
\end{array} \right|, \quad  ... \, , \quad   
\mathrm{d}_{16} = \left|
\begin{array}{cccc}
 0 \\
 0 \\
 0 \\
 0 \\
 0 \\
 0 \\
 0 \\
 0 \\
 0 \\
 0 \\
 0 \\
 0 \\
 0 \\
 0 \\
 0 \\
 1 \\
\end{array} \right|.
\end{equation}

The explicit form of the generators of the corresponding SU(2)-spin that satisfy the commutation relations (30) of the SU(2)
algebra is as follows

\begin{equation}
\label{eq212} \overrightarrow{s}_{16}=\left|
{{\begin{array}{*{20}c}
 \overrightarrow{s}_{8} \hfill & 0 \\
 0 \hfill &  -C\overrightarrow{s}_{8}C \\
\end{array} }} \right|,
\end{equation}

\noindent where the $8 \times 8$-matrices $\overrightarrow{s}_{8}$
are given by 

$$s^{1}_{8}= \frac{1}{\sqrt{2}}\left| {{\begin{array}{*{20}c}
 0 \hfill & \sqrt{2} \hfill & 0 \hfill & 0 \hfill & 0 \hfill  & 0 \hfill & 0 \hfill  & 0 \hfill \\
 \sqrt{2} \hfill & 0 \hfill & \sqrt{3} \hfill & 0 \hfill & 0 \hfill & 0 \hfill  & 0 \hfill  & 0 \hfill \\
 0 \hfill & \sqrt{3} \hfill & 0 \hfill & \sqrt{3} \hfill & 0 \hfill & 0 \hfill  & 0 \hfill  & 0 \hfill \\
 0 \hfill & 0 \hfill & \sqrt{3} \hfill & 0 \hfill & \sqrt{2} \hfill & 0 \hfill  & 0 \hfill  & 0 \hfill \\
 0 \hfill & 0 \hfill & 0 \hfill & \sqrt{2} \hfill & 0 \hfill & 0 \hfill  & 0 \hfill  & 0 \hfill \\
 0 \hfill & 0 \hfill & 0 \hfill & 0 \hfill & 0 \hfill & 0 \hfill  & 1 \hfill  & 0 \hfill \\
 0 \hfill & 0 \hfill & 0 \hfill & 0 \hfill & 0 \hfill & 1 \hfill  & 0 \hfill  & 1 \hfill \\
 0 \hfill & 0 \hfill & 0 \hfill & 0 \hfill & 0 \hfill & 0 \hfill  & 1 \hfill  & 0 \hfill \\
 \end{array} }} \right|,$$
\tiny
\begin{equation}
\label{eq213} s^{2}_{8}= \frac{i}{\sqrt{2}}\left| {{\begin{array}{*{20}c}
 0 \hfill & -\sqrt{2} \hfill & 0 \hfill & 0 \hfill & 0 \hfill  & 0 \hfill  & 0 \hfill  & 0 \hfill \\
 \sqrt{2} \hfill & 0 \hfill & -\sqrt{3} \hfill & 0 \hfill & 0 \hfill & 0 \hfill  & 0 \hfill  & 0 \hfill \\
 0 \hfill & \sqrt{3} \hfill & 0 \hfill & -\sqrt{3} \hfill & 0 \hfill & 0 \hfill  & 0 \hfill  & 0 \hfill\\
 0 \hfill & 0 \hfill & \sqrt{3} \hfill & 0 \hfill & -\sqrt{2} \hfill & 0 \hfill  & 0 \hfill  & 0 \hfill\\
 0 \hfill & 0 \hfill & 0 \hfill & \sqrt{2} \hfill & 0 \hfill & 0 \hfill  & 0 \hfill  & 0 \hfill\\
 0 \hfill & 0 \hfill & 0 \hfill & 0 \hfill & 0 \hfill & 0 \hfill & -1 \hfill  & 0 \hfill \\
 0 \hfill & 0 \hfill & 0 \hfill & 0 \hfill & 0 \hfill & 1 \hfill  & 0 \hfill  & -1 \hfill\\
 0 \hfill & 0 \hfill & 0 \hfill & 0 \hfill & 0 \hfill & 0 \hfill & 1 \hfill  & 0 \hfill \\
 \end{array} }} \right|,
\end{equation}
\normalsize
$$s^{3}_{8}= \left| {{\begin{array}{*{20}c}
 2 \hfill & 0 \hfill & 0 \hfill & 0 \hfill & 0 \hfill  & 0 \hfill & 0 \hfill  & 0 \hfill \\
 0 \hfill & 1 \hfill & 0 \hfill & 0 \hfill & 0 \hfill & 0 \hfill & 0 \hfill  & 0 \hfill \\
 0 \hfill & 0 \hfill & 0 \hfill & 0 \hfill & 0 \hfill & 0 \hfill & 0 \hfill  & 0 \hfill \\
 0 \hfill & 0 \hfill & 0 \hfill & -1 \hfill & 0 \hfill & 0 \hfill & 0 \hfill  & 0 \hfill \\
 0 \hfill & 0 \hfill & 0 \hfill & 0 \hfill & -2 \hfill & 0 \hfill & 0 \hfill  & 0 \hfill \\
 0 \hfill & 0 \hfill & 0 \hfill & 0 \hfill & 0 \hfill & 1 \hfill & 0 \hfill  & 0 \hfill \\
 0 \hfill & 0 \hfill & 0 \hfill & 0 \hfill & 0 \hfill & 0 \hfill & 0 \hfill  & 0 \hfill \\
 0 \hfill & 0 \hfill & 0 \hfill & 0 \hfill & 0 \hfill & 0 \hfill & 0 \hfill  & -1 \hfill \\
 \end{array} }} \right|,$$

\noindent and $C\mathrm{I}_{8}$ is $8 \times 8$ diagonal matrix operator of complex conjugation.

The Casimir operator is given by the following $16 \times 16$ diagonal matrix

\begin{equation}
\label{eq214} \overrightarrow{s}^{2}_{16}= \left|
{{\begin{array}{*{20}c}
 6\mathrm{I}_{5} \hfill & 0 \hfill & 0 \hfill & 0 \hfill\\
 0 \hfill & 2\mathrm{I}_{3} \hfill & 0 \hfill & 0 \hfill\\
 0 \hfill & 0 \hfill & 6\mathrm{I}_{5} \hfill & 0 \hfill\\
 0 \hfill & 0 \hfill & 0 \hfill & 2\mathrm{I}_{3} \hfill\\
\end{array} }} \right|,
\end{equation}

\noindent where $\mathrm{I}_{5}$ and $\mathrm{I}_{3}$ are  $5\times 5$ and $3\times 3$ unit matrices, respectively.

The stationary complete set of operators is given by $\overrightarrow{p}, \, s_{16}^{3}=s_{z}$ and the equations on the eigenvalues of the operators $\overrightarrow{p}$ and $s_{16}^{3}=s_{z}$ have the form

\begin{equation}
\label{eq215} \overrightarrow{p}e^{-ikx}\mathrm{d}_{\mathrm{\widetilde{A}}} =
\overrightarrow{k}e^{-ikx}\mathrm{d}_{\mathrm{\widetilde{A}}}, \quad \mathrm{\widetilde{A}}=\overline{1,16},
\end{equation}

$$s_{16}^{3}\mathrm{d}_{1} = 2\mathrm{d}_{1}, \, s_{16}^{3}\mathrm{d}_{2} = \mathrm{d}_{2}, \, s_{16}^{3}\mathrm{d}_{3} = 0, \,
s_{16}^{3}\mathrm{d}_{4} = -\mathrm{d}_{4},$$
\begin{equation}
\label{eq216} s_{16}^{3}\mathrm{d}_{5} = -2\mathrm{d}_{5}, \,
s_{16}^{3}\mathrm{d}_{6} = \mathrm{d}_{6}, \, s_{16}^{3}\mathrm{d}_{7} =
0, \, s_{16}^{3}\mathrm{d}_{8} = -\mathrm{d}_{8}.
\end{equation}
\small
$$s_{16}^{3}\mathrm{d}_{9} = -2\mathrm{d}_{9}, \, s_{16}^{3}\mathrm{d}_{10} = -\mathrm{d}_{10}, \, s_{16}^{3}\mathrm{d}_{11} = 0, \,
s_{16}^{3}\mathrm{d}_{12} = \mathrm{d}_{12},$$
$$s_{16}^{3}\mathrm{d}_{13} = 2\mathrm{d}_{13}, \, s_{16}^{3}\mathrm{d}_{14} = -\mathrm{d}_{14}, \, s_{16}^{3}\mathrm{d}_{15} =0, \,
s_{16}^{3}\mathrm{d}_{16} = \mathrm{d}_{16}.$$
\normalsize

Therefore, the functions $b^{1}(\overrightarrow{k})-b^{5}(\overrightarrow{k})$ in
solution (210) are the momentum-spin amplitudes of the massive
boson with the spin s=2 and the eigenvalues of the spin projection $(2,1,0,-1,-2)$,
respectively, $b^{6}(\overrightarrow{k})-b^{8}(\overrightarrow{k})$ are the amplitudes of the massive boson with the spin s=1 and the eigenvalues of the spin projection $(1,0,-1)$,
respectively,; $b^{9}(\overrightarrow{k})-b^{13}(\overrightarrow{k})$ in
solution (210) are the momentum-spin amplitudes of the massive
antiboson with the spin s=2 and the eigenvalues of the spin projection $(-2,-1,0,1,2)$,
respectively, $b^{14}(\overrightarrow{k})-b^{16}(\overrightarrow{k})$ are the amplitudes of the massive antiboson with the spin s=1 and the eigenvalues of the spin projection $(-1,0,1)$,
respectively.

The Schr$\mathrm{\ddot{o}}$dinger--Foldy equation (208) (and the
set $\{\mathrm{f}\}$ of its solutions (210)) is invariant with
respect to the reducible unitary bosonic representation (14) of
the Poincar$\mathrm{\acute{e}}$ group $\mathcal{P}$, whose
Hermitian $16 \times 16$ matrix-differential generators are given
by (12), (13), where the spin s=(2,1,2,1) SU(2) generators
$\overrightarrow{s}_{16}=(s_{16}^{\ell n})$ are given in (212).

The proof is similar to that given in section 2 after formula (14). The Casimir operators of this reducible bosonic spin s=(2,1,2,1) representation of the group $\mathcal{P}$ have the form

\begin{equation}
\label{eq217} p^{2}=\widehat{p}^{\mu}\widehat{p}_{\mu}=m^{2}\mathrm{I}_{16},
\end{equation}
$$W=w^{\mu}w_{\mu}=m^{2}\overrightarrow{s}_{8}^{2}=$$
\small
\begin{equation}
\label{eq218} m^{2}\left|
{{\begin{array}{*{20}c}
 2\left(2+1\right)\mathrm{I}_{5} \hfill & 0 \hfill & 0 \hfill & 0 \hfill\\
 0 \hfill & 1\left(1+1\right)\mathrm{I}_{3} \hfill & 0 \hfill & 0 \hfill\\
 0 \hfill & 0 \hfill & 2\left(2+1\right)\mathrm{I}_{5} \hfill & 0 \hfill\\
 0 \hfill & 0 \hfill & 0 \hfill & 1\left(1+1\right)\mathrm{I}_{3} \hfill\\
\end{array} }} \right|,
\end{equation}
\normalsize

\noindent where $\mathrm{I}_{16}$, $\mathrm{I}_{5}$ and $\mathrm{I}_{3}$ are $16\times 16$, $5\times 5$ and $3\times 3$ unit matrices, respectively.

Thus, above the foundations of the RCQM of the 16-component multiplet of two bosons with the spins s=(2,1) and their antiparticle doublet are considered briefly. It is the basis for the transition to the covariant local field theory of the spin s=(2,1,2,1) particle-antiparticle multiplet given below.

\section{On the relativistic canonical quantum mechanics of the arbitrary mass and spin particle multiplets}

Taking into account the step by step consideration of the different partial examples, which are given in the sections 7--17, the generalization for the particle multiplet of arbitrary spin can be formulated.

The theory is completely similar to the given in section 2, where the RCQM of the particle singlet of arbitrary spin is presented. The specification is only in application of the reducible representation of the SU(2) and the Poincar$\mathrm{\acute{e}}$ $\mathcal{P}$ groups.

Furthermore, the simple particle multiplets are constructed as the ordinary direct sum of the corresponding particle characteristics. The particle-antiparticle doublets and multiplets are constructed as the specific direct sum of the particle and antiparticle characteristics, in which the antiparticle is considered as the mirror reflection of the particle and the information about equal and positive masses of the particle and antiparticle is inserted. Therefore, and according to the Pauli principle, the third component of the antiparticle spin has the opposite sign to the third component of the particle spin. The partial examples are given in the formulae (64), (91), (153), (154), (172), (178), (179), (190), (191), (201), (212).

\section{On the transition to the nonrelativistic canonical quantum mechanics of the arbitrary mass and spin}

In the nonrelativistic limit the Schr$\mathrm{\ddot{o}}$dinger--Foldy equation (3) is transformed into the ordinary Schr$\mathrm{\ddot{o}}$dinger equation   

\begin{equation}
\label{eq219} \left(i\partial_{0}-\frac{\overrightarrow{p}^{2}}{2m}\right)\psi(x)=0,
\end{equation}

\noindent for the N-component wave function

\begin{equation}
\label{eq220}
\psi \equiv \mathrm{column}(\psi^{1},\psi^{2},...,\psi^{\mathrm {N}}), \quad \mathrm {N}=2\mathrm {s}+1.
\end{equation}

Each of the above considered equations of the RCQM (2-, 3-, 4-, 5-, 6-, 8-, 10-, 12-, 16-component) is transformed into the equation (219) with the corresponding number of the components.

Moreover, here the SU(2) spin operator is the same as in the RCQM and is given by the formulae (7). Therefore, here as in the RCQM, the SU(2) generators for the spin s=1/2 are given in (19), (20), for the spin s=1 in (29), for the spin s=1 in (29), for the spin s=3/2 in (39), for the spin s=2 in (48), e.t.c. for the multiplet SU(2) spins.

Therefore, the equation (219) is invariant with respect to same representations of the SU(2) group, with respect to which the relativistic equation (3) is invariant. The difference is in the application of the Galilean group and its representations instead of the Poincar$\mathrm{\acute{e}}$ $\mathcal{P}$ group and its representations.

For the models with interaction it is much more easier to solve the equation (219) with interaction potential $V(x)$ instead of pseudo-differential equation (3). Moreover, the solutions of the equation (219) with interaction can be useful for the obtaining of the corresponding solutions of the equation (3).

Thus, the equation (219) can be useful not only itself but for the different approximations of the relativistic equation (3) as well. 

\section{Transition from the nonlocal relativistic canonical quantum mechanics to the covariant local relativistic field theory}

The above formulated RCQM has the independent meaning as the useful model for the elementary particle physics. However, another application of the RCQM model has the important meaning as well. Each model of the quantum-mechanical particle singlet or multiplet considered above can be formulated also in the framework of the covariant local relativistic field theory. Moreover, it is not difficult to find the link between the RCQM and the covariant local relativistic field theory. For the partial case of spin s=(1/2,1/2) particle-antiparticle doublet such link is already given by $4 \times 4$ matrix-differential operators $v$ (102) multiply on $V$ (119), or resulting $W$ (120)--(123). Below for the case of higher spins doublets and multiplets the corresponding $6 \times 6$, $8 \times 8$, $10 \times 10$, $12 \times 12$, $16 \times 16$ transition operators are found. Therefore, below the relationship between the Schr$\mathrm{\ddot{o}}$dinger--Foldy equations (168) (and eq. (168) for the spin s=3/2), (197), (208) for different multiplets and the Dirac equation (or the Dirac-like equations) for these multiplets is introduced.

Thus, below the method of derivation of the different equations of the covariant local relativistic field theory is suggested. The start of such derivation is given from the corresponding Schr$\mathrm{\ddot{o}}$dinger--Foldy equations of RCQM. Note that derivation of the covariant particle equations from the equations in the FW representation (the so-called Foldy synthesis [13]) is well known. Here the new possibilities to fulfill the derivation of the covariant particle equations starting from the RCQM are open. The new covariant equations of the local relativistic field theory for the spin s=(3/2,3/2), s=(1,0,1,0), s=(2,0,2,0), s=(2,1,2,1) particle-antiparticle doublets and multiplets are found here by this method. The new equations for the spin s=(1,1) and s=(2,2) particle-antiparticle doublets in the FW representation are introduced as well.

These results are presented in the next sections.

\section{Field equation for the 6-component spin s=(1,1) particle-antiparticle doublet}

The RCQM of the spin s=(1,1) particle-antiparticle doublet is considered in section 11. Transition to the canonical field theory is fulfilled in a complete analogy with the method suggested in [3, 14] (see also sections 9, 10 here). The difference is only in the explicit form of the transition operator, which is now given by the $6 \times 6$ diagonal matrix with the $3 \times 3$ operator of the complex conjugation in the bottom part.

Thus, the Schr$\mathrm{\ddot{o}}$dinger--Foldy equation (149) and its solution (150) are linked with the canonical field theory equation

\begin{equation}
\label{eq221} (i\partial_{0}-
\Gamma_{6}^{0}\widehat{\omega})\phi(x)=0, \quad \Gamma_{6}^{0} =
\left| {{\begin{array}{*{20}c}
 \mathrm{I}_{3} \hfill &  0 \hfill\\
 0 \hfill & -\mathrm{I}_{3}  \hfill\\
 \end{array} }} \right|,
\end{equation}
$$\widehat{\omega}\equiv \sqrt{-\Delta+m^{2}},$$

\noindent and its solution

\begin{equation}
\label{eq222} \phi(x)=
\frac{1}{\left(2\pi\right)^{\frac{3}{2}}}\int
d^{3}k\left[e^{-ikx}c^{\ell}(\overrightarrow{k})\mathrm{d}_{\ell}+e^{ikx}c^{*\breve{\ell}}(\overrightarrow{k})\mathrm{d}_{{\breve{\ell}}}\right],
\end{equation}
$$\ell =1,2,3, \quad \breve{\ell} =4,5,6,$$

\noindent with the help of the following operator

\begin{equation}
\label{eq223} v_{6} = \left| {{\begin{array}{*{20}c}
 \mathrm{I}_{3} \hfill &  0 \hfill\\
 0 \hfill & C\mathrm{I}_{3}  \hfill\\
 \end{array} }} \right|, \quad v_{6}^{-1}= v_{6}^{\dag}= v_{6}, \quad v_{6}v_{6}=\mathrm{I}_{6},
\end{equation}

\noindent where the Cartesian orts $(\mathrm{d}_{\ell}, \, \mathrm{d}_{{\breve{\ell}}})$ are given in (151), $C$ is the operator of the complex conjugation and $\mathrm{I}_{3}$ is the $3\times 3$ unit matrix.

The operator $v_{6}$ (223) transforms an arbitrary operator $\hat{q}_{\mathrm{qm}}$ of the 6-component RCQM into the corresponding operator $\hat{q}_{\mathrm{cf}}$ of the canonical field theory and \textit{vice versa}:

\begin{equation}
\label{eq224} v_{6}\hat{q}_{\mathrm{qm}}^{\mathrm{anti-Herm}}v_{6}
= \hat{q}_{\mathrm{cf}}^{\mathrm{anti-Herm}},
\end{equation}
\begin{equation}
\label{eq225} v_{6}\hat{q}_{\mathrm{cf}}^{\mathrm{anti-Herm}}v_{6}
= \hat{q}_{\mathrm{qm}}^{\mathrm{anti-Herm}},
\end{equation}

\noindent where the operators $\hat{q}_{\mathrm{qm}}$ and $\hat{q}_{\mathrm{cf}}$ must be taken in the anti-Hermitian form. The corresponding link between solutions (150) and (222) is as follows

\begin{equation}
\label{eq226} \phi=v_{6} f, \quad f=v_{6}\phi.
\end{equation}

Note that formulae (224), (225) are valid for the anti-Hermitian (prime) operators only. For the goals stated in our previous papers [24--28], we often use the operators in the anti-Hermitian form (see also the comments in section 9). The mathematical correctness of such choice and the physical interpretation are explained in the books [29, 30]. Return to the Hermitian operators is very easy.

The important examples of the transitions (224), (225) are the transformations of the operators of equations (149), (221)

\begin{equation}
\label{eq227} v_{6}(\partial_{0}+i\widehat{\omega})v_{6}=
\partial_{0}+i \Gamma_{6}^{0}\widehat{\omega},
\end{equation}
\begin{equation}
\label{eq228}
v_{6}(\partial_{0}+i\Gamma_{6}^{0}\widehat{\omega})v_{6}=
\partial_{0}+i\widehat{\omega},
\end{equation}

\noindent and of the SU(2) spin operators. The spin operators of the canonical field theory found from the RCQM SU(2) spin (153) on the basis of the transformation (223) satisfy the commutation relations (30) and have the following form

\begin{equation}
\label{eq229} \overrightarrow{s}= \left| {{\begin{array}{*{20}c}
 \overrightarrow{s}_{3} \hfill  & 0 \hfill \\
 0 \hfill & \overrightarrow{s}_{3} \hfill \\
 \end{array} }} \right|,
\end{equation}

\noindent where the $3 \times 3$ spin s=1 SU(2) generators are denoted as $\overrightarrow{s}_{3}$ and are given in (29). The corresponding Casimir operator is given by

\begin{equation}
\label{eq230} \overrightarrow{s}^{2}= 2\mathrm{I}_{6}=
1(1+1)\mathrm{I}_{6}.
\end{equation}

\noindent where $\mathrm{I}_{6}$ is the $6 \times 6$- unit matrix.

The stationary complete set of operators is given by the operators $\overrightarrow{p}, \, s^{3}=s_{z}$ of the momentum and spin
projection. The equations on the eigenvectors and eigenvalues of the operators $\overrightarrow{p}$ and $s^{3}=s_{z}$ from (229) have the form

$$\overrightarrow{p}e^{-ikx}\mathrm{d}_{\ell} =
\overrightarrow{k}e^{-ikx}\mathrm{d}_{\ell}, \quad \ell =1,2,3,$$
$$\overrightarrow{p}e^{ikx}\mathrm{d}_{\breve{\ell}} =
-\overrightarrow{k}e^{ikx}\mathrm{d}_{\breve{\ell}}, \quad \breve{\ell} =4,5,6,$$
\begin{equation}
\label{eq231} s^{3}\mathrm{d}_{1} = \mathrm{d}_{1}, \, s^{3}\mathrm{d}_{2} = 0, \, s^{3}\mathrm{d}_{3} = - \mathrm{d}_{3},
\end{equation}
$$s^{3}\mathrm{d}_{4} =  \mathrm{d}_{4}, \,
s^{3}\mathrm{d}_{5} = 0, \, s^{3}\mathrm{d}_{6} = -
\mathrm{d}_{6},$$

\noindent and determine the interpretation of the amplitudes in the general solution (222). Note that the direct quantum-mechanical interpretation of the amplitudes $c^{\ell}(\overrightarrow{k}), \, c^{*\breve{\ell}}(\overrightarrow{k})$ in solution (222) should be taken from the quantum-mechanical equations (156), (157) and is given in section 11.

The relativistic invariance of the canonical field equation (221) follows from the corresponding invariance of the Schr$\mathrm{\ddot{o}}$dinger--Foldy equation (149) and transformation (223)--(226) (for the anti-Hermitian operators). The explicit form of the corresponding generators follows from of the explicit form the generators (12), (13) with the spin matrices (153), (154) and  transformation (223)--(225).

Thus, the canonical field equation (221) and the set
$\left\{\phi\right\}$ of its solutions (222) are invariant with respect
to the reducible unitary bosonic representation (14) of the
Poincar$\mathrm{\acute{e}}$ group $\mathcal{P}$, whose Hermitian  $6 \times 6$ matrix-differential generators are
given by

$$\widehat{p}^{0}=\Gamma_{6}^{0}\widehat{\omega}\equiv \Gamma_{6}^{0}\sqrt{-\Delta+m^{2}}, \quad \widehat{p}^{\ell}=-i\partial_{\ell},$$
\begin{equation}
\label{eq232} \widehat{j}^{\ell
n}=x^{\ell}\widehat{p}^{n}-x^{n}\widehat{p}^{\ell}+s^{\ell
n}\equiv \widehat{m}^{\ell n}+s^{\ell n},
\end{equation}
\begin{equation}
\label{eq233} \widehat{j}^{0 \ell}=-\widehat{j}^{\ell
0}=x^{0}\widehat{p}^{\ell}-\frac{1}{2}\Gamma_{6}^{0}\left\{x^{\ell},\widehat{\omega}\right\}+\Gamma_{6}^{0}\frac{(\overrightarrow{s}\times
\overrightarrow{p})^{\ell}}{\widehat{\omega}+m},
\end{equation}

\noindent where the spin s=(1,1) SU(2) generators
$\overrightarrow{s}=(s^{\ell n})$ have the form (229).

It is easy to prove by the direct verification that generators (232), (233) commute with the operator $(i\partial_{0}- \Gamma_{6}^{0}\widehat{\omega})$ of the canonical field equation (221) and satisfy the commutation relations (11) of the Lie algebra of the Poincar$\mathrm{\acute{e}}$ group $\mathcal{P}$.

The corresponding Casimir operators are given by

\begin{equation}
\label{eq234}
p^{2}=\widehat{p}^{\mu}\widehat{p}_{\mu}=m^{2}\mathrm{I}_{6},
\end{equation}
\begin{equation}
\label{eq235} W=w^{\mu}w_{\mu}=m^{2}\overrightarrow{s}^{2}=
1\left(1+1\right)m^{2}\mathrm{I}_{6},
\end{equation}

\noindent where $\mathrm{I}_{6}$ is the $6 \times 6$ unit matrix. Note that the difference between the Casimir operators (234), (235) and the corresponding expressions in RCQM, see the formulae (158), (159) above, is only in the explicit form of the operator $\overrightarrow{s}$.

Thus, due to the eigenvalues in equations (231), positive and
negative frequencies form of the solution (222) and the
Bargman--Wigner analysis of the Casimir operators (234), (235), one
can come to a conclusion that equation (221) describes the
canonical field (the bosonic particle-antiparticle doublet) with
the spins s=(1,1) and $m>0$. Transition to the $m=0$ limit leads
to the canonical field equation for the photon-antiphoton field.

Final transition to the covariant local field theory in the case of this 6-component doublet contains the specific features and is the subject of next publication. The result is the 6-component complex Maxwell-like equation for the spin s=(1,1) and $m>0$ classical field. In the case of $m=0$ the system of the 6-component complex equations for the electromagnetic field is found, which, after quantization, describes the photon and antiphoton states. Nevertheless, below in section 23 the corresponding equation, which follows from the RCQM of spin s=(1,1) particle antiparticle doublet, is found as the partial case of manifestly covariant equation for the spin s=(1,0,1,0) multiplet. 

\section{Covariant field equation for the 8-component spin s=(1,0,1,0) bosonic particle-antiparticle multiplet}

In this section, the field model of the spin s=(1,0) bosonic
multiplet and the corresponding spin s=(1,0) antiparticle
multiplet is constructed. At first, when the canonical field
equation is derived, the complete analogy with the case of the
spin s=(1,1) model of the previous section is used. Further, the
covariant local field equation for the spin s=(1,0,1,0) bosonic
particle-antiparticle multiplet is introduced. The analogy in the
beginning of this consideration with the previous section enabled
me to omit some details.

The start of this derivation is given in section 13, where the
RCQM of the 8-component bosonic spin s=(1,0,1,0)
particle-antiparticle multiplet is considered. The second step is
the transition from the Schr$\mathrm{\ddot{o}}$dinger--Foldy
equation (168) to the canonical field equation. This step, as
shown in previous section, is possible only for the
anti-Hermitian form of the operators. Nevertheless, the resulting
operators can be chosen in the standard Hermitian form and do not
contain the operator $C$ of complex conjugation. The last step of
the transition from the canonical field equation to the covariant
local field equation is fulfilled in analogy with the FW
transformation (119).

Thus, the corresponding canonical field equation is found in the
form

\begin{equation}
\label{eq236} (i\partial_{0}-
\Gamma_{8}^{0}\widehat{\omega})\phi(x)=0, \quad \phi=\left|
{{\begin{array}{*{20}c}
 \phi^{1} \hfill  \\
 \phi^{2} \hfill  \\
 \phi^{3} \hfill  \\
 \phi^{4} \hfill  \\
 \phi^{5} \hfill  \\
 \phi^{6} \hfill  \\
 \phi^{7} \hfill  \\
 \phi^{8} \hfill  \\
\end{array} }} \right|,
\end{equation}

where

$$\widehat{\omega}\equiv \sqrt{-\Delta+m^{2}},$$
\begin{equation}
\label{eq237} \Gamma_{8}^{0} = \left| {{\begin{array}{*{20}c}
 \mathrm{I}_{4} \hfill &  0 \hfill\\
 0 \hfill & -\mathrm{I}_{4}  \hfill\\
 \end{array} }} \right|, \quad \mathrm{I}_{4}=\left|
{{\begin{array}{*{20}c}
 1 \hfill & 0 \hfill & 0 \hfill & 0 \hfill\\
 0 \hfill & 1 \hfill & 0 \hfill & 0 \hfill\\
 0 \hfill & 0 \hfill & 1 \hfill & 0 \hfill\\
 0 \hfill & 0 \hfill & 0 \hfill & 1 \hfill\\
\end{array} }} \right|.
\end{equation}

The general solution of equation (236) is given by

$$\phi(x)= \frac{1}{\left(2\pi\right)^{\frac{3}{2}}}\int d^{3}k$$
\begin{equation}
\label{eq238}
\left[e^{-ikx}c^{\mathrm{A}}(\overrightarrow{k})\mathrm{d}_{\mathrm{A}}+e^{ikx}c^{*\mathrm{B}}(\overrightarrow{k})\mathrm{d}_{\mathrm{B}}\right],
\end{equation}

\noindent where $\mathrm{A}=\overline{1,4}, \,
\mathrm{B}=\overline{5,8}$ and the orts of the 8-component
Cartesian basis are given in (171).

The transition from the
Schr$\mathrm{\ddot{o}}$dinger--Foldy equation (168) and its
solution (170) to equation (236) and solution (238) is given by the
operator

\begin{equation}
\label{eq239} v_{8} = \left| {{\begin{array}{*{20}c}
 \mathrm{I}_{4} \hfill &  0 \hfill\\
 0 \hfill & C\mathrm{I}_{4}  \hfill\\
 \end{array} }} \right|, \quad v_{8}^{-1}= v_{8}^{\dag}= v_{8}, \quad v_{8}v_{8}=\mathrm{I}_{8},
\end{equation}
\begin{equation}
\label{eq240} \phi=v_{8} f, \quad f=v_{8}\phi,
\end{equation}
\begin{equation}
\label{eq241} v_{8}\hat{q}_{\mathrm{qm}}^{\mathrm{anti-Herm}}v_{8}
= \hat{q}_{\mathrm{cf}}^{\mathrm{anti-Herm}},
\end{equation}
\begin{equation}
\label{eq242} v_{8}\hat{q}_{\mathrm{cf}}^{\mathrm{anti-Herm}}v_{8}
= \hat{q}_{\mathrm{qm}}^{\mathrm{anti-Herm}}.
\end{equation}

\noindent Here $\hat{q}_{\mathrm{qm}}^{\mathrm{anti-Herm}}$ is an arbitrary operator from the RCQM of the 8-component particle-antiparticle doublet in the anti-Hermitian form, e. g., the operator $(\partial_{0}+i\widehat{\omega})$ of equation of motion, the operator of spin (172), etc.,  $\hat{q}_{\mathrm{cf}}^{\mathrm{anti-Herm}}$ is an arbitrary operator from the canonical field theory of the 8-component particle-antiparticle doublet in the anti-Hermitian form, $C\mathrm{I}_{4}$ is the $4\times 4$ operator of complex conjugation.

The SU(2) spin operators, which satisfy the commutation relations
(30) and commute with the operator $(i\partial_{0}-
\Gamma_{8}^{0}\widehat{\omega})$ of the equation of motion (98),
are derived from the RCQM operators (172) on the basis of
transformations (239), (241), (242). These canonical field spin
operators are given by

\begin{equation}
\label{eq243} \overrightarrow{s}_{8}=\left|
{{\begin{array}{*{20}c}
 \overrightarrow{s} \hfill & 0 \\
 0 \hfill & \overrightarrow{s} \\
\end{array} }} \right|,
\end{equation}

\noindent where the $4 \times 4$ operators $\overrightarrow{s}$ are
given in (163). The corresponding Casimir operator is given by the $8 \times 8$ diagonal matrix (173). Note that this Casimir operator is the same both for the spin (172) and for the spin (243).

The stationary complete set of operators is given by the operators $\overrightarrow{p}, \, s_{8}^{3}=s_{z}$ of the momentum and spin
projection. The equations on the eigenvectors and eigenvalues of the spin projection operator $s_{8}^{3}=s_{z}$ from (243) have the form

$$s_{8}^{3}\mathrm{d}_{1} = \mathrm{d}_{1}, \, s_{8}^{3}\mathrm{d}_{2} = 0, \, s_{8}^{3}\mathrm{d}_{3} = -\mathrm{d}_{3}, \,
s_{8}^{3}\mathrm{d}_{4} = 0,$$
\begin{equation}
\label{eq244} s_{8}^{3}\mathrm{d}_{5} = \mathrm{d}_{5}, \,
s_{8}^{3}\mathrm{d}_{6} = 0, \, s_{8}^{3}\mathrm{d}_{7} =
-\mathrm{d}_{7}, \, s_{8}^{3}\mathrm{d}_{8} = 0.
\end{equation}

Therefore, the functions $c^{1}(\overrightarrow{k}), \,
c^{2}(\overrightarrow{k}), \, c^{3}(\overrightarrow{k})$ in
solution (238) are the momentum-spin amplitudes of the massive
boson with the spin s=1 and the spin projection $(1,0,-1)$,
respectively, $c^{4}(\overrightarrow{k})$ is the amplitude of the
spinless boson; $c^{5}(\overrightarrow{k}), \,
c^{6}(\overrightarrow{k}), \, c^{7}(\overrightarrow{k})$ are the
momentum-spin amplitudes of the antiparticle (antiboson) with the
spin s=1 and the spin projection $(1,0,-1)$, respectively,
$c^{8}(\overrightarrow{k})$ is the amplitude of the spinless
antiboson.

Note that direct quantum-mechanical interpretation of the amplitudes in solution (238) should be given in the framework of the RCQM. Such interpretation is already given in section 13 in paragraph after equations (174). 

The generators of the reducible unitary bosonic spin s=(1,0,1,0)
multiplet representation of the Poincar$\mathrm{\acute{e}}$ group
$\mathcal{P}$, with respect to which the canonical field equation
(236) and the set $\left\{\phi\right\}$ of its solutions (238) are
invariant, are derived from the RCQM set of generators (12), (13)
with the spin (172) on the basis of the transformations (239),
(241), (242). These Hermitian  $8 \times 8$ matrix-differential
generators are given by

$$\widehat{p}^{0}=\Gamma_{8}^{0}\widehat{\omega}\equiv \Gamma_{8}^{0}\sqrt{-\Delta+m^{2}}, \quad \widehat{p}^{\ell}=-i\partial_{\ell},$$
\begin{equation}
\label{eq245} \widehat{j}^{\ell
n}=x^{\ell}\widehat{p}^{n}-x^{n}\widehat{p}^{\ell}+s_{8}^{\ell
n}\equiv \widehat{m}^{\ell n}+s_{8}^{\ell n},
\end{equation}
\begin{equation}
\label{eq246} \widehat{j}^{0 \ell}=-\widehat{j}^{\ell
0}=x^{0}\widehat{p}^{\ell}-\frac{1}{2}\Gamma_{8}^{0}\left\{x^{\ell},\widehat{\omega}\right\}+\Gamma_{8}^{0}\frac{(\overrightarrow{s}_{8}\times
\overrightarrow{p})^{\ell}}{\widehat{\omega}+m},
\end{equation}

\noindent where the spin s=(1,0,1,0) SU(2) generators
$\overrightarrow{s}_{8}=(s_{8}^{\ell n})$ have the form (243).

It is easy to prove by the direct verification that the generators (245), (246) commute with the operator $(i\partial_{0}- \Gamma_{8}^{0}\widehat{\omega})$ of the canonical field equation (236) and satisfy the commutation relations (11) of the Lie algebra of the Poincar$\mathrm{\acute{e}}$ group $\mathcal{P}$. The Casimir operators for the representation (245), (246) are given by

\begin{equation}
\label{eq247} p^{2}=\widehat{p}^{\mu}\widehat{p}_{\mu}=m^{2}\mathrm{I}_{8},
\end{equation}
$$W=w^{\mu}w_{\mu}=m^{2}\overrightarrow{s}_{8}^{2}=$$
\begin{equation}
\label{eq248} m^{2}\left| {{\begin{array}{*{20}c}
 1\left(1+1\right)\mathrm{I}_{3} \hfill & 0 \hfill & 0 \hfill & 0 \hfill\\
 0 \hfill & 0 \hfill & 0 \hfill & 0 \hfill\\
 0 \hfill & 0 \hfill & 1\left(1+1\right)\mathrm{I}_{3} \hfill & 0 \hfill\\
 0 \hfill & 0 \hfill & 0 \hfill & 0 \hfill\\
\end{array} }} \right|,
\end{equation}

\noindent where $\mathrm{I}_{8}$ and $\mathrm{I}_{3}$ are $8\times 8$ and $3\times 3$ unit matrices, respectively.

Thus, due to the eigenvalues in equations (244), positive and
negative frequencies form of solution (238) and the
Bargman--Wigner analysis of the Casimir operators (247), (248) one
can come to a conclusion that equation (244) describes the
8-component canonical field (the bosonic particle-antiparticle
doublet) with the spins s=(1,0,1,0) and $m>0$. Transition to the
$m=0$ limit leads to the canonical field equation for the
8-component (photon massless spinless boson)-(antiphoton-
masslees spinless antiboson) field.

The operator of the transition to the covariant local field theory
representation (the $8 \times 8$  analogy of the $4 \times 4$ FW
transformation operator (119)) is given by

\begin{equation}
\label{eq249} V^{\mp}=\frac{\mp
\overrightarrow{\Gamma}_{8}\cdot\overrightarrow{p}+\widehat{\omega}+m}{\sqrt{2\widehat{\omega}(\widehat{\omega}+m)}},
\quad V^{-}=(V^{+})^{\dag},
\end{equation}
$$V^{-}V^{+}=V^{+}V^{-}=\mathrm{I}_{8},$$
\begin{equation}
\label{eq250} \psi=V^{-}\phi, \, \phi = V^{+}\psi,
\end{equation}
\begin{equation}
\label{eq251} \hat{q}_{\mathrm{D}}=
V^{-}\hat{q}_{\mathrm{CF}}V^{+}, \, \hat{q}_{\mathrm{CF}}=
V^{+}\hat{q}_{\mathrm{D}}V^{-},
\end{equation}

\noindent where $\hat{q}_{\mathrm{D}}$ is an arbitrary operator
(both in the Hermitian and anti-Hermitian form) in the covariant
local field theory representation. The inverse transformation is
valid as well.

Thus, on the basis of transformation (249)--(251) the 8-component
Dirac-like equation is found from the canonical field equation
(236) in the form

\begin{equation}
\label{eq252}
\left[i\partial_{0}-\Gamma_{8}^{0}(\overrightarrow{\Gamma}_{8}\cdot\overrightarrow{p}+m)\right]\psi(x)=0.
\end{equation}

\noindent In formula (249) and in equation (252) the
$\Gamma_{8}^{\mu}$ matrices are given by

\begin{equation}
\label{eq253} \Gamma_{8}^{0}=\left| {{\begin{array}{*{20}c}
 \mathrm{I}_{4} \hfill & 0 \\
 0 \hfill & -\mathrm{I}_{4} \\
\end{array} }} \right|, \quad \Gamma_{8}^{j}=\left| {{\begin{array}{*{20}c}
 0 \hfill & \Sigma^{j} \\
 -\Sigma^{j} \hfill & 0 \\
\end{array} }} \right|,
\end{equation}

\noindent where $\Sigma^{j}$ are the $4\times 4$ Pauli matrices

\begin{equation}
\label{eq254} \Sigma^{j}=\left| {{\begin{array}{*{20}c}
 \sigma^{j} \hfill & 0 \\
 0 \hfill & \sigma^{j} \\
\end{array} }} \right|,
\end{equation}

\noindent and $\sigma^{j}$ are the standard $2\times 2$ Pauli matrices (20).

The matrices $\Sigma^{j}$ satisfy the similar commutation
relations as the $2\times 2$ Pauli matrices (20) and have other
similar properties. The matrices $\Gamma_{8}^{\mu}$ (253) satisfy
the anticommutation relations of the Clifford--Dirac algebra in
the form

\begin{equation}
\label{eq255}
\Gamma_{8}^{\mu}\Gamma_{8}^{\nu}+\Gamma_{8}^{\nu}\Gamma_{8}^{\mu}=2g^{\mu\nu}.
\end{equation}

It is evident that equation (252) is not the ordinary direct sum
of the two Dirac equations. Therefore, it is not the complex
Dirac--Kahler equation [34]. Moreover, it is not the standard
Dirac--Kahler equation [35].

The solution of equation (252) is derived from solution (238) of
this equation in the canonical representation (236) on the basis
of transformation (249), (250) and is given by

$$\psi(x)=V^{-}\phi(x)= \frac{1}{\left(2\pi\right)^{\frac{3}{2}}}\int d^{3}k$$
\begin{equation}
\label{eq256}
\left[e^{-ikx}c^{\mathrm{A}}(\overrightarrow{k})\mathrm{v}^{-}_{\mathrm{A}}(\overrightarrow{k})+e^{ikx}c^{*\mathrm{B}}(\overrightarrow{k})\mathrm{v}^{+}_{\mathrm{B}}(\overrightarrow{k})\right],
\end{equation}

\noindent where $\mathrm{A}=\overline{1,4}, \, \mathrm{B}=\overline{5,8}$ and the 8-component spinors $(\mathrm{v}^{-}_{\mathrm{A}}(\overrightarrow{k}), \, \mathrm{v}^{+}_{\mathrm{B}}(\overrightarrow{k}))$ are given by

$$\mathrm{v}^{-}_{1}(\overrightarrow{k}) = N\left|
\begin{array}{cccc}
 \tilde{\omega}+m \\
 0 \\
 0 \\
 0 \\
 k^{3} \\
 k^{1}+ik^{2} \\
 0 \\
 0 \\
\end{array} \right|, \, \mathrm{v}^{-}_{2}(\overrightarrow{k}) = N\left|
\begin{array}{cccc}
 0 \\
 \tilde{\omega}+m \\
 0 \\
 0 \\
 k^{1}-ik^{2} \\
 -k^{3} \\
 0 \\
 0 \\
\end{array} \right|,$$
$$ \mathrm{v}^{-}_{3}(\overrightarrow{k}) = N \left|
\begin{array}{cccc}
 0 \\
 0 \\
 \tilde{\omega}+m \\
 0 \\
 0 \\
 0 \\
 k^{3} \\
 k^{1}+ik^{2} \\
\end{array} \right|,
\, \mathrm{v}^{-}_{4}(\overrightarrow{k}) = N\left|
\begin{array}{cccc}
 0 \\
 0 \\
 0 \\
 \tilde{\omega}+m \\
 0 \\
 0 \\
 k^{1}-ik^{2} \\
 -k^{3} \\
\end{array} \right|,$$
\begin{equation}
\label{eq257} \mathrm{v}^{+}_{5}(\overrightarrow{k}) = N\left|
\begin{array}{cccc}
 k^{3} \\
 k^{1}+ik^{2} \\
 0 \\
 0 \\
 \tilde{\omega}+m \\
 0 \\
 0 \\
 0 \\
\end{array} \right|, \,
\mathrm{v}^{+}_{6}(\overrightarrow{k}) = N\left|
\begin{array}{cccc}
 k^{1}-ik^{2} \\
 -k^{3} \\
 0 \\
 0 \\
 0 \\
 \tilde{\omega}+m \\
 0 \\
 0 \\
\end{array} \right|,
\end{equation}
$$\mathrm{v}^{+}_{7}(\overrightarrow{k}) = N\left|
\begin{array}{cccc}
 0 \\
 0 \\
 k^{3} \\
 k^{1}+ik^{2} \\
 0 \\
 0 \\
 \tilde{\omega}+m \\
 0 \\
\end{array} \right|,
\mathrm{v}^{+}_{8}(\overrightarrow{k}) = N\left|
\begin{array}{cccc}
 0 \\
 0 \\
 k^{1}-ik^{2} \\
 -k^{3} \\
 0 \\
 0 \\
 0 \\
 \tilde{\omega}+m \\
\end{array} \right|,$$

\noindent where

\begin{equation}
\label{eq258} N\equiv
\frac{1}{\sqrt{2\tilde{\omega}(\tilde{\omega}+m)}}, \quad
\tilde{\omega}\equiv \sqrt{\overrightarrow{k}^{2}+m^{2}}.
\end{equation}

The spinors (257) are derived from the orts
$\left\{\mathrm{d}_{\alpha}\right\}$ of the Cartesian basis (171)
with the help of the transformation (249), (250). The spinors
(257) satisfy the relations of the orthonormalization and
completeness similar to the corresponding relations for the
standard 4-component Dirac spinors, see, e. g., [33]. In the
covariant local field theory, the operators of the SU(2) spin,
which satisfy the corresponding commutation relations $
\left[s_{8\mathrm{D}}^{j},s_{8\mathrm{D}}^{\ell}\right]=i\varepsilon^{j
\ell n}s_{8\mathrm{D}}^{n}$ and commute with the operator
$\left[i\partial_{0}-\Gamma_{8}^{0}(\overrightarrow{\Gamma}_{8}\cdot\overrightarrow{p}+m)\right]$
of equation (252), are derived from the pure matrix operators
(243) with the help of operator (249), (251). The explicit form
of these SU(2) generators is given by

\begin{equation}
\label{eq259}
s_{8\mathrm{D}}^{1}=\frac{1}{2\sqrt{2}\widehat{\omega}\Omega}\left|
{{\begin{array}{*{20}c}
 11 \hfill & 12 \hfill & 13 \hfill & 14 \hfill & 15 \hfill  & 16 \hfill & 17 \hfill & 18 \hfill \\
 21 \hfill & 22 \hfill & 23 \hfill & 24 \hfill & 25 \hfill & 26 \hfill & 27 \hfill & 28 \hfill \\
 31 \hfill & 32 \hfill & 33 \hfill & 34 \hfill & 35 \hfill & 36 \hfill & 37 \hfill & 38 \hfill \\
 41 \hfill & 42 \hfill & 43 \hfill & 44 \hfill & 45 \hfill & 46 \hfill & 47 \hfill & 48 \hfill \\
 51 \hfill & 52 \hfill & 53 \hfill & 54 \hfill & 55 \hfill & 56 \hfill & 57 \hfill & 58 \hfill \\
 61 \hfill & 62 \hfill & 63 \hfill & 64 \hfill & 65 \hfill & 66 \hfill & 67 \hfill & 68 \hfill \\
 71 \hfill & 72 \hfill & 73 \hfill & 74 \hfill & 75 \hfill & 76 \hfill & 77 \hfill & 78 \hfill \\
 81 \hfill & 82 \hfill & 83 \hfill & 84 \hfill & 85 \hfill & 86 \hfill & 87 \hfill & 88 \hfill \\
\end{array} }} \right|,
\end{equation}

\noindent where the matrix elements have the form

$$11=2p^{13}, \, 12=2p^{1}z^{*}m\Omega, \,
13=p^{3}z^{*}, \, 14=z^{*2},$$
$$15=2ip^{2}\Omega, \, 16=-2p^{3}\Omega, \,17=-z\Omega, \, 18=0,$$
$$21=2p^{1}zm\Omega, \, 22=-2p^{13}, \, 23=\Omega^{2}-p^{33},$$
$$24=-p^{3}z^{*},$$
$$25=2p^{3}\Omega, \, 26=-2ip^{2}\Omega, \, 27=2p^{3}\Omega, \, 28=z^{*}\Omega,$$
$$31=p^{3}z, \, 32=\Omega^{2}-p^{33}, \, 33=0, \, 34=0,$$
$$35=z\Omega, \, 36=-2p^{3}\Omega, 37=0, \, 38=0,$$
$$41=z^{2}, \, 42=-p^{3}z, \, 43=0, \, 44=0,$$
$$45=0, \, 46=-z\Omega, \, 47=0, 48=0,$$
$$51=-2ip^{2}\Omega, \, 52=2p^{3}\Omega, \, 53=z^{*}\Omega, \, 54=0$$
$$55=2p^{13}, \, 56=2p^{1}z^{*}m\Omega, \, 57=p^{3}z^{*}, \, 58=z^{*2},$$
$$61=-2p^{3}\Omega, \, 62=2ip^{2}\Omega, \, 63=-2p^{3}\Omega,$$
$$64=-z^{*}\Omega,$$
$$65=2p^{1}zm\Omega, \,66=-2p^{13}, \, 67=\Omega^{2}-p^{33},$$
$$68=-p^{3}z^{*},$$
$$71=-z\Omega, \,72=2p^{3}\Omega, \, 73=0, \, 74=0,$$
$$75=p^{3}z, 76=\Omega^{2}-p^{33}, \, 77=0, \, 78=0,$$
$$81=0, \, 82=z\Omega, \, 83=0, \, 84=0,$$
$$85=z^{2}, \, 86=-p^{3}z, \, 87=0, \, 88=0,$$

\begin{equation}
\label{eq260}
s_{8\mathrm{D}}^{2}=\frac{i}{2\sqrt{2}\widehat{\omega}\Omega}\left|
{{\begin{array}{*{20}c}
 11 \hfill & 12 \hfill & 13 \hfill & 14 \hfill & 15 \hfill  & 16 \hfill & 17 \hfill & 18 \hfill \\
 21 \hfill & 22 \hfill & 23 \hfill & 24 \hfill & 25 \hfill & 26 \hfill & 27 \hfill & 28 \hfill \\
 31 \hfill & 32 \hfill & 33 \hfill & 34 \hfill & 35 \hfill & 36 \hfill & 37 \hfill & 38 \hfill \\
 41 \hfill & 42 \hfill & 43 \hfill & 44 \hfill & 45 \hfill & 46 \hfill & 47 \hfill & 48 \hfill \\
 51 \hfill & 52 \hfill & 53 \hfill & 54 \hfill & 55 \hfill & 56 \hfill & 57 \hfill & 58 \hfill \\
 61 \hfill & 62 \hfill & 63 \hfill & 64 \hfill & 65 \hfill & 66 \hfill & 67 \hfill & 68 \hfill \\
 71 \hfill & 72 \hfill & 73 \hfill & 74 \hfill & 75 \hfill & 76 \hfill & 77 \hfill & 78 \hfill \\
 81 \hfill & 82 \hfill & 83 \hfill & 84 \hfill & 85 \hfill & 86 \hfill & 87 \hfill & 88 \hfill \\
\end{array} }} \right|,
\end{equation}

\noindent where the matrix elements are given by

$$11=-2i{p}^{33}, \, 12=-2p^{3}z^{*}m\Omega, \, 13=-p^{3}z^{*},$$
$$14=-z^{*2},$$
$$15=-2p^{1}\Omega, \, 16=2p^{3}\Omega, \,  17=z^{*}\Omega, \, 18=0$$
$$21=2p^{2}zm\Omega, \, 22=2ip^{23}, \, 23=-\Omega^{2}+p^{33},$$
$$24=p^{3}z^{*},$$
$$25=2p^{3}\Omega, \, 26=2p^{1}\Omega, \, 27=-2p^{3}\Omega, \, 28=-z^{*}\Omega,$$
$$31=p^{3}z, \, 32=\Omega^{2}-p^{33}, \, 33=0, \, 34=0,$$
$$35=z\Omega, \, 36=-2p^{3}\Omega, \, 37=0, \, 38=0,$$
$$41=z^{2}, \, 42=-p^{3}z, \, 43=0, \, 44=0,$$
$$45=0, \, 46=-z\Omega, \, 47=0, \, 48=0,$$
$$51=2p^{1}\Omega, \, 52=-2p^{3}\Omega, \, 53=-z^{*}\Omega, \, 54=0,$$
$$55=-2ip^{23}, \, 56=-2p^{2}z^{*}m\Omega, \, 57=-p^{3}z^{*},$$
$$58=-z^{*2},$$
$$61=-2p^{3}\Omega, \, 62=-2p^{1}\Omega, \, 63=2p^{3}\Omega, \, 64=z^{*}\Omega,$$
$$65=2p^{2}zm\Omega, \, 66=2ip^{23}, \, 67=-\Omega^{2}+p^{33},$$
$$68=p^{3}z^{*},$$
$$71=-z\Omega, \, 72=2p^{3}\Omega, \, 73=0, \, 74=0,$$
$$75=p^{3}z, \, 76=\Omega^{2}-p^{33}, \, 77=0, \, 78=0,$$
$$81=0, \, 82=z\Omega, \, 83=0, \, 84=0,$$
$$85=z^{2}, \, 86=-p^{3}z, \, 87=0, \, 88=0,$$

\begin{equation}
\label{eq261}
s_{8\mathrm{D}}^{3}=\frac{1}{2\widehat{\omega}\Omega}\left|
{{\begin{array}{*{20}c}
 11 \hfill & 12 \hfill & 13 \hfill & 14 \hfill & 15 \hfill  & 16 \hfill & 17 \hfill & 18 \hfill \\
 21 \hfill & 22 \hfill & 23 \hfill & 24 \hfill & 25 \hfill & 26 \hfill & 27 \hfill & 28 \hfill \\
 31 \hfill & 32 \hfill & 33 \hfill & 34 \hfill & 35 \hfill & 36 \hfill & 37 \hfill & 38 \hfill \\
 41 \hfill & 42 \hfill & 43 \hfill & 44 \hfill & 45 \hfill & 46 \hfill & 47 \hfill & 48 \hfill \\
 51 \hfill & 52 \hfill & 53 \hfill & 54 \hfill & 55 \hfill & 56 \hfill & 57 \hfill & 58 \hfill \\
 61 \hfill & 62 \hfill & 63 \hfill & 64 \hfill & 65 \hfill & 66 \hfill & 67 \hfill & 68 \hfill \\
 71 \hfill & 72 \hfill & 73 \hfill & 74 \hfill & 75 \hfill & 76 \hfill & 77 \hfill & 78 \hfill \\
 81 \hfill & 82 \hfill & 83 \hfill & 84 \hfill & 85 \hfill & 86 \hfill & 87 \hfill & 88 \hfill \\
\end{array} }} \right|,
\end{equation}

\noindent where the matrix elements have the form

$$11=\Omega^{2}+p^{33}, \, 12=p^{3}z^{*}, \, 13=0, \, 14=0,$$
$$15=0, \, 16=z^{*}\Omega, \, 17=0, \, 18=0,$$
$$21=p^{3}z, \, 22=p^{11}+p^{22}, \, 23=0, \, 24=0,$$
$$25=-z\Omega, \, 26=0, \, 27=0, \, 28=0,$$
$$31=0, \, 32=0, \, 33=-\Omega^{2}-p^{33}, \, 34=-p^{3}z^{*},$$
$$35=0, \, 36=0, \, 37=0, \, 38=-z^{*}\Omega,$$
$$41=0, \, 42=0, \, 43=-p^{3}z, \, 44=-p^{11}-p^{22},$$
$$45=0, \, 46=0, 47=z\Omega, \, 48=0,$$
$$51=0, \, 52=-z^{*}\Omega, \, 53=0, \, 54=0,$$
$$55=\Omega^{2}+p^{33}, \, 56=p^{3}z^{*}, \, 57=0, \, 58=0,$$
$$61=z\Omega, \, 62=0, \, 63=0, \, 64=0, \,$$
$$65=p^{3}z, \, 66=p^{11}+p^{22}, \, 67=0, \, 68=0,$$
$$71=0, \, 72=0, \, 73=0, \, 74=z^{*}\Omega,$$
$$75=0, \, 76=0, \, 77=-\Omega^{2}-p^{33}, \, 78=-p^{3}z^{*},$$
$$81=0, \, 82=0, \, 83=-z\Omega, \, 84=0,$$
$$85=0, \, 86=0, \, 87=-p^{3}z, \, 88=-p^{11}-p^{22}.$$

\noindent Above, in the formulae for matrix elements of the spin operators (259)--(261) the following notations are used

$$p^{1}zm\Omega\equiv p^{1}z+m\Omega, \, p^{1}z^{*}m\Omega\equiv p^{1}z^{*}+m\Omega,$$
$$p^{2}zm\Omega\equiv -ip^{2}z+m\Omega, \, p^{2}z^{*}m\Omega\equiv ip^{2}z^{*}+m\Omega,$$
\begin{equation}
\label{eq262} p^{12}\equiv p^{1}p^{2}, \,p^{13}\equiv p^{1}p^{3},
\, p^{23}\equiv p^{2}p^{3},
\end{equation}
$$p^{11}\equiv p^{1}p^{1}, \, p^{22}\equiv p^{2}p^{2}, \, p^{33}\equiv p^{3}p^{3},$$
$$\Omega \equiv \omega +m, \, \omega \equiv \sqrt{\overrightarrow{p}+m^{2}},$$
$$z\equiv p^{1}+ip^{2}, \, z^{*}\equiv p^{1}-ip^{2}, \, z^{*2} \equiv (z^{*})^{2}.$$

The equations on eigenvectors and eigenvalues of the operator
$s_{8\mathrm{D}}^{3}$ (261) follow from the equations (244) and
the transformation (249)--(251). In addition to it, the action of
the operator $s_{8\mathrm{D}}^{3}$ (261)  on the spinors
$(\mathrm{v}^{-}_{\mathrm{A}}(\overrightarrow{k}), \,
\mathrm{v}^{+}_{\mathrm{B}}(\overrightarrow{k}))$ (257) also
leads to the result

$$s_{8\mathrm{D}}^{3}\mathrm{v}^{-}_{1}(\overrightarrow{k})= \mathrm{v}^{-}_{1}(\overrightarrow{k}), \, s_{8\mathrm{D}}^{3}\mathrm{v}^{-}_{2}(\overrightarrow{k}) = 0,$$
$$s_{8\mathrm{D}}^{3}\mathrm{v}^{-}_{3}(\overrightarrow{k}) = -\mathrm{v}^{-}_{3}(\overrightarrow{k}), \,
s_{8\mathrm{D}}^{3}\mathrm{v}^{-}_{4}(\overrightarrow{k}) = 0,$$
\begin{equation}
\label{eq263}
s_{8\mathrm{D}}^{3}\mathrm{v}^{+}_{5}(\overrightarrow{k}) =
\mathrm{v}^{+}_{5}(\overrightarrow{k}), \,
s_{8\mathrm{D}}^{3}\mathrm{v}^{+}_{6}(\overrightarrow{k}) = 0,
\end{equation}
$$s_{8\mathrm{D}}^{3}\mathrm{v}^{+}_{7}(\overrightarrow{k}) = -\mathrm{v}^{+}_{7}(\overrightarrow{k}), \, s_{8\mathrm{D}}^{3}\mathrm{v}^{+}_{8}(\overrightarrow{k}) = 0.$$

In order to verify equations (263) the following identity

\begin{equation}
\label{eq264}
(\tilde{\omega}+m)^{2}+(\overrightarrow{k})^{2}=2\tilde{\omega}(\tilde{\omega}+m),
\end{equation}

\noindent should be used. In the case
$\mathrm{v}^{+}_{\mathrm{B}}(\overrightarrow{k})$ in the
expression $s_{8\mathrm{D}}^{3}(\overrightarrow{k})$ (261) the
substitution $\overrightarrow{k}\rightarrow - \overrightarrow{k}$
is made.

The equations (263) determine the interpretation of the
amplitudes in solution (256). This interpretation is similar to the given above in the paragraph after equations (244). Nevertheless, the direct quantum-mechanical interpretation of the amplitudes should be made in the
framework of the RCQM, see the section 13 above.

The explicit form of the $\mathcal{P}$-generators of the bosonic
representation of the Poincar$\mathrm{\acute{e}}$ group
$\mathcal{P}$, with respect to which the covariant equation (252)
and the set $\left\{\psi\right\}$ of its solutions (256) are
invariant, is derived from the generators (245), (246) on the
basis of the transformation (249), (251). The corresponding
generators are given by

$$\widehat{p}^{0}=\Gamma_{8}^{0}(\overrightarrow{\Gamma}_{8}\cdot\overrightarrow{p}+m), \quad \widehat{p}^{\ell}=-i\partial_{\ell},$$
\begin{equation}
\label{eq265} \widehat{j}^{\ell
n}=x_{\mathrm{D}}^{\ell}\widehat{p}^{n}-x_{\mathrm{D}}^{n}\widehat{p}^{\ell}+s_{8\mathrm{D}}^{\ell
n}\equiv \widehat{m}^{\ell n}+s_{8\mathrm{D}}^{\ell n},
\end{equation}
\begin{equation}
\label{eq266} \widehat{j}^{0 \ell}=-\widehat{j}^{\ell
0}=x^{0}\widehat{p}^{\ell}-\frac{1}{2}\left\{x_{\mathrm{D}}^{\ell},\widehat{p}^{0}\right\}+\frac{\widehat{p}^{0}(\overrightarrow{s}_{8\mathrm{D}}\times
\overrightarrow{p})^{\ell}}{\widehat{\omega}(\widehat{\omega}+m)},
\end{equation}

\noindent where the spin matrices
$\overrightarrow{s}_{8\mathrm{D}}=(s_{8\mathrm{D}}^{\ell n})$ are
given in (259)--(261)  and the operator
$\overrightarrow{x}_{\mathrm{D}}$ has the form

\begin{equation}
\label{eq267}
\overrightarrow{x}_{\mathrm{D}}=\overrightarrow{x}+\frac{i\overrightarrow{\Gamma}_{8}}{2\widehat{\omega}}-\frac{\overrightarrow{s}_{8}\times \overrightarrow{p}}{\widehat{\omega}(\widehat{\omega}+m)}-\frac{i\overrightarrow{p} (\overrightarrow{\Gamma}_{8}\cdot \overrightarrow{p})}{2\widehat{\omega}^{2}(\widehat{\omega}+m)},
\end{equation}

\noindent where the spin matrices $\overrightarrow{s}_{8}$ are given in (243).

It is easy to verify that the generators (265), (266) commute with
the operator
$\left[i\partial_{0}-\Gamma_{8}^{0}(\overrightarrow{\Gamma}_{8}\cdot\overrightarrow{p}+m)\right]$
of equation (252), satisfy the commutation relations (11) of the
Lie algebra of the Poincar$\mathrm{\acute{e}}$ group and the
corresponding Casimir operators are given by

\begin{equation}
\label{eq268} p^{2}=\widehat{p}^{\mu}\widehat{p}_{\mu}=m^{2}\mathrm{I}_{8},
\end{equation}
$$W=w^{\mu}w_{\mu}=m^{2}\overrightarrow{s}_{8\mathrm{D}}^{2}=$$
\begin{equation}
\label{eq269} m^{2}\left| {{\begin{array}{*{20}c}
 1\left(1+1\right)\mathrm{I}_{3} \hfill & 0 \hfill & 0 \hfill & 0 \hfill\\
 0 \hfill & 0 \hfill & 0 \hfill & 0 \hfill\\
 0 \hfill & 0 \hfill & 1\left(1+1\right)\mathrm{I}_{3} \hfill & 0 \hfill\\
 0 \hfill & 0 \hfill & 0 \hfill & 0 \hfill\\
\end{array} }} \right|.
\end{equation}

\noindent where $\mathrm{I}_{8}$ and $\mathrm{I}_{3}$ are $8\times 8$ and $3\times 3$ unit matrices, respectively.

As it was already explained in details in the previous sections,
the conclusion that equation (252) describes the bosonic
particle-antiparticle multiplet of the spin s=(1,0,1,0) and mass
$m>0$ (and its solution (256) is the bosonic field of the above
mentioned spin and nonzero mass) follows from the analysis of
equations (263) and the Casimir operators (268), (269). Moreover, the external argument in the validity of such interpretation is the link with the corresponding RCQM of spin s=(1,0,1,0) particle-antiparticle multiplet, where the quantum-mechanical interpretation is direct and evident.
Therefore, the bosonic spin s=(1,0,1,0) properties of equation (252) are proved. In particular, the equation (252) can be used for the $W^{\mp}$-boson.

Contrary to the above found bosonic properties of the equation  (252), the fermionic properties of this equation are evident. The fact that equation (252) describes the multiplet of two fermions with the spin s=1/2 and two antifermions with that spin can be proved much more easier then the above bosonic consideration. The proof is similar to that given in the standard 4-component Dirac model (sections 7, 9, 10 above). Moreover, it is easy to show that equation (252) can describe the spin s=(3/2,3/2) particle-antiparticle doublet and the corresponding spinor field (section 24 below). Therefore, equation (252) has more extended property of the Fermi--Bose duality then the standard Dirac equation [24--28].

\section{New Maxwell-like equations for the spin s=1 boson and the generalized Maxwell electrodynamics}

In the partial case of $m=0$ equation (252) describes (among the
other possibilities) the system of the 8-component
\textit{(photon massless spinless boson)-(antiphoton masslees
spinless antiboson)} field. The 4-component photon massless
spinless boson subsystem was described in [36--38]. In other
possible interpretation [36--38], it is the system of the Maxwell
equations with gradient type current and charge densities (the
magnetic gradient type current and charge densities also are included). It was
shown [36--38] that this 4-component model can describe the
relativistic hydrogen spectrum (i. e. has the coupled finite
solutions) and the longitudinal electromagnetic waves. The
8-component model given here above has the additional
possibilities related to the antiphoton presence. The antiphoton can be considered as the photon with opposite sign of helicity (or right-handed in comparison with left-handed photon).

The link between the standard 4-component Dirac equation and the original Maxwell  equations is
the direct consequence of the substitutions like

\begin{equation}
\label{eq270}
\psi = \left|
\begin{array}{cccc}
 E^{3}+iH^{0} \\
 E^{1}+iE^{2} \\
 iH^{3}+E^{0} \\
 -H^{2}+iH^{1} \\
 \end{array} \right|,
\end{equation}

\noindent or

\begin{equation}
\label{eq271}
\psi = \left|
\begin{array}{cccc}
\overrightarrow{E}-i\overrightarrow{H} \\
 E^{0}-iH^{0} \\
  \end{array} \right|,
\end{equation}

\noindent into the massless Dirac equation (see [36--38] for details). In formulae (270), (271)
the functions $\overrightarrow{E}=(E^{1},E^{2},E^{3}), \,
\overrightarrow{H}=(H^{1},H^{2},H^{3})$ are the field strengths
of electromagnetic field and the functions $E^{0}, \, H^{0}$ are
the corresponding characteristics of the spinless bosonic field.
The necessary substitutions like (270), (271) for the 4-component spinors have been
found step by step by many authors, see, e. g., [36--38] and
references therein.

The link between the 8-component Dirac-like equation (252) and the original Maxwell  equations is
the direct consequence of the substitution

\begin{equation}
\label{eq272}
\psi = \left|
\begin{array}{cccc}
 E^{3}+iH^{0} \\
 E^{1}+iE^{2} \\
 iH^{3}+E^{0} \\
 -H^{2}+iH^{1} \\
 -iH^{3}-E^{0} \\
 H^{2}-iH^{1} \\
 -E^{3}-iH^{0} \\
 -E^{1}-iE^{2} \\
 \end{array} \right|
\end{equation}

\noindent into equation (252). The resulting system of Maxwell-like equations is given by

$$\partial_{0}E^{0}+\partial_{j}E^{j}\mp mH^{3}=0,$$
$$\partial_{0}E^{1}-(\mathrm{curl}\overrightarrow{H})^{1}+\partial_{1}E^{0} \mp mE^{2}=0,$$
$$\partial_{0}E^{2}-(\mathrm{curl}\overrightarrow{H})^{2}+\partial_{2}E^{0} \pm mE^{1}=0,$$
$$\partial_{0}E^{3}-(\mathrm{curl}\overrightarrow{H})^{3}+\partial_{3}E^{0} \mp mH^{0}=0,$$
\begin{equation}
\label{eq273}
\partial_{0}H^{0}+\partial_{j}H^{j}\pm mE^{3}=0,
\end{equation}
$$\partial_{0}H^{1}+(\mathrm{curl}\overrightarrow{E})^{1}+\partial_{1}H^{0} \mp mH^{2}=0,$$
$$\partial_{0}H^{2}+(\mathrm{curl}\overrightarrow{E})^{2}+\partial_{2}H^{0} \pm mH^{1}=0,$$
$$\partial_{0}H^{3}+(\mathrm{curl}\overrightarrow{E})^{3}+\partial_{3}H^{0} \pm mE^{0}=0.$$

The system of equations (273) contains 16 equations for the 8 real functions $E^{\alpha}=(E^{0},E^{1},E^{2},E^{3}), \,
H^{\alpha}=(H^{0},H^{1},H^{2},H^{3})$. The subsystem of 8 equations with upper sign near mass member is the consequence of the four upper equations from (252), while the subsystem of 8 equations with lower sign near mass member is the consequence of the four lower equations from (252). The situation is similar to the Dirac equation, wherever from the form of 8 equations with $\pm m$

\begin{equation}
\label{eq274}
(i\gamma^{\mu}\partial_{\mu}-m)(i\gamma^{\nu}\partial_{\nu}+m)=-\partial_{0}\partial_{0}+\Delta -m^{2}
\end{equation}
  
\noindent the system of 4 equations with $-m$ $(i\gamma^{\mu}\partial_{\mu}-m)\psi(x)=0$ for 4 functions is taken. In complete analogy here from the equations (273) the system of 8 equations with upper sign near mass for 8 functions $E^{\alpha}=(E^{0},E^{1},E^{2},E^{3}), \,
H^{\alpha}=(H^{0},H^{1},H^{2},H^{3})$ is taken. It is the system (273) with upper sign near mass. 

The system of equations (273) (upper sign near mass) describes the bosonic particle multiplet of the spin s=(1,0) and mass
$m>0$ (the coupled electromagnetic and scalar fields in the case of zero mass). Contrary to (252), here (in this form of equation) the relation to the Maxwell electrodynamics is evident. In partial case $m=0$ equations (273) (8 real or 4 complex equations) coincide with the Maxwell equations with gradient-like current and charge densities considered in [36--38].

Furthermore, the partial case $E^{0}=H^{0}=0$ in (273) leads to the system of equations

$$\partial_{j}E^{j} - mH^{3}=0,$$
$$\partial_{0}E^{1}-(\mathrm{curl}\overrightarrow{H})^{1} - mE^{2}=0,$$
$$\partial_{0}E^{2}-(\mathrm{curl}\overrightarrow{H})^{2}+ mE^{1}=0,$$
$$\partial_{0}E^{3}-(\mathrm{curl}\overrightarrow{H})^{3}=0,$$
\begin{equation}
\label{eq275}
\partial_{j}H^{j} + mE^{3}=0,
\end{equation}
$$\partial_{0}H^{1}+(\mathrm{curl}\overrightarrow{E})^{1} - mH^{2}=0,$$
$$\partial_{0}H^{2}+(\mathrm{curl}\overrightarrow{E})^{2}+ mH^{1}=0,$$
$$\partial_{0}H^{3}+(\mathrm{curl}\overrightarrow{E})^{3}=0.$$

\noindent Equations (275) can be considered as the generalized electrodynamics for the electromagnetic field of the photon with mass. In general, equations (275) describes the field of arbitrary boson with spin s=1 and mass $m>0$. 

The system of equations (275) is derived from the equations (149), (221) similarly to the way, in which the system of equations (252), and further (273) (with upper sign near mass), is derived from the equations (168), (236). The corresponding 6 component transformation of FW type will be the subject of special consideration. Hence, the system (275) is the direct consequence of the equations (149), (221) for the spin s=(1,1) particle-antiparticle doublet.

Moreover, equations (273) with upper sign near mass are linked directly with the RCQM equation (160) for the spin s=(1,0) multiplet. This link is based on the anti-Hermitian transformations, essential application of complex conjugation operator C and the corresponding equations for eigenvectors lead to complex number eigenvalues. Therefore, such links and transformations will be given in special publication.

Finally, the substitution $m=0$ into the system of equations (275) leads to the Maxwell equations     

$$\mathrm{div}\overrightarrow{E}=0, \quad \partial_{0}\overrightarrow{E}-\mathrm{curl}\overrightarrow{H}=0,$$
\begin{equation}
\label{eq276}
\mathrm{div}\overrightarrow{H}=0, \quad \partial_{0}\overrightarrow{H}+\mathrm{curl}\overrightarrow{E}=0,
\end{equation}

\noindent for free electromagnetic field. Thus, the link between the RCQM and the Maxwell electrodynamics is finished.

\section{Covariant field equation for the 8-component spin s=(3/2,3/2) fermionic particle-antiparticle doublet}

The model is constructed in complete analogy with the consideration in section 22.

The start of this derivation is given in section 14, where the
RCQM of the 8-component fermionic spin s=(3/2,3/2)
particle-antiparticle doublet is considered. The second step is
the transition from the Schr$\mathrm{\ddot{o}}$dinger--Foldy
equation (168) to the canonical field equation. This step, as
shown in section 22, is possible only for the
anti-Hermitian form of the operators. Nevertheless, the resulting
operators can be chosen in the standard Hermitian form and do not
contain the operator $C$ of complex conjugation. The last step of
the transition from the canonical field equation to the covariant
local field equation is fulfilled in analogy with the FW
transformation (119) with the help of the transformation (249).

Thus, the canonical field equation for the 8-component spin s=(3/2,3/2) fermionic particle-antiparticle doublet (8 component analogy of the FW equation) is found from the the Schr$\mathrm{\ddot{o}}$dinger--Foldy
equation (168) on the basis of the transformation $v_{8}$ (239)--(242) and is given in (236).

The general solution of this equation in the case of spin s=(3/2,3/2) fermionic particle-antiparticle doublet is found with the help of the transformation $v_{8}$ (239), (240) from the general solution (177) of the Schr$\mathrm{\ddot{o}}$dinger--Foldy
equation (168) and is given by

$$\phi(x)= \frac{1}{\left(2\pi\right)^{\frac{3}{2}}}\int d^{3}k$$
\begin{equation}
\label{eq277}
\left[e^{-ikx}b^{\mathrm{A}}(\overrightarrow{k})\mathrm{d}_{\mathrm{A}}+e^{ikx}b^{*\mathrm{B}}(\overrightarrow{k})\mathrm{d}_{\mathrm{B}}\right],
\end{equation}

\noindent where $\mathrm{A}=\overline{1,4}, \,
\mathrm{B}=\overline{5,8}$, the orts of the 8-component
Cartesian basis are given in (171) and the quantum-mechanical interpretation of the amplitudes $(b^{\mathrm{A}}(\overrightarrow{k}), \, b^{*\mathrm{B}}(\overrightarrow{k}))$ is given according to (182), (183).

The SU(2) spin operators, which satisfy the commutation relations
(30) and commute with the operator $(i\partial_{0}-
\Gamma_{8}^{0}\widehat{\omega})$ of the equation of motion (236),
are derived from the corresponding RCQM operators (178), (179) on the basis of
transformations (239), (241), (242). These canonical field spin
operators are given by

\begin{equation}
\label{eq278} \overrightarrow{s}_{8}=\left|
{{\begin{array}{*{20}c}
 \overrightarrow{s} \hfill & 0 \\
 0 \hfill & \overrightarrow{s} \\
\end{array} }} \right|, \quad \overrightarrow{s}_{8}^{2}=\frac{3}{2}\left(\frac{3}{2}+1\right)\mathrm{I}_{8},
\end{equation}

\noindent where the $4 \times 4$ operators $\overrightarrow{s}$ are
given in (39). In the explicit form the SU(2) spin operators (278) are given by

$${s}^{1}_{8}= \frac{1}{2}\cdot$$
$$\left| {{\begin{array}{*{20}c}
 0 \hfill & \sqrt{3} \hfill & 0 \hfill & 0 \hfill & 0 \hfill  & 0 \hfill & 0 \hfill  & 0 \hfill \\
 \sqrt{3} \hfill & 0 \hfill & 2 \hfill & 0 \hfill & 0 \hfill & 0 \hfill & 0 \hfill  & 0 \hfill \\
 0 \hfill & 2 \hfill & 0 \hfill & \sqrt{3} \hfill & 0 \hfill & 0 \hfill & 0 \hfill  & 0 \hfill \\
 0 \hfill & 0 \hfill & \sqrt{3} \hfill & 0 \hfill & 0 \hfill & 0 \hfill & 0 \hfill  & 0 \hfill \\
 0 \hfill & 0 \hfill & 0 \hfill & 0 \hfill & 0 \hfill & \sqrt{3} \hfill & 0 \hfill  & 0 \hfill \\
 0 \hfill & 0 \hfill & 0 \hfill & 0 \hfill & \sqrt{3} \hfill & 0 \hfill& 2 \hfill  & 0 \hfill \\
 0 \hfill & 0 \hfill & 0 \hfill & 0 \hfill & 0 \hfill & 2 \hfill& 0 \hfill  & \sqrt{3} \hfill \\
 0 \hfill & 0 \hfill & 0 \hfill & 0 \hfill & 0 \hfill & 0 \hfill& \sqrt{3} \hfill  & 0 \hfill \\
\end{array} }} \right|,$$

$${s}^{2}_{8}= \frac{i}{2}\cdot$$
\begin{equation}
\label{eq279}\left| {{\begin{array}{*{20}c}
 0 \hfill & -\sqrt{3} \hfill & 0 \hfill & 0 \hfill & 0 \hfill  & 0 \hfill & 0 \hfill  & 0 \hfill \\
 \sqrt{3} \hfill & 0 \hfill & -2 \hfill & 0 \hfill & 0 \hfill & 0 \hfill & 0 \hfill  & 0 \hfill \\
 0 \hfill & 2 \hfill & 0 \hfill & -\sqrt{3} \hfill & 0 \hfill & 0 \hfill & 0 \hfill  & 0 \hfill \\
 0 \hfill & 0 \hfill & \sqrt{3} \hfill & 0 \hfill & 0 \hfill & 0 \hfill & 0 \hfill  & 0 \hfill \\
 0 \hfill & 0 \hfill & 0 \hfill & 0 \hfill & 0 \hfill & -\sqrt{3} \hfill & 0 \hfill  & 0 \hfill \\
 0 \hfill & 0 \hfill & 0 \hfill & 0 \hfill & \sqrt{3} \hfill & 0 \hfill& -2 \hfill  & 0 \hfill \\
 0 \hfill & 0 \hfill & 0 \hfill & 0 \hfill & 0 \hfill & 2 \hfill& 0 \hfill  & -\sqrt{3} \hfill \\
 0 \hfill & 0 \hfill & 0 \hfill & 0 \hfill & 0 \hfill & 0 \hfill& \sqrt{3} \hfill  & 0 \hfill \\
\end{array} }} \right|,
\end{equation}
$${s}^{3}_{8}= \frac{1}{2}\left| {{\begin{array}{*{20}c}
 3 \hfill & 0 \hfill & 0 \hfill & 0 \hfill & 0 \hfill  & 0 \hfill & 0 \hfill  & 0 \hfill \\
 0 \hfill & 1 \hfill & 0 \hfill & 0 \hfill & 0 \hfill & 0 \hfill & 0 \hfill  & 0 \hfill \\
 0 \hfill & 0 \hfill & -1 \hfill & 0 \hfill & 0 \hfill & 0 \hfill & 0 \hfill  & 0 \hfill \\
 0 \hfill & 0 \hfill & 0 \hfill & -3 \hfill & 0 \hfill & 0 \hfill & 0 \hfill  & 0 \hfill \\
 0 \hfill & 0 \hfill & 0 \hfill & 0 \hfill & 3 \hfill & 0 \hfill & 0 \hfill  & 0 \hfill \\
 0 \hfill & 0 \hfill & 0 \hfill & 0 \hfill & 0 \hfill & 1 \hfill& 0 \hfill  & 0 \hfill \\
 0 \hfill & 0 \hfill & 0 \hfill & 0 \hfill & 0 \hfill & 0 \hfill& -1 \hfill  & 0 \hfill \\
 0 \hfill & 0 \hfill & 0 \hfill & 0 \hfill & 0 \hfill & 0 \hfill& 0 \hfill  & -3 \hfill \\
\end{array} }} \right|.$$

The stationary complete set of operators is given by the operators $g, \overrightarrow{p}, \, s_{8}^{3}=s_{z}$ of the charge sign, momentum and spin projection, respectively (see section 14 for details). The equations on the eigenvectors and eigenvalues of the spin projection operator $s_{8}^{3}=s_{z}$ from (279) have the form

$$s_{8}^{3}\mathrm{d}_{1} = \frac{3}{2}\mathrm{d}_{1}, \, s_{8}^{3}\mathrm{d}_{2} = \frac{1}{2}\mathrm{d}_{2},$$
$$s_{8}^{3}\mathrm{d}_{3} = -\frac{1}{2}\mathrm{d}_{3}, \,
s_{8}^{3}\mathrm{d}_{4} = -\frac{3}{2}\mathrm{d}_{4},$$
\begin{equation}
\label{eq280} s_{8}^{3}\mathrm{d}_{5} = \frac{3}{2}\mathrm{d}_{5}, \, s_{8}^{3}\mathrm{d}_{6} = \frac{1}{2}\mathrm{d}_{6},
\end{equation}
$$s_{8}^{3}\mathrm{d}_{7} = -\frac{1}{2}\mathrm{d}_{7}, \,
s_{8}^{3}\mathrm{d}_{8} = -\frac{3}{2}\mathrm{d}_{8}.$$

Therefore, the functions $b^{1}(\overrightarrow{k}), \,
b^{2}(\overrightarrow{k}), \, b^{3}(\overrightarrow{k}), \, b^{4}(\overrightarrow{k})$ in
solution (277) are the momentum-spin amplitudes of the massive
fermion with the spin s=3/2 and the spin projection $(3/2,1/2,-1/2,-3/2)$,
respectively; $b^{5}(\overrightarrow{k}), \,
b^{6}(\overrightarrow{k}), \, b^{7}(\overrightarrow{k}), \, b^{8}(\overrightarrow{k})$ are the
momentum-spin amplitudes of the antiparticle (antifermion) with the
spin s=3/2 and the spin projection $(3/2,1/2,-1/2,-3/2)$, respectively.

Note that direct quantum-mechanical interpretation of the amplitudes in solution (277) should be given in the framework of the RCQM. Such interpretation is already given in section 14 in paragraph after equations (183). 

The generators of the reducible unitary fermionic spin s=(3/2,3/2)
doublet representation of the Poincar$\mathrm{\acute{e}}$ group
$\mathcal{P}$, with respect to which the canonical field equation
(236) and the set $\left\{\phi\right\}$ of its solutions (277) are
invariant, are derived from the RCQM set of generators (12), (13)
with the spin (178), (179) on the basis of the transformations (239),
(241), (242). These Hermitian  $8 \times 8$ matrix-differential
generators are given by (245), (246), where the explicit form of SU(2) spin s=(3/2,3/2) is given in (279).

It is easy to prove by the direct verification that the generators (245), (246) commute with the operator $(i\partial_{0}- \Gamma_{8}^{0}\widehat{\omega})$ of the canonical field equation (236) and satisfy the commutation relations (11) of the Lie algebra of the Poincar$\mathrm{\acute{e}}$ group $\mathcal{P}$. The Casimir operators for the representation (245), (246) with SU(2) spin (279) are given by

\begin{equation}
\label{eq281} p^{2}=\widehat{p}^{\mu}\widehat{p}_{\mu}=m^{2}\mathrm{I}_{8},
\end{equation}
\begin{equation}
\label{eq282} W=w^{\mu}w_{\mu}=m^{2}\overrightarrow{s}_{8}^{2}=\frac{3}{2}\left(\frac{3}{2}+1\right)\mathrm{I}_{8}.
\end{equation}

Thus, due to the eigenvalues in equations (280), positive and
negative frequencies form of solution (277) and the
Bargman--Wigner analysis of the Casimir operators (281), (282) one
can come to a conclusion that equation (236) describes the
8-component canonical field (the fermionic particle-antiparticle
doublet) with the spins s=(3/2,3/2) and $m>0$.

The operator of the transition to the covariant local field theory
representation (the $8 \times 8$  analogy of the $4 \times 4$ FW
transformation operator (119)) is the same as in the section 22 and is given by (249)--(251). Thus, on the basis of transformation (249)--(251) the 8-component
Dirac-like equation is found from the canonical field equation
(236) in the form (252).

Note that equation (252) is not the ordinary direct sum
of the two Dirac equations. Therefore, it is not the complex
Dirac--Kahler equation [34]. Moreover, it is not the standard
Dirac--Kahler equation [35].

The solution of equation (252) is derived from solution (277) of
this equation in the canonical representation (236) on the basis
of transformation (249), (250) and is given by

$$\psi(x)=V^{-}\phi(x)= \frac{1}{\left(2\pi\right)^{\frac{3}{2}}}\int d^{3}k$$
\begin{equation}
\label{eq283}
\left[e^{-ikx}b^{\mathrm{A}}(\overrightarrow{k})\mathrm{v}^{-}_{\mathrm{A}}(\overrightarrow{k})+e^{ikx}b^{*\mathrm{B}}(\overrightarrow{k})\mathrm{v}^{+}_{\mathrm{B}}(\overrightarrow{k})\right],
\end{equation}

\noindent where $\mathrm{A}=\overline{1,4}, \, \mathrm{B}=\overline{5,8}$ and the 8-component spinors $(\mathrm{v}^{-}_{\mathrm{A}}(\overrightarrow{k}), \, \mathrm{v}^{+}_{\mathrm{B}}(\overrightarrow{k}))$ are given in (257).

The spinors (257) are derived from the orts
$\left\{\mathrm{d}_{\alpha}\right\}$ of the Cartesian basis (171)
with the help of the transformation (249), (250). The spinors
(257) satisfy the relations of the orthonormalization and
completeness similar to the corresponding relations for the
standard 4-component Dirac spinors, see, e. g., [33]. 

The direct quantum-mechanical interpretation of the amplitudes in solution (283) should be given in the framework of the RCQM. Such interpretation is already given in section 14 in paragraph after equations (183).

In the covariant local field theory, the operators of the SU(2) spin,
which satisfy the corresponding commutation relations $
\left[s_{8\mathrm{D}}^{j},s_{8\mathrm{D}}^{\ell}\right]=i\varepsilon^{j
\ell n}s_{8\mathrm{D}}^{n}$ and commute with the operator
$\left[i\partial_{0}-\Gamma_{8}^{0}(\overrightarrow{\Gamma}_{8}\cdot\overrightarrow{p}+m)\right]$
of equation (252), are derived from the pure matrix operators
(279) with the help of transition operator (249), (251) $\overrightarrow{s}_{8\mathrm{D}}=V^{-}\overrightarrow{s}_{8}V^{+}$. The explicit form
of these s=(3/2,3/2) SU(2) generators is given by

\begin{equation}
\label{eq284}
s_{8\mathrm{D}}^{1}=k\left|
{{\begin{array}{*{20}c}
 11 \hfill & 12 \hfill & 13 \hfill & 14 \hfill & 15 \hfill  & 16 \hfill & 17 \hfill & 18 \hfill \\
 21 \hfill & 22 \hfill & 23 \hfill & 24 \hfill & 25 \hfill & 26 \hfill & 27 \hfill & 28 \hfill \\
 31 \hfill & 32 \hfill & 33 \hfill & 34 \hfill & 35 \hfill & 36 \hfill & 37 \hfill & 38 \hfill \\
 41 \hfill & 42 \hfill & 43 \hfill & 44 \hfill & 45 \hfill & 46 \hfill & 47 \hfill & 48 \hfill \\
 51 \hfill & 52 \hfill & 53 \hfill & 54 \hfill & 55 \hfill & 56 \hfill & 57 \hfill & 58 \hfill \\
 61 \hfill & 62 \hfill & 63 \hfill & 64 \hfill & 65 \hfill & 66 \hfill & 67 \hfill & 68 \hfill \\
 71 \hfill & 72 \hfill & 73 \hfill & 74 \hfill & 75 \hfill & 76 \hfill & 77 \hfill & 78 \hfill \\
 81 \hfill & 82 \hfill & 83 \hfill & 84 \hfill & 85 \hfill & 86 \hfill & 87 \hfill & 88 \hfill \\
\end{array} }} \right|,
\end{equation}

\noindent where the matrix elements have the form

$$11=\sqrt{3}p^{13}, \, 12=\sqrt{3}(p^{1}z^{*}+m\Omega), \,
13=p^{3}z^{*},$$
$$14=z^{*2},$$
$$15=i\sqrt{3}p^{2}\Omega, \, 16=-\sqrt{3}p^{3}\Omega, \, 17=-\Omega z^{*}, \, 18=0,$$
$$21=\sqrt{3} (p^{1}z+m\Omega), \, 22=-\sqrt{3} p^{13},$$
$$23=2m\Omega + p^{11}_{22}, \, 24=-p^{3}z^{*},$$
$$25=\sqrt{3}p^{3}\Omega, \, 26=-i\sqrt{3}p^{2}\Omega, \, 27=2p^{3}\Omega, \, 28=z^{*}\Omega,$$
$$31=p^{3}z, \, 32=2m\Omega+p^{11}_{22}, \, 33=\sqrt{3} p^{13},$$
$$34=\sqrt{3}(p^{1}z^{*}+m\Omega),$$
$$35=2\Omega z, \, 36=-2p^{3}\Omega, 37=\sqrt{3}p^{1}\Omega, \, 38=-\sqrt{3}p^{3}\Omega,$$
$$41=z^{2}, \, 42=-p^{3}z, \, 43=\sqrt{3} (p^{1}z+m\Omega), $$
$$44=-\sqrt{3} p^{13},$$
$$45=0, \, 46=-\Omega z, \, 47=\sqrt{3}p^{3}\Omega, 48=-i\sqrt{3}p^{2}\Omega,$$
$$51=-i\sqrt{3}\Omega, \, 52=\sqrt{3}p^{3}\Omega, \, 53=\Omega z^{*}, \, 54=0$$
$$55=\sqrt{3}p^{13}, \, 56=\sqrt{3}(p^{1}z^{*}+m\Omega), \, 57=p^{3}z^{*},$$
$$58=z^{*2},$$
$$61=-\sqrt{3}p^{3}\Omega, \, 62=i\sqrt{3}p^{2}\Omega, \, 63=-2p^{3}\Omega,$$
$$64=-\Omega z^{*},$$
$$65=\sqrt{3}(p^{1}z+m\Omega), \,66=-\sqrt{3}p^{13},$$
$$67=2m\Omega+p^{11}_{22}, \, 68=-p^{3}z^{*},$$
$$71=-\Omega z, \, 72=2p^{3}\Omega, \, 73=-i\sqrt{3}p^{2}\Omega,$$
$$74=\sqrt{3}p^{3}\Omega,$$
$$75=p^{3}z^{*}, 76=2m\Omega^{2}+p^{11}_{22}, \, 77=\sqrt{3}p^{13},$$
$$78=\sqrt{3}(p^{1}z^{*}+m\Omega),$$
$$81=0, \, 82=\Omega z, \, 83=-\sqrt{3}p^{3}\Omega, \, 84=i\sqrt{3}p^{2}\Omega,$$
$$85=z^{2}, \, 86=-p^{3} z, \, 87=\sqrt{3} (p^{1}z+m\Omega),$$
$$88=-\sqrt{3}p^{13},$$

\begin{equation}
\label{eq285}
s_{8\mathrm{D}}^{2}=k\left|
{{\begin{array}{*{20}c}
 11 \hfill & 12 \hfill & 13 \hfill & 14 \hfill & 15 \hfill  & 16 \hfill & 17 \hfill & 18 \hfill \\
 21 \hfill & 22 \hfill & 23 \hfill & 24 \hfill & 25 \hfill & 26 \hfill & 27 \hfill & 28 \hfill \\
 31 \hfill & 32 \hfill & 33 \hfill & 34 \hfill & 35 \hfill & 36 \hfill & 37 \hfill & 38 \hfill \\
 41 \hfill & 42 \hfill & 43 \hfill & 44 \hfill & 45 \hfill & 46 \hfill & 47 \hfill & 48 \hfill \\
 51 \hfill & 52 \hfill & 53 \hfill & 54 \hfill & 55 \hfill & 56 \hfill & 57 \hfill & 58 \hfill \\
 61 \hfill & 62 \hfill & 63 \hfill & 64 \hfill & 65 \hfill & 66 \hfill & 67 \hfill & 68 \hfill \\
 71 \hfill & 72 \hfill & 73 \hfill & 74 \hfill & 75 \hfill & 76 \hfill & 77 \hfill & 78 \hfill \\
 81 \hfill & 82 \hfill & 83 \hfill & 84 \hfill & 85 \hfill & 86 \hfill & 87 \hfill & 88 \hfill \\
\end{array} }} \right|,
\end{equation}

\noindent where the matrix elements are given by

$$11=\sqrt{3}{p}^{23}, \, 12=-i\sqrt{3}(m\Omega+ip^{2}z^{*}),$$
$$13=--ip^{3}z^{*}, \,  14=-iz^{*2},$$
$$15=-i\sqrt{3}p^{1}\Omega, \, 16=i\sqrt{3}p^{3}\Omega, \,  17=iz^{*}\Omega, \, 18=0,$$
$$21=i\sqrt{3}(m\Omega-ip^{2}z), \, 22=\sqrt{3}p^{23},$$
$$23=-i(\Omega^{2}-p^{33}), \, 24=ip^{3}z^{*},$$
$$25=i\sqrt{3}p^{3}\Omega, \, 26=i\sqrt{3}p^{1}\Omega, \, 27=-2ip^{3}\Omega,$$
$$28=-iz^{*}\Omega,$$
$$31=ip^{3}z, \, 32=i(\Omega^{2}-p^{33}), \, 33=\sqrt{3}p^{23},$$
$$34=-i\sqrt{3}(m\Omega+ip^{2}z^{*}),$$
$$35=iz\Omega, \, 36=-2ip^{3}\Omega, \, 37=-i\sqrt{3}p^{1}\Omega,$$
$$38=i\sqrt{3}p^{3}\Omega,$$
$$41=iz^{2}, \, 42=-ip^{3}z, \, 43=i\sqrt{3}(m\Omega-ip^{2}z),$$
$$44=-\sqrt{3}{p}^{23},$$
$$45=0, \, 46=-iz\Omega, \, 47=i\sqrt{3}{p}^{3}\Omega, 48=i\sqrt{3}{p}^{1}\Omega,$$
$$51=i\sqrt{3}p^{1}\Omega, \, 52=-i\sqrt{3}p^{3}\Omega,$$
$$ 53=-2iz^{*}\Omega, \, 54=0,$$
$$55=\sqrt{3}p^{23}, \, 56=-i\sqrt{3}(m\Omega+ip^{2}z^{*}),$$
$$57=-ip^{3}z^{*}, \, 58=-iz^{*2},$$
$$61=-i\sqrt{3}p^{3}\Omega, \, 62=-i\sqrt{3}p^{1}\Omega, \, 63=2ip^{3}\Omega,$$
$$64=iz^{*}\Omega,$$
$$65=i\sqrt{3}(m\Omega-ip^{2}z), \, 66=-\sqrt{3}p^{23},$$
$$67=-i(\Omega^{2}-p^{33}), \, 68=ip^{3}z^{*},$$
$$71=-iz\Omega, \, 72=2ip^{3}\Omega, \, 73=i\sqrt{3}p^{1}\Omega,$$
$$74=-i\sqrt{3}p^{3}\Omega,$$
$$75=ip^{3}z, \, 76=i(\Omega^{2}-p^{33}), \, 77=\sqrt{3}p^{23},$$
$$78=-i\sqrt{3}(m\Omega+ip^{2}z^{*}),$$
$$81=0, \, 82=iz\Omega, \, 83=-i\sqrt{3}p^{3}\Omega,$$
$$84=-i\sqrt{3}p^{1}\Omega,$$
$$85=iz^{2}, 86=-ip^{3}z, 87=i\sqrt{3}(m\Omega-ip^{2}z),$$
$$88=-\sqrt{3}p^{23},$$

\begin{equation}
\label{eq286}
s_{8\mathrm{D}}^{3}=k\left|
{{\begin{array}{*{20}c}
 11 \hfill & 12 \hfill & 0 \hfill & 0 \hfill & 0 \hfill  & 16 \hfill & 0 \hfill & 0 \hfill \\
 21 \hfill & 22 \hfill & 0 \hfill & 0 \hfill & 25 \hfill & 0 \hfill & 0 \hfill & 0 \hfill \\
 0 \hfill & 0 \hfill & 33 \hfill & 34 \hfill & 0 \hfill & 0 \hfill & 0 \hfill & 38 \hfill \\
 0 \hfill & 0 \hfill & 43 \hfill & 44 \hfill & 0 \hfill & 0 \hfill & 47 \hfill & 0 \hfill \\
 0 \hfill & 52 \hfill & 0 \hfill & 0 \hfill & 55 \hfill & 56 \hfill & 0 \hfill & 0 \hfill \\
 61 \hfill & 0 \hfill & 0 \hfill & 0 \hfill & 65 \hfill & 66 \hfill & 0 \hfill & 0 \hfill \\
 0 \hfill & 0 \hfill & 0 \hfill & 74 \hfill & 0 \hfill & 0 \hfill & 77 \hfill & 78 \hfill \\
 0 \hfill & 0 \hfill & 83 \hfill & 0 \hfill & 0 \hfill & 0 \hfill & 87 \hfill & 88 \hfill \\
\end{array} }} \right|,
\end{equation}

\noindent where the nonzero matrix elements have the form

$$11=3\omega\Omega -p^{11}_{22}, \, 12=p^{3}z^{*}, \, 16=z^{*}\Omega,$$
$$21=p^{3}z, \, 22=\omega\Omega + p^{11}_{22}, \, 25=-z\Omega,$$
$$33=-\omega\Omega - p^{11}_{22}, \, 34=p^{3}z^{*}, \, 38=z^{*}\Omega,$$
$$43=p^{3}z, \, 44=-3\omega\Omega +p^{11}_{22}, \, 47=-z\Omega,$$
$$52=-z^{*}\Omega, \, 55=3\omega\Omega -p^{11}_{22}, \, 56=p^{3}z^{*},$$
$$61=z\Omega, \, 65=p^{3}z, \, 66=\omega\Omega +p^{11}_{22},$$
$$74=-z^{*}\Omega, \, 77=-\omega\Omega -p^{11}_{22}, \, 78=p^{3}z^{*},$$
$$83=z\Omega, \, 87=p^{3}z, \, 88=-3\omega\Omega +p^{11}_{22}.$$

\noindent Above, in the formulae (284)--(286) and in corresponding matrix elements the following notations are used

\begin{equation}
\label{eq287} k \equiv \frac{1}{2\omega\Omega}, \quad p^{11}_{22}\equiv p^{1}p^{1}+p^{2}p^{2}.
\end{equation}

\noindent Other notations are already given in (262).

The equations on eigenvectors and eigenvalues of the operator
$s_{8\mathrm{D}}^{3}$ (286) follow from the equations (280) and
the transformation (249)--(251). In addition to it, the action of
the operator $s_{8\mathrm{D}}^{3}$ (286)  on the spinors
$(\mathrm{v}^{-}_{\mathrm{A}}(\overrightarrow{k}), \,
\mathrm{v}^{+}_{\mathrm{B}}(\overrightarrow{k}))$ (257) also
leads to the result

$$s_{8\mathrm{D}}^{3}\mathrm{v}^{-}_{1}(\overrightarrow{k})= \frac{3}{2}\mathrm{v}^{-}_{1}(\overrightarrow{k}), \, s_{8\mathrm{D}}^{3}\mathrm{v}^{-}_{2}(\overrightarrow{k}) = \frac{1}{2}\mathrm{v}^{-}_{2}(\overrightarrow{k}),$$
$$s_{8\mathrm{D}}^{3}\mathrm{v}^{-}_{3}(\overrightarrow{k}) = -\frac{1}{2}\mathrm{v}^{-}_{3}(\overrightarrow{k}), \,
s_{8\mathrm{D}}^{3}\mathrm{v}^{-}_{4}(\overrightarrow{k}) = -\frac{3}{2}\mathrm{v}^{-}_{4}(\overrightarrow{k}),$$
\begin{equation}
\label{eq288}
s_{8\mathrm{D}}^{3}\mathrm{v}^{+}_{5}(\overrightarrow{k}) =
\frac{3}{2}\mathrm{v}^{+}_{5}(\overrightarrow{k}), \,
s_{8\mathrm{D}}^{3}\mathrm{v}^{+}_{6}(\overrightarrow{k}) = \frac{1}{2}\mathrm{v}^{+}_{6}(\overrightarrow{k}),
\end{equation}
$$s_{8\mathrm{D}}^{3}\mathrm{v}^{+}_{7}(\overrightarrow{k}) = -\frac{1}{2}\mathrm{v}^{+}_{7}(\overrightarrow{k}), \, s_{8\mathrm{D}}^{3}\mathrm{v}^{+}_{8}(\overrightarrow{k}) = -\frac{3}{2}\mathrm{v}^{+}_{8}(\overrightarrow{k}).$$

In order to verify equations (288) the identity (264) is used. In the case
$\mathrm{v}^{+}_{\mathrm{B}}(\overrightarrow{k})$ in the
expression $s_{8\mathrm{D}}^{3}(\overrightarrow{k})$ (286) the
substitution $\overrightarrow{k}\rightarrow - \overrightarrow{k}$
is made.

The equations (288) determine the interpretation of the
amplitudes in solution (283). This interpretation is similar to the given above in the paragraph after equations (280). Nevertheless, the direct quantum-mechanical interpretation of the amplitudes should be made in the
framework of the RCQM, see the section 14 above.

The explicit form of the $\mathcal{P}$-generators of the fermionic
representation of the Poincar$\mathrm{\acute{e}}$ group
$\mathcal{P}$, with respect to which the covariant equation (252)
and the set $\left\{\psi\right\}$ of its solutions (283) are
invariant, is derived from the generators (245), (246) with SU(2) spin (279) on the
basis of the transformation (249), (251). The corresponding
generators are given by (265), (266) where the spin matrices
$\overrightarrow{s}_{8\mathrm{D}}=(s_{8\mathrm{D}}^{\ell n})$ are
given in (284)--(286)  and the operator
$\overrightarrow{x}_{\mathrm{D}}$ has the form (267) with spin matrices $\overrightarrow{s}_{8}$ (278), (279).

It is easy to verify that the generators (265), (266) with SU(2) spin (284)--(286) commute with
the operator $\left[i\partial_{0}-\Gamma_{8}^{0}(\overrightarrow{\Gamma}_{8}\cdot\overrightarrow{p}+m)\right]$
of equation (252), satisfy the commutation relations (11) of the
Lie algebra of the Poincar$\mathrm{\acute{e}}$ group and the
corresponding Casimir operators are given by

\begin{equation}
\label{eq289} p^{2}=\widehat{p}^{\mu}\widehat{p}_{\mu}=m^{2}\mathrm{I}_{8},
\end{equation}
\begin{equation}
\label{eq290} W=w^{\mu}w_{\mu}=m^{2}\overrightarrow{s}_{8\mathrm{D}}^{2}=\frac{3}{2}\left(\frac{3}{2}+1\right)m^{2}\mathrm{I}_{8}.
\end{equation}

As it was already explained in details in the previous sections,
the conclusion that equation (252) describes the local field of fermionic
particle-antiparticle doublet of the spin s=(3/2,3/2) and mass
$m>0$ (and its solution (283) is the local fermionic field of the above
mentioned spin and nonzero mass) follows from the analysis of
equations (288) and the Casimir operators (289), (290).
This proof is completely similar to given in sections 7, 9, 10 in order to link with RCQM and to interprate the standard Dirac equation.
Hence, the equation (252) describes the spin s=(3/2,3/2) particle-antiparticle doublet on the same level, on which the standard 4-component Dirac equation describes the spin s=(1/2,1/2) particle-antiparticle doublet. Moreover, the external argument in the validity of such interpretation is the link with the corresponding RCQM of spin s=(3/2,3/2) particle-antiparticle doublet, where the quantum-mechanical interpretation is direct and evident. Therefore, the fermionic spin s=(3/2,3/2) properties of equation (252) are proved.

Contrary to the bosonic spin s=(1,0,1,0) properties of the equation  (252) found in section 22, the fermionic spin s=(1/2,1/2,1/2,1/2) properties of this equation are evident. The fact that equation (252) describes the multiplet of two fermions with the spin s=1/2 and two antifermions with that spin can be proved much more easier then the above given consideration. The proof is similar to that given in the standard 4-component Dirac model. The detailed consideration can be found in sections 7, 9, 10 here. Moreover, in this section equation (252) has been proved to describe the spin s=(3/2,3/2) particle-antiparticle doublet and the corresponding spinor field. Therefore, equation (252) has more extended property of the Fermi--Bose duality then the standard Dirac equation [24--28]. This equation has the property of the Fermi--Bose triality. The property of the Fermi--Bose triality of the manifestly covariant equation (252) means that this equation describes on equal level (i) the spin s=(1/2,1/2,1/2,1/2) multiplet of two spin s=(1/2,1/2) fermions and two spin s=(1/2,1/2) antifermions, (ii) the spin s=(1,0,1,0) multiplet of the vector and scalar bosons together with their antiparticles, (iii) the spin s=(3/2,3/2) particle-antiparticle doublet.

Interaction, quantization and Lagrange approach in the given spin s=(3/2,3/2) model are completely similar to the Dirac 4-component theory and standard quantum electrodynamics. For example, the Lagrange function of the system of interacting 8 component spinor and electromagnetic fields (in the terms of 4-vector potential $A^{\mu}(x)$) is given by

$$L=-\frac{1}{4}F^{\mu\nu}F_{\mu\nu}+$$
\begin{equation}
\label{eq291} \frac{i}{2} \left(\overline{\psi}(x)\Gamma_{8}^{\mu}\frac{\partial \psi (x)}{\partial x^{\mu}}-\frac{\partial \overline{\psi}(x)}{\partial x^{\mu}}\Gamma_{8}^{\mu}\psi (x)\right)-
\end{equation}
$$m\overline{\psi}(x)\psi (x)+q\overline{\psi}(x)\Gamma_{8}^{\mu}\psi (x)A_{\mu}(x),$$

\noindent where $\overline{\psi}(x)$ is the independent Lagrange variable and $\overline{\psi}=\psi^{\dag}\Gamma_{8}^{0}$ in the space of solutions $\left\{\psi\right\}$. In Lagrangian (291) $F_{\mu\nu}=\partial _{\mu}A_{\nu}-\partial _{\nu}A_{\mu}$ is the electromagnetic field tensor in the terms of potentials, which play the role of variational variables in this Lagrange approach.

Therefore, the covariant local quantum field theory model for the interacting particles with spin s=3/2 and photons can be constructed in complete analogy to the construction of the modern quantum electrodynamics. This model can be useful for the investigations of processes with interacting hyperons and photons.  

\section{Field equation for the 10-component spin s=(2,2) particle-antiparticle doublet}

This model is constructed in complete analogy for the consideration of the spin s=(1,1) particle-antiparticle doublet given in section 21.

The RCQM of the spin s=(2,2) particle-antiparticle doublet is considered in section 15. Transition to the canonical field theory is fulfilled in a complete analogy with the method suggested in [3, 14] (see details in sections 9, 10 here). The difference is only in the explicit form of the transition operator, which is now given by the $10 \times 10$ diagonal matrix with the $5 \times 5$ operator of the complex conjugation in the bottom part.

Thus, the Schr$\mathrm{\ddot{o}}$dinger--Foldy equation (186) and its solution (187) are linked with the canonical field theory equation

\begin{equation}
\label{eq292} (i\partial_{0}-
\Gamma^{0}_{10}  \widehat{\omega})\phi(x)=0, \quad \Gamma^{0}_{10} =
\left| {{\begin{array}{*{20}c}
 \mathrm{I}_{5} \hfill &  0 \hfill\\
 0 \hfill & -\mathrm{I}_{5}  \hfill\\
 \end{array} }} \right|,
\end{equation}
$$\widehat{\omega}\equiv \sqrt{-\Delta+m^{2}},$$

\noindent and its solution

$$\phi(x)=\frac{1}{\left(2\pi\right)^{\frac{3}{2}}}\int d^{3}k$$
\begin{equation}
\label{eq293} \left[e^{-ikx}g^{\mathrm{A}}(\overrightarrow{k})\mathrm{d}_{\mathrm{A}}+e^{ikx}g^{*\mathrm{B}}(\overrightarrow{k})\mathrm{d}_{{\mathrm{B}}}\right],
\end{equation}
$$\mathrm{A} = 1,2,3,4,5, \quad \mathrm{B} = 6,7,8,9,10,$$

\noindent with the help of the following operator

\begin{equation}
\label{eq294} v_{10} = \left| {{\begin{array}{*{20}c}
 \mathrm{I}_{5} \hfill &  0 \hfill\\
 0 \hfill & C\mathrm{I}_{5}  \hfill\\
 \end{array} }} \right|, \quad v_{10}^{-1}= v_{10}^{\dag}= v_{10}, \, v_{10}v_{10}=\mathrm{I}_{10},
\end{equation}

\noindent where the Cartesian orts $(\mathrm{d}_{\mathrm{A}}, \, \mathrm{d}_{\mathrm{B}})$ are given in (188), $C$ is the operator of the complex conjugation and $\mathrm{I}_{5}$ is the $5\times 5$ unit matrix.

The operator $v_{10}$ (294) transforms an arbitrary operator $\hat{q}_{\mathrm{qm}}$ of the 10-component RCQM from section 15 into the corresponding operator $\hat{q}_{\mathrm{cf}}$ of the canonical field theory and \textit{vice versa}:

\begin{equation}
\label{eq295} v_{10}\hat{q}_{\mathrm{qm}}^{\mathrm{anti-Herm}}v_{10}
= \hat{q}_{\mathrm{cf}}^{\mathrm{anti-Herm}},
\end{equation}
\begin{equation}
\label{eq296} v_{10}\hat{q}_{\mathrm{cf}}^{\mathrm{anti-Herm}}v_{10}
= \hat{q}_{\mathrm{qm}}^{\mathrm{anti-Herm}},
\end{equation}

\noindent where the operators $\hat{q}_{\mathrm{qm}}$ and $\hat{q}_{\mathrm{cf}}$ must be taken in the anti-Hermitian form. The corresponding link between solutions (187) and (293) is as follows

\begin{equation}
\label{eq297} \phi=v_{10} f, \quad f=v_{10}\phi.
\end{equation}

Note that formulae (295), (296) are valid for the anti-Hermitian (prime) operators only. For the goals stated in our previous papers [24--28], we often use the operators in the anti-Hermitian form (see also the comments in section 9). The mathematical correctness of such choice and the physical interpretation are explained in the books [29, 30]. Return to the Hermitian operators is very easy.

The important examples of the transitions (295), (296) are the transformations of the operators of equations (186), (292)

\begin{equation}
\label{eq298} v_{10}(\partial_{0}+i\widehat{\omega})v_{10}=
\partial_{0}+i \Gamma_{10}^{0}\widehat{\omega},
\end{equation}
\begin{equation}
\label{eq299}
v_{10}(\partial_{0}+i\Gamma_{10}^{0}\widehat{\omega})v_{10}=
\partial_{0}+i\widehat{\omega},
\end{equation}

\noindent and of the SU(2) spin operators. The spin operators of the canonical field theory found from the RCQM SU(2) spin (190) on the basis of the transformation (294) satisfy the commutation relations (30) and have the following form

\begin{equation}
\label{eq300} \overrightarrow{s}_{10}= \left| {{\begin{array}{*{20}c}
 \overrightarrow{s}_{5} \hfill  & 0 \hfill \\
 0 \hfill & \overrightarrow{s}_{5} \hfill \\
 \end{array} }} \right|,
\end{equation}

\noindent where the $5 \times 5$ spin s=2 SU(2) generators are denoted as $\overrightarrow{s}_{5}$ and are given in (48). The corresponding Casimir operator is given by

\begin{equation}
\label{eq301} \overrightarrow{s}_{10}^{2}= 6\mathrm{I}_{10}=
2(2+1)\mathrm{I}_{10},
\end{equation}

\noindent where $\mathrm{I}_{10}$ is the $10 \times 10$- unit matrix.

The stationary complete set of operators is given by the operators $\overrightarrow{p}, \, s^{3}_{10}=s_{z}$ of the momentum and spin
projection. The equations on the eigenvectors and eigenvalues of the operators $\overrightarrow{p}$ and $s^{3}_{10}=s_{z}$ from (300) have the form

$$\overrightarrow{p}e^{-ikx}\mathrm{d}_{\mathrm{A}} =
\overrightarrow{k}e^{-ikx}\mathrm{d}_{\mathrm{A}}, \quad \mathrm{A} =\overline{1,5},$$
$$\overrightarrow{p}e^{ikx}\mathrm{d}_{\mathrm{B}} =
-\overrightarrow{k}e^{ikx}\mathrm{d}_{\mathrm{B}}, \quad \mathrm{B} =\overline{6,10},$$
$$s^{3}_{10}\mathrm{d}_{1} = 2\mathrm{d}_{1}, \, s^{3}_{10}\mathrm{d}_{2} = \mathrm{d}_{2}, \, s^{3}_{10}\mathrm{d}_{3} =0, \, s^{3}_{10}\mathrm{d}_{4} =  -\mathrm{d}_{4},$$
\begin{equation}
\label{eq302} s^{3}_{10}\mathrm{d}_{5} = -2\mathrm{d}_{5}, \, s^{3}_{10}\mathrm{d}_{6} = 2\mathrm{d}_{6}, \, s^{3}_{10}\mathrm{d}_{7} = \mathrm{d}_{7},
\end{equation}
$$s^{3}_{10}\mathrm{d}_{8} = 0, \, s^{3}_{10}\mathrm{d}_{9} = -\mathrm{d}_{9}, \,  s^{3}_{10}\mathrm{d}_{10} =  -2\mathrm{d}_{10},$$

\noindent and determine the interpretation of the amplitudes in the general solution (293). Note that the direct quantum-mechanical interpretation of the amplitudes $g^{\mathrm{A}} (\overrightarrow{k}), \, g^{\mathrm{B}} (\overrightarrow{k})$ in solution (293) should be taken from the quantum-mechanical equations (193), (194) and is given in section 15.

The relativistic invariance of the canonical field equation (292) follows from the corresponding invariance of the Schr$\mathrm{\ddot{o}}$dinger--Foldy equation (186) and transformation (294)--(297) (for the anti-Hermitian operators). The explicit form of the corresponding generators follows from of the explicit form the generators (12), (13) with the spin matrices (190), (191) and  transformation (294)--(297).

Thus, the canonical field equation (292) and the set
$\left\{\phi\right\}$ of its solutions (293) are invariant with respect
to the reducible unitary bosonic representation (14) of the
Poincar$\mathrm{\acute{e}}$ group $\mathcal{P}$, whose Hermitian  $10 \times 10$ matrix-differential generators are
given by

$$\widehat{p}^{0}=\Gamma_{10}^{0}\widehat{\omega}\equiv \Gamma_{10}^{0}\sqrt{-\Delta+m^{2}}, \quad \widehat{p}^{\ell}=-i\partial_{\ell},$$
\begin{equation}
\label{eq303} \widehat{j}^{\ell
n}=x^{\ell}\widehat{p}^{n}-x^{n}\widehat{p}^{\ell}+s^{\ell
n}_{10}\equiv \widehat{m}^{\ell n}+s^{\ell n}_{10},
\end{equation}
\begin{equation}
\label{eq304} \widehat{j}^{0 \ell}=-\widehat{j}^{\ell
0}=x^{0}\widehat{p}^{\ell}-\frac{1}{2}\Gamma_{10}^{0}\left\{x^{\ell},\widehat{\omega}\right\}+\Gamma_{10}^{0}\frac{(\overrightarrow{s}_{10}\times
\overrightarrow{p})^{\ell}}{\widehat{\omega}+m},
\end{equation}

\noindent where the spin s=(2,2) SU(2) generators
$\overrightarrow{s}_{10}=(s^{\ell n}_{10})$ have the form (300).

It is easy to prove by the direct verification that generators (303), (304) commute with the operator $(i\partial_{0}- \Gamma_{6}^{0}\widehat{\omega})$ of the canonical field equation (292) and satisfy the commutation relations (11) of the Lie algebra of the Poincar$\mathrm{\acute{e}}$ group $\mathcal{P}$.

The corresponding Casimir operators are given by

\begin{equation}
\label{eq305}
p^{2}=\widehat{p}^{\mu}\widehat{p}_{\mu}=m^{2}\mathrm{I}_{10},
\end{equation}
\begin{equation}
\label{eq306} W=w^{\mu}w_{\mu}=m^{2}\overrightarrow{s}^{2}_{10}=
2\left(2+1\right)m^{2}\mathrm{I}_{10},
\end{equation}

\noindent where $\mathrm{I}_{10}$ is the $10 \times 10$ unit matrix. Note that the difference between the Casimir operators (305), (306) and the corresponding expressions in RCQM, see the formulae (195), (196) above, is only in the explicit form of the operator $\overrightarrow{s}_{10}$.

Thus, due to the eigenvalues in equations (302), positive and
negative frequencies form of the solution (293) and the
Bargman--Wigner analysis of the Casimir operators (305), (306), one
can come to a conclusion that equation (292) describes the
canonical field (the bosonic particle-antiparticle doublet) with
the spins s=(2,2) and $m>0$. Transition to the $m=0$ limit leads
to the canonical field equation for the graviton-antigraviton field.

Final transition to the covariant local field theory in the case of this 10-component doublet contains the specific features and is the subject of next publication. The result of such transition can be the Duffin--Kemmer--Petiau equation or the 10-component Dirac-like equation for the spin s=(2,2) and $m>0$ classical field. The resulting covariant local field theory equation can be found in an analogy with the consideration of the spin s=(1,1) doublet given above. 

\section{Covariant field equation for the 12-component spin s=(2,0,2,0) bosonic particle-antiparticle multiplet}

In this section, the field model of the spin s=(2,0) bosonic
multiplet and the corresponding spin s=(2,0) antiparticle
multiplet is constructed. The complete analogy with the case of the
spin s=(1,0,1,0) model of section 22 is used.

The start of this derivation is given in section 16, where the
RCQM of the 12-component bosonic spin s=(2,0,2,0)
particle-antiparticle multiplet is considered. The second step is
the transition from the Schr$\mathrm{\ddot{o}}$dinger--Foldy
equation (197) to the canonical field equation. This step, as
shown above, is possible only for the
anti-Hermitian form of the operators. Nevertheless, the resulting
operators can be chosen in the standard Hermitian form and do not
contain the operator $C$ of complex conjugation. The last step of
the transition from the canonical field equation to the covariant
local field equation is fulfilled in analogy with the FW
transformation (119).

Thus, the corresponding canonical field equation is found in the
form

\begin{equation}
\label{eq307} (i\partial_{0}-
\Gamma_{12}^{0}\widehat{\omega})\phi(x)=0, \quad \phi=\left|
{{\begin{array}{*{20}c}
 \phi^{1} \hfill  \\
 \phi^{2} \hfill  \\
   . \hfill  \\
   . \hfill  \\
   . \hfill  \\
 \phi^{12} \hfill  \\
\end{array} }} \right|,
\end{equation}

where

$$\widehat{\omega}\equiv \sqrt{-\Delta+m^{2}},$$
\begin{equation}
\label{eq308} \Gamma_{12}^{0} = \left| {{\begin{array}{*{20}c}
 \mathrm{I}_{6} \hfill &  0 \hfill\\
 0 \hfill & -\mathrm{I}_{6}  \hfill\\
 \end{array} }} \right|, \quad \mathrm{I}_{6}=\left|
{{\begin{array}{*{20}c}
 1 \hfill & 0 \hfill & 0 \hfill & 0 \hfill & 0 \hfill & 0 \hfill\\
 0 \hfill & 1 \hfill & 0 \hfill & 0 \hfill & 0 \hfill & 0 \hfill\\
 0 \hfill & 0 \hfill & 1 \hfill & 0 \hfill & 0 \hfill & 0 \hfill\\
 0 \hfill & 0 \hfill & 0 \hfill & 1 \hfill & 0 \hfill & 0 \hfill\\
 0 \hfill & 0 \hfill & 0 \hfill & 0 \hfill & 1 \hfill & 0 \hfill\\
 0 \hfill & 0 \hfill & 0 \hfill & 0 \hfill & 0 \hfill & 1 \hfill\\
\end{array} }} \right|.
\end{equation}

The general solution of equation (307) is given by

$$\phi(x)= \frac{1}{\left(2\pi\right)^{\frac{3}{2}}}\int d^{3}k$$
\begin{equation}
\label{eq309}
\left[e^{-ikx}g^{\mathrm{A}}(\overrightarrow{k})\mathrm{d}_{\mathrm{A}}+e^{ikx}g^{*\mathrm{B}}(\overrightarrow{k})\mathrm{d}_{\mathrm{B}}\right],
\end{equation}

\noindent where $\mathrm{A}=\overline{1,6}, \,
\mathrm{B}=\overline{5,12}$ and the orts of the 12-component
Cartesian basis are given in (199).

The transition from the
Schr$\mathrm{\ddot{o}}$dinger--Foldy equation (197) and its
solution (198) to canonical field equation (307) and solution (309) is given by the
operator

\begin{equation}
\label{eq310} v_{12} = \left| {{\begin{array}{*{20}c}
 \mathrm{I}_{6} \hfill &  0 \hfill\\
 0 \hfill & C\mathrm{I}_{6}  \hfill\\
 \end{array} }} \right|, \quad v_{12}^{-1}= v_{12}^{\dag}= v_{12}, \, v_{12}v_{12}=\mathrm{I}_{12},
\end{equation}
\begin{equation}
\label{eq311} \phi=v_{12} f, \quad f=v_{12}\phi,
\end{equation}
\begin{equation}
\label{eq312} v_{12}\hat{q}_{\mathrm{qm}}^{\mathrm{anti-Herm}}v_{12}
= \hat{q}_{\mathrm{cf}}^{\mathrm{anti-Herm}},
\end{equation}
\begin{equation}
\label{eq313} v_{12}\hat{q}_{\mathrm{cf}}^{\mathrm{anti-Herm}}v_{12}
= \hat{q}_{\mathrm{qm}}^{\mathrm{anti-Herm}}.
\end{equation}

\noindent Here $\hat{q}_{\mathrm{qm}}^{\mathrm{anti-Herm}}$ is an arbitrary operator from the RCQM of the 12-component particle-antiparticle doublet in the anti-Hermitian form, e. g., the operator $(\partial_{0}+i\widehat{\omega})$ of equation of motion, the operator of spin (201), etc.,  $\hat{q}_{\mathrm{cf}}^{\mathrm{anti-Herm}}$ is an arbitrary operator from the canonical field theory of the 12-component particle-antiparticle doublet in the anti-Hermitian form, $C\mathrm{I}_{6}$ is the $6\times 6$ operator of complex conjugation.

The SU(2) spin operators, which satisfy the commutation relations
(30) and commute with the operator $(i\partial_{0}-
\Gamma_{12}^{0}\widehat{\omega})$ of the equation of motion (307),
are derived from the RCQM operators (201) on the basis of
transformations (310), (312). These canonical field spin
operators are given by

\begin{equation}
\label{eq314} \overrightarrow{s}_{12}=\left|
{{\begin{array}{*{20}c}
 \overrightarrow{s}_{6} \hfill & 0 \\
 0 \hfill & \overrightarrow{s}_{6} \\
\end{array} }} \right|,
\end{equation}

\noindent where the $6 \times 6$ operators $\overrightarrow{s}_{6}$ are
given in (202). The corresponding Casimir operator is given by the $12 \times 12$ diagonal matrix (203). Note that this Casimir operator is the same both for the spin (201) and for the spin (314).

The stationary complete set of operators is given by the operators $\overrightarrow{p}, \, s_{12}^{3}=s_{z}$ of the momentum and spin
projection. The equations on the eigenvectors and eigenvalues of the spin projection operator $s_{12}^{3}=s_{z}$ from (314) have the form

$$s^{3}_{12}\mathrm{d}_{1} = 2\mathrm{d}_{1}, \, s^{3}_{12}\mathrm{d}_{2} = \mathrm{d}_{2}, \, s^{3}_{12}\mathrm{d}_{3} = 0, \, s^{3}_{12}\mathrm{d}_{4} = -\mathrm{d}_{4},$$
\small
\begin{equation}
\label{eq315} s^{3}_{12}\mathrm{d}_{5} = - 2\mathrm{d}_{5}, \, s^{3}_{12}\mathrm{d}_{6} =0, \, s^{3}_{12}\mathrm{d}_{7} = 2\mathrm{d}_{7}, \, 
s^{3}_{12}\mathrm{d}_{8} = \mathrm{d}_{8},
\end{equation}
\normalsize
$$s^{3}_{12}\mathrm{d}_{9} = 0, \, s^{3}_{12}\mathrm{d}_{10}=-\mathrm{d}_{10}, \, s^{3}_{12}\mathrm{d}_{11} = -2\mathrm{d}_{11},$$
$$s^{3}_{12}\mathrm{d}_{12} =0.$$

Therefore, the functions $g^{1}(\overrightarrow{k}), \,
g^{2}(\overrightarrow{k}), \, g^{3}(\overrightarrow{k}), \, g^{4}(\overrightarrow{k}), \, g^{5}(\overrightarrow{k})$ in
solution (309) are the momentum-spin amplitudes of the massive
boson with the spin s=2 and the spin projection $(2,1,0,-1,-2)$,
respectively, $g^{6}(\overrightarrow{k})$ is the amplitude of the
spinless boson; $g^{7}(\overrightarrow{k}), \,
g^{8}(\overrightarrow{k}), \, g^{9}(\overrightarrow{k}), \, \, g^{10}(\overrightarrow{k}), \, g^{11}(\overrightarrow{k})$ are the
momentum-spin amplitudes of the antiparticle (antiboson) with the
spin s=2 and the spin projection $(2,1,0,-1,-2)$, respectively,
$g^{12}(\overrightarrow{k})$ is the amplitude of the spinless
antiboson.

Note that direct quantum-mechanical interpretation of the amplitudes in solution (309) should be given in the framework of the RCQM. Such interpretation is already given in section 16 in paragraph after equations (205). 

The generators of the reducible unitary bosonic spin s=(2,0,2,0)
multiplet representation of the Poincar$\mathrm{\acute{e}}$ group
$\mathcal{P}$, with respect to which the canonical field equation
(307) and the set $\left\{\phi\right\}$ of its solutions (309) are
invariant, are derived from the RCQM set of generators (12), (13)
with the spin (201) on the basis of the transformations (310),
(312). These Hermitian  $12 \times 12$ matrix-differential
generators are given by

$$\widehat{p}^{0}=\Gamma_{12}^{0}\widehat{\omega}\equiv \Gamma_{12}^{0}\sqrt{-\Delta+m^{2}}, \quad \widehat{p}^{\ell}=-i\partial_{\ell},$$
\begin{equation}
\label{eq316} \widehat{j}^{\ell
n}=x^{\ell}\widehat{p}^{n}-x^{n}\widehat{p}^{\ell}+s_{12}^{\ell
n}\equiv \widehat{m}^{\ell n}+s_{12}^{\ell n},
\end{equation}
\begin{equation}
\label{eq317} \widehat{j}^{0 \ell}=-\widehat{j}^{\ell
0}=x^{0}\widehat{p}^{\ell}-\frac{1}{2}\Gamma_{12}^{0}\left\{x^{\ell},\widehat{\omega}\right\}+\Gamma_{12}^{0}\frac{(\overrightarrow{s}_{12}\times
\overrightarrow{p})^{\ell}}{\widehat{\omega}+m},
\end{equation}

\noindent where the spin s=(2,0,2,0) SU(2) generators
$\overrightarrow{s}_{12}=(s_{12}^{\ell n})$ have the form (314).

It is easy to prove by the direct verification that the generators (316), (317) commute with the operator $(i\partial_{0}- \Gamma_{12}^{0}\widehat{\omega})$ of the canonical field equation (307) and satisfy the commutation relations (11) of the Lie algebra of the Poincar$\mathrm{\acute{e}}$ group $\mathcal{P}$. The Casimir operators for the representation (316), (317) are given by

\begin{equation}
\label{eq318} p^{2}=\widehat{p}^{\mu}\widehat{p}_{\mu}=m^{2}\mathrm{I}_{12},
\end{equation}
$$W=w^{\mu}w_{\mu}=m^{2}\overrightarrow{s}_{12}^{2}=$$
\begin{equation}
\label{eq319} m^{2}\left| {{\begin{array}{*{20}c}
 2\left(2+1\right)\mathrm{I}_{5} \hfill & 0 \hfill & 0 \hfill & 0 \hfill\\
 0 \hfill & 0 \hfill & 0 \hfill & 0 \hfill\\
 0 \hfill & 0 \hfill & 2\left(2+1\right)\mathrm{I}_{5} \hfill & 0 \hfill\\
 0 \hfill & 0 \hfill & 0 \hfill & 0 \hfill\\
\end{array} }} \right|,
\end{equation}

\noindent where $\mathrm{I}_{12}$ is $12 \times 12$ unit matrix and $\mathrm{I}_{5}$ is $5 \times 5$ unit matrix.

Thus, due to the eigenvalues in equations (315), positive and
negative frequencies form of solution (309) and the
Bargman--Wigner analysis of the Casimir operators (318), (319) one
can come to a conclusion that equation (307) describes the
12-component canonical field (the bosonic particle-antiparticle
doublet) with the spins s=(2,0,2,0) and $m>0$. Transition to the
$m=0$ limit leads to the canonical field equation for the
12-component (graviton massless spinless boson)-(antigraviton-
masslees spinless antiboson) field.

The operator of the transition to the covariant local field theory
representation (the $12 \times 12$  analogy of the $4 \times 4$ FW
transformation operator (119)) is given by

\begin{equation}
\label{eq320} V^{\mp}=\frac{\mp
\overrightarrow{\Gamma}_{12}\cdot\overrightarrow{p}+\widehat{\omega}+m}{\sqrt{2\widehat{\omega}(\widehat{\omega}+m)}},
\quad V^{-}=(V^{+})^{\dag},
\end{equation}
$$V^{-}V^{+}=V^{+}V^{-}=\mathrm{I}_{12},$$
\begin{equation}
\label{eq321} \psi=V^{-}\phi, \, \phi = V^{+}\psi,
\end{equation}
\begin{equation}
\label{eq322} \hat{q}_{\mathrm{D}}=
V^{-}\hat{q}_{\mathrm{CF}}V^{+}, \, \hat{q}_{\mathrm{CF}}=
V^{+}\hat{q}_{\mathrm{D}}V^{-},
\end{equation}

\noindent where $\hat{q}_{\mathrm{D}}$ is an arbitrary operator
(both in the Hermitian and anti-Hermitian form) in the covariant
local field theory representation. The inverse transformation is
valid as well.

Thus, on the basis of transformation (320)--(322) the 12-component
Dirac-like equation is found from the canonical field equation
(307) in the form

\begin{equation}
\label{eq323}
\left[i\partial_{0}-\Gamma_{12}^{0}(\overrightarrow{\Gamma}_{12}\cdot\overrightarrow{p}+m)\right]\psi(x)=0.
\end{equation}

\noindent In formula (320) and in equation (323) the
$\Gamma_{12}^{\mu}$ matrices are given by

\begin{equation}
\label{eq324} \Gamma_{12}^{0}=\left| {{\begin{array}{*{20}c}
 \mathrm{I}_{6} \hfill & 0 \\
 0 \hfill & -\mathrm{I}_{6} \\
\end{array} }} \right|, \quad \Gamma_{12}^{j}=\left| {{\begin{array}{*{20}c}
 0 \hfill & \Sigma^{j}_{6} \\
 -\Sigma^{j}_{6} \hfill & 0 \\
\end{array} }} \right|,
\end{equation}

\noindent where $\Sigma^{j}_{6}$ are the $6\times 6$ Pauli matrices

\begin{equation}
\label{eq325} \Sigma^{1}_{6}=\left| {{\begin{array}{*{20}c}
 0 \hfill & \mathrm{I}_{3} \\
 \mathrm{I}_{3} \hfill &  0 \\
\end{array} }} \right|, \quad \Sigma^{2}_{6}=\left| {{\begin{array}{*{20}c}
 0 \hfill & -i\mathrm{I}_{3} \\
 i\mathrm{I}_{3} \hfill &  0 \\
\end{array} }} \right|,
\end{equation}
$$\Sigma^{3}_{6}=\left| {{\begin{array}{*{20}c}
 \mathrm{I}_{3} \hfill & 0 \\
 0 \hfill &  -\mathrm{I}_{3}\\
\end{array} }} \right|,$$

\noindent and $\mathrm{I}_{3}$ is the $3\times 3$ unit matrix.

The matrices $\Sigma^{j}_{6}$ satisfy the similar commutation
relations as the $2\times 2$ Pauli matrices (20) and have other
similar properties. The matrices $\Gamma_{12}^{\mu}$ (324) satisfy
the anticommutation relations of the Clifford--Dirac algebra in
the form

\begin{equation}
\label{eq326}
\Gamma_{12}^{\mu}\Gamma_{12}^{\nu}+\Gamma_{12}^{\nu}\Gamma_{12}^{\mu}=2g^{\mu\nu}.
\end{equation}

The solution of equation (323) is derived from solution (309) of
this equation in the canonical representation (307) on the basis
of transformation (320), (321) and is given by

$$\psi(x)=V^{-}\phi(x)= \frac{1}{\left(2\pi\right)^{\frac{3}{2}}}\int d^{3}k$$
\begin{equation}
\label{eq327}
\left[e^{-ikx}g^{\mathrm{A}}(\overrightarrow{k})\mathrm{v}^{-}_{\mathrm{A}}(\overrightarrow{k})+e^{ikx}g^{*\mathrm{B}}(\overrightarrow{k})\mathrm{v}^{+}_{\mathrm{B}}(\overrightarrow{k})\right],
\end{equation}

\noindent where $\mathrm{A}=\overline{1,6}, \, \mathrm{B}=\overline{7,12}$ and the 12-component spinors $(\mathrm{v}^{-}_{\mathrm{A}}(\overrightarrow{k}), \, \mathrm{v}^{+}_{\mathrm{B}}(\overrightarrow{k}))$ are given by

$$\mathrm{v}^{-}_{1}(\overrightarrow{k}) = N\left|
\begin{array}{cccc}
 \tilde{\omega}+m \\
 0 \\
 0 \\
 0 \\
 0 \\
 0 \\
 k^{3} \\
 0 \\
 0 \\ 
 k^{1}+ik^{2} \\
 0 \\
 0 \\
\end{array} \right|, \, \mathrm{v}^{-}_{2}(\overrightarrow{k}) = N\left|
\begin{array}{cccc}
 0 \\
 \tilde{\omega}+m \\
 0 \\
 0 \\
 0 \\
 0 \\
 0 \\
 k^{3} \\
 0 \\
 0 \\ 
 k^{1}+ik^{2} \\
 0 \\
\end{array} \right|,$$
$$ \mathrm{v}^{-}_{3}(\overrightarrow{k}) = N \left|
\begin{array}{cccc}
 0 \\
 0 \\
 \tilde{\omega}+m \\
 0 \\
 0 \\
 0 \\
 0 \\
 0 \\
 k^{3} \\
 0 \\
 0 \\ 
k^{1}+ik^{2} \\
\end{array} \right|,
\, \mathrm{v}^{-}_{4}(\overrightarrow{k}) = N\left|
\begin{array}{cccc}
 0 \\
 0 \\
 0 \\
 \tilde{\omega}+m \\
 0 \\
 0 \\
 k^{1}-ik^{2} \\
 0 \\
 0 \\ 
-k^{3} \\
 0 \\
 0 \\
\end{array} \right|,$$
\begin{equation}
\label{eq328} \mathrm{v}^{-}_{5}(\overrightarrow{k}) = N\left|
\begin{array}{cccc}
 0 \\
 0 \\
 0 \\
 0 \\
 \tilde{\omega}+m \\
 0 \\
 0 \\
 k^{1}-ik^{2} \\
 0 \\
 0 \\ 
-k^{3} \\
 0 \\
\end{array} \right|, \,
\mathrm{v}^{-}_{6}(\overrightarrow{k}) = N\left|
\begin{array}{cccc}
 0 \\
 0 \\
 0 \\
 0 \\
 0 \\
 \tilde{\omega}+m \\
 0 \\
 0 \\
 k^{1}-ik^{2} \\
 0 \\
 0 \\ 
-k^{3} \\
\end{array} \right|,
\end{equation}
$$\mathrm{v}^{+}_{7}(\overrightarrow{k}) = N\left|
\begin{array}{cccc}
 k^{3} \\
 0 \\
 0 \\
 k^{1}+ik^{2} \\
 0 \\
 0 \\
 \tilde{\omega}+m \\
 0 \\
 0 \\
 0 \\
 0 \\
 0 \\
\end{array} \right|, \,
\mathrm{v}^{+}_{8}(\overrightarrow{k}) = N\left|
\begin{array}{cccc}
 0 \\
 k^{3} \\
 0 \\
 0 \\
 k^{1}+ik^{2} \\
 0 \\
 0 \\
 \tilde{\omega}+m \\
 0 \\
 0 \\
 0 \\
 0 \\
\end{array} \right|,$$
$$\mathrm{v}^{+}_{9}(\overrightarrow{k}) = N\left|
\begin{array}{cccc}
 0 \\
 0 \\
 k^{3} \\
 0 \\
 0 \\
 k^{1}+ik^{2} \\
 0 \\
 0 \\
 \tilde{\omega}+m \\
 0 \\
 0 \\
 0 \\
 \end{array} \right|, \,
\mathrm{v}^{+}_{10}(\overrightarrow{k}) = N\left|
\begin{array}{cccc}
 k^{1}-ik^{2} \\ 
 0 \\
 0 \\
 -k^{3} \\
 0 \\
 0 \\
 0 \\
 0 \\
 0 \\
 \tilde{\omega}+m \\
 0 \\
 0 \\
\end{array} \right|,$$
$$\mathrm{v}^{+}_{11}(\overrightarrow{k}) = N\left|
\begin{array}{cccc}
 0 \\
 k^{1}-ik^{2} \\
 0 \\
 0 \\
 -k^{3} \\
 0 \\
 0 \\
 0 \\
 0 \\
 0 \\
 \tilde{\omega}+m \\
 0 \\
 \end{array} \right|, \,
\mathrm{v}^{+}_{12}(\overrightarrow{k}) = N\left|
\begin{array}{cccc}
 0 \\
 0 \\ 
 k^{1}-ik^{2} \\ 
 0 \\
 0 \\
 -k^{3} \\
 0 \\
 0 \\
 0 \\
 0 \\
 0 \\
 \tilde{\omega}+m \\
 \end{array} \right|,$$

\noindent where $N$ is given in (258).

The spinors (328) are derived from the orts
$\left\{\mathrm{d}\right\}$ of the Cartesian basis (199)
with the help of the transformation (320), (321). The spinors
(328) satisfy the relations of the orthonormalization and
completeness similar to the corresponding relations for the
standard 4-component Dirac spinors, see, e. g., [33]. 

In the covariant local field theory, the operators of the SU(2) spin,
which satisfy the corresponding commutation relations $
\left[s_{12\mathrm{D}}^{j},s_{12\mathrm{D}}^{\ell}\right]=i\varepsilon^{j
\ell n}s_{12\mathrm{D}}^{n}$ and commute with the operator
$\left[i\partial_{0}-\Gamma_{12}^{0}(\overrightarrow{\Gamma}_{12}\cdot\overrightarrow{p}+m)\right]$
of equation (323), are derived from the pure matrix operators
(314) with the help of operator (320), (322):

\begin{equation}
\label{eq329}
\overrightarrow{s}_{12\mathrm{D}} =
V^{-}\overrightarrow{s}_{12}V^{+}.
\end{equation}

\noindent The explicit form can be presented in analogy with previous sections.

The equations on eigenvectors and eigenvalues of the operator
$s_{12\mathrm{D}}^{3}$ (329) follow from the equations (315) and
the transformation (320)--(322). In addition to it, the action of
the operator $s_{12\mathrm{D}}^{3}$ (329)  on the spinors
$(\mathrm{v}^{-}_{\mathrm{A}}(\overrightarrow{k}), \,
\mathrm{v}^{+}_{\mathrm{B}}(\overrightarrow{k}))$ (328) also
leads to the result

$$s_{12\mathrm{D}}^{3}\mathrm{v}^{-}_{1}(\overrightarrow{k})= 2\mathrm{v}^{-}_{1}(\overrightarrow{k}), \, s_{12\mathrm{D}}^{3}\mathrm{v}^{-}_{2}(\overrightarrow{k}) = \mathrm{v}^{-}_{2}(\overrightarrow{k}),$$
$$s_{12\mathrm{D}}^{3}\mathrm{v}^{-}_{3}(\overrightarrow{k}) = 0, \,
s_{12\mathrm{D}}^{3}\mathrm{v}^{-}_{4}(\overrightarrow{k}) = -\mathrm{v}^{-}_{4}(\overrightarrow{k}),$$
\begin{equation}
\label{eq330}
s_{12\mathrm{D}}^{3}\mathrm{v}^{-}_{5}(\overrightarrow{k}) =
-2\mathrm{v}^{-}_{5}(\overrightarrow{k}), \,
s_{12\mathrm{D}}^{3}\mathrm{v}^{-}_{6}(\overrightarrow{k}) = 0,
\end{equation}
$$s_{12\mathrm{D}}^{3}\mathrm{v}^{+}_{7}(\overrightarrow{k}) = 2\mathrm{v}^{+}_{7}(\overrightarrow{k}), \, s_{12\mathrm{D}}^{3}\mathrm{v}^{+}_{8}(\overrightarrow{k}) = \mathrm{v}^{+}_{8}(\overrightarrow{k}),$$
$$s_{12\mathrm{D}}^{3}\mathrm{v}^{+}_{9}(\overrightarrow{k}) = 0, \, s_{12\mathrm{D}}^{3}\mathrm{v}^{+}_{10}(\overrightarrow{k}) = -\mathrm{v}^{+}_{10}(\overrightarrow{k}),$$
$$s_{12\mathrm{D}}^{3}\mathrm{v}^{+}_{11}(\overrightarrow{k}) = -2\mathrm{v}^{+}_{11}(\overrightarrow{k}), \, s_{12\mathrm{D}}^{3}\mathrm{v}^{+}_{12}(\overrightarrow{k}) = 0.$$

In order to verify equations (330) the identity (264)
should be used. In the case $\mathrm{v}^{+}_{\mathrm{B}}(\overrightarrow{k})$ in the
expression $s_{12\mathrm{D}}^{3}(\overrightarrow{k})$ (329) the
substitution $\overrightarrow{k}\rightarrow - \overrightarrow{k}$
is made.

The equations (330) determine the interpretation of the
amplitudes in solution (327). This interpretation is similar to the given above in the paragraph after equations (315). Nevertheless, the direct quantum-mechanical interpretation of the amplitudes should be made in the
framework of the RCQM, see the section 16 above.

The explicit form of the $\mathcal{P}$-generators of the bosonic
representation of the Poincar$\mathrm{\acute{e}}$ group
$\mathcal{P}$, with respect to which the covariant equation (323)
and the set $\left\{\psi\right\}$ of its solutions (327) are
invariant, is derived from the generators (316), (317) on the
basis of the transformation (320), (322). The corresponding
generators are given by

$$\widehat{p}^{0}=\Gamma_{12}^{0}(\overrightarrow{\Gamma}_{12}\cdot\overrightarrow{p}+m), \quad \widehat{p}^{\ell}=-i\partial_{\ell},$$
\begin{equation}
\label{eq331} \widehat{j}^{\ell
n}=x_{\mathrm{D}}^{\ell}\widehat{p}^{n}-x_{\mathrm{D}}^{n}\widehat{p}^{\ell}+s_{12\mathrm{D}}^{\ell
n}\equiv \widehat{m}^{\ell n}+s_{12\mathrm{D}}^{\ell n},
\end{equation}
\begin{equation}
\label{eq332} \widehat{j}^{0 \ell}=-\widehat{j}^{\ell
0}=x^{0}\widehat{p}^{\ell}-\frac{1}{2}\left\{x_{\mathrm{D}}^{\ell},\widehat{p}^{0}\right\}+\frac{\widehat{p}^{0}(\overrightarrow{s}_{12\mathrm{D}}\times
\overrightarrow{p})^{\ell}}{\widehat{\omega}(\widehat{\omega}+m)},
\end{equation}

\noindent where the spin matrices
$\overrightarrow{s}_{12\mathrm{D}}=(s_{12\mathrm{D}}^{\ell n})$ are
given in (329)  and the operator
$\overrightarrow{x}_{\mathrm{D}}$ has the form

\begin{equation}
\label{eq333}
\overrightarrow{x}_{\mathrm{D}}=\overrightarrow{x}+\frac{i\overrightarrow{\Gamma}_{12}}{2\widehat{\omega}}-\frac{\overrightarrow{s}_{12}\times \overrightarrow{p}}{\widehat{\omega}(\widehat{\omega}+m)}-\frac{i\overrightarrow{p} (\overrightarrow{\Gamma}_{12}\cdot \overrightarrow{p})}{2\widehat{\omega}^{2}(\widehat{\omega}+m)},
\end{equation}

\noindent where the spin matrices $\overrightarrow{s}_{12}$ are given in (314).

It is easy to verify that the generators (331), (332) commute with
the operator
$\left[i\partial_{0}-\Gamma_{12}^{0}(\overrightarrow{\Gamma}_{12}\cdot\overrightarrow{p}+m)\right]$
of equation (323), satisfy the commutation relations (11) of the
Lie algebra of the Poincar$\mathrm{\acute{e}}$ group and the
corresponding Casimir operators are given by

\begin{equation}
\label{eq334} p^{2}=\widehat{p}^{\mu}\widehat{p}_{\mu}=m^{2}\mathrm{I}_{12},
\end{equation}
$$W=w^{\mu}w_{\mu}=m^{2}\overrightarrow{s}_{12\mathrm{D}}^{2}=$$
\begin{equation}
\label{eq335} m^{2}\left| {{\begin{array}{*{20}c}
 2\left(2+1\right)\mathrm{I}_{5} \hfill & 0 \hfill & 0 \hfill & 0 \hfill\\
 0 \hfill & 0 \hfill & 0 \hfill & 0 \hfill\\
 0 \hfill & 0 \hfill & 2\left(2+1\right)\mathrm{I}_{5} \hfill & 0 \hfill\\
 0 \hfill & 0 \hfill & 0 \hfill & 0 \hfill\\
\end{array} }} \right|,
\end{equation}

\noindent where $\mathrm{I}_{12}$ is $12 \times 12$ unit matrix and $\mathrm{I}_{5}$ is $5 \times 5$ unit matrix.

As it was already explained in details in the previous sections,
the conclusion that equation (323) describes the bosonic
particle-antiparticle multiplet of the spin s=(2,0,2,0) and mass
$m>0$ (and its solution (327) is the bosonic field of the above
mentioned spin and nonzero mass) follows from the analysis of
equations (330) and the Casimir operators (334), (335). Moreover, the external argument in the validity of such interpretation is the link with the corresponding RCQM of spin s=(2,0,2,0) particle-antiparticle multiplet, where the quantum-mechanical interpretation is direct and evident.
Therefore, the bosonic spin s=(2,0,2,0) properties of equation (323) are proved.

\section{Covariant field equation for the 16-component spin s=(2,1,2,1) bosonic particle-antiparticle multiplet}

In this section, the field model of the spin s=(2,1) bosonic
multiplet and the corresponding spin s=(2,1) antiparticle
multiplet is constructed. The complete analogy with the case of the
spin s=(1,0,1,0) model of section 22 and the spin s=(2,0,2,0) model of the previous section is used.

The start of this derivation is given in section 17, where the
RCQM of the 16-component bosonic spin s=(2,1,2,1)
particle-antiparticle multiplet is considered. The second step is
the transition from the Schr$\mathrm{\ddot{o}}$dinger--Foldy
equation (208) to the canonical field equation. This step, as
shown above, is possible only for the
anti-Hermitian form of the operators. Nevertheless, the resulting
operators can be chosen in the standard Hermitian form and do not
contain the operator $C$ of complex conjugation. The last step of
the transition from the canonical field equation to the covariant
local field equation is fulfilled in analogy with the FW
transformation (119).

Thus, the corresponding canonical field equation is found in the
form

\begin{equation}
\label{eq336} (i\partial_{0}-
\Gamma_{16}^{0}\widehat{\omega})\phi(x)=0, \quad \phi=\left|
{{\begin{array}{*{20}c}
 \phi^{1} \hfill  \\
 \phi^{2} \hfill  \\
   . \hfill  \\
   . \hfill  \\
   . \hfill  \\
 \phi^{16} \hfill  \\
\end{array} }} \right|,
\end{equation}

where

\small
$$\widehat{\omega}\equiv \sqrt{-\Delta+m^{2}},$$
\begin{equation}
\label{eq337} \Gamma_{16}^{0} = \left| {{\begin{array}{*{20}c}
 \mathrm{I}_{8} \hfill &  0 \hfill\\
 0 \hfill & -\mathrm{I}_{8}  \hfill\\
 \end{array} }} \right|, \quad \mathrm{I}_{8}=\left|
{{\begin{array}{*{20}c}
 1 \hfill & 0 \hfill & 0 \hfill & 0 \hfill & 0 \hfill & 0 \hfill & 0 \hfill & 0 \hfill\\
 0 \hfill & 1 \hfill & 0 \hfill & 0 \hfill & 0 \hfill & 0 \hfill & 0 \hfill & 0 \hfill\\
 0 \hfill & 0 \hfill & 1 \hfill & 0 \hfill & 0 \hfill & 0 \hfill & 0 \hfill & 0 \hfill\\
 0 \hfill & 0 \hfill & 0 \hfill & 1 \hfill & 0 \hfill & 0 \hfill & 0 \hfill & 0 \hfill\\
 0 \hfill & 0 \hfill & 0 \hfill & 0 \hfill & 1 \hfill & 0 \hfill & 0 \hfill & 0 \hfill\\
 0 \hfill & 0 \hfill & 0 \hfill & 0 \hfill & 0 \hfill & 1 \hfill & 0 \hfill & 0 \hfill\\
 0 \hfill & 0 \hfill & 0 \hfill & 0 \hfill & 0 \hfill & 0 \hfill & 1 \hfill & 0 \hfill\\
 0 \hfill & 0 \hfill & 0 \hfill & 0 \hfill & 0 \hfill & 0 \hfill & 0 \hfill & 1 \hfill\\
\end{array} }} \right|.
\end{equation}
\normalsize

The general solution of equation (336) is given by

$$\phi(x)= \frac{1}{\left(2\pi\right)^{\frac{3}{2}}}\int d^{3}k$$
\begin{equation}
\label{eq338}
\left[e^{-ikx}g^{\mathrm{A}}(\overrightarrow{k})\mathrm{d}_{\mathrm{A}}+e^{ikx}g^{*\mathrm{B}}(\overrightarrow{k})\mathrm{d}_{\mathrm{B}}\right],
\end{equation}

\noindent where $\mathrm{A}=\overline{1,8}, \,
\mathrm{B}=\overline{9,16}$ and the orts of the 16-component
Cartesian basis are given in (211).

The transition from the
Schr$\mathrm{\ddot{o}}$dinger--Foldy equation (208) and its
solution (210) to canonical field equation (336) and solution (338) is given by the
operator

\begin{equation}
\label{eq339} v_{16} = \left| {{\begin{array}{*{20}c}
 \mathrm{I}_{8} \hfill &  0 \hfill\\
 0 \hfill & C\mathrm{I}_{8}  \hfill\\
 \end{array} }} \right|, \quad v_{16}^{-1}= v_{16}^{\dag}= v_{16}, \, v_{16}v_{16}=\mathrm{I}_{16},
\end{equation}
\begin{equation}
\label{eq340} \phi=v_{16} f, \quad f=v_{16}\phi,
\end{equation}
\begin{equation}
\label{eq341} v_{16}\hat{q}_{\mathrm{qm}}^{\mathrm{anti-Herm}}v_{16}
= \hat{q}_{\mathrm{cf}}^{\mathrm{anti-Herm}},
\end{equation}
\begin{equation}
\label{eq342} v_{16}\hat{q}_{\mathrm{cf}}^{\mathrm{anti-Herm}}v_{16}
= \hat{q}_{\mathrm{qm}}^{\mathrm{anti-Herm}}.
\end{equation}

\noindent Here $\hat{q}_{\mathrm{qm}}^{\mathrm{anti-Herm}}$ is an arbitrary operator from the RCQM of the 16-component particle-antiparticle doublet in the anti-Hermitian form, e. g., the operator $(\partial_{0}+i\widehat{\omega})$ of equation of motion, the operator of spin (212), etc.,  $\hat{q}_{\mathrm{cf}}^{\mathrm{anti-Herm}}$ is an arbitrary operator from the canonical field theory of the 16-component particle-antiparticle doublet in the anti-Hermitian form, $C\mathrm{I}_{8}$ is the $8\times 8$ operator of complex conjugation.

The SU(2) spin operators, which satisfy the commutation relations
(30) and commute with the operator $(i\partial_{0}-
\Gamma_{16}^{0}\widehat{\omega})$ of the equation of motion (336),
are derived from the RCQM operators (212) on the basis of
transformations (339), (341). These canonical field spin
operators are given by

\begin{equation}
\label{eq343} \overrightarrow{s}_{16}=\left|
{{\begin{array}{*{20}c}
 \overrightarrow{s}_{8} \hfill & 0 \\
 0 \hfill & \overrightarrow{s}_{8} \\
\end{array} }} \right|,
\end{equation}

\noindent where the $8 \times 8$ operators $\overrightarrow{s}_{8}$ are
given in (213). The corresponding Casimir operator is given by the $16 \times 16$ diagonal matrix (214). Note that this Casimir operator is the same both for the spin (212) and for the spin (343).

The stationary complete set of operators is given by the operators $\overrightarrow{p}, \, s_{16}^{3}=s_{z}$ of the momentum and spin
projection. The equations on the eigenvectors and eigenvalues of the spin projection operator $s_{16}^{3}=s_{z}$ from (343) have the form

$$s^{3}_{16}\mathrm{d}_{1} = 2\mathrm{d}_{1}, \, s^{3}_{16}\mathrm{d}_{2} = \mathrm{d}_{2}, \, s^{3}_{16}\mathrm{d}_{3} = 0, \, s^{3}_{16}\mathrm{d}_{4} = -\mathrm{d}_{4},$$
\small
\begin{equation}
\label{eq344} s^{3}_{16}\mathrm{d}_{5} = - 2\mathrm{d}_{5}, \, s^{3}_{16}\mathrm{d}_{6} =\mathrm{d}_{6}, \, s^{3}_{16}\mathrm{d}_{7} = 0, \,
s^{3}_{16}\mathrm{d}_{8} = -\mathrm{d}_{8},
\end{equation}
\normalsize
$$s^{3}_{16}\mathrm{d}_{9} = 2\mathrm{d}_{9}, \, s^{3}_{16}\mathrm{d}_{10}=\mathrm{d}_{10}, \, s^{3}_{16}\mathrm{d}_{11} = 0,$$
$$s^{3}_{16}\mathrm{d}_{12} = -\mathrm{d}_{12}, \, s^{3}_{16}\mathrm{d}_{13} = -2\mathrm{d}_{13}, \, s^{3}_{16}\mathrm{d}_{14} = \mathrm{d}_{14},$$
$$s^{3}_{16}\mathrm{d}_{15} = 0, \, s^{3}_{16}\mathrm{d}_{16} = -\mathrm{d}_{16}.$$

Therefore, the functions $g^{1}(\overrightarrow{k}), \,
g^{2}(\overrightarrow{k}), \, g^{3}(\overrightarrow{k}), \, g^{4}(\overrightarrow{k}), \, g^{5}(\overrightarrow{k})$ in
solution (309) are the momentum-spin amplitudes of the massive
boson with the spin s=2 and the spin projection $(2,1,0,-1,-2)$,
respectively, $g^{6}(\overrightarrow{k}), \, g^{7}(\overrightarrow{k}), \, g^{8}(\overrightarrow{k})$ are the momentum-spin amplitudes of the
massive boson with the spin s=1 and the spin projection $(1,0,-1)$; the functions $g^{9}(\overrightarrow{k}), \,
g^{10}(\overrightarrow{k}), \, g^{11}(\overrightarrow{k}), \, \, g^{12}(\overrightarrow{k}), \, g^{13}(\overrightarrow{k})$ are the
momentum-spin amplitudes of the antiparticle (antiboson) with the
spin s=2 and the spin projection $(2,1,0,-1,-2)$, respectively,
$g^{14}(\overrightarrow{k}), \, g^{15}(\overrightarrow{k}), \, g^{16}(\overrightarrow{k})$ are the momentum-spin amplitudes of the antiparticle
(massive antiboson) with the spin s=1 and the spin projection $(1,0,-1)$.

Note that direct quantum-mechanical interpretation of the amplitudes in solution (338) should be given in the framework of the RCQM. Such interpretation is already given in section 17 in paragraph after equations (216). 

The generators of the reducible unitary bosonic spin s=(2,1,2,1)
multiplet representation of the Poincar$\mathrm{\acute{e}}$ group
$\mathcal{P}$, with respect to which the canonical field equation
(336) and the set $\left\{\phi\right\}$ of its solutions (338) are
invariant, are derived from the RCQM set of generators (12), (13)
with the spin (212) on the basis of the transformations (339),
(341). These Hermitian  $16 \times 16$ matrix-differential
generators are given by

$$\widehat{p}^{0}=\Gamma_{16}^{0}\widehat{\omega}\equiv \Gamma_{16}^{0}\sqrt{-\Delta+m^{2}}, \quad \widehat{p}^{\ell}=-i\partial_{\ell},$$
\begin{equation}
\label{eq345} \widehat{j}^{\ell
n}=x^{\ell}\widehat{p}^{n}-x^{n}\widehat{p}^{\ell}+s_{16}^{\ell
n}\equiv \widehat{m}^{\ell n}+s_{16}^{\ell n},
\end{equation}
\begin{equation}
\label{eq346} \widehat{j}^{0 \ell}=-\widehat{j}^{\ell
0}=x^{0}\widehat{p}^{\ell}-\frac{1}{2}\Gamma_{16}^{0}\left\{x^{\ell},\widehat{\omega}\right\}+\Gamma_{16}^{0}\frac{(\overrightarrow{s}_{16}\times
\overrightarrow{p})^{\ell}}{\widehat{\omega}+m},
\end{equation}

\noindent where the spin s=(2,1,2,1) SU(2) generators
$\overrightarrow{s}_{16}=(s_{16}^{\ell n})$ have the form (343).

It is easy to prove by the direct verification that the generators (345), (346) commute with the operator $(i\partial_{0}- \Gamma_{16}^{0}\widehat{\omega})$ of the canonical field equation (336) and satisfy the commutation relations (11) of the Lie algebra of the Poincar$\mathrm{\acute{e}}$ group $\mathcal{P}$. The Casimir operators for the representation (345), (346) are given by

\begin{equation}
\label{eq347} p^{2}=\widehat{p}^{\mu}\widehat{p}_{\mu}=m^{2}\mathrm{I}_{16},
\end{equation}
$$W=w^{\mu}w_{\mu}=m^{2}\overrightarrow{s}_{16}^{2}=$$
\small
\begin{equation}
\label{eq348} m^{2}\left|
{{\begin{array}{*{20}c}
 2(2+1)\mathrm{I}_{5} \hfill & 0 \hfill & 0 \hfill & 0 \hfill\\
 0 \hfill & 1(1+1)\mathrm{I}_{3} \hfill & 0 \hfill & 0 \hfill\\
 0 \hfill & 0 \hfill & 2(2+1)\mathrm{I}_{5} \hfill & 0 \hfill\\
 0 \hfill & 0 \hfill & 0 \hfill & 1(1+1)\mathrm{I}_{3} \hfill\\
\end{array} }} \right|,
\end{equation}
\normalsize

\noindent where $\mathrm{I}_{16}$ is $16 \times 16$ unit matrix, $\mathrm{I}_{5}$ is $5 \times 5$ unit matrix and $\mathrm{I}_{3}$ is $3 \times 3$ unit matrix, respectively.

Thus, due to the eigenvalues in equations (344), positive and
negative frequencies form of solution (338) and the
Bargman--Wigner analysis of the Casimir operators (347), (348) one
can come to a conclusion that equation (336) describes the
16-component canonical field (the bosonic particle-antiparticle
doublet) with the spins s=(2,1,2,1) and $m>0$. Transition to the
$m=0$ limit leads to the canonical field equation for the
16-component (graviton - photon)-(antigraviton -
antiphoton) field.

The operator of the transition to the covariant local field theory
representation (the $16 \times 16$  analogy of the $4 \times 4$ FW
transformation operator (119)) is given by

\begin{equation}
\label{eq349} V^{\mp}=\frac{\mp
\overrightarrow{\Gamma}_{16}\cdot\overrightarrow{p}+\widehat{\omega}+m}{\sqrt{2\widehat{\omega}(\widehat{\omega}+m)}},
\quad V^{-}=(V^{+})^{\dag},
\end{equation}
$$V^{-}V^{+}=V^{+}V^{-}=\mathrm{I}_{16},$$
\begin{equation}
\label{eq350} \psi=V^{-}\phi, \, \phi = V^{+}\psi,
\end{equation}
\begin{equation}
\label{eq351} \hat{q}_{\mathrm{D}}=
V^{-}\hat{q}_{\mathrm{CF}}V^{+}, \, \hat{q}_{\mathrm{CF}}=
V^{+}\hat{q}_{\mathrm{D}}V^{-},
\end{equation}

\noindent where $\hat{q}_{\mathrm{D}}$ is an arbitrary operator
(both in the Hermitian and anti-Hermitian form) in the covariant
local field theory representation. The inverse transformation is
valid as well.

Thus, on the basis of transformation (349)--(351) the 16-component
Dirac-like equation is found from the canonical field equation
(336) in the form

\begin{equation}
\label{eq352}
\left[i\partial_{0}-\Gamma_{16}^{0}(\overrightarrow{\Gamma}_{16}\cdot\overrightarrow{p}+m)\right]\psi(x)=0.
\end{equation}

\noindent In operator (349) and in equation (352) the
$\Gamma_{16}^{\mu}$ matrices are given by

\begin{equation}
\label{eq353} \Gamma_{16}^{0}=\left| {{\begin{array}{*{20}c}
 \mathrm{I}_{8} \hfill & 0 \\
 0 \hfill & -\mathrm{I}_{8} \\
\end{array} }} \right|, \quad \Gamma_{16}^{j}=\left| {{\begin{array}{*{20}c}
 0 \hfill & \Sigma^{j}_{8} \\
 -\Sigma^{j}_{8} \hfill & 0 \\
\end{array} }} \right|,
\end{equation}

\noindent where $\Sigma^{j}_{8}$ are the $8\times 8$ Pauli matrices

\begin{equation}
\label{eq354} \Sigma^{j}_{8}=\left|
{{\begin{array}{*{20}c}
 \sigma^{j} \hfill & 0 \hfill & 0 \hfill & 0 \hfill\\
 0 \hfill & \sigma^{j} \hfill & 0 \hfill & 0 \hfill\\
 0 \hfill & 0 \hfill & \sigma^{j} \hfill & 0 \hfill\\
 0 \hfill & 0 \hfill & 0 \hfill & \sigma^{j} \hfill\\
\end{array} }} \right|,
\end{equation}

\noindent and the standard Pauli matrices $\sigma^{j}$ are given in (20).

The matrices $\Sigma^{j}_{8}$ satisfy the similar commutation
relations as the $2\times 2$ Pauli matrices (20) and have other
similar properties. The matrices $\Gamma_{16}^{\mu}$ (353) satisfy
the anticommutation relations of the Clifford--Dirac algebra in
the form

\begin{equation}
\label{eq355}
\Gamma_{16}^{\mu}\Gamma_{16}^{\nu}+\Gamma_{16}^{\nu}\Gamma_{16}^{\mu}=2g^{\mu\nu}.
\end{equation}

The solution of equation (352) is derived from solution (338) of
this equation in the canonical representation (336) on the basis
of transformation (349), (350) and is given by

$$\psi(x)=V^{-}\phi(x)= \frac{1}{\left(2\pi\right)^{\frac{3}{2}}}\int d^{3}k$$
\begin{equation}
\label{eq356}
\left[e^{-ikx}g^{\mathrm{A}}(\overrightarrow{k})\mathrm{v}^{-}_{\mathrm{A}}(\overrightarrow{k})+e^{ikx}g^{*\mathrm{B}}(\overrightarrow{k})\mathrm{v}^{+}_{\mathrm{B}}(\overrightarrow{k})\right],
\end{equation}

\noindent where $\mathrm{A}=\overline{1,8}, \, \mathrm{B}=\overline{9,16}$ and the 16-component spinors $(\mathrm{v}^{-}_{\mathrm{A}}(\overrightarrow{k}), \, \mathrm{v}^{+}_{\mathrm{B}}(\overrightarrow{k}))$ are given by

$$\mathrm{v}^{-}_{1}(\overrightarrow{k}) = N\left|
\begin{array}{cccc}
 \tilde{\omega}+m \\
 0 \\
 0 \\
 0 \\
 0 \\
 0 \\
 0 \\
 0 \\
 k^{3} \\
 k^{1}+ik^{2} \\
 0 \\
 0 \\
 0 \\
 0 \\
 0 \\
 0 \\
\end{array} \right|, \, \mathrm{v}^{-}_{2}(\overrightarrow{k}) = N\left|
\begin{array}{cccc}
 0 \\
 \tilde{\omega}+m \\
 0 \\
 0 \\
 0 \\
 0 \\
 0 \\
 0 \\
 k^{1}-ik^{2} \\
 -k^{3} \\
 0 \\
 0 \\
 0 \\
 0 \\
 0 \\
 0 \\
\end{array} \right|,$$
$$ \mathrm{v}^{-}_{3}(\overrightarrow{k}) = N \left|
\begin{array}{cccc}
 0 \\
 0 \\
 \tilde{\omega}+m \\
 0 \\
 0 \\
 0 \\
 0 \\
 0 \\
 0 \\
 0 \\ 
 k^{3} \\
 k^{1}+ik^{2} \\
 0 \\
 0 \\
 0 \\
 0 \\
\end{array} \right|,
\, \mathrm{v}^{-}_{4}(\overrightarrow{k}) = N\left|
\begin{array}{cccc}
 0 \\
 0 \\
 0 \\
 \tilde{\omega}+m \\
 0 \\
 0 \\
 0 \\
 0 \\
 0 \\
 0 \\
 k^{1}-ik^{2} \\
 -k^{3} \\
 0 \\
 0 \\
 0 \\
 0 \\
\end{array} \right|,$$
$$ \mathrm{v}^{-}_{5}(\overrightarrow{k}) = N \left|
\begin{array}{cccc}
 0 \\
 0 \\
 0 \\
 0 \\
 \tilde{\omega}+m \\
 0 \\
 0 \\
 0 \\
 0 \\
 0 \\
 0 \\
 0 \\ 
 k^{3} \\
 k^{1}+ik^{2} \\
 0 \\
 0 \\
 \end{array} \right|,
\, \mathrm{v}^{-}_{6}(\overrightarrow{k}) = N\left|
\begin{array}{cccc}
 0 \\
 0 \\
 0 \\
 0 \\
 0 \\
 \tilde{\omega}+m \\
 0 \\
 0 \\
 0 \\
 0 \\
 0 \\
 0 \\
 k^{1}-ik^{2} \\
 -k^{3} \\
 0 \\
 0 \\
 \end{array} \right|,$$
\begin{equation}
\label{eq357} \mathrm{v}^{-}_{7}(\overrightarrow{k}) = N\left|
\begin{array}{cccc}
 0 \\
 0 \\
 0 \\
 0 \\
 0 \\
 0 \\
 \tilde{\omega}+m \\
 0 \\
 0 \\
 0 \\
 0 \\
 0 \\
 0 \\
 0 \\
 k^{3} \\
 k^{1}+ik^{2} \\
 \end{array} \right|, \,
\mathrm{v}^{-}_{8}(\overrightarrow{k}) = N\left|
\begin{array}{cccc}
 0 \\
 0 \\
 0 \\
 0 \\
 0 \\
 0 \\
 0 \\
 \tilde{\omega}+m \\
 0 \\
 0 \\
 0 \\
 0 \\
 0 \\
 0 \\
 k^{1}-ik^{2} \\
 -k^{3} \\
\end{array} \right|,
\end{equation}
$$\mathrm{v}^{+}_{9}(\overrightarrow{k}) = N\left|
\begin{array}{cccc}
 k^{3} \\
 k^{1}+ik^{2} \\
 0 \\
 0 \\
 0 \\
 0 \\
 0 \\
 0 \\
 \tilde{\omega}+m \\
 0 \\
 0 \\
 0 \\
 0 \\
 0 \\
 0 \\
 0 \\
\end{array} \right|,
\mathrm{v}^{+}_{10}(\overrightarrow{k}) = N\left|
\begin{array}{cccc}
 k^{1}-ik^{2} \\
 -k^{3} \\
 0 \\
 0 \\
 0 \\
 0 \\
 0 \\
 0 \\
 0 \\
 \tilde{\omega}+m \\
 0 \\
 0 \\
 0 \\
 0 \\
 0 \\
 0 \\
\end{array} \right|,$$
$$\mathrm{v}^{+}_{11}(\overrightarrow{k}) = N\left|
\begin{array}{cccc}
 0 \\
 0 \\
 k^{3} \\
 k^{1}+ik^{2} \\
 0 \\
 0 \\
 0 \\
 0 \\
 0 \\
 0 \\
 \tilde{\omega}+m \\
 0 \\
 0 \\
 0 \\
 0 \\
 0 \\
 \end{array} \right|,
\mathrm{v}^{+}_{12}(\overrightarrow{k}) = N\left|
\begin{array}{cccc}
 0 \\
 0 \\
 k^{1}-ik^{2} \\ 
 -k^{3} \\
 0 \\
 0 \\
 0 \\
 0 \\
 0 \\
 0 \\
 0 \\
 \tilde{\omega}+m \\
 0 \\
 0 \\
 0 \\
 0 \\
\end{array} \right|,$$
$$\mathrm{v}^{+}_{13}(\overrightarrow{k}) = N\left|
\begin{array}{cccc}
 0 \\
 0 \\
 0 \\
 0 \\
 k^{3} \\
 k^{1}+ik^{2}
 0 \\
 0 \\
 0 \\
 0 \\
 0 \\
 0 \\
 \tilde{\omega}+m \\
 0 \\
 0 \\
 0 \\
 \end{array} \right|,
\mathrm{v}^{+}_{14}(\overrightarrow{k}) = N\left|
\begin{array}{cccc}
 0 \\
 0 \\
 0 \\
 0 \\ 
 k^{1}-ik^{2} \\ 
 -k^{3} \\
 0 \\
 0 \\
 0 \\
 0 \\
 0 \\
 0 \\
 0 \\
 \tilde{\omega}+m \\
 0 \\
 0 \\
 \end{array} \right|,$$
$$\mathrm{v}^{+}_{15}(\overrightarrow{k}) = N\left|
\begin{array}{cccc}
 0 \\
 0 \\
 0 \\
 0 \\
 0 \\
 0 \\
 k^{3} \\
 k^{1}+ik^{2} \\
 0 \\
 0 \\
 0 \\
 0 \\
 0 \\
 0 \\
 \tilde{\omega}+m \\
 0 \\
 \end{array} \right|,
\mathrm{v}^{+}_{16}(\overrightarrow{k}) = N\left|
\begin{array}{cccc}
 0 \\
 0 \\
 0 \\
 0 \\
 0 \\
 0 \\
 k^{1}-ik^{2} \\
 -k^{3} \\
 0 \\
 0 \\
 0 \\
 0 \\
 0 \\
 0 \\
 0 \\
 \tilde{\omega}+m \\
 \end{array} \right|,$$

\noindent where $N$ and $\tilde{\omega}$ are given in (258).

The spinors (357) are derived from the orts
$\left\{\mathrm{d}\right\}$ of the Cartesian basis (211)
with the help of the transformation (349), (350). The spinors
(357) satisfy the relations of the orthonormalization and
completeness similar to the corresponding relations for the
standard 4-component Dirac spinors, see, e. g., [33]. 

In the covariant local field theory, the operators of the SU(2) spin,
which satisfy the corresponding commutation relations $
\left[s_{16\mathrm{D}}^{j},s_{16\mathrm{D}}^{\ell}\right]=i\varepsilon^{j
\ell n}s_{16\mathrm{D}}^{n}$ and commute with the operator
$\left[i\partial_{0}-\Gamma_{16}^{0}(\overrightarrow{\Gamma}_{16}\cdot\overrightarrow{p}+m)\right]$
of equation (352), are derived from the pure matrix operators
(343) with the help of operator (349), (351):

\begin{equation}
\label{eq358}
\overrightarrow{s}_{16\mathrm{D}} =
V^{-}\overrightarrow{s}_{16}V^{+}.
\end{equation}

\noindent The explicit form can be presented in analogy with previous sections. The third component of this spin has the explicit form

\begin{equation}
\label{eq359}
s_{16\mathrm{D}}^{3}=\frac{1}{2\omega\Omega}\cdot\left|16 \times 16\right|,
\end{equation}

\noindent where the nonzero matrix elements $a,b$ of $16 \times 16$ matrix are given by

$$1,1=4\omega\Omega-p^{11}_{22}, \, 1,2=p^{3}z^{*}, \, 1,10=z^{*}\Omega,$$
$$2,1=p^{3}z, \, 2,2=2\omega\Omega+p^{11}_{22}, \, 2,9=-z\Omega,$$
$$3,3=-p^{11}_{22}, \, 3,4=p^{3}z^{*}, \, 3,12=z^{*}\Omega,$$
$$4,3=p^{3}z, \, 4,4=\Omega^{2}-p^{33}, \, 4,11=-z\Omega,$$
\small
$$5,5=-4\omega\Omega+3p^{11}_{22}, \, 5,6=-3p^{3}z^{*}, \, 5,14=-3z^{*}\Omega,$$
\normalsize
$$6,5=-3p^{3}z, \, 6,6=2\omega\Omega-3p^{11}_{22}, \, 6,13=3z\Omega,$$
$$7,7=-p^{11}_{22}, \, 7,8=p^{3}z^{*}, \, 7,16=z^{*}\Omega,$$
$$8,7=p^{3}z, \, 8,8=-\Omega^{2}-p^{33}, \, 8,15=-z\Omega,$$
$$9,2=-z^{*}\Omega, \, 9,9=4\omega\Omega-p^{11}_{22}, \, 9,10=p^{3}z^{*},$$
$$10,1=z\Omega, \, 10,9=p^{3}z, \, 10,10=2\omega\Omega+p^{11}_{22},$$
$$11,4=-z^{*}\Omega, \, 11,11=-p^{11}_{22}, \, 11,12=p^{3}z^{*},$$
$$12,3=z\Omega, \, 12,11=p^{3}z, \, 12,12=-\Omega^{2}-p^{33},$$
\small
$$13,6=3z^{*}\Omega, \, 13,13=-4\omega\Omega+3p^{11}_{22}, \, 13,14=-3p^{3}z^{*},$$
$$14,5=-3z\Omega, \, 14,13=-3p^{3}z, \, 14,14=2\omega\Omega-3p^{11}_{22},$$
\normalsize
$$15,8=-z^{*}\Omega, \, 15,15=-p^{11}_{22}, \, 15,16=p^{3}z^{*},$$
$$16,7=z\Omega, \, 16,15=p^{3}z, \, 16,16=-\Omega^{2}-p^{33}.$$

\noindent Here in matrix elements the notations (262), (287) are used.

The equations on eigenvectors and eigenvalues of the operator
$s_{16\mathrm{D}}^{3}$ (359) follow from the equations (344) and
the transformation (349)--(351). In addition to it, the action of
the operator $s_{16\mathrm{D}}^{3}$ (359)  on the spinors
$(\mathrm{v}^{-}_{\mathrm{A}}(\overrightarrow{k}), \,
\mathrm{v}^{+}_{\mathrm{B}}(\overrightarrow{k}))$ (357) also
leads to the result

$$s_{16\mathrm{D}}^{3}\mathrm{v}^{-}_{1}(\overrightarrow{k})= 2\mathrm{v}^{-}_{1}(\overrightarrow{k}), \, s_{16\mathrm{D}}^{3}\mathrm{v}^{-}_{2}(\overrightarrow{k}) = \mathrm{v}^{-}_{2}(\overrightarrow{k}),$$
$$s_{16\mathrm{D}}^{3}\mathrm{v}^{-}_{3}(\overrightarrow{k}) = 0, \,
s_{16\mathrm{D}}^{3}\mathrm{v}^{-}_{4}(\overrightarrow{k}) = -\mathrm{v}^{-}_{4}(\overrightarrow{k}),$$
$$s_{16\mathrm{D}}^{3}\mathrm{v}^{-}_{5}(\overrightarrow{k})= -2\mathrm{v}^{-}_{5}(\overrightarrow{k}), \, s_{16\mathrm{D}}^{3}\mathrm{v}^{-}_{6}(\overrightarrow{k}) = \mathrm{v}^{-}_{6}(\overrightarrow{k}),$$
\begin{equation}
\label{eq360}
s_{16\mathrm{D}}^{3}\mathrm{v}^{-}_{7}(\overrightarrow{k}) =0, \,
s_{16\mathrm{D}}^{3}\mathrm{v}^{-}_{8}(\overrightarrow{k}) = -\mathrm{v}^{-}_{8}(\overrightarrow{k}),
\end{equation}
$$s_{16\mathrm{D}}^{3}\mathrm{v}^{+}_{9}(\overrightarrow{k}) = 2\mathrm{v}^{+}_{9}(\overrightarrow{k}), \, s_{16\mathrm{D}}^{3}\mathrm{v}^{+}_{10}(\overrightarrow{k}) = \mathrm{v}^{+}_{10}(\overrightarrow{k}),$$
$$s_{16\mathrm{D}}^{3}\mathrm{v}^{+}_{11}(\overrightarrow{k}) = 0, \, s_{16\mathrm{D}}^{3}\mathrm{v}^{+}_{12}(\overrightarrow{k}) = -\mathrm{v}^{+}_{12}(\overrightarrow{k}),$$
$$s_{16\mathrm{D}}^{3}\mathrm{v}^{+}_{13}(\overrightarrow{k}) = -2\mathrm{v}^{+}_{13}(\overrightarrow{k}), \, s_{16\mathrm{D}}^{3}\mathrm{v}^{+}_{14}(\overrightarrow{k}) = \mathrm{v}^{+}_{14}(\overrightarrow{k}),$$
$$s_{16\mathrm{D}}^{3}\mathrm{v}^{+}_{15}(\overrightarrow{k}) = 0, \, s_{16\mathrm{D}}^{3}\mathrm{v}^{+}_{16}(\overrightarrow{k}) = -\mathrm{v}^{+}_{16}(\overrightarrow{k}).$$

In order to verify equations (360) the identity (264)
should be used. In the case $\mathrm{v}^{+}_{\mathrm{B}}(\overrightarrow{k})$ in the
expression $s_{12\mathrm{D}}^{3}(\overrightarrow{k})$ (359) the
substitution $\overrightarrow{k}\rightarrow - \overrightarrow{k}$
is made.

The equations (360) determine the interpretation of the
amplitudes in solution (356). This interpretation is similar to the given above in the paragraph after equations (344). Nevertheless, the direct quantum-mechanical interpretation of the amplitudes should be made in the
framework of the RCQM, see the section 17 above.

The explicit form of the $\mathcal{P}$-generators of the bosonic
representation of the Poincar$\mathrm{\acute{e}}$ group
$\mathcal{P}$, with respect to which the covariant equation (352)
and the set $\left\{\psi\right\}$ of its solutions (356) are
invariant, is derived from the generators (345), (346) on the
basis of the transformation (349), (351). The corresponding
generators are given by

$$\widehat{p}^{0}=\Gamma_{16}^{0}(\overrightarrow{\Gamma}_{16}\cdot\overrightarrow{p}+m), \quad \widehat{p}^{\ell}=-i\partial_{\ell},$$
\begin{equation}
\label{eq361} \widehat{j}^{\ell
n}=x_{\mathrm{D}}^{\ell}\widehat{p}^{n}-x_{\mathrm{D}}^{n}\widehat{p}^{\ell}+s_{16\mathrm{D}}^{\ell
n}\equiv \widehat{m}^{\ell n}+s_{16\mathrm{D}}^{\ell n},
\end{equation}
\begin{equation}
\label{eq362} \widehat{j}^{0 \ell}=-\widehat{j}^{\ell
0}=x^{0}\widehat{p}^{\ell}-\frac{1}{2}\left\{x_{\mathrm{D}}^{\ell},\widehat{p}^{0}\right\}+\frac{\widehat{p}^{0}(\overrightarrow{s}_{16\mathrm{D}}\times
\overrightarrow{p})^{\ell}}{\widehat{\omega}(\widehat{\omega}+m)},
\end{equation}

\noindent where the spin matrices
$\overrightarrow{s}_{16\mathrm{D}}=(s_{16\mathrm{D}}^{\ell n})$ are
given in (358), (359),  and the operator
$\overrightarrow{x}_{\mathrm{D}}$ has the form

\begin{equation}
\label{eq363}
\overrightarrow{x}_{\mathrm{D}}=\overrightarrow{x}+\frac{i\overrightarrow{\Gamma}_{16}}{2\widehat{\omega}}-\frac{\overrightarrow{s}_{16}\times \overrightarrow{p}}{\widehat{\omega}(\widehat{\omega}+m)}-\frac{i\overrightarrow{p} (\overrightarrow{\Gamma}_{16}\cdot \overrightarrow{p})}{2\widehat{\omega}^{2}(\widehat{\omega}+m)},
\end{equation}

\noindent where the spin matrices $\overrightarrow{s}_{16}$ are given in (343)

It is easy to verify that the generators (361), (362) commute with
the operator
$\left[i\partial_{0}-\Gamma_{16}^{0}(\overrightarrow{\Gamma}_{12}\cdot\overrightarrow{p}+m)\right]$
of equation (352), satisfy the commutation relations (11) of the
Lie algebra of the Poincar$\mathrm{\acute{e}}$ group and the
corresponding Casimir operators are given by

\begin{equation}
\label{eq364} p^{2}=\widehat{p}^{\mu}\widehat{p}_{\mu}=m^{2}\mathrm{I}_{16},
\end{equation}
$$W=w^{\mu}w_{\mu}=m^{2}\overrightarrow{s}_{16\mathrm{D}}^{2}=$$
\small
\begin{equation}
\label{eq365} m^{2}\left|
{{\begin{array}{*{20}c}
 2(2+1)\mathrm{I}_{5} \hfill & 0 \hfill & 0 \hfill & 0 \hfill\\
 0 \hfill & 1(1+1)\mathrm{I}_{3} \hfill & 0 \hfill & 0 \hfill\\
 0 \hfill & 0 \hfill & 2(2+1)\mathrm{I}_{5} \hfill & 0 \hfill\\
 0 \hfill & 0 \hfill & 0 \hfill & 1(1+1)\mathrm{I}_{3} \hfill\\
\end{array} }} \right|,
\end{equation}
\normalsize

\noindent where $\mathrm{I}_{16}$, $\mathrm{I}_{5}$ and $\mathrm{I}_{3}$ are $16 \times 16$, $5 \times 5$ and $3 \times 3$ unit matrices, respectively.

As it was already explained in details in the previous sections,
the conclusion that equation (352) describes the bosonic
particle-antiparticle multiplet of the spin s=(2,1,2,1) and mass
$m>0$ (and its solution (356) is the bosonic field of the above
mentioned spin and nonzero mass) follows from the analysis of
equations (360) and the Casimir operators (364), (365). Moreover, the external argument in the validity of such interpretation is the link with the corresponding RCQM of spin s=(2,1,2,1) particle-antiparticle multiplet, where the quantum-mechanical interpretation is direct and evident.
Therefore, the bosonic spin s=(2,1,2,1) properties of equation (352) are proved.

It is easy to prove that equation (352) describes four spin s=(1/2,1/2) fermionic particle-antiparticle doublets, two spin s=(3/2,3/2) fermionic particle-antiparticle doublets and two spin s=(1,0,1,0) bosonic particle-antiparticle multiplets as well. Therefore, this equation has the extended property of Fermi--Bose duality. Such characteristic can be called the quadro Fermi--Bose property.

\section{The partial case of zero mass}

All equations of motion considered here are valid for the partial case $m=0$. The corresponding expressions are found by the transition $m\rightarrow 0$. This assertion is valid both for the RCQM and for the covariant local field theory.

In the case $m=0$ the above considered equations of motion are invariant with respect to all representations of the SU(2) group, which are mentioned in this article. The SU(2) symmetry does not depend on the difference between $m=0$ and $m\neq 0$. 

The difference between $m=0$ and $m\neq 0$ is observed only for the symmetry with respect to the Poincar$\mathrm{\acute{e}}$ group
$\mathcal{P}$. In the case of $m=0$ equations of motion are invariant with respect to all Poincar$\mathrm{\acute{e}}$ generators given above in the paper. Nevertheless, the Casimir operators are equal to zero in this case. Therefore, another class of Poincar$\mathrm{\acute{e}}$ group representations, using the Casimir operator $\overrightarrow{s}\cdot \overrightarrow{p}$ related to the helicity of elementary particles, is preferable. Hence, in the case $m=0$ the momentum-helicity basis vectors are preferable in comparison with momentum-spin basis, which is applied in this paper.

After this warning the theory for $m=0$ follows from the theory with $m\neq 0$.  

\section{Brief conclusions}

The foundations of the relativistic canonical quantum mechanics of the arbitrary mass and spin have been formulated.  The possibilities of the model have been demonstrated and explained on the examples of the spin s=1/2, s=1, s=3/2, s=2 singlets, spin s=(1,1) particle-antiparticle doublet, spin s=(1,0) multiplet, spin s=(1,1) particle-antiparticle doublet, spin s=(1,0,1,0) particle-antiparticle multiplet, spin s=(3/2,3/2) particle-antiparticle doublet, spin s=(2,2) particle-antiparticle doublet, spin s=(2,0,2,0) particle-antiparticle multiplet  and spin s=(2,1,2,1) particle-antiparticle multiplet. The link between the relativistic canonical quantum mechanics of the arbitrary spin and the covariant local field theory has been found. The corresponding transition operators have been visualized. The test example of the relativistic canonical quantum mechanics of the spin s=(1/2,1/2) particle-antiparticle doublet and its link with the standard Dirac equation has been considered in details. 

On this basis the covariant local field theory equations for spin s=(1,1) particle-antiparticle doublet, spin s=(1,0,1,0) particle-antiparticle multiplet, spin s=(3/2,3/2) particle-antiparticle doublet, spin s=(2,2) particle-antiparticle doublet, spin s=(2,0,2,0) particle-antiparticle multiplet  and spin s=(2,1,2,1) particle-antiparticle multiplet have been derived.

The 8-component manifestly covariant equation for the spin s=3/2 field found here is the s=3/2 analogy of the 4-component Dirac equation for the spin s=1/2 doublet, while the 4-component Rarita-Schwinger (Pauli-Fierz) equation is the spin 3/2 analogy of the 2-component Pauli equation for the spin s=1/2 singlet.

In the case of manifestly covariant equations for the s=(1,0,1,0) particle-antiparticle multiplet the limit $m\rightarrow 0$ leads to the corresponding equations for the photonic and massless bosonic fields including the antiparticle fields. Thus, the electromagnetic field equations that follow from the corresponding relativistic quantum mechanical equations have been found. The new electrodynamical equations containing the hypothetical antiphoton and massless spinless antiboson have been introduced. The Maxwell-like equations for the boson with spin s=1 and $m>0$ (W-boson) have been introduced as well. In other words, the Maxwell equations for the field with nonzero mass have been introduced.

The properties of the Fermi--Bose duality, triality and quadro Fermi--Bose properties of equations found have been discussed briefly.


%


\ifCLASSOPTIONcaptionsoff
  \newpage
\fi




\begin{thebibliography}{99}

\bibitem{1} P.A.M. Dirac, "The quantum theory of the electron," \textit{Proc. R. Soc. Lond. A},  vol. 117, no. 778, pp. 610-624, 1928. \\
\bibitem{2} L.L. Foldy and S.A. Wouthuysen, "On the Dirac theory of spin 1/2 particles and its non-relativistic limit," \textit{Phys. Rev.},  vol. 78, no. 1, pp. 29-36, 1950. \\
\bibitem{3} V.M. Simulik and I.Yu. Krivsky, "Link between the relativistic canonical quantum mechanics and the Dirac equation," \textit{Univ. J. Phys. Appl.},  vol. 2, no. 2, pp. 115-128, 2014. \\
\bibitem{4} P.A.M. Dirac, "Relativistic wave equations," \textit{Proc. R. Soc. Lond. A},  vol. 155, no. 886, pp. 447-459, 1936. \\
\bibitem{5} H.J. Bhabha, "Relativistic wave equations for the elementary particles," \textit{Rev. Mod. Phys.},  vol. 17, no. 2-3, pp. 200-216, 1945. \\
\bibitem{6} D.L. Weaver, C.L. Hammer and R.H. Good Jr., "Description of a particle with arbitrary mass and spin," \textit{Phys. Rev.},  vol. 135, no. 1 B, pp. 241-248, 1964. \\
\bibitem{7} P.M. Mathews, "Relativistic Schr$\mathrm{\ddot{o}}$dinger equations for particles of arbitrary spin," \textit{Phys. Rev.},  vol. 143, no. 4, pp. 978-985, 1966. \\
\bibitem{8} T.J. Nelson and R.H. Good Jr., "Second-quantization process for particles with any spin and with internal symmetry," \textit{Rev. Mod. Phys.},  vol. 40, no. 3, pp. 508-522, 1968. \\
\bibitem{9} R.F. Guertin, "Relativistic Hamiltonian equations for any spin," \textit{Ann. Phys. (USA)},  vol. 88, pp. 504-553, 1974. \\
\bibitem{10} R.A. Kraicik and M.M. Nieto, "Bhabha first-order wave equations. VI. Exact, closed-form, Foldy-Wouthuysen transformations and solutions," \textit{Phys. Rev. D},  vol. 15, no. 2, pp. 433-444, 1977. \\
\bibitem{11} R-K. Loide, I. Ots and R. Saar, "Bhabha relativistic wave equations," \textit{J. Phys. A},  vol. 30, no. 11, pp. 4005-4017, 1997. \\
\bibitem{12} D.M. Gitman and A.L. Shelepin, "Fields on the Poincar$\mathrm{\acute{e}}$ group: Arbitrary spin description and relativistic wave equations," \textit{Inter. J. Theor. Phys.},  vol. 40, no. 3, pp. 603-684, 2001. \\
\bibitem{13} L.L. Foldy, "Synthesis of covariant particle equations," \textit{Phys. Rev.},  vol. 102, no. 2, pp. 568-581, 1956. \\
\bibitem{14} V.M. Simulik and I.Yu. Krivsky, "Quantum-mechanical description of the fermionic doublet and its link with the Dirac equation," \textit{Ukr. J. Phys.},  vol. 58, no. 12, pp. 1192-1203, 2013. \\
\bibitem{15} von J. Neumann, \textit{Mathematische Grundlagen der Quantenmechanik,} Berlin: Verlag von Julius Springer, 1932.\\
\bibitem{16} L.L. Foldy, "Relativistic particle systems with interaction," \textit{Phys. Rev.},  vol. 122, no. 1, pp. 275-288, 1961. \\
\bibitem{17} E.E. Salpeter, "Mass corrections to the fine structure of hydrogen-like atoms," \textit{Phys. Rev.},  vol. 87, no. 2, pp. 328-343, 1952. \\
\bibitem{18} W. Lucha and F.F. Schobert, "Relativistic Coulomb problems: Analitic upper bounds on energy levels," \textit{Phys. Rev. A},  vol. 54, no. 5, pp. 3790-3794, 1996. \\
\bibitem{19} Y. Chargui and A. Trabelsi, "The zero-mass spinless Salpeter equation with a regularized inverse square potential," \textit{Phys. Lett. A},  vol. 377, no. 3-4, pp. 158-168, 2013. \\
\bibitem{20} V.S. Vladimirov, \textit{Methods of the theory of generalized functions,} London: Taylor and Francis, 2002.\\
\bibitem{21} N.N. Bogoliubov, A.A. Logunov and I.T. Todorov, \textit{Introduction to axiomatic quantum field theory,} Berlin: Springer-Verlag, 1992.\\
\bibitem{22} G. Bluman and S. Anco, \textit{Symmetry and integration methods for differential equations,} New York: Springer, 2002.\\
\bibitem{23} P. Garbaczewski, "Boson - Fermion duality in four dimensions: comments on the paper of Luther and Schotte," \textit{Intern. J. Theor. Phys.},  vol. 25, no. 11, pp. 1193-1208, 1986. \\
\bibitem{24} V.M. Simulik and I.Yu. Krivsky, "On the extended real Clifford--Dirac algebra and new physically meaningful symmetries of the Dirac equation with nonzero mass," \textit{Reports of the National Academy of Sciences of Ukraine}, no. 5, pp. 82-88, 2010 (in Ukrainian). \\
\bibitem{25} V.M. Simulik and I.Yu. Krivsky, "Bosonic symmetries of the Dirac equation," \textit{Phys. Lett. A},  vol. 375, no. 25, pp. 2479-2483, 2011. \\
\bibitem{26} V.M. Simulik, I.Yu. Krivsky and I.L. Lamer, "Some statistical aspects of the spinor field Fermi--Bose duality," \textit{Cond. Matt. Phys.},  vol. 15, no. 4, pp. 43101(1-10), 2012. \\
\bibitem{27} V.M. Simulik, I.Yu. Krivsky and I.L. Lamer, "Application of the generalized Clifford--Dirac algebra to the proof of the Dirac equation Fermi--Bose duality," \textit{TWMS J. App. Eng. Math.},  vol. 3, no. 1, pp. 46-61, 2013. \\
\bibitem{28} V.M. Simulik, I.Yu. Krivsky and I.L. Lamer, "Bosonic symmetries, solutions and conservation laws for the Dirac equations with nonzero mass," \textit{Ukr. J. Phys.},  vol. 58, no. 6, pp. 523-533, 2013. \\
\bibitem{29} J. Elliott and P. Dawber, \textit{Symmetry in Physics, vol.1,} London: Macmillian Press, 1979.\\
\bibitem{30} B. Wybourne, \textit{Classical groups for Physicists,} New York: John Wiley and sons, 1974.\\
\bibitem{31} B. Thaller, \textit{The Dirac equation,} Berlin: Springer, 1992.\\
\bibitem{32} I.Yu. Krivsky and V.M. Simulik, "Fermi-Bose duality of the Dirac equation and extended real Clifford-Dirac algebra," \textit{Cond. Matt. Phys.},  vol. 13, no. 4, pp. 43101(1-15), 2010. \\
\bibitem{33} N.N. Bogoliubov and D.V. Shirkov, \textit{Introduction to the theory of
quantized fields,} New York: John Wiley and Sons, 1980.\\
\bibitem{34} I.Yu. Krivsky, R.R. Lompay and V.M. Simulik, "Symmetries of the complex Dirac--Kahler equation," \textit{Theor. Math. Phys.},  vol. 143, no. 1, pp. 541-558, 2005. \\
\bibitem{35} S.I. Kruglov, "Dirac--Kahler equation," \textit{Inter. J. Theor. Phys},  vol. 41, no. 4, pp. 653-687, 2002. \\
\bibitem{36} V.M. Simulik and I.Yu. Krivsky, "Relationship between the Maxwell and Dirac equations: symmetries, quantization, models of atom," \textit{Rep. Math. Phys.},  vol. 50, no. 3, pp. 315-328, 2002. \\
\bibitem{37} V.M. Simulik and I.Yu. Krivsky, "Classical electrodynamical aspect of the Dirac equation," \textit{Electromagnetic Phenomena},  vol. 3, no. 1(9), pp. 103-114, 2003. \\
\bibitem{38} V.M. Simulik, In \textit{What is the electron?} edit. V. Simulik, Montreal: Apeiron, 2005.\\





\end{thebibliography}
%

%

\end{document}